\documentclass[a4paper,11pt,oneside]{book}
\usepackage[left=4cm,right=2.5cm,top=2.5cm,bottom=2.5cm]{geometry} 

\usepackage{blindtext} 
\usepackage{graphicx}
\usepackage{verbatim}
\usepackage{latexsym}
\usepackage{mathchars}
\usepackage{setspace}
\usepackage{titlesec}
\usepackage{changepage}
\usepackage[hyperfootnotes=false,hypertexnames=false]{hyperref} % for hyperlinks to sections etc.
\usepackage{amsmath}
\usepackage{amsfonts}
\usepackage{amsthm}
\usepackage{amssymb}
\usepackage{cancel}
\usepackage{caption}
\usepackage{dirtytalk}
\usepackage{nonumonpart}
\usepackage{tikz}
\usetikzlibrary{tqft}
\usepackage{subcaption}
\usepackage[toc,page]{appendix}
\usepackage{bbding}
\usepackage{slashed}
\usepackage[sort&compress, numbers]{natbib} % bibliography package
\usepackage[T1]{fontenc} % for \dj
\usepackage{tocbibind} % adds bibliography to table of contents
\usepackage{bookmark} % for side pannel when viewing pdf
\usepackage[bottom]{footmisc}
\usepackage{fancyhdr}
\usepackage{comment}
\usepackage[compat=1.0.0]{tikz-feynman}
\setcitestyle{square} %This makes the citations appear as [NUMBER] rather than (NUMBER)

\usepackage{tocloft,calc}

\AtBeginDocument{\addtolength\cftchapnumwidth{\widthof{Chapter}}}

\makeatletter
\g@addto@macro\appendix{%
  \addtocontents{toc}{%
    \protect%
  }%
}

\usepackage{minitoc}
\mtcsetoffset{minitoc}{0pt} 
\mtcsetdepth{minitoc}{2} \mtcsetfont{minitoc}{*}{\small\rmfamily\upshape\mdseries} \mtcsetfont{minitoc}{section}{\small\rmfamily\upshape\bfseries}
\fancypagestyle{front}{% style for TOC, LOF, LOT
  \fancyhf{}
  
  \cfoot{\thepage}
}
\fancypagestyle{main}{% style for the mainmatter
  \fancyhf{}
  \fancyhead[C]{\slshape \rightmark}
  \fancyfoot[C]{\thepage}
  
}
\makeatletter
  \newcommand\frontpagestyle{\cleardoublepage\pagestyle{front}\let\ps@plain\ps@front}
  \newcommand\mainpagestyle{\cleardoublepage\pagestyle{main}\let\ps@plain\ps@main}
\makeatother

\input{blocked.sty}
\input{uhead.sty}
\input{boxit.sty}
\input{uon_thesis.sty}

\newcommand\figref{Fig.~\ref}
\newcommand\chpref{Chap.~\ref}
\newcommand\secref{Sec.~\ref}
\newcommand\ssecref{Subsec.~\ref}
\newcommand\appref{App.~\ref}
\newcommand\tabref{Tab.~\ref}
\renewcommand{\eqref}[1]{Eq.~(\ref{#1})}
\newcommand{\kah}{K\"ahler }
\newcommand{\barj}{\bar{\jmath}}
\newcommand{\bari}{\bar{\imath}}
\renewcommand{\Im}{\text{Im}\ }
\renewcommand{\Re}{\text{Re}\ }
\newcommand{\x}{\relax\ifmmode \mathcal{X} \else $\mathcal{X}$ \fi}
\newcommand{\tmax}{\text{max}}

\def\vo{\mathcal{V}}
\def\la{\langle}
\def\ra{\rangle}

\newcommand{\mc}[1]{\ensuremath{\mathcal #1}}

\newcommand{\normallinespacing}{\renewcommand{\baselinestretch}{1.5} \normalsize}

\newcommand\blankpage{% comando pagina vuota
    \clearpage
    \null
    \thispagestyle{empty}%
    \addtocounter{page}{-1}%
    \clearpage
}

\newtheorem{conjecture}{Conjecture}
\newtheorem*{statement_nat}{Naturalness}

\def\be{\begin{equation}}
\def\ee{\end{equation}}
\def\ben{\begin{equation*}}
\def\een{\end{equation*}}
\def\um#1{{\"#1}}

\def\del{\partial}

\newcommand{\vol}{\relax\ifmmode \mathcal{V} \else $\mathcal{V}$ \fi}
\newcommand{\order}[1]{{$\mathcal{O}\left(#1\right)$}}
\let\mpl\mp
\renewcommand{\mp}{\relax\ifmmode M_{\text{pl}} \else $M_{\text{pl}}$ \fi}
\newcommand{\vmin}{\relax\ifmmode V^{(n)}_{\text{min}}\else $V^{(n)}_{\text{min}}$ \fi}

\DeclareMathAlphabet\mathbfcal{OMS}{cmsy}{b}{n}

\begin{document}

\title{\LARGE {\bf Dark energy: EFTs and supergravity}\\
 \vspace*{6mm}
}

\author{Francesc Cunillera}
\uni{Centre for Astronomy and Particle Theory\\Nottingham Centre for Gravity}
\submitdate{March 2022}
\degree{Doctor of Philosophy}
\supervisor{Prof. Antonio Padilla\\
			Prof. Ed Copeland}

\maketitle
\normallinespacing
\blankpage
\preface
\addstarredchapter{Summary}
\chaptermark{Summary}

\chapter*{\vspace{-3.5em}Summary}
\vspace{-4.5em}
\noindent\makebox[\linewidth]{\rule{\textwidth}{0.4pt}}
\vspace{1em}\\

The subject of this thesis is cosmological implications of string compactifications understood in a broad sense. In the first half of the thesis, we will begin by reviewing the four-dimensional description of the tree-level perturbative type IIB action. We will then introduce a number of open questions in cosmology and their relevance with regards to the remainder of the thesis. We will first explore some of these questions from the perspective of effective field theories motivated by supergravity. In particular, we provide a description of a naturally light dark energy field in terms of the clockwork mechanism and the Dvali-Kaloper-Sorbo four-form mixing. We study its possible UV completion and show a no-go for its embedding within perturbative type IIA supergravity. We also discuss the coincidence problem for dynamical models of dark energy consistent with a quintessence field slowly rolling down a potential slope, of the type one would expect from the asymptotics of moduli space. As it rolls, a tower of heavy states will generically descend, triggering a phase transition in the low energy cosmological dynamics after at most a few hundred Hubble times. As a result, dark energy domination cannot continue indefinitely and there is at least a percentage chance that we find ourselves in the first Hubble epoch.

In the second half of the thesis, we introduce the effects of perturbative and non-perturbative corrections to the tree-level type IIB action. We then focus on obtaining a viable model of quintessence from the type IIB effective field theory. However, we are able to show that such a model must have a non-supersymmetric Minkowski vacuum at leading order. Furthermore, it must necessarily take the form of axion hilltop quintessence. When we consider the effects of quantum fluctuations during the early Universe, we see that such models must have extremely fine-tuned initial conditions to describe a slow-rolling scalar field at present times. We conclude that quintessence faces more challenges than a true cosmological constant, to the point that quintessence is very unattractive for model building modulo a ruling out of the cosmological constant by observations. Following this line of reasoning, we consider whether other perturbative corrections can generate de Sitter solutions in an appropriate setting. In particular, we consider the effects of higher curvature corrections in the Gauss-Bonnet term. Remarkably, we are able to show that, for the particular setting of a fluxed runaway potential motivated by heterotic supergravity, the curvature corrections reduce the space of solutions.

\addstarredchapter{Declaration}
\chaptermark{Declaration}
 
\chapter*{\vspace{-3.5em}Declaration}
\vspace{-4.5em}
\noindent\makebox[\linewidth]{\rule{\textwidth}{0.4pt}}
\vspace{1em}\\

This dissertation is the result of my own work and includes nothing which is the outcome of work done in collaboration except where specifically indicated in the text. No part of this thesis has previously been submitted for a degree or other qualification at this or any other university.\par

The following chapters contain original research presented in the following papers:
\begin{itemize}
\item \chpref{chap:clockwork}: {\bf  A natural theory of dark energy}, Lorenzo Bordin, FC, Antoine Leh\'ebel and Antonio Padilla, \emph{Phys.Rev.D} (2020),  arXiv:1912.04905 [hep-th]
\item \chpref{chap:coincidence}: {\bf  A stringy perspective on the coincidence problem}, FC and Antonio Padilla, \emph{JHEP} (2021),  arXiv:2105.03426 [hep-th]
\item \chpref{chap:intermezzo}: {\bf  Quintessence and the Swampland. The parametrically controlled regime of moduli space.}, Michele Cicoli, FC, Antonio Padilla and Francisco G. Pedro, \emph{Fortsch.Phys.} (2022), arXiv:2112.10779 [hep-th]
\item \chpref{chap:model_building}: {\bf  Quintessence and the Swampland. The numerically controlled regime of moduli space.}, Michele Cicoli, FC, Antonio Padilla and Francisco G. Pedro, \emph{Fortsch.Phys.} (2022), arXiv:2112.10783 [hep-th]
\item \chpref{chap:gbds}: {\bf  Quadratic curvature corrections and the absence of de Sitter vacua}, FC, Will Emond, Antoine Leh\'ebel and Antonio Padilla, \emph{JHEP} (2022), arXiv:2112.05771 [hep-th]
\end{itemize}
These papers are references \cite{1912.04905, 2105.03426, Cicoli:2021fsd, Cunillera:2021fbc, Cicoli:2021skd} in the bibliography, respectively.

\vfill
\begin{flushright}
Francesc Cunillera\par
Nottingham, \today
\end{flushright}
\vfill

\addstarredchapter{Acknowledgements}
\chaptermark{Acknowledgements}

\chapter*{\vspace{-3.5em}Acknowledgements}
\vspace{-4.5em}
\noindent\makebox[\linewidth]{\rule{\textwidth}{0.4pt}}
\vspace{1em}\\

The first and biggest thanks goes to Tony. For some reason you decided that I was worth your time and put up with 3 years of my painstaking mistakes and dumb ideas. You gave me enough freedom to go on a random supergravity journey for my PhD and taught me so much about Physics and what it means to be a good physicist during that time. I will always be thankful for all the time you have dedicated to nurture me as a budding physicist and, hopefully, some of your innate intuition about Physics will have rubbed off. 

I would also like to give a special thanks to Ed Copeland, Tasos Avgoustidis, Michele Cicoli and Francisco Gil Pedro for many long discussions about strings and physics in general. Ed, you also acted as a supervisor for me, making sure the less obvious admin duties where fulfilled and entertained long discussions about everything cosmological when I would randomly pop into your office. Tasos, our chats and coffees at Choulia's will be one of the things I miss the most from my times at Nottingham. Michele and Francisco, you guys patiently put up with many discussions over Zoom while I was getting up to speed with model building aspects of supergravity. You also made both Tony and I jealous of the beautiful Italian countryside, especially when comparing it to our bleak English weather. 

I would also like to thank Arthur Hebecker, Chad Briddon, Joe Conlon, Lorenzo Bordin, Clare Burrage, Will Emond, Cristiano Germani, Oli Gould, Antoine Leh\'ebel, Pete Millington, Adam Moss, Paul Saffin and Marco Scalisi for many useful and insightful discussions over the years.

\blankpage
\body
%\input{chapter_folder/chapter_name}
%%Add \blankpage after chapters and parts
\blankpage
\chapter{Introduction and motivations}

Let us begin by considering the age-old question: {\it why string theory?} To answer it, we first consider the two underlying low-energy theories that string theory aims to reproduce at the four-dimensional classical level: the Standard Model (SM) and General Relativity (GR). These two theories have been extremely successful in the many decades since their inception. Almost all of the physical probes\footnote{As far as we are aware, the only measurement that requires an explicit modification of the SM are the non-zero neutrino masses. The anomalous dipole moment of the muon might also indicate some new physics, but its interpretation is not as clear cut as the three-generations of massive neutrinos, at the moment.} currently available to physicists are consistent with a description in terms of GR+SM, from gravitational waves to the low energy particle spectrum of our Universe, at least to scales of order TeV\cite{PDG}. It is astounding that these two effective low energy theories can explain such a range of physical phenomena to such a degree of accuracy. In fact, one of the current problems within particle physics is the inability to design experiments able to probe regimes where we expect the SM description to break down\footnote{Modulo the discovery of supersymmetry at a collider experiment, given that the supersymmetry breaking scale for the matter sector could lie anywhere between the mass of the top quark and \mp.}\footnote{Between the writing and defence of this thesis, the CDF collaboration \cite{CDF:2022hxs} claimed a $\sim7\sigma$ tension between their experimental results and the SM expectation for the mass of the W boson. If confirmed by other experiments, this would be the strongest signal for beyond the SM physics to date.}. 

Nonetheless, despite the many successes, both the Standard Model and General Relativity falter in some aspects. Regarding the SM, there are issues when one considers it as a theory with some twenty free parameters to be fixed by experiments. Although these parameters can be fixed by experiments, one can ask the following two questions. First, what is the origin of the parameters in the SM Lagrangian? Second, why do some of these parameters take unnatural values, {\it i.e.} values that are much smaller than unity? Even if the physical origin and smallness of the parameters in the SM Lagrangian were not a worry, other issues arise when considering quantum effects \cite{0801.2562}. Most notably the Higgs, or any scalar not protected by some symmetry for that matter, is unstable against radiative corrections. This is to say that loop corrections couple the Higgs mass to the heaviest mass available in the theory, one of the incarnations of the so-called hierarchy problem \cite{Gildener:1976ai}. Therefore, unless new physics comes in to protect the Higgs at some scale above the current collider energies, there would be no reason to expect the Higgs to have its observed mass.  

On the other hand, the addition of gravity presents even more problems. Setting aside the usual considerations about singularities of GR, like in the interior of a black hole, the mismatch between theoretical and observational predictions for the value of the vacuum energy density presents one of the biggest open challenges in modern physics \cite{1502.05296}. Indeed, in terms of phenomenology we can think of the vacuum energy as a scalar field, which is either a cosmological constant or a very slowly varying scalar dubbed quintessence. Again, as a scalar field which is not protected by a symmetry at the level of GR, the cosmological constant is radiatively unstable and will tend to couple to the heavier masses available in the theory. In this case, the presence of the SM particles alone would naively induce a correction to the cosmological constant of $\mathcal{O}(\text{TeV}^4)$, many orders of magnitude above its observed value at $\mathcal{O}(\text{meV}^4)$ scales. This is the so-called cosmological constant problem \cite{Weinberg:1988cp}.

Turning on gravity also introduces technical difficulties, which are not entirely unrelated to the previous issues, in that the quantum field theory of GR+SM becomes non-renormalisable. This is easily seen as soon as one realises that the Ricci term in the Einstein-Hilbert action is parametrising the strength of graviton interactions, with a coupling strength of $\mp^{-2}$. This is an irrelevant operator, growing weaker at low energies and strong at large energies. It is certainly weak enough that at the level of collider physics one seems to be able to turn gravity off and accurately predict the scattering amplitudes of any interaction. The problem with non-renormalisability becomes clear when considering internal graviton loops. Indeed, in that case the integral over momentum space becomes divergent for arbitrarily high energies. This divergence only becomes worse with higher orders in the loop calculation \cite{1502.05296}. This divergence can be mapped to position space as the limit where graviton vertices become coincident\cite{Polchinski:1998rq}.

String theory provides a beautiful description of graviton scatterings that gets rid of these divergences. Heuristically, due to the extended nature of strings, the scattering diagrams for gravitons are no longer coincident but spread out over the worldsheet of the string. Incidentally, string theory\footnote{Henceforth, by string theory we loosely refer to {\it a complete theory of extended objects as fundamental blocks of nature for which supergravity is a low energy description}.} also has many desirables properties for a would-be unification theory of the SM with gravity. Some of these are \cite{Polchinski:1998rq}
\begin{enumerate}
\item String theory reproduces the low energy description of gravity, {\it i.e.} string theory contains a massless spin-2 particle whose interactions reduce to general relativity at the four-dimensional low energy level \cite{Scherk:1974ca}.
\item Supersymmetry of the full theory follows from consistency requirements and it provides a natural way of introducing fermions and bosons into the particle spectrum of the theory, with a very clear physical interpretation in terms of periodic and anti-periodic boundary conditions on the string worldsheet \cite{hep-ph/9511313}. It could also aid in solving the SM hierarchy problem and ameliorate the cosmological constant problem, to some extent \cite{0801.2562}.
\item Further consistency considerations imply that the supersymmetric string theory requires a definite number of spacetime dimensions. Supersymmetric string compactifications allow a low-energy description of the spacetime in terms of 4 large external dimensions, that can host physics similar to those of the SM, and 6 small internal dimensions, that can be {\it integrated out} at collider energies \cite{Ibanez:2012zz}.
\item Supersymmetry can be partially broken so that the low energy dynamics of the string theory are chiral, in agreement with the SM \cite{Ibanez:2012zz}.
\item String theory, in its final incarnation as described in the footnote, contains no free parameters and is unique \cite{1009.3497}.
\end{enumerate}
Of course, the lack of supersymmetry found at the LHC presents some subtleties in engineering a phenomenologically viable low-energy description of string theory. Nonetheless, these issues can, {\it a priori}, be fixed by softly breaking supersymmetry in the matter sector at some scale between the energy limit at collider searches and \mp. It is quite striking that the consistency considerations of string theory can lead to such powerful constraints and still yield a theory that is remarkably capable of describing our Universe. 

Clearly, string theory represents a powerful framework\footnote{And is possibly the only one.} to unify SM and GR. However, there must be some work left to be done, otherwise theoretical physics would be solved. Most of the open questions in this line of research can be recast as a lack of understanding of the full non-perturbative description of string theory. Indeed, the work in this thesis will concentrate on supergravity, {\it i.e.} the low-energy physics of string theory. The supergravity action only contains the massless sector of string theory, that is we will drop the dependence on oscillator modes and keep only the string zero modes. This allows an exploration of the perturbative aspects of string theory, where we see supergravity as an expansion to some order in the genus of the worldsheet, parametrised by the string coupling $g_s$, and to some order in worldsheet loops, parametrised by the string length $l_s$, or equivalently by the $\alpha'$-parameter with $l_s:=\sqrt{2\pi\alpha'}$. Thus, supergravity inherits two expansion parameters\footnote{Note that, on top of the expansion parameters, ten-dimensional supergravities admit a number of fluxes being turned on, which at the low-energy effective level will effectively act as continuous free parameters.} in $g_s$ and $\alpha'$. We also note that our knowledge of non-perturbative corrections is quite limited. Nonetheless, when trying to describe the low-energy dynamics of our four-dimensional world, we would expect that in most instances supergravity provides a {\it close enough description}. Keeping the perturbative expansion under control, so that the supergravity predictions can be trusted, will be a running theme in this thesis. 

Within the supergravity approximation, we can begin to ask phenomenological question about our Universe. The central question that we tackle in this work is regarding the microscopic origin of the vacuum energy and its implications.  A number of the technical questions about naturalness and hierarchies that concerned the SM, will also apply to the vacuum energy. Indeed, the ten-dimensional supergravity action has a characteristic energy scale of $\mathcal{O}(l_s^{-1})$, with $l_s^{-1}\lesssim \mp$, whereas the observational dark energy scale lies at $\mathcal{O}(10^{-30}\mp)$. Thus, if supergravity is to explain the scale of dark energy we will require a theory able to generate huge scale hierarchies. Can these large scales hierarchies be generated in the first place? If so, can it be done while avoiding extreme fine-tuning of the supergravity parameters? 

These are some important considerations that will play a key role in the latter stages of this thesis. Before moving onto the bulk of our work, we would like to touch upon two more points: EFTs and the Swampland (for a review see \cite{1903.06239, 2102.01111}). Computations in string theory and, by proxy, supergravity are complex. We have a relatively small toolbox to compute quantities in supergravity in comparison to, say, GR. This means that it is sometimes convenient to explore an idea in a simplified framework within effective field theories that are in a more or less rigorous sense motivated by supergravity. In some sense, by working with supergravity we are doing just so. String theory is too complicated to be treated in full generality, so we concentrate on the massless sector and accept the limitations that this carries. Even then, certain concepts might be far from the reach of our current understanding of supergravity. It is then convenient to study such concepts within effective field theory toy models acknowledging its limitations. When a toy model proves promising, we can then try to find its supergravity embedding. 

An immediate question that can be raised is whether any self-consistent EFT inspired by supergravity can find itself embedded in one corner or another of the supergravity parameter space. The reasonable expectation is that this would not be true. For example, one could imagine an EFT with more free parameters than a particular supergravity theory, such that supergravity imposes some non-trivial constraints on the free parameters of the EFT which cannot be satisfied. These constraints are not, in general, accessible from the EFT perspective as they require some knowledge of the UV embedding above the EFT cut-off. The Swampland programme aims to provide  conditions for any EFT to be consistently embedded within a UV parent theory\footnote{This is usually thought to be string theory/supergravity, although some aspects of the Swampland are claimed to apply more generally.}. This programme has become an important part of the research done by the community in the past decade. Its more fundamental parts concern M-theory/F-theory compactifications and, in general, have been quite successful in improving our understanding of the ten-dimensional description of supergravity (for example \cite{Lee:2020blx, 1808.09427, 1908.04788}). On the other hand, there exists another section of the Swampland programme concerned with deriving EFT constraints from the asymptotics of the ten-dimensional theory. In this regime, the supergravity description becomes quite simplified and strong phenomenological bounds on the resulting EFTs can be imposed \cite{Obied:2018sgi, Bedroya:2019snp}. Critically, these results hold only in the asymptotic regime and claiming that they hold anywhere else is a matter of speculation. However, due to the powerful phenomenological constraints, this part of the Swampland has attracted considerable research and statements about the asymptotic behaviour of supergravity have turned into general conjectures anywhere in the parameter space of the theory. In this thesis, we will briefly discuss the phenomenological branch of the Swampland programme as some of its results overlap with the research presented here. We will often refer to it as the {\it Swampland programme} for convenience but we remark that our discussion concerns only this second aspect of the Swampland.

Finally, one could worry whether our current description of our effective low-energetic world, in the SM and GR, is at all consistent with a supergravity embedding. We have already pointed out that string theory has many desirable properties that would make us optimistic that such an embedding exists, we also have constructions that closely resemble the low-energy dynamics of SM+GR, and we have no evidence to the contrary, but one can never disregard the possibility that this could not be the case. However, to conclude that such embedding does not exists with any certain degree of confidence we will require a more complete picture of supergravity. It is certainly not enough to explore the asymptotics of the ten-dimensional description to claim anything about the bulk of the theory.

\section{Outline}
The contents of this thesis are organised as follows. In \chpref{chap:pert_II}, we will obtain the effective field theory derived from the tree-level type IIB supergravity action. In \chpref{chap:openqs}, we introduce a number of open question in physics and cosmology that will be relevant to the contents of the following sections. After this, in \chpref{chap:clockwork} and \chpref{chap:coincidence} concentrate on studying EFTs motivated by supergravity that aim at answering some of the questions in \chpref{chap:openqs}. We conclude the first half of the thesis with a brief discussion about perturbative control and the Swampland in \chpref{chap:intermezzo}. 

The second half of the thesis is mostly devoted to the implications that the microscopic nature of dark energy has on supergravity model building. In \chpref{chap:non_pert}, we introduce a number of perturbative and non-perturbative corrections to the tree-level type IIB action that will be necessary to build viable models of dark energy. Then, in \chpref{chap:model_building}, we try to construct viable dynamical models of dark energy and contrast them with a true cosmological constant in terms of model building. We remark that the effects of higher order corrections might prove crucial and begin to study whether this can be relevant in a fluxed runaway scenario with Gauss-Bonnet corrections in \chpref{chap:gbds}. Finally, in \chpref{chap:future}, we conclude by reviewing our findings and provide some general thoughts as to possible directions for future research.

\section{Units and conventions}
Unless explicitly stated, we take the usual unit conventions where $\hbar=c=\mp=1$. We will also set $\alpha'=1$; this will in turn fix the string scale to $M_s:=l_s^{-1}=1/\sqrt{2\pi}$. For the metric, we take the mostly positive convention with Latin uppercase indices ($M,N...$) making reference to the ten-dimensional metric, Greek indices ($\mu,\nu,...$) referring to the external four-dimensional metric and Latin lowercase indices ($i,j,...$) will represent the internal six-dimensional metric. Note that this index convention only refers to metric indices and not to the (co)homology indices which will follow particular conventions set out in the pertinent chapters.

Where ambiguities may arise, we will also ``hat'' the ten-dimensional quantities, for example $\hat{\phi}$ will be the ten-dimensional dilaton field and $\phi$ will be its four-dimensional analogue.
\blankpage
\part{String inspired EFTs}
\blankpage
\chapter{Perturbative supergravity in type IIB theories\label{chap:pert_II}}
In this chapter we explore the fundamentals of Calabi-Yau compactifications. We begin in \secref{sec:special_geo} by motivating the use and describing the geometry of Calabi-Yau manifolds. In \secref{sec:n2}, we derive the moduli spectrum and scalar potential for $\mathcal{N}=2$ type IIB Calabi-Yau compactifications. This leaves us with 2 supersymmetry generators and a non-chiral theory. Phenomenologically, we are interested in a chiral $\mathcal{N}=1$ supersymmetric theory. To achieve this, in \secref{sec:orientifold}, we orientifold our $\mathcal{N}=2$ action and provide the $\mathcal{N}=1$ spectrum as well as its effective scalar potential. 

Finally, we would like to also discuss how to stabilise the moduli fields of the theory. To do so, in \secref{sec:fluxes}, we introduce background fluxes through the three-form flux piece in type IIB. This allows us to stabilise the axio-dilaton and the complex structure sector. The stabilisation of the \kah sector is a challenging question that we will discuss in detail in \chpref{chap:non_pert}, as it requires non-perturbative corrections. We also briefly discuss localised sources and tadpole cancellation conditions. In \secref{sec:swamp}, we discuss the Swampland programme and, in particular, the refined versions of the Swampland distance conjecture and the de Sitter conjecture. 

\section{Special geometry of K\"ahler manifolds}\label{sec:special_geo}
Let \x be a Calabi-Yau (CY) three-fold, {\it i.e.} a \kah manifold in 3 complex dimensions with $SU(3)$ holonomy. By the famous theorem conjectured by Calabi and proved by Yau \cite{Yau:1977ms}, this description for a CY three-fold is equivalent to that of a \kah manifold in 3 complex dimensions that is Ricci-flat, since a vanishing first Chern class implies an n-fold \kah manifold admits a \kah metric with $SU(n)$ holonomy. A \kah manifold is a hermitian manifold with metric $g_{i\barj}$, {\it i.e.} the purely holomorphic and anti-holomorphic components of the metric vanish, whose  \kah form
\be
J=-i g_{i\barj} dz^i\wedge d\bar{z}^{\barj}
\label{eq:kah_form}
\ee
is closed, $dJ=0$, under the exterior derivative $d$, where $z^i$ denote complex local coordinates and $\bar{z}^{\bari}$ are their complex conjugates. 

There are a few properties that make studying CY three-folds attractive. First, the critical superstring lives in a ten-dimensional spacetime. By compactifying a six-dimensional subspace on a CY three-fold, we will be left with an effective four-dimensional description of the low-energy physics of the superstring, matching the low-energy dimensionality of our Universe. Second, a compactification on a CY three-fold gets rid of $3/4$ of the underlying supersymmetry of the theory, when compared to a toroidal compactification. The number of preserved supersymmetries is related to the decomposition of the spinor representation $\mathbf{16}\in SO(1,9)$. The existence of a singlet in the decomposition of the internal spinor representation will lead to the conservation of a supersymmetry in the 4-dimensional spectrum \cite{Polchinski:1998rr}. For a generic compact manifold $\mathcal{M}$ and requiring 4-dimensional Poincar\'e invariance, the ten-dimensional Lorentz group decomposes into
\be
SO(1,9)\longrightarrow SO(1,3)\times SO(6)\ ,
\ee
and the spinor representation becomes
\be
\mathbf{16}\longrightarrow (\mathbf{2},\mathbf{4})\oplus (\mathbf{\bar 2},\mathbf{\bar 4})\ ,
\ee
where $\mathbf{2}$ is the Weyl spinor of $SO(1,3)$ and $\mathbf{4}$ is the corresponding spinor of $SO(6)\cong SU(4)$, with $(\mathbf{\bar 2},\mathbf{\bar 4})$ their respective conjugates. Since $SU(4)$ does not contain a singlet in its representation, this would lead to a non-supersymmetric four-dimensional theory. To preserve some supersymmetry we require $\mathcal{M}$ to have $SU(n)$ holonomy with $n<4$, that is under parallel transport around a closed loop the spinor singlet comes back to itself with a rotation in $SU(n)\subset SO(6)\cong SU(4)$. For example, in the case of a CY three-fold, which by definition has $SU(3)$ holonomy, the internal spinor representation decomposes as
\be
\mathbf{4}\longrightarrow \mathbf{3}\oplus\mathbf{1}\ ,
\ee
such that a singlet is left invariant and an $\mathcal{N}=1$ ten-dimensional theory compactified on a CY three-fold will preserve $\mathcal{N}=1$ supersymmetry in four dimensions\footnote{Note that type IIB supergravity has $\mathcal{N}=2$ in ten dimensions and thus preserves $\mathcal{N}=2$ supersymmetry after compactification on a CY three-fold. This is by virtue of type IIB having both left and right moving states, which yield a spinor decomposition in terms of $\mathbf{4}_{LR} \oplus \mathbf{4}_{RL}$.}. Finally, Ricci-flatness, or equivalently $SU(3)$ holonomy, allows to find a one-to-one mapping between metric perturbations and the cohomology of the CY three-folds. We explore this relation in the remainder of this section.

The {\it p}-th de Rham cohomology associated to the exterior derivative of the three-fold $H^p(\x)$ is defined to be the quotient space
\be
H^p(\x) := C^p(\x)/Z^p(\x)\ ,
\ee
where $C^p(\x)$ is the space of closed p-forms, {\it i.e.} $dA_p=0$, and $Z^p(\x)$ is the space of exact p-forms, {\it i.e.} $A_p=dA_{p-1}$. Then, $H^p(\x)$ is composed by the representatives of equivalency classes of p-forms whose elements differ at most by an exact p-form. The dimensionality of the $H^p(\x)$ group is given by the Betti number $b^p$, which is related to the Euler characteristic\footnote{This relation holds even more generally for a CY n-fold $\x_n$ where the Euler characteristic is then given by 
\ben
\chi(\x_n) = \sum_{p=0}^{2n} (-1)^p b^p \ .
\een} of \x by
\be
\chi = \sum_{p=0}^6 (-1)^p b^p \ .
\ee
Since \x is a complex manifold, it will be useful to adopt a holomorphic/anti-holomorphic index notation and to split the p-forms and exterior derivative in holomorphic/anti-holomorphic pieces. Indeed, we will define a (p,q)-form to be a form with p holomorphic and q anti-holomorphic indices
\be
A_{p,q}= \frac{1}{p!q!} A_{m1,...,m_p,\bar{n}_1,...,\bar{n}_q}dz^{m_1}\wedge...\wedge dz^{m_p} \wedge d\bar{z}^{\bar{n}_1} \wedge ...\wedge d\bar{z}^{\bar{n}_q}\ .
\ee
We also expand the exterior derivative in Dolbeault operators
\be
d = \partial +\bar{\partial}\ , \qquad \partial=dz^a{\partial\over\partial z^a}\ , \quad \bar{\partial}=d\bar{z}^{\bar{a}}{\partial\over\partial \bar{z}^{\bar{a}}}\ ,
\ee
where $(\partial,\bar{\partial})$ map (p,q)-forms to (p+1,q)-forms and (p,q+1)-forms, respectively. The Dolbeault cohomology\footnote{In a slight abuse of nomenclature, for the rest of the thesis we will use ``cohomology'' to refer to either de Rham cohomology or Dolbeault cohomology equally.}  group $H^{p,q}(\x)$ follows from the definition of the operator\footnote{Note that the Dolbeault cohomology of \x can be defined with respect to $\bar{\partial}$, since the {\it two} cohomologies will be mapped one-to-one by the complex conjugation of their elements. This is by virtue of \x being a \kah manifold.} ${\partial}$. Indeed, $H^{p,q}(\x)$ is composed by equivalency classes of closed (p,q)-forms under $\partial$, whose elements differ at most by an exact (p,q)-form. The dimension of $H^{p,q}(\x)$ is the Hodge number $h^{p,q}$ and it follows that
\be
b^q = \sum_{p=0}^{q} h^{p,q-p}\ . \label{eq:euler_bp}
\ee
It is convenient to represent cohomology structure of \x through the so-called the Hodge diamond, since this makes the CY symmetries manifest. The Hodge diamond of any CY three-fold is given by
\begin{equation}
\begin{array}{ccccccc}
&&& 1 &&&\\
&& 0 && 0 &&\\
& 0 && h_{1,1}&& 0 &\\
1 && h_{1,2}&& h_{1,2}&& 1 \\
& 0 && h_{1,1}&& 0 &\\
&& 0 && 0 &&\\
&&& 1 &&&
\end{array}\ .
\label{eq:hodge_dia}
\end{equation}
Note that the Hodge diamond, or the cohomology classes to be more precise, are not a complete characterisation of a Calabi-Yau. Indeed, non-homeomorphic Calabi-Yaus can have the same Hodge diamond. This can be better understood with a simple example. Consider a strip with periodic boundary conditions (a band) and one with anti-periodic boundary conditions (a M\um{o}bius strip). Both of them are homotopically equivalent to a circle, and also cohomologically equivalent, however there exists no one-to-one bijective map between the two topologies due to the torsion. Therefore, the two topologies are cohomologically equivalent but not homeomorphic. Although this is far from a proof for generic Calabi-Yaus, we find the intuitive picture useful.

The following symmetries have been exploited to eliminate redundant groups in the Hodge diamond:
\begin{itemize}
\item Isomorphism of $H^p(\x)$ and $H^{3-p}(\x)$ requires $h^{p,\,0}=h^{3-p,\,0}$.
\item Complex conjugation implies $h^{p,\,q}=h^{q,\,p}$.
\item Poincar\'e duality leads to $h^{p,\,q}=h^{3-p,\,3-q}$;
\item Simply connected CYs will have vanishing first homology group as every one-cycle can be infinitesimally shrunk to a point such that $h^{0,1}=h^{1,0}=0$.
\end{itemize}
After applying these symmetries, we can see that the cohomological structure of \x will be determined by a pair of Hodge numbers $h^{1,1},\,h^{2,1}$ which are related to its Euler characteristic through \eqref{eq:euler_bp}
\be
\chi = 2\left(h^{1,1}-h^{2,1}\right)\ .
\ee
One can then introduce a base of harmonic p-forms for each cohomology class. We denote the basis of harmonic (1,1)-forms with $\omega_A$, its Poincar\'e dual being the basis of (2,2)-forms $\tilde{\omega}^A$, with $A=1,...,h^{1,1}$. Introducing the basis of $H^{3}(\x)$ to be $\left(\alpha_{\hat{K}},\beta^{\hat{L}}\right)$, which are related to each other by complex conjugation and $\hat{K},\hat{L}=0,...,h^{1,2}$, the only non-trivial intersection numbers are
\be
\int_\x\omega_A\wedge\tilde{\omega}^B=\delta_A^B\ , \qquad \int_\x\alpha_{\hat{K}}\wedge\beta^{\hat{L}}=\delta_{\hat{K}}^{\hat{L}}\ .
\label{eq:int_num_2}
\ee

It is also useful to breakdown the $H^{3}(\x)$ basis in its components as follows $\left(\alpha_{\hat{K}},\beta^{\hat{L}}\right)=(\alpha_0,\chi_K,\beta^0,\chi^L)$ with $K,L=1,...,h^{2,1}$.

\begin{table}[h!]
\centering
\begin{tabular}{|c|c|c|}
\hline
Cohomology group & Dimension    & Basis                                           \\ \hline
$H^{1,1}$        & $h^{1,1}$    & $\omega_A$                                      \\ \hline
$H^{2,2}$        & $h^{1,1}$    & $\tilde{\omega}^A$                              \\ \hline
$H^3$            & $2+2h^{2,1}$ & $\left(\alpha_{\hat{K}},\beta^{\hat{L}}\right)$ \\ \hline
$H^{2,1}$        & $h^{2,1}$    & $\chi_K$                                        \\ \hline
$H^{3,3}$        & $1$          & $\text{dvol}(\x)$                                       \\ \hline
\end{tabular}
\caption{Bases for a Calabi-Yau threefold.}
\end{table}

At this point, one can expand any p-form of \x as a product of moduli fields and cohomology basis elements. For instance, a 2-form field $B_2$ can generically be expanded as
\be
B_2=\bar{B}_2(x) + b^A(x)\omega_A \ ,
\ee
where $x$ denotes external coordinates. 

Aside from the field content present in the theory, one is free to consider deformations $\delta g_{mn}$ of the CY metric $g_{mn}$ on \x, where the indices $(m,n)$ cover both the holomorphic and anti-holomorphic indices, that is $m=(i,\bari)$. In particular, we consider deformations that preserve the Ricci flatness of the manifold, {\it i.e.} $R_{mn}(g+\delta g)=0$. In this way, even after deforming the metric, the manifold is guaranteed to be Calabi-Yau \cite{Grimm:2004uq, Grimm:2004ua, Grimm:2005fa}.

What are the set of deformations that preserve the Ricci flatness of \x? First of all, notice that if the metric of the manifold $g_{mn}$ is Ricci-flat so will be any other $g'_{mn}$ related to the original metric by diffeomorphisms. This diffeomorphism invariance allows us to gauge fix $\nabla^m\delta g_{mn}= {1\over2} \nabla_n \delta g$, where $\delta g=g^{mn}\delta g_{mn}$, such that the linearised Lichnerowicz equation governing the deformations becomes
\be
\Delta \delta g_{mn}:= \nabla^p\nabla_p \delta g_{mn} + 2 R_{m\ n}^{\ \ p\ \, q}\delta g_{pq}  = 0\ ,
\ee
and we can study the equations for the pure and mixed indices independently as these decouple.
\begin{itemize}
\item Mixed components: The Lichnerowicz equations reads
\be
\Delta \delta g_{i\barj}=0.
\ee
Therefore, $\delta g_{i\barj}$ is a harmonic form of the Laplacian and will admit an expansion in terms of $\omega_A$. These modes are related to non-trivial deformations of the \kah form $J$ through \eqref{eq:kah_form}.

\item  Pure components: In this case, the Lichnerowicz equations reads
\be
\Delta \delta g_{ij}=0,
\ee
implying an expansion in (2,0)-forms. However, since we would like to recover a CY manifold after the deformations, there should exist a one-to-one mapping between the $\delta g_{ij}$ and some combination of mixed deformations. Indeed, one can generate such a mapping between pure deformations and a new set of mixed component deformations through the unique holomorphic (3,0)-form $\Omega$ of \x (cf. \ssecref{ssec:cs_kahler}). 
\end{itemize}
All in all, one finds that the moduli space of deformations is spanned by \cite{Polchinski:1998rr, hep-th/0403067}
\begin{gather}
\delta g_{i\barj} = -iv^A(\omega_A)_{i\barj}\ , \label{eq:mixed_def}\\
\delta g_{i j} = \frac{i}{||\Omega||^2}\overline{(z^K \chi_K)}_{i\bari\barj}\Omega_{jkl} g^{\bari k}g^{\barj l}:=\frac{i}{||\Omega||^2}\bar{z}^K(\bar{\chi}_K)_{i\bari\barj}\Omega^{\bari \barj}_j\ , \label{eq:pure_def} 
\end{gather}
where the last equality in \eqref{eq:pure_def} simply introduces shorthand notation and we have defined\footnote{Again, in this shorthand notation, this definition really means $||\Omega||^2:=\frac{1}{3!}\Omega_{ijk}\bar{\Omega}_{\bari\barj\bar{k}}g^{\bari i}g^{\barj j}g^{\bar{k} k}$.} $||\Omega||^2:=\frac{1}{3!}\Omega_{ijk}\bar{\Omega}^{ijk}$. At this level, the moduli space is locally a direct product of two \kah manifolds
\begin{equation*}
\mathcal{M}^{ks}\times\mathcal{M}^{cs}\ ,
\end{equation*}
this will continue being true at the 4D level so the potential generated by either sector will decouple from the other and we can study them independently\footnote{Note that the existence of this direct product of two manifolds will be directly related to the no-scale structure of type IIB, as we will discuss in \ssecref{ssec:n1}. Breaking the no-scale structure will mean that the \kah and complex structure sectors will not decouple anymore.}. Furthermore, this implies that (at least locally) the metrics of both sectors are the metrics of \kah manifolds independently, {\it i.e.} there will exist a pre-potential function for each sector such that the metric can be expressed as a derivative of this pre-potential. In the following two section we explore the structure of $\mathcal{M}^{ks}$ and $\mathcal{M}^{cs}$.

\subsection{The \kah structure of $\mathcal{M}^{ks}$}
The first of the two sectors is given by the deformations to the \kah structure. Intuitively these deformations can be understood as changes in the size of the internal space. For example, in the simple case of a $T^2$ the \kah structure deformations are parametrised by the modulus $\rho=i R_1R_2$ where $R_i$ are the two radii of the torus. A change in $\rho$ corresponds to a change in the size of the torus.

Back to the more complicated case of a CY, we have $h^{1,1}$ \kah moduli. It is convenient to package the moduli coming from the Kalb-Ramond 2-form together with the real scalar moduli of \eqref{eq:mixed_def} in a new set of imaginary scalars given by
\be
t^A:= v^A+ i b^A\ .
\ee
The metric of the moduli space is defined as
\be
G_{AB}:=\frac{1}{4\mathcal{V}}\int\omega_A\wedge\star\omega_B=-\frac{1}{4\mathcal{V}}\left(\mathcal{K}_{AB}-\frac{\mathcal{K}_A\mathcal{K}_B}{4\mathcal{V}}\right)=\partial_{t^A}\partial_{\bar{t}^B} K^{\text{ks}}\ , \label{eq:ks_metric}
\ee
where we find the \kah potential of the \kah structure metric $K^{\text{ks}}$ to be
\be
K^{\text{ks}}=-\ln\left[8\mathcal{V}\right]=-\ln\left[\frac{1}{6}\mathcal{K}_{ABC}(t+\bar{t})^A(t+\bar{t})^B(t+\bar{t})^C\right] \label{eq:kks}
\ee
with $\mathcal{V}$ the volume of \x and we define the triple intersection numbers $\mathcal{K}_{ABC}$ and their {\it contractions} as
\begin{gather}
\mathcal{K}_{ABC}:=\int\omega_A\wedge\omega_B\wedge\omega_C \ , \qquad \mathcal{K}_{AB}:=\int\omega_A\wedge\omega_B\wedge J =\mathcal{K}_{ABC} v^C\ , \nonumber \\
\mathcal{K}_{A}:=\int\omega_A\wedge J \wedge J =\mathcal{K}_{ABC} v^B v^C\ , \qquad \mathcal{K}:=\int J\wedge J\wedge J =\mathcal{K}_{ABC}v^A v^B v^C = 6\mathcal{V} \nonumber \ .
\end{gather}
The metric $G_{AB}$ is fully encoded in a single function $K^{\text{ks}}$. Even more so, the moduli space is a {\it special \kah}manifold since the \kah potential itself is given in terms of a prepopential 
\begin{gather}
K^{\text{ks}}=-\ln \left[2(f+\bar{f})-(t+\bar{t})^A(\partial_{t^A}f+\partial_{\bar{t}^A}\bar{f})\right]\ , \\
f(t)=-\frac{1}{6}\mathcal{K}_{ABC}t^At^Bt^C\ ,
\end{gather}
To ensure that the metric is positive definite, even after the deformations, we require $\mathcal{K}_{AB},\mathcal{K}_{A},\mathcal{K}>0\ \forall(A,B)$. These conditions define the \kah cone spanned by the moduli $t^A$.

Finally, every {\it special \kah}manifold has a complex matrix defined as 
\be
\mathcal{N}_{\hat{A}\hat{B}}:=\bar{\mathcal{F}}_{\hat{A}\hat{B}}+2i\frac{(\text{Im}\ \mathcal{F})_{\hat{A}\hat{C}}\; t^{\hat{C}}(\text{Im}\ \mathcal{F})_{\hat{B}\hat{D}}\; t^{\hat{D}}}{t^{\hat{C}}(\text{Im}\ \mathcal{F})_{\hat{C}\hat{D}}t^{\hat{D}}}\ , \label{eq:gc}
\ee
where we use adapted coordinates $t^{\hat{A}}=\{1,t^A\}$, $\mathcal{F}=f(t)$ and $\mathcal{F}_{\hat{A}\hat{B}}=\{0,\partial_{t^A}\partial_{t^B}\mathcal{F}\}$. Plugging in the form of $f(t)$ we obtain 
\begin{gather}
\text{Re} \mathcal{N} = \begin{pmatrix}
-\frac{1}{3}\mathcal{K}_{ABC}b^A b^B b^C & \frac{1}{2}\mathcal{K}_{ABC}b^A b^B \\
 \frac{1}{2}\mathcal{K}_{ABC}b^A b^B & -\mathcal{K}_{ABC} b^C
\end{pmatrix}\ ,\label{eq:gk_coupling1}\\
\text{Im} \mathcal{N} = -\mathcal{V}\begin{pmatrix}
1+4G_{AB}b^A b^B & -4G_{AB}b^B\\
-4G_{AB}b^B & 4G_{AB}
\end{pmatrix}\ ,\label{eq:gk_coupling2}\\
(\text{Im} \mathcal{N})^{-1} = -\frac{1}{\mathcal{V}}\begin{pmatrix}
1 & b^A\\
b^A & \frac{1}{4}G^{AB}+b^A b^B
\end{pmatrix}\ .
\label{eq:gk_coupling3}
\end{gather}

\subsection{The complex structure of $\mathcal{M}^{cs}$\label{ssec:cs_kahler}}
The second sector is given by the deformations to the complex structure. Back to the analogy with a $T^2$, these deformations are parametrised by the modulus $\tau=i\frac{R_2}{R_1}$ and control the changes in shape of the torus.

In the case of the CY, there will be $h^{2,1}$ complex structure fields. The metric of the complex structure moduli space is defined as
\be
G_{K\bar{L}}:=-\frac{\int \chi_K\wedge\bar{\chi}_L}{\int \Omega\wedge\bar{\Omega}}\ . \label{eq:cs_metric}
\ee
Kodaira's formula gives the basis $\chi_L$ in terms of the holomorphic (3,0)-form $\Omega$
\be
\chi_K(z,\bar{z})=\partial_{z^K}\Omega(z)+\Omega(z)\partial_{z^K}\left[-\ln\left(i\int \Omega\wedge\bar{\Omega}\right) \right]\ .\label{eq:kodaira_form}
\ee
The bracket is the \kah potential for the complex structure metric since
\be
G_{K\bar{L}}=\partial_{z^K}\partial_{\bar{z}^L}K^{\text{cs}}\ , \qquad K^{\text{cs}}=-\ln\left(i\int \Omega\wedge\bar{\Omega}\right)=-\ln \ i\left(\mathcal{Z}^{\hat{K}}\bar{\mathcal{F}}_{\hat{K}}-\bar{\mathcal{Z}}^{\hat{K}}\mathcal{F}_{\hat{K}}\right)\ . \label{eq:kcs}
\ee
Thus, the complex structure moduli space is \kah as well. In the last relation of \eqref{eq:kcs} we have used that the harmonic (3,0)-form $\Omega$ benefits from an expansion in terms of the harmonic bases
\be
\Omega(z)=\mathcal{Z}^{\hat{K}}(z)\alpha_{\hat{K}}-\mathcal{F}_{\hat{K}}(z)\beta^{\hat{K}}\ .
\ee
Since the \kah potential is only defined up to a complex rescaling of $\Omega$ by a holomorphic function
\be
\Omega\rightarrow \tilde{\Omega}=\Omega e^{-h(z)}\ , \qquad K^{\text{cs}}\rightarrow \tilde{K}^{\text{cs}}=K^{\text{cs}}+h+\bar{h}\ , \label{eq:gauge_omega}
\ee
we are allowed to gauge fix $\mathcal{Z}^0=1$; this is the so-called \kah gauge. This leaves $h^{2,1}$ free components for the periods of $\Omega$ and one can identify these periods with the $z^K$ moduli fields by defining $z^K=\mathcal{Z}^K/\mathcal{Z}^0$. Furthermore, the manifold $\mathcal{M}^{cs}$ is also a {\it special \kah}manifold. 

Indeed, one can choose $\mathcal{F}=\frac{1}{2}\mathcal{Z}^{\hat{K}}\mathcal{F}_{\hat{K}}$ and the relations set out in the previous section between {\it special} manifolds and its {\it special} matrix still hold (now adapted to the {\it special} coordinates $z^K$ instead of $t^A$). Namely\footnote{We keep the $\mathcal{Z}^0$ dependence explicit for the sake of generality}
\begin{gather}
K^{\text{cs}}=-\ln i\left|\mathcal{Z}^0\right|^2 \left[2(f-\bar{f})-(z-\bar{z})^K(\partial_{z^K}f+\partial_{\bar{z}^K}\bar{f})\right]\ , \\
f(z)={1\over \left(\mathcal{Z}^0\right)^2}\mathcal{F}\ ,
\end{gather}
and the special matrix (also known in this case as the period matrix) is given by
\be
\mathcal{M}_{\hat{K}\hat{L}}:=\bar{\mathcal{F}}_{\hat{K}\hat{L}}+2i\frac{(\text{Im}\ \mathcal{F})_{\hat{K}\hat{M}}\; \mathcal{Z}^{\hat{M}}(\text{Im}\ \mathcal{F})_{\hat{L}\hat{N}}\; \mathcal{Z}^{\hat{N}}}{\mathcal{Z}^{\hat{M}}(\text{Im}\ \mathcal{F})_{\hat{M}\hat{N}}\mathcal{Z}^{\hat{N}}}\ . \label{eq:period_matrix}
\ee
This matrix will appear in the dimensionally reduced type II supergravity action as the kinetic metric for the complex structure vectors  and is most conveniently written down as
\begin{gather}
\int \alpha_{\hat{K}}\wedge\star\alpha_{\hat{L}}=-\left[\text{Im}\ \mathcal{M} + (\text{Re}\ \mathcal{M})(\text{Im}\ \mathcal{M})^{-1}(\text{Re}\ \mathcal{M})\right]_{\hat{K}\hat{L}} \ , \\
\int \beta^{\hat{K}}\wedge\star\beta^{\hat{L}}=-\left(\text{Im}\ \mathcal{M}\right)^{-1\ \hat{K}\hat{L}} \ , \\
\int \alpha_{\hat{K}}\wedge\star\beta^{\hat{L}}=-\left[(\text{Re}\ \mathcal{M})(\text{Im}\ \mathcal{M})^{-1}\right]_{\hat{K}}^{\hat{L}} \ .
\end{gather}
 
With this, we finish our review on the geometry of moduli space. We will now proceed to study the $\mathcal{N}=2$ bosonic sector for the type II supergravity actions.

\section{Perturbative $\mathcal{N}=2$ supergravity\label{sec:n2}}
The massless bosonic spectrum of type IIB supergravity is composed of two sectors: the NSNS sector comprising of the metric $\hat{g}_{mn}$, the dilaton $\hat{\phi}$ and the Kalb-Ramond two-form $\hat{B}_2$; and the RR sector containing the even p-form gauge fields $\hat{C}_0$, $\hat{C}_2$, $\hat{C}_4$.
 
Our starting point for this section is the ten dimensional type IIB action in Einstein frame \cite{Polchinski:1998rr}
\begin{gather}
\mathcal{S}_{IIB}=-\int\left({1\over2}\hat{R} \star \mathbf{1} + {1\over 4}d\hat{\phi}\wedge\star d\hat{\phi}+{1\over4}e^{-\hat{\phi}} \hat{H}_3\wedge\star\hat{H}_3\right)\\
-\int\frac{1}{4}\left(e^{-2\hat{\phi}}\hat{F}_1\wedge\star\hat{F}_1+e^{\hat{\phi}}\hat{F}_3\wedge\star\hat{F}_3+{1\over 2}\hat{F}_5\wedge\star\hat{F}_5+\hat{C}_4\wedge\hat{H}_3\wedge \hat{F}_3\right)\ ,\nonumber
\label{eq:iib_n2}
\end{gather}
where the field strength are defined as follows
\begin{gather}
\hat{H}_3=d\hat{B}_2\ , \quad \hat{F}_1=d\hat{C}_0\ , \quad \hat{F}_3=d\hat{C}_2-\hat{C}_0 \hat{H}_3\ , \nonumber \\
\hat{F}_5=d\hat{C}_4 - {1\over2}\hat{H}_3\wedge\hat{C}_2 + {1\over2}\hat{B}_2\wedge d\hat{C}_2\ , \quad  
\label{eq:f_strength}
\end{gather}
and the self-duality of $\hat{F}_5$ must be imposed by hand at the level of the equations of motion.

By making use of the harmonic bases, we can expand the field content of the theory as
\begin{gather}
\hat{B}_2=B_2(x)+b^A(x)\omega_A\ , \quad \hat{C}_2 = C_2(x) + c_2^A\omega_A \ , \nonumber \\
\hat{C}_4= c_4^A(x)\omega_A + \xi^{\hat{K}}(x)\wedge\alpha_{\hat{K}}-\tilde{\xi}_{\hat{K}}(x)\wedge\beta^{\hat{K}}+\rho_A(x)\tilde{\omega}^A\ . \label{eq:field_content_iib}
\end{gather}
We can further make use of the self-duality of $\hat{F}_5$ to get rid of half of the degrees of freedom of $\hat{C}_4$ --- we choose to keep the $\xi^{\hat{K}}$ and $\rho_A$ fields. On top of these, we will also obtain a number of moduli fields from the deformations of the metric as discussed in the previous section.

Taking the metric ansatz to be
\be
ds^2 = g_{\mu\nu}(x)dx^\mu dx^\nu + g_{i\barj}(x,z)dz^id\bar{z}^{\barj}\ , \label{eq:metric_ansatz}
\ee
where $g_{\mu\nu}$ (with $\mu,\nu=0,...,3$) is the external metric and $g_{i\barj}$ (with $i,\barj=1,...,3$) is the CY metric of \x and plugging the definitions of \eqref{eq:f_strength} in \eqref{eq:iib_n2} one finds the tree-level four-dimensional action\footnote{Where a further Weyl rescaling has been made to reabsorb the internal volume $\vol$, so that the action is in Einstein frame.} \cite{hep-th/9908007, hep-th/0507153}  (see \appref{app:dim_redux_iib} for further details)
\begin{gather}
\mathcal{S}^{(4)}_{IIB}=\int -{1\over2} R\star{\mathbf 1}+{1\over 4} \text{Re}\ \mathcal{M}_{\hat{K}\hat{L}} F^{\hat{K}}\wedge F^{\hat{L}}+{1\over 4} \text{Im}\ \mathcal{M}_{\hat{K}\hat{L}} F^{\hat{K}}\wedge\star F^{\hat{L}}\nonumber \\
- G_{K\bar{L}} dz^K\wedge\star d\bar{z}^{\bar{L}} - h_{pq} dq^p\wedge\star dq^q\ ,
\label{eq:iib_dual}
\end{gather}
where the real and imaginary parts of $\mathcal{M}_{\hat{K}\hat{L}}$ are found from \eqref{eq:period_matrix}, $F^{\hat{K}}:=d\xi^{\hat{K}}$, the complex structure metric $G_{K\bar{L}}$ is given in \eqref{eq:kcs} and we define the four-dimensional dilaton to be
\be
e^\phi := {1\over \mathcal{V}^{1/2}}e^{\hat{\phi}} \ , \label{eq:4dim_dilaton}
\ee
such that the quaternionic metric $h_{pq}$ becomes\footnote{Where we have dualised the $\left(C_2,\,B_2\right)$ gauge fields into the scalars $\left(h,\,\tilde{h}\right)$, as demonstrated in \appref{app:dual_axions}.} \cite{Ferrara:1988ff, Ferrara:1989ik}
\begin{gather}
h_{pq} dq^p\wedge\star dq^q := (d\phi)^2 + G_{AB} dT^A \wedge\star dT^B + {1\over 4} e^{2\phi} \mathcal{V} (dC_0)^2 \nonumber\\ 
+e^{2\phi} \mathcal{V} G_{AB}\left(dc_2^A-C_0db^A\right)\wedge\star\left(dc_2^B-C_0db^B\right) \nonumber\\
+{1\over 16\mathcal{V}} e^{2\phi} G^{AD}\left(d\rho_A-\mathcal{K}_{ABC}c_2^Bdb^C\right)\wedge\star\left(d\rho_D-\mathcal{K}_{DEF}c_2^Edb^F\right)\nonumber\\
+{1\over 4\mathcal{V}} e^{2\phi} \left[dh-{1\over2}\left(\rho_A db^A - b^A d\rho_A\right)\right]^2\nonumber\\
+{1\over 2} e^{4\phi}\left[d\tilde{h}+C_0dh+c_2^Ad\rho_A+{1\over2}C_0\left(\rho_A db^A - b^A d\rho_A\right)-{1\over4}\mathcal{K}_{ABC}c_2^Ac_2^Bdb^C\right]^2\ . 
\end{gather}
In \cite{Ferrara:1988ff, Ferrara:1989ik}, it was shown that the dimensionally reduced type IIB theory in \eqref{eq:iib_dual} admits a description in terms of a local direct product $\mathcal{M}^Q\times\mathcal{M}^{\text{cs}}$, where $\mathcal{M}^\text{Q}$ is the quaternionic manifold described by the metric $h_{pq}$ and spanned by the coordinates $q^p$, with $\mathcal{M}^{\text{ks}}$ a submanifold of $\mathcal{M}^\text{Q}$, and $\mathcal{M}^\text{cs}$ the complex structure manifold described in \ssecref{ssec:cs_kahler}. The ordered multiplets of \eqref{eq:iib_dual} are given in \tabref{tab:field_content}. 
\begin{table}[h!]
\centering
\begin{tabular}{|c|c|c|}
\hline
Multiplet & Dimension    & Field content                                           \\ \hline
Gravity multiplet       & 1 & $\left(g_{\mu\nu},\ \xi^0\right)$                                      \\ \hline
Vector multiplet        & $h^{2,1}$    & $\left(\xi^K,\ z^K\right)$                              \\ \hline
Hypermultiplet        & $h^{1,1}+1$ & $\left(v^A,b^A,c_2^A,\rho_A\right)+\left(h,\tilde{h},\phi,C_0\right)$ \\ \hline
\end{tabular}
\caption{Multiplets for the four-dimensional type IIB supergravity spectrum.}
\label{tab:field_content}
\end{table}
The four-dimensional effective theory described by \eqref{eq:iib_dual} preserves $\mathcal{N}=2$ supersymmetry. However, phenomenologically we are interested in $\mathcal{N}=1$ theories. Half of the supersymmetry can be broken by introducing an orientifold projection in an appropriate manner, as we will discuss in the next section. Furthermore, the scalar potential for the four dimensional scalars is flat. To be able to stabilise the moduli fields we will require the use of background fluxes, which we describe in \secref{sec:fluxes}.

\section{Orientifolding the supergravity action\label{sec:orientifold}}
In this section we will briefly discuss the orientifolded truncation of type IIB supergravity. For a more complete description, we point the interested reader to \cite{hep-th/0403067, hep-th/0412277, hep-th/0507153} and references therein.

We would like to consider the most generic field content invariant under the action of the $\mathcal{N}=1$ orientifold operator\footnote{A second choice of orientifold projection is possible in $\tilde{\mathcal{O}}=\Omega_p\sigma$; this would lead to the introduction of $O5$ and $O9$-planes in the theory.}
\be
\mathcal{O}=\Omega_p(-1)^{\text{F}_\text{L}}\sigma\ ,
\ee
which is the combined action of the world-sheet parity transformation $\Omega_p$, {\it i.e.} the operator that transforms the world-sheet coordinates $\left(\tau_0,\tau_1\right)\rightarrow\left(\tau_0,-\tau_1\right)$, the left-moving fermion number operator $(-1)^{\text{F}_\text{L}}$ and an isometric and holomorphic involution $\sigma$ acting solely on \x $\left(\sigma^2=\mathbf{1}\right)$. Furthermore, $\sigma$ leaves the metric and complex structure of \x intact \cite{hep-th/0507153}. This implies that the \kah form will also be invariant under the involution $\sigma^*J=J$, where $\sigma^*$ is the pullback of $\sigma$. Holomorphicity of $\sigma$ implies that the Dolbeault decomposition for the cohomologies is respected, and in particular $\sigma^*H^{3,\,0}=H^{3,\,0}$; together with its idempotency we choose for the holomorphic (3,0)-form $\Omega$
\be
\sigma^*\Omega=-\Omega \ . \label{eq:omega_choice}
\ee
This choice for the orientifold projection is consistent with the introduction of $O3$ and $O7$-planes into the effective action \cite{hep-th/0507153}.

Another effect of the involution is that the cohomology of \x will split in even and odd cohomologies under its action, by virtue of $\sigma$ being an involutive symmetry of \x, and thus we write
\be
H^{p,\,q}=H^{p,\,q}_+ \oplus H^{p,\,q}_-\ ,
\ee
such that an element $\omega_{\pm}\in H^{p,\,q}_\pm$ transforms like $\sigma^*\omega_\pm=\pm\omega_\pm$. Furthermore, we can see the following \cite{hep-th/0507153}
\begin{itemize}
\item The involution being isometric means that the metric and orientation of \x are preserved. The Hodge-star operator $\star$ then commutes with the involution and $h^{1,\,1}_\pm=h^{2,\,2}_\pm$.
\item The holomorphicity of $\sigma$ implies that there will not be any mixing between holomorphic and antiholomorphic groups $h^{1,\,2}_\pm=h^{2,\,1}_\pm$.
\item The choice \eqref{eq:omega_choice} leads to $h^{3,\,0}_-=h^{0,\,3}_-=1$ and $h^{3,\,0}_+=h^{0,\,3}_+=0$.
\item The previous two points telll us that $\Omega\wedge\bar{\Omega}$ is invariant under $\sigma^*$ such that $h^{3,\,3}_-=h^{0,\,0}_-=0$ and $h^{3,\,3}_+=h^{0,\,0}_+=1$.
\end{itemize}
In \tabref{tab:bases_orient} we summarise the notation for the truncated harmonic bases after orientifolding.
\begin{table}[h!]
\centering
\begin{tabular}{|c|c|c|}
\hline
Cohomology group & Dimension & Basis  \\ \hline
$H^{1,\,1}_-$     & $h^{1,\,1}_- $ & $\omega_a$                                   \\ \hline
$H^{1,\,1}_+   $   & $h^{1,\,1}_+ $     & $\omega_\alpha$                             \\ \hline
$H^{2,\,2}_-     $ & $h^{1,\,1}_- $  & $\tilde{\omega}^a$   \\ \hline
$H^{2,\,2}_+    $  & $h^{1,\,1}_+ $  &$\tilde{\omega}^\alpha$    \\ \hline
$H^{2,\,1}_-   $  & $h^{2,\,1}_- $  & $\chi_k$    \\ \hline
$H^{2,\,1}_+   $    & $h^{2,\,1}_+ $  & $\chi_\kappa$    \\ \hline
$H^{3}_-     $  & $2h^{2,\,1}_- +2$  & $\left(\alpha_{\hat{k}},\beta^{\hat{l}}\right)$    \\ \hline
$H^{3}_+     $  & $2h^{2,\,1}_+ $  & $\left(\alpha_{\kappa},\beta^{\lambda}\right)$    \\ \hline
\end{tabular}
\caption{Harmonic bases for the orientifolded \x.}
\label{tab:bases_orient}
\end{table}

Finally, the non-trivial intersection numbers are given by truncation of \eqref{eq:int_num_2} as
\be
\int_\x\omega_a\wedge\tilde{\omega}^b=\delta_a^b\ , \qquad \int_\x\omega_\alpha\wedge\tilde{\omega}^\beta=\delta_\alpha^\beta\ , \qquad \int_\x\alpha_{\hat{k}}\wedge\beta^{\hat{l}}=\delta_{\hat{k}}^{\hat{l}}\ , \qquad \int_\x\alpha_{\kappa}\wedge\beta^{\lambda}=\delta_{\kappa}^{\lambda}\ .
\ee
The action of the parity operator and the fermion number operator on the ten-dimensional fields is summarised in \tabref{tab:orient_1} \cite{hep-th/9804208}. It is clear from it that the invariant orientifolded states must obey
\begin{gather}
\sigma^*\hat{\phi}=+\hat{\phi}\ ,\qquad \sigma^*\hat{g}=+\hat{g}\ ,\qquad \sigma^*\hat{B_2}=-\hat{B_2}\ , \nonumber \\
\sigma^*\hat{C_0}=+\hat{C_0}\ , \qquad\sigma^*\hat{C_2}=-\hat{C_2}\ , \qquad\sigma^*\hat{C_4}=+\hat{C_4}\ .
\end{gather}
\begin{table}[h!]
\centering
\begin{tabular}{|c|c|c|c|}
\hline
Field & Under $\Omega_p$    & Under $(-1)^{\text{F}_\text{L}}$       & Under  $\Omega_p\,(-1)^{\text{F}_\text{L}}$  \\ \hline
$\hat{\phi}$      & $+$ & $+$    & $+$                                      \\ \hline
$\hat{g}$        & $+$    & $+$      & $+$                            \\ \hline
$\hat{B_2}$       & $-$ & $+$ & $-$    \\ \hline
$\hat{C_0}$       & $-$ & $-$  & $+$   \\ \hline
$\hat{C_2}$       & $+$ & $-$  & $-$   \\ \hline
$\hat{C_4}$       & $-$ & $-$ & $+$    \\ \hline
\end{tabular}
\caption{Actions of the parity and fermion number operators on the ten-dimensional spectrum.}
\label{tab:orient_1}
\end{table}

We are now able to expand the ten-dimensional spectrum in the new truncated harmonic bases, much like we did in the $\mathcal{N}=2$ case. In the following subsections we briefly study the truncation of the $\mathcal{N}=2$ \kah manifolds into their $\mathcal{N}=1$ analogues.
\subsection{Truncating the \kah geometry}
After orientifolding, we can expand the \kah form and the Kalb-Ramond 2-form as\footnote{Note that the orientifold invariant states of $\hat{B}_2$ have to be odd under $\sigma^*$. However, the external piece $B_2(x)$ is even since $\sigma$ leaves the external space intact. Therefore, $B_2(x)$ gets projected out by the orientifold projection.} 
\be
\sigma^*J=+J\ \longrightarrow \ J=v^\alpha\omega_\alpha \qquad \text{and} \qquad  \sigma^*\hat{B}_2=-\hat{B}_2\ \longrightarrow \hat{B}_2=b^a \omega_a\ ,
\ee
where $\alpha=1,...,h^{1,\,1}_+$ and $a=1,...,h^{1,\,1}_-$. The invariance of the \kah form under orientifolding, together with the condition that the metric must remain semi-positive definite after orientifolding, implies that some intersection numbers have to vanish. In particular, any intersection number (or its contraction) with an odd number of odd 2-cycles will have to vanish since orientifolding will flip the sign of the intersection number, breaking the \kah cone conditions. This means that
\be
\mathcal{K}_{abc}=\mathcal{K}_{a\alpha\beta}=\mathcal{K}_{a\alpha}=\mathcal{K}_{a}=0\ .
\label{eq:vanish_triple}
\ee
Plugging these conditions in the definition of the metric in \eqref{eq:ks_metric}, we find the $\mathcal{N}=1$ truncated metric for the \kah sector
\be
G_{\alpha\beta}=-\frac{1}{4\mathcal{V}}\left(\mathcal{K}_{\alpha\beta}-\frac{\mathcal{K}_\alpha\mathcal{K}_\beta}{4\mathcal{V}}\right)\ , \quad G_{ab}=-{1\over 4\mathcal{V}} \mathcal{K}_{ab}\ , \quad G_{\alpha b}=G_{a\beta}=0 \ , 
\label{eq:ks_metric_1}
\ee
where $\mathcal{V}:={1\over 6}\mathcal{K}_{\alpha\beta\gamma}v^\alpha v^\beta v^\gamma$ is the volume of the truncated \x and there is no ambiguity in the contractions of the triple intersection numbers since $J\in H^{1,\,1}_+$. It will be convenient for later calculations to also introduce the inverse metrics
\be
G^{\alpha\beta}=-4\mathcal{V}\mathcal{K}^{\alpha\beta}+2v^\alpha v^\beta\ , \quad G^{ab}=-4\mathcal{V}\mathcal{K}^{ab}\ .
\label{eq:ks_inverse_1}
\ee
\begin{comment}
Similarly, the intersection number constraints can be applied to the complex matrix \eqref{eq:gc} to find\footnote{We remind the reader that we are using adapted coordinates $T^A=\left(1,\left\{T^\alpha,T^a\right\}\right)$.}
\begin{gather}
\text{Re} \mathcal{N}_{0\alpha} = \frac{1}{2}\mathcal{K}_{ab\alpha}b^a b^b\ ,\label{eq:gk_coupling1_1}\\
\text{Im} \mathcal{N}_{00}=-\mathcal{V}\left(1+4G_{ab}b^a b^b\right)\ , \quad \text{Im} \mathcal{N}_{ab}=-4\mathcal{V}G_{ab}\ , \quad  \text{Im} \mathcal{N}_{\alpha\beta}=-4\mathcal{V}G_{\alpha\beta}\ ,\label{eq:gk_coupling2_1}
\end{gather}
with all other components vanishing.
\end{comment}
We now concentrate on the complex structure sector. After orientifolding, the unique holomorphic form will be expanded as
\be
\sigma^*\Omega=-\Omega\ \longrightarrow \ \Omega=\mathcal{Z}^{\hat{k}}\alpha_{\hat{k}}-\mathcal{F}_{\hat{k}}\beta^{\hat{k}}\ ,
\ee
where $\hat{k}=0,...,h^{1,2}_-$. By proxy of \eqref{eq:kodaira_form}, the even deformations of the complex structure sector will be projected out, $z_k$ being the only ones left. The \kah potential for the truncated complex structure sector will become
\be
G_{k\bar{l}}=\partial_{z^k}\partial_{\bar{z}^l}K^{\text{cs}}\ , \qquad K^{\text{cs}}=-\ln\left(i\int \Omega\wedge\bar{\Omega}\right)=-\ln \ i\left(\mathcal{Z}^{\hat{k}}\bar{\mathcal{F}}_{\hat{k}}-\bar{\mathcal{Z}}^{\hat{k}}\mathcal{F}_{\hat{k}}\right)\ . \label{eq:kcs_1}
\ee
\subsection{The effective $\mathcal{N}=1$ action \label{ssec:n1}}
Let us begin by writing down the orientifold invariant spectrum
\begin{gather}
\hat{B}_2=b^a(x)\omega_a\ , \quad \hat{C}_2 =c_2^a\omega_a \ , \nonumber \\
\hat{C}_4= \xi^{k}(x)\wedge\alpha_{k}+\rho_\alpha(x)\tilde{\omega}^\alpha\ .
\end{gather}
The truncated effective action is directly inherited from the parent $\mathcal{N}=2$ action, based on the considerations of the previous subsection,
\begin{gather}
\mathcal{S}^{(4)}_{IIB,\,\mathcal{N}=1}=\int -{1\over2} R\star{\mathbf 1}+{1\over 4} \text{Re}\ \mathcal{M}_{\kappa\lambda} F^{\kappa}\wedge F^{\lambda}+{1\over 4} \text{Im}\ \mathcal{M}_{\kappa\lambda} F^{\kappa}\wedge\star F^{\lambda}\nonumber \\
- G_{k\bar{l}} dz^k\wedge\star d\bar{z}^{\bar{l}} - h_{\tilde{p}\tilde{q}} dq^{\tilde{p}}\wedge\star dq^{\tilde{q}}\ ,
\label{eq:iib_dual_1}
\end{gather}
where the quaternionic sector becomes
\begin{gather}
h_{\tilde{p}\tilde{q}} dq^{\tilde{p}}\wedge\star dq^{\tilde{q}} := (d\phi)^2 + G_{\alpha\beta} dv^\alpha \wedge\star dv^\beta + G_{ab} db^a \wedge\star db^b + {1\over 4} e^{2\phi} \mathcal{V} (dC_0)^2 \nonumber\\ 
+e^{2\phi} \mathcal{V} G_{ab}\left(dc_2^a-C_0db^a\right)\wedge\star\left(dc_2^b-C_0db^b\right) \nonumber\\
+{1\over 16\mathcal{V}} e^{2\phi} G^{\alpha\beta}\left(d\rho_\alpha-\mathcal{K}_{\alpha bc}c_2^bdb^c\right)\wedge\star\left(d\rho_\beta-\mathcal{K}_{\beta de}c_2^ddb^e\right)\ .
\end{gather}
We can see that due to the effect of the orientifold projection, while the $z^k$ coordinates are still the natural candidate that spans $\mathcal{M}^{\text{cs}}$, the coordinates $t^A$ are no longer adequate to describe $\mathcal{M}^{\text{ks}}$, and $\mathcal{M}^{\text{Q}}$ by proxy. Finding the correct choice of coordinates is quite involved and we will only present the result in the following. For $O3/O7$-planes we have \cite{hep-th/0507153, hep-th/0103068, hep-th/0107044, 0907.0665}
\begin{itemize}
\item Axio-dilaton: $S=e^{-\phi}-iC_0$.
\item Complex structure: $U^k:=z^k$, with $k=1,...,h^{1,\,2}_-(\x)$.
\item Odd \kah moduli: $G^a=c^a-iSb^a$, with $a=1,...,h^{1,\,1}_-(\x)$
\item Even \kah moduli: 
\be
T_\alpha=\tau_\alpha-{1\over 2\left(S+\bar{S}\right)}\mathcal{K}_{\alpha ab}G^a\left(G-\bar{G}\right)^b+i\rho_\alpha\ ,
\label{eq:T_coords}
\ee
with $\alpha=1,...,h^{1,\,1}_+$ and where $\tau_\alpha$ are given as an implicit function of $v^\alpha$ by
\be
\tau_\alpha = {1\over2}\int_\x \omega_\alpha\wedge J\wedge J={1\over 2} \mathcal{K}_{\alpha \beta\gamma}v^\beta v^\gamma\ ,
\label{eq:tau_def}
\ee
in this way, $\tau_\alpha$ is the volume of a four-cycle $\tilde{\omega}^\alpha\in H^{2,\,2}_+(\x)$, the Poincar\'e dual to the two-cycle ${\omega}_\alpha\in H^{1,\,1}_+(\x)$ with volume $v^\alpha$. We note that in terms of the four-cycles, the volume has become an implicit function of $\tau_\alpha$. In general, we would require to invert the relation $\tau_\alpha={1\over 2} \mathcal{K}_{\alpha \beta\gamma}v^\beta v^\gamma$ but this cannot be done in full generality for a Calabi-Yau with a large number $h^{1,\,1}$ of \kah moduli. In \cite{2005.11329}, some advances have been made in terms of studying \kah moduli stabilisation in the two-cycle frame allowing for a more general approach to model building with, a priori, any number of \kah moduli.
\end{itemize}
We also note that based on the definition of $\tau_\alpha$ in \eqref{eq:tau_def}, we have that the internal volume is a homogeneous function of degree $3/2$ on the $\tau_\alpha$, or equivalently on $(T+\bar T)_\alpha$. This is a powerful result that holds for any CY based on the definition of the internal volume derived from the \kah geometry. This also implies, due to a nice theorem by Euler, the useful relations
\be
{1\over \vol}{\partial \vol\over \partial T_\alpha} = {3\over 2}\left[\left(T+\bar T\right)^{-1}\right]^\alpha \ , \qquad {1\over \vol}{\partial^2 \vol\over \partial T_\alpha \partial T_\beta } = {3\over 4} \left[\left(T+\bar T\right)^{-1}\right]^\alpha\left[\left(T+\bar T\right)^{-1}\right]^\beta \ , \label{eq:useful_rels}
\ee
where $\left[\left(T+\bar T\right)^{-1}\right]^\alpha$ is taken to be the inverse of $\left(T+\bar T\right)_\alpha$. The \kah potential of the truncated theory is then given by
\be
K=-\ln\left(S+\bar{S}\right) - \ln\left(-i\int_\x \Omega\wedge\bar{\Omega}\right) - 2\ln\left(\mathcal{V}\right)\ . \label{eq:kah_1}
\ee 
Here, and in the following, the volume is taken to be in the natural coordinate frame in terms of four-cycles, {\it i.e.} $\vol=\vol(T+\bar{T})$, and the $\mathcal{N}=1$ moduli space is given locally by the direct product
\be
\mathcal{M}^{\text{cs}}_{h^{1,\,2}_-}\times\mathcal{M}^{\text{Q}}_{h^{1,\,1}+1}\ .
\ee
In these coordinates, the $\mathcal{N}=1$ effective action, in terms of the \kah form $K$ and superpotential $W$, takes the well-known form \cite{Wess:1992cp, Gates:1983nr}
\be
\mathcal{S}^{(4)}_{IIB,\,\mathcal{N}=1}=-\int {1\over2} R\star{\mathbf 1} + K_{I\bar{J}} dM^I\wedge\star d\bar{M}^{\bar{J}} +{1\over 4} \text{Re}\ \mathcal{M}_{\kappa\lambda} F^{\kappa}\wedge F^{\lambda}+{1\over 4} \text{Im}\ \mathcal{M}_{\kappa\lambda} F^{\kappa}\wedge\star F^{\lambda} + V\star\mathbf{1}\ , \label{eq:iib_1_W}
\ee
where the scalar potential is given by
\be
V=e^K\left(K^{I\bar{J}}D_IWD_{\bar{J}}\bar{W}-3\left|W\right|^2\right)+{1\over 4} \text{Re}\ \mathcal{M}^{-1 \ \kappa\lambda} D_\kappa D_\lambda\ , \label{eq:scalar_pot_1}
\ee
where $D_\kappa$ are the would-be D-term contributions to the scalar potential. We have used $M^I$ to denote all complex scalars in chiral multiplets in the theory, $K_{I\bar{J}}$ is the \kah metric obtained from the \kah potential of \eqref{eq:kah_1}, {\it i.e.} $\partial_I\partial_{\bar{J}}K$; and we have introduced the \kah derivative $D_I W = \partial_I W+W\partial_I K$.

Comparison of the actions in \eqref{eq:iib_dual_1} and \eqref{eq:iib_1_W} immediately implies that the scalar potential for the moduli fields is flat, $V=0$, or in other words that the superpotential and D-terms vanish $W=D_\kappa=0$. We will have to generate a potential to be able to stabilise the moduli fields. This is no trivial task and, among other things, it will require the introduction of background fluxes in our theory. We will dedicate the next section to the study of flux compactifications. Before doing so, however, it is important to remark a couple of things.

Orientifolding requires the introduction of $O$-planes. These objects, as well as the presence of $D$-branes and background fluxes, will backreact onto the Calabi-Yau geometry and the sourced solutions will, in general, not admit a metric ansatz of the form \eqref{eq:metric_ansatz}. A warp factor must be included to account for the backreaction and the new metric ansatz is \cite{hep-th/0009211, hep-th/0105097, hep-th/0106014}
\be
ds^2=e^{2A(z)}g_{\mu\nu}(x) dx^\mu dx^\nu + e^{-2A(z)} g_{i\barj}(x,z)dz^id\bar{z}^{\barj}\ ,
\ee
In general, the effects of the warping can be dramatic, to the point where the resultant geometry is not necessarily conformally Calabi-Yau anymore. In the remainder of this thesis, however, we will concentrate on compactifications at large volume, relative to the string length, where the warp factor tends to unity and the warped and unwarped metric coincide \cite{hep-th/0105097}. In this limit, the formulae derived so far hold. This is the limit of interest, in any case, since we understand supergravity as a low-energy effective theory that is well-defined only in the corner of parameter space with small string coupling and large volume. Furthermore, the perturbative and non-perturbative corrections that will become important in the second half of the thesis will be only under control at large volume. It is important to note that although we use the words ``small'' and ``large'' we really mean {\it finitely} small and {\it finitely} large, not necessarily the strict limits $g_s\rightarrow0$ and $\mathcal{V}\rightarrow\infty$ that take us to the boundary of moduli space, more on this later.

Finally, the addition of $Op$-planes and $Dp$-branes introduces a term into the ten-dimensional action that is localized in $(p+1)$-dimensions, which we have neglected in \eqref{eq:iib_1_W}. This is consistent with considering a small number of $O3/O7$-planes to be coincident with an equal number of $D3/D7$-branes, such that their contributions to \eqref{eq:iib_1_W} cancel locally. At the level of the perturbative action in \eqref{eq:iib_1_W}, the main use of localised objects is to satisfy tadpole cancellation conditions, allowing non-trivial background fluxes, and to avoid a number of no-go theorems \cite{hep-th/9910053, hep-th/0007018}. We will briefly mention these in some more detail in the next section. 

Two other applications that will become relevant in the remainder of the thesis are the introduction of non-perturbative corrections through a stack of $N$ $D$-branes \cite{hep-th/9910053, Hamada:2021ryq} and the addition of uplifting contributions to the potential through anti $D3$-branes \cite{Kachru:2003aw}. In the former, a stack of $N$ $D$-branes will give rise to non-Abelian $U(N)$ gauge theories. The \kah moduli can then be charged under the action of such non-Abelian groups obtaining a non-flat potential, this will be discussed in more detail in \chpref{chap:non_pert}. In the latter, anti $D$-branes at the tip of a Klebanov-Strassler throat are used to uplift four-dimensional anti-de Sitter (AdS$_4$) vacua to de Sitter (dS$_4$) (for example, see \cite{Kachru:2019dvo} for an overview of the uplift mechanism). 

Although the ten-dimensional description of such objects is very interesting and of paramount importance, within the scope of this thesis we will only treat these at an effective four-dimensional level. We will therefore not dedicate more time to the ten-dimensional description of localised sources and we point the interested reader to \cite{hep-th/9602052, Johnson:2003glb, Becker:2006dvp, Green:1987mn} for fundamentals and to \cite{hep-th/0007191, 0708.2865, 0710.2951, 0808.2811, 0802.2916, 0810.5001} for phenomenological applications.

 \section{Flux compactifications\label{sec:fluxes}} 
 As we have seen in the previous section, the scalar potential for the moduli fields is flat. The introduction of background fluxes allows us to generate a non-trivial superpotential at tree-level. The scalar potential obtained from this non-zero superpotential will be able to stabilise the complex structure sector as well as the axio-dilaton. However, we will also show that due to the no-scale structure, the \kah sector does not receive any contribution from background fluxes and it remains flat. The stabilisation of \kah moduli will be the motivation to introduce corrections to the tree-level action in the second half of the thesis. Finally, let us note that the introduction of fluxes provides a way to spontaneously break the leftover $\mathcal{N}=1$ supersymmetry.
 
Take a p-form gauge field $A_p$ with field strength $F_{p+1}=dA_p$. We will say, by analogy with the case of electromagnetism in 4-dimensions, that the gauge field $A_p$ gives rise to a {\it magnetic flux} of the form
\be
\int_{\Sigma_{p+1}} F_{p+1} = m\in\mathbb{Z}\ ,
\ee
where $\Sigma_{p+1}\subset\x$ is the (p+1)-cycle wrapping the (p+1)-form field strength $F$. The same gauge field will also give rise to an {\it electric flux} of the form
\be
\int_{\Sigma_{D-p-1}} \star F_{p+1} = n\in\mathbb{Z}\ ,
\ee
where $D=\text{dim}(\x)$ and  $\Sigma_{D-p-1}\in\x$ is the (D-p-1)-cycle wrapping the electric dual of $F$. We note that these fluxes are quantised in units of ${1\over 2\pi\alpha'}$ as a consequence of a generalised Dirac quantisation condition \cite{hep-th/0105097}, much like the electric and magnetic fluxes are quantised in the case of electromagnetism. However, within the context of the low energy supergravity description, these fluxes will be treated as continuous parameters in the effective theory.

In type IIB supergravity, one can turn on background fluxes for the 3-form field strengths in the NSNS and RR sector. For the background fluxes to be consistent with our orientifold projection we see that the flux pieces must be odd under the action of the involution $\sigma$. Thus, we expand the flux pieces as
\be
\tilde{H}_3 = m_{(H)}^{\hat{k}} \alpha_{\hat{k}} + n^{(H)}_{\hat{k}} \beta^{\hat{k}}\ , \quad \tilde{F}_3 = m_{(F)}^{\hat{k}} \alpha_{\hat{k}} + n^{(F)}_{\hat{k}} \beta^{\hat{k}}\ , \qquad \hat{k}=0,...,h^{1,\,2}_-\ , 
\ee
and the fluxed effective action can be obtained by substituting \cite{hep-th/0507153}
\be
\hat{H}_3 \rightarrow \hat{H}_3^{(F)}=\hat{H}_3 +\tilde{H}_3\ , \qquad \hat{F}_3 \rightarrow  \hat{F}_3^{(F)}=\hat{F}_3 +\tilde{F}_3\ , 
\ee
in the ten-dimensional action since the $\hat{F}_5$ terms that would involve the 3-form fluxes are projected out due to the orientifold constraints. It is convenient to repackage the two 3-form field strength in the imaginary self-dual three-form field strength \cite{hep-th/0105097}
\be
\hat{G}_3= \hat{F}_3^{(F)} - i S \hat{H}_3^{(F)} \ , \qquad \star\hat{G}_3=i \hat{G}_3 \ ,
\ee
and we define the shorthand notation
\be
\tilde{G}_3 = (m_{(F)}-i S m_{(H)})^{\hat{k}}\alpha_{\hat{k}} + (n_{(F)}-i Sn^{(H)})_{\hat{k}} \beta^{\hat{k}} :=  m^{\hat{k}}\alpha_{\hat{k}} + n_{\hat{k}} \beta^{\hat{k}}\ .
\ee
The fluxed scalar potential is given by \cite{hep-th/9912152, hep-th/0105097, hep-th/0204254, hep-th/0208123, hep-th/0403067, hep-th/0507153}
\be
V=e^K\left(\int\Omega\wedge\bar{\tilde{G}}_3\int\bar{\Omega}\wedge\tilde{G}_3+G^{kl}\int\chi_k\wedge\tilde{G}_3\int\bar{\chi}_l\wedge\bar{\tilde{G}}_3\right)\ .
\ee
From this expression, it is clear to see that setting the background fluxes to zero will make the scalar potential vanish. To make a connection with \eqref{eq:iib_1_W}, it is also shown, for orientifolds without an odd-\kah sector, the GVW superpotential 
\be
W_{GVW}=\int \Omega\wedge\tilde{G}_3\ ,
\ee
generates the same scalar potential when making use of \eqref{eq:scalar_pot_1}. For the case of orientifold with an odd \kah sector this same results holds as long as the no-scale structure is unbroken. Generating a potential for the \kah sector directly relates to breaking this no-scale structure, since (at least) the volume mode will couple to the other sectors as well as the \kah moduli. Given that our objective is to discuss four-dimensional phenomenology, we will have to eventually break no-scale. In the remainder of this thesis, we will concentrate in orientifolds that project out the odd \kah sector, {\it i.e.} $h^{1,1}=h^{1,1}_+$, and the results discussed so far will apply straightforwardly. Now that we have generated a scalar potential, we would like to see what sectors can be stabilised from it. As already anticipated, only the axio-dilaton and complex structure moduli will receive a non-zero potential. Let us see this now.

The idea is to show that the scalar potential obtained from GVW superpotential is independent of the \kah coordinates $T_\alpha$ \eqref{eq:T_coords}, up to an overall factor. We begin by noting that
\be
W_{GVW}=W_{GVW}(S,U)\ ,
\ee
since $\Omega=\Omega(U)$ and $\tilde{G}_3=\tilde{G}_3(S)$. The F-terms for $T_\alpha$ are
\be
D_T W_{GVW} = W_{GVW} K_T = -2W_{GVW} {\vol_T\over \vol}=-3 W_{GVW}\left[(T+\bar T)^{-1}\right]^\alpha \ , \label{eq:kah_fterm}
\ee
where $X_T:={\partial X\over \partial T_\alpha}$, we have also made use of the form of the \kah potential given in \eqref{eq:kah_1}. The \kah metric will be block diagonal in the $\left(S,U^k,T_\alpha\right)$ sectors such that the scalar potential takes the form
\begin{gather}
V = e^{K}\left[K^{S\bar{S}}D_SWD_{\bar{S}}\bar{W}+K^{U\bar{U}}D_UWD_{\bar{U}}\bar{W}+K^{T\bar{T}}D_T W D_{\bar{T}}\bar{W}-3\left|W\right|^2\right] \notag \\
= e^{K}\left[K^{S\bar{S}}D_SWD_{\bar{S}}\bar{W}+K^{U\bar{U}}D_UWD_{\bar{U}}\bar{W} + \left(K^{T\bar{T}} \partial_T K \partial_{\bar{T}} K-3\right)\left|W\right|^2\right]\ , \label{eq:scalar_pot_kah}
\end{gather}
where we remark that the first two terms in the brackets are independent of $T_\alpha$. 
\begin{comment}
From the definition of the internal volume and \eqref{eq:tau_def}, we note that \vol is a homogeneous function of degree $3/2$ in the $\tau_\alpha$ coordinates. This also implies that under a rescaling $\tau\rightarrow\lambda\tau$, the \kah potential gets shifted as $K\rightarrow K-3\ln(\lambda)$. Together, this implies the no-scale identity
\end{comment}
The \kah metric for the $T_\alpha$ sector and its inverse are given by
\begin{gather}
K_{T\bar{T}}={2\over\vol}\left({\vol_T\vol_{\bar T}\over \vol}-\vol_{T\bar T}\right)=3\left[(T+\bar T)^{-1}\right]^\alpha\left[(T+\bar T)^{-1}\right]^\beta\ ,
\end{gather}
where we have made use of the relations in \eqref{eq:useful_rels}. Contracting the inverse metric we find
\be
K^{T\bar{T}}K_TK_{\bar{T}}=3\ ,
\ee
so that the parentheses in \eqref{eq:scalar_pot_kah} vanish identically and the scalar potential has the no-scale structure
\be
V=e^K\left(K^{S\bar{S}}D_SWD_{\bar{S}}\bar{W}+K^{U\bar{U}}D_UWD_{\bar{U}}\bar{W}\right)\ ,\label{eq:no_scale_pot}
\ee
with the only $T_\alpha$ dependence in the overall exponential factor. From the previous expression, it is clear that the \kah sector remains flat at tree-level. 

Breaking the no-scale structure can be done in one of two ways, either the superpotential gains a dependence on the \kah moduli $W(S,U) \rightarrow W(S,U,T)$ or by breaking the block diagonal structure of the \kah metric. The first option will be related to the introduction of non-perturbative corrections to the superpotential. The second one can be associated to perturbative corrections in the large volume/small coupling expansion. We will dedicate \chpref{chap:non_pert} to the study of these effects.

By looking at \eqref{eq:no_scale_pot} we see that, since any supersymmetric vacua must have vanishing F-terms $D_UW=D_SW=0$, only supersymmetric Minkowski vacua are allowed. Non-supersymmetric solutions with non vanishing F-terms are also allowed by appropriate choices of fluxes. We can see this by turning once again to the F-terms.

We already found the F-term for the \kah sector in \eqref{eq:kah_fterm}. Preserving supersymmetry will require $W_{\text{GVW}}=0$, in other words the (0,3)-component of $\tilde{G}_3$ must vanish. The remaining F-terms read
\be
D_S W = {1\over S-\bar{S}} \int_\x \Omega\wedge\bar{\tilde{G}}_3 = 0 \ , \qquad D_U W = \int_{\x} \chi_k \wedge \tilde{G}_3 =0\ .
\ee
These F-terms imply that the (3,0)-component and the (1,2)-component need to vanish. Therefore, we have the following: if $\tilde{G}_3\in H^{2,\,1}_-$ all conditions are satisfied and supersymmetry is preserved, if $\tilde{G}_3\in H^{0,\,3}_-$ we find some broken supersymmetry in $D_TW\neq 0$ with $V=0$, and if $\tilde{G}_3\in H^{3,\,0}_-\oplus H^{1,\,2}_-$ none of the above F-terms would vanish and we only find unstable vacuum configurations. 

Since the axio-dilaton and the complex structure sector are fixed at tree-level, one is safe to consider them fixed by their vacuum expectations values. We will take the tree-level superpotential to be
\be
W_{\text{tree}}:=W_0=\left\langle\int_\x \Omega\wedge\tilde{G}_3 \right\rangle\ ,
\ee
for the remainder of the thesis, effectively integrating out the axio-dilaton and the complex structure moduli. Any quantum corrections to their vevs in the large volume limit will be subleading to the tree-level contributions and we will neglect them safely. This also implies that the tree-level \kah potential will take the form
\be
K_{\text{tree}}:=-2 \ln\left(\vol\right) + K_0 = -2 \ln\left(\vol\right) -\ln\left({2\over g_s}\right) - \ln \left\langle-i\int_\x\Omega\wedge\bar{\Omega}\right\rangle\ .
\ee
Finally, before moving on, we would like to discuss the tadpole cancellation conditions in the presence of fluxes and localised sources. The choice of non-trivial fluxes induces a new term in the Bianchi identity for the $\hat{F}_5$ field strength. In particular, 
\be
d\hat{F}_5=\tilde{H}_3\wedge \tilde{F}_3\ , \label{eq:BianchiF5}
\ee 
which leads to the no-go theorem of \cite{deWit:1986mwo}. Since the fluxes are the only contribution to the $\hat{C}_4$ tadpole, one would have to consider compactifications with trivial fluxes only. However, the discovery of D-branes \cite{hep-th/9510017} led to the realisation that charge densities from $D3/D7$ branes, and also from orientifold planes, contribute to the tadpole \cite{hep-th/0105097}. To satisfy the Bianchi identity and guarantee the absence of anomalies in the effective four-dimensional theory, it is then required that the following tadpole cancellation condition holds true
\be
N_{D3}-N_{\overline{D3}}+\int_\x \tilde{H}_3\wedge \tilde{F}_3 = {\chi(\x_4)\over 24} \ ,
\ee
where $\left(N_{D3},\, N_{\overline{D3}}\right)$ are the charge contributions from $D3$-branes and $\overline{D3}$-branes, respectively, and $\chi(\x_4)$ is the contribution from $D7$-branes and orientifold planes, which gains a geometrical interpretation in terms of the Euler characteristic of the corresponding four-fold $\x_4$ \cite{0803.1194}.

\section{The Swampland programme\label{sec:swamp}}
In the last section we achieved the stabilisation of the non-\kah moduli through background fluxes. We highlighted that stabilising the \kah sector will generically require the introduction of perturbative and non-perturbative correction to the tree-level \kah potential and superpotential discussed previously. These corrections are quintessential to supergravity model building, as we will explore in \chpref{chap:non_pert}. It turns out that obtaining a systematic understanding of the stabilisation of the \kah moduli is a highly non-trivial task. Furthermore, since we lack a complete picture of non-perturbative supergravity, it is complicated to define frameworks where computations can be carried out under complete control. Defining {\it complete control} is itself a hotly debated topic in the community. Another reason behind the difficulty of the task at hand is that supergravity inherits two perturbation parameter from string theory: $\alpha'$ and the string coupling $g_s$. In writing down quantum corrections to the tree-level action, we will inevitably encounter terms of $\mathcal{O}(\alpha^p\cdot g_s^q)$, for some $p,q>0$. This can make it so that, at times, it can become unclear which terms enter at a given order in perturbation theory. For example, comparing terms $\mathcal{O}(\alpha'^3 g_s^2)$ and $\mathcal{O}(\alpha'^2 g_s^3)$ can be hard. A very important part of ongoing supergravity research is to systematically analyse these corrections at all orders and ensure that \kah moduli stabilisation is not spoiled in the bulk of moduli space (see for example \cite{Cicoli:2007xp, Cicoli:2008va, Kachru:2019dvo, AbdusSalam:2020ywo, Cicoli:2021rub}).

The lack of a completely satisfactory answer to the challenges presented above led to the creation of the so-called {\it Swampland programme} (see \cite{Palti:2019pca} for an in-depth review). Its core goal is to differentiate between low energy effective field theories that have a consistent completion within a prospective quantum gravity UV theory and those that do not. In doing so, the Swampland programme has formed a {\it web of conjectures} that attempt to delineate the space between the consistent and inconsistent theories. In the next section we will briefly describe two of these conjectures, the particular ones that concern the contents of this thesis. One should remark that, although the ultimate goal is for the programme to be independent of a prospective UV completion, most of the conjectures will draw from string theoretical intuitions.
\begin{figure}[h!]
\centering
\includegraphics[width=0.7\textwidth]{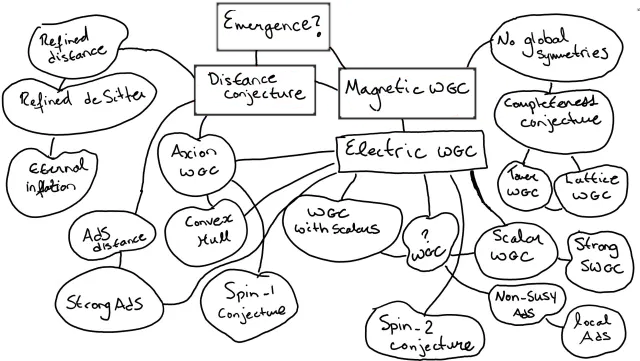}
\caption{A pictorial representation of the web of conjectures in the Swampland programme \cite{robert}}\label{fig:web_conjectures}
\end{figure}

\subsection{The Swampland distance conjecture \label{ssec:sdc}}
In \cite{hep-th/0509212, hep-th/0605264}, the authors proposed that a number of conjectures. Among them was what would eventually become to be known as the Swampland distance conjecture (SDC). A refined version of the SDC (rDC) claims the following \cite{Palti:2019pca, 1610.00010, 1602.06517}, 
\begin{adjustwidth}{2em}{2em}
\begin{conjecture}\label{rDC}
Consider a theory coupled to gravity with some moduli space parametrised by the expectation values of some fields with no potential. Let the geodesic distance between any two points $P\in\mathcal{M}$ and $Q\in\mathcal{M}$ in moduli space be denoted by $d(P,Q)$. Then, there exists a tower of states in theory with mass scale $M$ and scaling
\be
M(Q)<M(P) e^{-\alpha {d(P,Q)\over \mp}}\ ,
\ee
if $d(P<Q)\gtrsim\mp$ and where $\alpha\sim\mathcal{O}(1)$ is a positive model dependent coefficient.
\end{conjecture}
\end{adjustwidth}
A second claim of the rDC is that this statement holds true even for fields with a potential by exchanging the moduli space with the field space of the effective field theory.

The motivation for the existence of a light tower of states can be understood from the mass spectrum in $D$ dimensions of a closed string when compactifying one direction on a circle of radius $R$. Indeed, the $(D-1)$-dimensional mass for the for the string is \cite{1903.06239}
\be
M^2 \sim n^2 \left({1\over R^2}\right)^{{d-1\over d-2}} + w^2 \left({R^2\over \alpha'^2}\right)^{{d-1\over d-2}} \ ,
\ee
where $\left(n,w\right)\in\mathbb{N}$ are the Kaluza-Klein number and winding number, respectively. Indeed, in the infinite distance limit $R\rightarrow \infty$, a infinite tower of Kaluza-Klein modes labelled by $n$ will become massless. The existence of the winding modes ensures that in the T-dual frame, where $R\rightarrow \tilde{R}=1/R$, there also exists another tower becoming massless, in this case related to the winding modes with labels $w$. 

More detailed analysis of the conjecture has been carried out in a number of string scenarios \cite{1610.00010, 1602.06517, 1611.00394, 1611.07084, 1703.05776, 1708.06761, 1801.05434, 1904.05379}. Generically, the analyses focus on finding solutions and then identifying the objects in the theory that introduce an infinite tower of states which becomes light at the boundary of moduli space. For the volume modulus, it is quite straightforward to show that the tower of states can be associated with the KK modes like we did in the simple example above. In more generality, the tower of states can be harder to identify and related to, for example, D-branes becoming tensionless.

\subsection{The refined de Sitter conjecture\label{ssec:dsc}}
A second conjecture is connected to the existence of (meta)stable dS backgrounds in an effective field theory obtained from string theory compactifications. In \cite{1806.09718}, the authors introduced the following claim that a potential of any theory coupled to gravity would have to satisfy
 \be
\left|\nabla V\right|\geq {c\over \mp} V\ .
\ee
It was pointed out in \cite{1807.06581}, that this claim was in tension with the Higgs potential. The previous statement was refined to its present form in \cite{1807.05193, 1810.05506}. This refined de Sitter conjecture (rdSC) claims the following
\begin{adjustwidth}{2em}{2em}
\begin{conjecture}
Given an effective field theory consistent with a theory of quantum gravity, its potential must satisfy either
\be
\left|\nabla V\right|\geq {c\over \mp} V\ , \qquad \text{or} \qquad \text{min}\left(\nabla_i\nabla_j V\right) \geq -{c'\over \mp^2} V\ ,
\ee
where $\left(c,\ c'\right)$ are positive constants of $\mathcal{O}(1)$ and $\text{min}\left(\nabla_i\nabla_j V\right)$ is the minimum of the eigenvalue of the corresponding Hessian.
\end{conjecture}
\end{adjustwidth}
This conjecture forbids the existence of stable de Sitter vacua in string theory. Importantly, metastable vacua are allowed as long as a tachyonic instability exists such that the lifetime of the vacua is bounded by its Hubble time $H^{-1}\sim\mp / \sqrt{V}$.

Whereas the refined distance conjecture has been well studied, the UV origin of the refined de Sitter conjecture is more opaque. In part, this is related to the fact that the de Sitter conjecture is entirely supported by cherry-picked examples and difficulty of obtaining de Sitter vacua in string theory. To some extent, it has been argued that the rdSC is consistent with the rSDC \cite{1810.05506}, and thus the rdSC could be understood from an infinite distance limit perspective in the boundary of moduli space. 

If the statement of the de Sitter conjecture was the inability of obtaining deSitter vacua in the boundary of moduli space, this would not be a surprise. Indeed, it has been known for quite some time \cite{Dine:1985he} that non-supersymmetric degrees of freedom (like those of de Sitter vacua) are notably hard to access from the point of view of weak coupling (perturbative) supergravity. In the infinite distance limit all we have left are perturbative degrees of freedom. Therefore, it would be na\um{i}ve to expect that de Sitter vacua are hiding in that corner of parameter space. However, the dS conjecture forbids dS vacua existing also in the bulk of moduli space, where perturbative and non-perturbative corrections become important.

\begin{figure}[h!]
\centering
\includegraphics[width=0.9\textwidth]{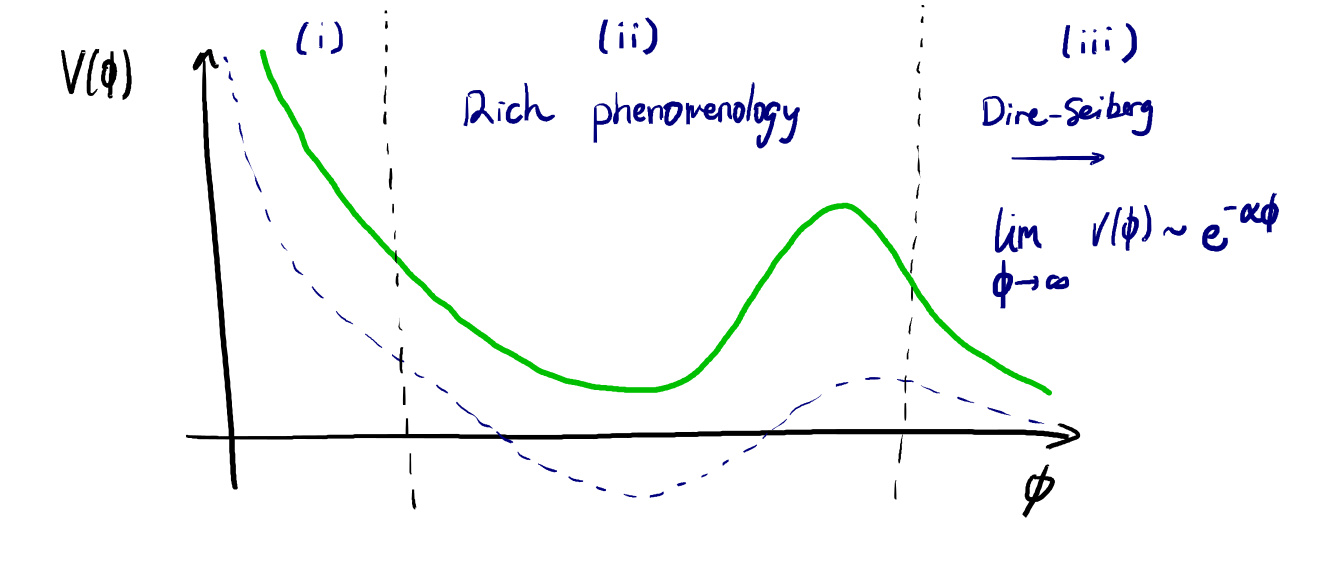}
\caption{A heuristic example of a typical profile for the scalar potential of a canonical saxion modulus $\phi$. The potential consists of three regions: (i) the deep bulk of moduli space where the perturbative expansion breaks down, (ii) the asymptotic bulk where numeric control can be retained and (iii) the boundary of moduli space where the perturbative expansion can be made exact.}\label{fig:pheno_swamp_1}
\end{figure}

Indeed, let us consider a heuristic scenario to exemplify this point ---we will make this point more formal in \chpref{chap:intermezzo} and when discussing corrections to the tree-level action in \chpref{chap:non_pert}---. In \cite{Dine:1985he}, the authors describe the Dine-Seiberg Minkowski vacua, the decompactification limit, for the dilaton modulus. A similar argument for the canonical volume modulus, $\varphi$, can also be made. Indeed, let us imagine that we have some de Sitter vacua generated through corrections to the tree-level action as depicted in \figref{fig:pheno_swamp_1}. 
At large field values, the potential decreases exponentially tending to the Dine-Seiberg type vacuum, in the language of the canonical field $\varphi$. This behaviour at large field values is quite generic since, in this limit, we expect the terms in the action \eqref{eq:iib_dual_1} to dominate over any corrections, and the non-canonical saxions $(s)$ enter the theory as $d\phi \sim s^{-1} ds$, based on the form of the block diagonal \kah metric obtained from \eqref{eq:kah_1}. 

We can observe two things. Moving the minimum in the bulk towards asymptotic larger values of the field would imply tuning the corrections to the point where the leading perturbative expansion is comparable with the tree-level terms, at which point we lose computational control. In \chpref{chap:intermezzo} we revisit the discussion of the Swampland programme from the point of view of phenomenology and point out that there exists a disconnect between the Swampland programme and phenomenology. Secondly, an interesting phenomenological region exists in the asymptotic bulk of moduli space, where the effective theory can remain under numeric control. That region is where all the known de Sitter proposals exists and where the corrections to the tree-level action remain under control. This is where we will concentrate our efforts in the second half of the thesis. Finally, a third region exists, the deep bulk of moduli space, where the effective theory becomes strongly coupled and the perturbative expansion breaks down. This regions maps to small volumes (in units of string length) and/or a string coupling that becomes larger than unity.

\blankpage
%\input{2_Swampland/swamp}
%\blankpage
 \chapter{Some open questions in modern cosmology\label{chap:openqs}}
Throughout this thesis we will concern ourselves with the consequences of a number of important open questions in modern physics and more particularly, its implications for cosmology. In \chpref{chap:clockwork} we will discuss how to generate a naturally light dark energy field. In \chpref{chap:coincidence}, we discuss the implications of the coincidence problem and a possible stringy solution. Naturalness (in the context of cosmology) and the coincidence problem are intimately related to the cosmological constant problem. A problem which is, to a large extent, the underlying motivation for a lot of the research we have undertaken.

In this chapter, we will begin by giving an overview of the naturalness and hierarchy problem in \secref{sec:nat_hie}. We will then shortly review some problems in cosmology, diving into the well-known cosmological constant problem in \secref{sec:ccp}. Finally, we will discuss the coincidence problem in \secref{sec:coincidence_problem}.

\section{The hierarchy problem and naturalness\label{sec:nat_hie}}
In theoretical physics, the original setting for the hierarchy problem is often thought of to be in the context of grand unified theories \cite{Gildener:1976ai, Weinberg:1978ym}. To see the hierarchy problem in action, we can take the much simpler example of a scalar $\phi^4$ effective theory with some UV cut-off $\Lambda$ given by the Lagrangian
\be
\mathcal{L}=-{1\over 2}(\partial_\mu \phi)^2 - {1\over 2} m^2 \phi^2 - {\lambda \over 4!} \phi^4\ . \label{eq:L_quartic}
\ee
We are interested in calculating the effects of one-loop quantum corrections to the above Lagrangian. The tree-level Feynman diagrams for the theory are given by
\be
\begin{tikzpicture}[baseline=-\the\dimexpr\fontdimen22\textfont2\relax]
    \begin{feynman}
    \vertex (a) at (-2,0) ;
     \vertex (b) at ( 2,0)  ;
    \diagram* {
        (a) -- [scalar] (b),
      };
    \end{feynman}
  \end{tikzpicture}=D^{(2)}\qquad \qquad
  \begin{tikzpicture}[baseline=-\the\dimexpr\fontdimen22\textfont2\relax]
    \begin{feynman}
    \vertex[dot]  (m)  at (0, 0);
    \vertex (a) at (-2,-1) ;
     \vertex (b) at ( 2,-1)  ;
      \vertex (c) at (-2, 1) ;
      \vertex (d) at ( 2, 1)  ;
    \diagram* {
        (d) -- [scalar] (m) -- [scalar] (c),
        (b) -- [scalar] (m) -- [scalar] (a),
      }; 
    \end{feynman}
  \end{tikzpicture}  = D^{(4)} \ .
\ee
By standard calculation techniques \cite{Peskin:1995ev}, the two-point and four-point propagators in Fourier space are given by
\be
D^{(2)} = {-i \over p^2 + m^2 + i\epsilon} \ , \qquad D^{(4)} =- i\lambda\ ,
\ee
and the one loop-diagrams are
\be
\begin{tikzpicture}[baseline=-\the\dimexpr\fontdimen22\textfont2\relax]
    \begin{feynman}
    \vertex (m)  at (0, -1);
    \vertex (a) at (-2,-1) ;
     \vertex (b) at ( 2,-1)  ;
      \vertex (ll) at ( -1,0)  ;
       \vertex (lr) at ( 1,0)  ;
     \vertex (lu) at (0,1) ;
    \diagram* {
        (a) -- [scalar] (m) -- [scalar] (b),
        (m) -- [scalar, quarter left] (ll) -- [scalar, quarter left] (lu) -- [scalar,  quarter left] (lr) -- [scalar, quarter left] (m),
      };
    \end{feynman}
  \end{tikzpicture}  \qquad \qquad
  \begin{tikzpicture}[baseline=-\the\dimexpr\fontdimen22\textfont2\relax]
    \begin{feynman}
    \vertex  (ll)  at (-1,0);
     \vertex  (lu)  at (0,1);
    \vertex  (lr)  at (1,0);
     \vertex  (ld)  at (0,-1);
    \vertex (a) at (2,1) ;
     \vertex (b) at ( 2,-1)  ;
      \vertex (c) at (-2, 1) ;
      \vertex (d) at (-2, -1)  ;
    \diagram* {
        (d) -- [scalar] (ll)  -- [scalar] (c),
        (ll) -- [scalar, quarter left] (lu) -- [scalar, quarter left] (lr) -- [scalar, quarter left] (ld) -- [scalar, quarter left] (ll),
        (b) -- [scalar] (lr)  -- [scalar] (a),
      };
    \end{feynman}
  \end{tikzpicture} \ .
\ee
The introduction of the one-loop diagrams corrects the tree-level expressions such that the integral along a loop carrying internal momenta $q^\mu$ is
\be
\int_0^{\Lambda} {d^4 q\over (2\pi)^4} {-i \over q^2 + m^2 + i\epsilon}{ -i \over (p-q)^2 + m^2 + i\epsilon} \propto ln \left({\Lambda \over p}\right)\ .
\ee
We can see that in the limit $\Lambda \rightarrow \infty$ the integral diverges. More importantly for our discussion, the theory and the physical observables now have knowledge of the cut-off scale $\Lambda$. For example, the scattering amplitude for the 2-to-2 particle scattering diagrams is now $i\mathcal{M}\propto -i\lambda + i \lambda^2 \ln\left({\Lambda^2\over m^2}\right)$. Another related problem is that the self-energy diagram contributes to the two point correlation function with a term
\be
\Gamma_{(2)}^{1-\text{loop}} \propto \lambda \int_0^{\Lambda} {d^4 q\over (2\pi)^4} {1 \over q^2 + m^2 + i\epsilon} \propto \lambda \left[\Lambda^2 - m^2 \ln \left({\Lambda^2\over m^2}\right)\right]\ .
\ee
The canonical solution to these issues is to introduce counterterms to renormalise the theory. After doing so, the {\it renormalised} observables of the theory are cut-off independent. For example, introducing the mass counterterm
\be
\mathcal{L}_{\delta m} = -{\delta m^2\over 2}\phi^2\ , \qquad \delta m^2 = -{\lambda \over 32\pi^2} \left[\Lambda^2-m^2\ln\left({\Lambda^2\over m^2}\right)\right] ,
\ee
turns the two-point correlation function into its renormalised version
\be
\bar{\Gamma}_{(2)} = i \left(p^2 + m^2\right) \ ,
\ee
to one-loop, with higher orders in loop corrections being systematically reabsorbed into the definition of the counterterms.

This has allowed us to decouple the effective description of the low-energy physics from the cut-off of the theory, and got rid of the UV divergence that appear by sending the cut-off to arbitrarily large values. This will require an infinite number of counterterms but, {\it a priori}, we have cured the theory from UV divergences. We would like to go one step further now and ask what happens for a theory with a light scalar $\phi$ and a heavy scalar $\varphi$. Let us now consider a theory with two scalar sectors coupled to each other
\be
\mathcal{L}=-{1\over 2}(\partial_\mu \phi)^2 - {1\over 2} m^2 \phi^2 - {\lambda_\phi \over 4!} \phi^4 -{1\over 2}(\partial_\mu \varphi)^2 - {1\over 2} M^2 \varphi^2 - {\lambda_\varphi \over 4!} \varphi^4 - {g\over 4} \varphi^2 \phi^2\ , \label{eq:l_h_L}
\ee
here $m\ll M$. Much like in the previous case, one must regularise the theory. After doing so, the tree-level one-to-one scatterings are decoupled from one another since there is no quadratic term mixing $\phi$ and $\varphi$. However, the one-loop diagrams for the light scalar now contain an extra contribution
\be
\begin{tikzpicture}[baseline=-\the\dimexpr\fontdimen22\textfont2\relax]
    \begin{feynman}
    \vertex (m)  at (0, -1);
    \vertex (a) at (-2,-1) ;
     \vertex (b) at ( 2,-1)  ;
      \vertex (ll) at ( -1,0)  ;
       \vertex (lr) at ( 1,0)  ;
     \vertex (lu) at (0,1) ;
    \diagram* {
        (a) -- [scalar] (m) -- [scalar] (b),
        (m) -- [quarter left] (ll) -- [quarter left] (lu) -- [quarter left] (lr) -- [quarter left] (m),
      };
    \end{feynman}
  \end{tikzpicture} \qquad \qquad ,
\ee
where the solid line represents the heavy scalar $\varphi$. By reproducing the calculations for the renormalised two-point correlator, we find
\be
\bar{\Gamma}_{(2)} \sim i \left(p^2 + m^2\right) - {i \lambda_\phi m^2\over 32\pi^2} - {i g M^2\over 32\pi^2}\ .
\ee
The physics associated to the light sector now depend on the heavy sector, {\it i.e.} the light and heavy physics only decouple at tree-level. We could try to retune the counterterms to reabsorb the heavy sector dependence. However, by continuing our perturbative expansion to two-loops we would find diagrams of the type
\be
\begin{tikzpicture}[baseline=-\the\dimexpr\fontdimen22\textfont2\relax]
    \begin{feynman}
    \vertex (m1)  at (-2, -1);
    \vertex (a) at (-4,-1) ;
     \vertex (b) at ( 4,-1)  ;
      \vertex (ll1) at ( -3,0)  ;
       \vertex (lr1) at ( -1,0)  ;
     \vertex (lu1) at (-2,1) ;
     \vertex (m2) at (2,-1);
      \vertex (ll2) at ( 1,0)  ;
       \vertex (lr2) at ( 3,0)  ;
     \vertex (lu2) at (2,1) ;
    \diagram* {
        (a) -- [scalar] (m1) -- [scalar] (m2) -- [scalar] (b),
        (m1) -- [quarter left] (ll1) -- [quarter left] (lu1) -- [quarter left] (lr1) -- [quarter left] (m1),
        (m2) -- [quarter left] (ll2) -- [quarter left] (lu2) -- [quarter left] (lr2) -- [quarter left] (m2),
      };
    \end{feynman}
  \end{tikzpicture} \qquad \qquad ,
\ee
which would again contribute a term of $\mathcal{O}(gM^2)$ to the two-point correlation function. We would have to fine-tune to the same accuracy at every order in perturbation theory, and we would say that the light scalar is unstable against radiative corrections. This is effectively an infinite fine-tuning and the theory loses all power of predictability. The hierarchy problem can then be restated as: {\it how can we keep the heavy physics decoupled from the light physics?} 

A simple solution would be to argue that $g$ is strictly vanishing. Then the theory with two scalars turns into two theories with one scalar each, analogous to the one in \eqref{eq:L_quartic}.This implies the existence of an extra symmetry in the theory. However, consider the case of the Higgs, for example, where the interactions between different sectors are crucial to describe our Universe, then $g\neq0$ by necessity. 

Another possibility is that the symmetry does not fix $g=0$ but rather introduces new degrees of freedom such that the dependence on the heavy physics drops off from the observables. Of course, this is the idea behind supersymmetry \cite{Dimopoulos:1981zb} where each component in a supersymmetric multiplet contributes with opposite sign to the n-point correlators, by virtue of supersymmetry pairing fermions to bosons and vice versa. However, no supersymmetric partner to the Standard Model particles has been found at energies $\mathcal{O}(\text{TeV})\sim 10^{-15} \mp$, and supersymmetry must then be softly broken at some low-energy scale $E_{SUSY} \gtrsim  10^{-15}\mp$. 

Finally, one could just set $g\ll 1$ by hand such that, even if the two different sectors are coupled, the effects due to heavy physics are subleading to the light sector physics. However, two considerations arise. One is that the usual theoretical physics' lore would tell us that the smallness of $g$ has to be protected by some approximate symmetry in the theory. Otherwise, the RG flow running for $g$ towards arbitrarily high UV physics can spoil the decoupling. A second one is that, in Physics, we do not find arbitrarily tuned constants. Usually, couplings are $\mathcal{O}(1)$; if the couplings are very small there is an underlying reason. This sentiment is the idea behind naturalness. We say that a theory is natural when it contains only couplings which are $\mathcal{O}(1)$. Setting $g\ll 1$ by hand would be unnatural.

More concretely, we introduce `t Hooft's definition of naturalness \cite{tHooft:1979rat}, 
\begin{adjustwidth}{2em}{2em}\vspace{-2em}
\begin{statement_nat}
Consider an effective theory valid up to some cut-off $\Lambda$. Make any free parameters dimensionless by measuring them in units of the cut-off. A theory is natural if the only parameters that are much smaller than unity imply an enhancement of the symmetry group of theory in the limit of those parameters vanishing. Otherwise, the theory is unnatural. 
\end{statement_nat}
\end{adjustwidth}
To exemplify the meaning of an enhanced symmetry, we take our simple one-field $\phi^4$ model and follow the discussion of \cite{tHooft:1979rat}. Fixing $\lambda=0$ leads to a non-interacting theory, a freely propagating massive particle with a conserved particle number, $\lambda \neq0$ breaks particle number conservation. Therefore, a theory with $\lambda=0$ is natural despite having a vanishingly small parameter since $\lambda=0$ induces a new symmetry in the theory. On the other hand, $m=0$ does not enhance the symmetries of the theory in any way, and thus fixing $m/\Lambda\ll1$ by hand would render the theory unnatural. Similarly, in the light-heavy Lagrangian \eqref{eq:l_h_L} setting $g=0$ does not recover any symmetries, as the total number of particles is still not conserved due to $\lambda_\phi, \, \lambda_\varphi\neq0$. 

As discussed, within the framework of particle physics the hierarchy problem and naturalness issues appear under the guise of the Higgs. The Higgs mass is much smaller than the mass of the top gauge boson, the heaviest Standard Model particle. As mentioned, supersymmetry aims to solve this. Another proposed solution is technicolour \cite{Susskind:1978ms, Dimopoulos:1979za, Farhi:1979zx}, where the Higgs is a composite particle of two new fermions and the underlying theory in terms of the two fermions is natural. A more in-depth review of the historical developments and philosophy around naturalness and the hierarchy problem can be found in \cite{0801.2562}.

We can also think of the hierarchy problem and naturalness in more general terms as being a question regarding how to introduce parametrically large scale differences in a theory. For example, the cosmological constant energy scale is given by $E_{cc}\sim\sqrt{\Lambda} \sim 10^{-60} \mp$ whereas top collider physics have an associated energy scale $E_{\text{top}}\sim10^{-16}\mp$ while quantum gravity is often associated with $E_{QG}\sim\mathcal{O}(\mp)$ physics. A full theory describing all of these phenomena, and anything in between, would have to span $10^{60}$ orders of magnitude. If string theory is to describe our Universe, it will certainly have to contain small numbers and whether it can do so in a natural way is non-trivial. 

\section{The cosmological constant problem\label{sec:ccp}}
In the previous section, we reviewed how light modes in the effective theory can become unstable due to radiative corrections. This statement also affects the cosmological constant, making the theoretical cosmological constant orders of magnitude larger than the observed value for the vacuum energy of our de Sitter Universe. This problem goes by the name of the cosmological constant problem \cite{Weinberg:1988cp}, for a modern formulation of this problem, and possible resolutions, we direct the interested reader to \cite{1309.4133, 1502.05296}, as well as references therein.  

To see the problem in action, we consider the one-loop vacuum diagrams for the $\phi^4$-theory of \eqref{eq:L_quartic} in a flat spacetime
\be
\begin{tikzpicture}[baseline=-\the\dimexpr\fontdimen22\textfont2\relax]
    \begin{feynman}
    \vertex (m)  at (0, -1);
      \vertex (ll) at ( -1,0)  ;
       \vertex (lr) at ( 1,0)  ;
     \vertex (lu) at (0,1) ;
    \diagram* {
        (m) -- [scalar, quarter left] (ll) -- [scalar, quarter left] (lu) -- [scalar,  quarter left] (lr) -- [scalar, quarter left] (m),
      };
    \end{feynman}
  \end{tikzpicture}\  = \int {d^4 p\over (2\pi)^4} \log\left(p^2+m^2\right) \propto m^4 \int d^4 x  \subset -V_{\text{vac}} \int d^4 x\ , 
  \ee
where we have renormalised the ${\epsilon}$ pole, obtained by calculating the loop integral through the usual dimensional renormalisation methods, as this will not play a role in our discussion. In a theory with $N$ particle species, we might expect the vacuum energy to be given by
\be
V_{\text{vac}} \sim \sum_{0<i\leq N} c_i m^4_i\ ,
\ee
where $c_i$ are arbitrary coefficients that by naturalness we would expect to be $\mathcal{O}(1)$ and $m_i$ denotes the renormalised mass of the i-th particle species. When gravity kicks in, the vacuum energy will gravitate in the usual way 
\be
V_{\text{vac}} \int d^4 x \rightarrow V_{\text{vac}} \int d^4 x \sqrt{-g}\ ,
\ee
and in terms of Feynman diagrams this can be thought of as being given by the sum of vacuum diagrams with an arbitrary number of graviton legs
\be
\begin{tikzpicture}[baseline=-\the\dimexpr\fontdimen22\textfont2\relax]
    \begin{feynman}
    \vertex (m)  at (0, -1);
      \vertex (ll) at ( -1,0)  ;
       \vertex (lr) at ( 1,0)  ;
     \vertex (lu) at (0,1) ;
    \diagram* {
        (m) -- [scalar, quarter left] (ll) -- [scalar, quarter left] (lu) -- [scalar,  quarter left] (lr) -- [scalar, quarter left] (m),
      };
    \end{feynman}
  \end{tikzpicture} \ + \
  \begin{tikzpicture}[baseline=-\the\dimexpr\fontdimen22\textfont2\relax]
   \begin{feynman}
    \vertex (m)  at (0, -1);
      \vertex (ll) at ( -1,0)  ;
       \vertex (lr) at ( 1,0)  ;
     \vertex (lu) at (0,1) ;
     \vertex (ol) at (-2,0) ;
    \diagram* {
        (m) -- [scalar, quarter left] (ll) -- [scalar, quarter left] (lu) -- [scalar,  quarter left] (lr) -- [scalar, quarter left] (m),
        (ol) -- [photon] (ll),
      };
    \end{feynman}
  \end{tikzpicture} \ + \
    \begin{tikzpicture}[baseline=-\the\dimexpr\fontdimen22\textfont2\relax]
   \begin{feynman}
    \vertex (m)  at (0, -1);
      \vertex (ll) at ( -1,0)  ;
       \vertex (lr) at ( 1,0)  ;
     \vertex (lu) at (0,1) ;
     \vertex (ol) at (-2,0) ;
     \vertex (or) at (2,0) ;
    \diagram* {
        (m) -- [scalar, quarter left] (ll) -- [scalar, quarter left] (lu) -- [scalar,  quarter left] (lr) -- [scalar, quarter left] (m),
        (ol) -- [photon] (ll),
        (lr) -- [photon] (or),
      };
    \end{feynman}
  \end{tikzpicture} \ + \ ... \sim -V_{\text{vac}} \int d^4 x \sqrt{-g}\ .
\ee
After renormalisation, the one-loop contribution to the vacuum energy for this scalar field will again be of $\mathcal{O}(m^4)$. If we consider the matter sector to be given by the Standard model, we would expect that $V_{\text{vac}} \sim m^4_{\text{top}}\sim \mathcal{O}(\text{TeV})^4$ which is much heavier than the observed value $V_{\text{vac}}^{\text{observed}}\sim \mathcal{O}(\text{meV})^4$, this constitutes a tuning to an accuracy of 60 orders of magnitude at one-loop. However, we know that the problems do not stop here. Analogously to the discussion of the last section, let us imagine that we do fine-tune the counterterms to allow the theoretical value of the cosmological constant to match the observational one at one-loop. Then, we consider the effects of two-loop diagrams like
\be
\begin{tikzpicture}[baseline=-\the\dimexpr\fontdimen22\textfont2\relax]
    \begin{feynman}
    \vertex (m)  at (-1, -1);
      \vertex (ll) at ( -2,0)  ;
       \vertex (lr) at ( 0,0)  ;
     \vertex (lu) at (-1,1) ;
     \vertex (m2)  at (1, -1);
       \vertex (lr2) at ( 2,0)  ;
     \vertex (lu2) at (1,1) ;
    \diagram* {
        (m) -- [scalar, quarter left] (ll) -- [scalar, quarter left] (lu) -- [scalar,  quarter left] (lr) -- [scalar, quarter left] (m),
        (m2) -- [scalar, quarter left] (lr) -- [scalar, quarter left] (lu2) -- [scalar,  quarter left] (lr2) -- [scalar, quarter left] (m2),
      };
    \end{feynman}
  \end{tikzpicture} \ = \ -\lambda \left(\int {d^4 p\over (2\pi)^4} {1\over p^2+m^2+i\epsilon}\right)^2 \propto -\lambda m^4
  \ee
which contributes to the vacuum energy of the theory at the same order as the one-loop diagrams. This will continue to hold true for higher orders loop diagrams. Again, we find that the cosmological constant is unstable against radiative corrections and that, to match the observations, we must fine-tune the effective field theory at every level in the loop expansion. The cosmological constant problem remains one of the most important and challenging open questions in modern physics. Below, we briefly review some of the proposed resolutions.

\subsection{The anthropic principle}

The failure to obtain a microscopic explanation for the smallness of the cosmological constant $\Lambda$ led researchers to look for a solution in the form of the anthropic principle. The philosophy behind the anthropic principle is best encapsulated in the statement: ``$\Lambda$ should be small enough to allow the Universe to evolve for long enough such that there are scientists that would ask about its smallness''. If $\Lambda$ is much larger in magnitude than its observed value, dark energy domination would start sooner and structure formation could be affected. For concreteness, let us imagine a Universe with $\Lambda \sim \pm \mp^4$. In this Universe, we would not fret about small numbers regarding $\Lambda$ and thus the cosmological problem would be no more. However, no galaxy formation could have happened and no civilisation could ask questions about the cosmological constant in the first place. On the other hand, if galaxy formation bounds the cosmological constant to small enough values, then the cosmological constant problem can be solved by arguing that a small cosmological constant is the only acceptable value that would sustain intelligent life, which would then go on to worry about its smallness.

In \cite{Weinberg:1987dv}, Weinberg builds upon previous anthropic considerations\footnote{See \cite{Barrow:1986nmg} for a review of the work on anthropics prior to \cite{Weinberg:1987dv}.} to provide a bound on the available window for the cosmological constant. The anthropic bound on the cosmological constant is given as
\be
-10^{-120} \mp^4 < \Lambda < 10^{-118}\mp^4 \ ,
\ee
where the upper bound is obtained from demanding that dark energy domination must not begin until sufficiently late times such that structure formation can go through and the lower bound follows from demanding that the Universe does not collapse during matter domination, while stars and galaxies form. Subsequent work has tried to strengthen the bound \cite{Weinberg:1996xe, astro-ph/9701099, astro-ph/9908115}. Work on anthropics has also inspired other resolutions to the cosmological constant problem and the coincidence problem, which we discuss in the following. 

Although anthropic considerations might seem anti-climactic in nature, these are by no means trivial statements. A number of underlying assumptions about the true nature of our Universe are at work. For anthropics to solve the cosmological constant problem we require that our Universe can select from different values of the cosmological constant, that at least one of such values is in the observational range and that there exists a dynamical mechanism to allow some regions to attain such value. Since no concrete example of such theories exists, anthropic considerations remain as a proposed resolution.

\subsection{Sequestering the vacuum energy}

A qualitatively different approach to solving the cosmological constant problem relies on vacuum energy sequestering \cite{Kaloper:2013zca, Kaloper:2014dqa, Kaloper:2014fca, Kaloper:2015jra, Kaloper:2016yfa, Kaloper:2016jsd}. In this case, the radiative corrections are reabsorbed in new rigid degrees of freedom. The theory remains locally flat while global dynamics are modified. The mechanism itself is reminiscent of decapitation \cite{hep-th/0209226, 1706.04778}.

Let us consider a semiclassical action of the form
\be
\mathcal{S}=\int d^4 x \sqrt{-g} \left[{1\over 2} R -\lambda^4 \mathcal{L}_m(\lambda^{-2} g^{\mu\nu}, \Psi) - \Lambda\right] + \sigma\left({\Lambda\over \lambda^4\mu^4}\right)\ ,
\ee
with $\mu$ some mass scale and where the classical gravitation sector has been coupled to a quantum matter sector $\mathcal{L}_m$, which we take to contain the Standard model. The global parameter $\lambda$ is introduced by hand to couple matter $\Psi$ to the cosmological constant $\Lambda$, which has been promoted to a global dynamical variable. Finally, $\sigma$ is a global interaction term, which can be seen to be the dual of a top-form.

The equations of motion for the global variables $\lambda, \Lambda$ and the metric are
\begin{gather}
\sigma'\left({\Lambda\over \lambda^4\mu^4}\right) = \lambda^4 \mu^4 \int d^4x \sqrt{-g}\ , \\
\sigma'\left({\Lambda\over \lambda^4\mu^4}\right) = {\lambda^4 \mu^4\over 4\Lambda} \int d^4x \sqrt{-g} T^{\mu}_{\mu} , \\
G_{\mu\nu} = - \Lambda g_{\mu\nu} + T_{\mu\nu}\ , \label{eq:GR}
\end{gather}
where $T_{\mu\nu}$ is the energy momentum tensor for the matter sector with respect to the metric $g_{\mu\nu}$. The last equation is just Einstein's equations, the other two equations allow us to fix the cosmological constant counterterm. By combining the first two equations of motion we find
\be
\Lambda = {1\over 4} {\int d^4x \sqrt{-g} T^{\mu}_{\mu} \over  \int d^4x \sqrt{-g}} := {1\over 4} \langle T^{\mu}_{\mu}  \rangle\ , \label{eq:lambda_constraint}
\ee
where $\langle T^{\mu}_{\mu}  \rangle$ is understood as a spacetime average. Furthermore, we write the energy-momentum tensor as $T_{\mu\nu}=-V_{\text{vac}} \,g_{\mu\nu} + \tau_{\mu\nu}$ where $\tau_{\mu\nu}$ describes local excitations and $V_{\text{vac}}$ is the vacuum energy coming from Standard model loops. Plugging this expansion for the energy-momentum tensor and \eqref{eq:lambda_constraint} into \eqref{eq:GR} we find
\be
G_{\mu\nu} = \tau_{\mu\nu} - {1\over 4} g_{\mu\nu} \langle \tau_{\mu\nu} \rangle\ .
\ee
meaning that the radiative corrections due to Standard model loops always drop out from the effective field theory. This holds true at any order in loop expansion due to the diffeomorphism invariance of the action and the fact that matter couples universally to $\lambda$ through the metric $g_{\mu\nu}$, ensuring that at any order in loops the vacuum energy contribution scales like $\lambda^4$.

Sequestering resolves the cosmological constant problem at an effective field theory level, but the action has to take a very special form. Without a better understanding of why the effective field theory would have to take that particular form or what kind of UV embedding would yield a low-energy effective action of this kind, we cannot justify sequestering as the one panacea solving all our radiative problems. 

\subsection{Supersymmetry}

Another possible resolution could be supersymmetry. After all, supersymmetry came to the rescue when we considered the hierarchy problem for the Standard model. The problem here is that a de Sitter Universe is necessarily non-supersymmetric\footnote{We note that supersymmetry may be realised non-linearly on de Sitter space whenever the symmetry is broken spontaneously, as in \cite{Bergshoeff:2015tra, 1507.07842, 1511.03024, 1910.14349, 2010.13758}.}, meaning that supersymmetry must be (at least) softly broken below some scale $E_{\text{SUSY}}$. In this case, the bosonic-fermionic superpartners have a split in their mass spectrum that goes like
\be
m_B^2-m_F^2 \simeq g_s^2 E_{\text{SUSY}}^2\ ,
\ee
and this ruins the cancellation for the vacuum loop diagrams giving raise to vacuum energy contribution of $\mathcal{O}(g_s^2 E_{\text{SUSY}}^2)$. Since we have not observed supersymmetry in the LHC, we can at most hope to explain why the cosmological constant does not have an $\mp^4$ energy scale, assuming $E_{\text{SUSY}}<\mp$, but the coupling problems to the standard model sector remain. It should be noted that one could try to engineer a situation where supersymmetry is broken at two different scales $E_{\text{brane}}\gg E_{\text{bulk}}$, so that the matter sector living in a D3-brane has $E_{\text{SUSY}}=E_{\text{brane}}$ and gravity propagating in the bulk has $E_{\text{SUSY}}=E_{\text{bulk}}$. This type of scenario was already considered in braneworld constructions where the extra dimensions had infinite volume \cite{hep-th/0002190, hep-ph/0002297} and later in supersymmetric large extra dimensions brane models \cite{hep-th/0304256, 1108.0345, 1101.0152}. It remains unclear how to achieve this from a UV perspective. The problem stems from demanding a (nearly) flat four-dimensional solution in the brane. This requires the brane couplings to preserve bulk scale invariance and leads to a fine-tuning in the parameters of the model to achieve the observed curvature scales \cite{1508.01124, 1512.03800}.

\subsection{The flux landscape}

To some extend motivated by the anthropic prospects discussed previously, Bousso and Polchinski \cite{hep-th/0004134} proposed how we may achieve a small enough cosmological constant through four-form quantisation. Indeed, consider the Einstein-Hilbert action minimally coupled to a four-form
\be
\mathcal{S} = \int d^4 x \sqrt{-g}\left({1\over 2} R - \Lambda - {Z\over 48} F_4^2\right)\ , \label{eq:4theory}
\ee
where $Z$ is a normalisation constant. The four-form equations of motion fix $F_4 = c \epsilon_4$, where $\epsilon_4$ is the four-dimensional Levi-Civita symbol and $c$ is a constant related to the Dirac quantisation condition of the four-form, as we will see below. The effective cosmological constant is the given by \cite{hep-th/0004134}
\be
\Lambda_{\text{eff}} = \Lambda - {Z\over 48} F_4^2 = \Lambda + {Z c^2\over 2}\ .
\ee
If $c$ could take any value, the bare cosmological constant $\Lambda$ could be cancelled to any degree of accuracy. However, in string theory both the four-form and its dual are quantised and for the dual of the four-form the quantisation condition reads
\be
\star F_4 = F_0 = {e n \over Z} \ , \quad n\in \mathbb{Z}\ ,
\ee
where $e$ is the charge of the membrane coupled to the four-form. In turn, this fixes our arbitrary constant $c$ to take values in the integer set, in units of ${e\over Z}$. For the effective cosmological constant to be efficiently cancelled by the four-form contribution, we require that 
\be
\Lambda + {e^2 n^2\over 2Z} \sim 0 \qquad n\sim {\sqrt{2 Z\left|\Lambda\right|}\over e}\ ,
\ee
and that the spacing between different effective cosmological constants, associated to different values of $n$, is below the observed value $\delta \Lambda \sim {e^2 n\over Z} < 10^{-120}$, so that a number of effective cosmological constants can fit observations. All together, we find
\be 
\sqrt{\left|\Lambda\right|\over Z} e \lesssim 10^{-120}\ .
\ee
The problem is that the bare cosmological constant $\Lambda$ will see Standard model loops, meaning that $\Lambda\sim \mathcal{O}(\text{TeV})^4$. This implies that $e \lesssim \sqrt{Z} 10^{-108}$, for the action in \eqref{eq:4theory} we would expect $Z\sim \mathcal{O}(1)$ which would require $e\lesssim \mathcal{O}(10^{-108})$. One four-form is not enough since this is just exchanging a fine-tuning on the cosmological constant by a slightly smaller fine-tuning on the value of the membrane charge $e$. 

However, in typical compactifications, we expect a large number of these four-forms to be present. In \cite{hep-th/0004134}, it was also shown that if we have some $\mathcal{O}(100)$ of these four-forms arising from a string compactification, then the individual charges only need to be of $\mathcal{O}(1)$ even if $\Lambda$ couples to the Planck scale. Then, by anthropic considerations, our Universe is one such that the cosmological constant picked the right value to allow for structure formation. Overall, this argument is based on the idea that flux compactifications generate a very large number of vacua such that even exponentially small cosmological constant may exists in the landscape of theories. 

In supergravity models of our Universe, the smallness of the cosmological constant is often related to a balance between an uplift contribution and a tuned tree-level superpotential $W_0$. In \cite{Ashok:2003gk, Denef:2004ze}, it was argued that statistics of classical flux vacua do indeed agree with the intuition that an exponentially-tuned $W_0$ is achievable in the landscape such that the cosmological constant can be tuned to a similar degree. For example, in the case of KKLT models which we will discuss in \chpref{chap:non_pert}, \cite{Demirtas:2021ote, Demirtas:2021nlu} provide realisations of AdS$_4$ vacua with exponentially small $W_0$, fitting observations in magnitude but not in sign as an uplift is still required to obtain dS$_4$. The idea of using the flux landscape to solve the cosmological constant problem remains attractive but no concrete ten-dimensional realisation of a dS$_4$ vacua has been achieved yet.

\section{Coincidence problem \label{sec:coincidence_problem}}
A related but qualitatively distinct puzzle to the cosmological constant problem is the {\it cosmological coincidence problem}  \cite{astro-ph/9807002}. This  is usually presented as a simple question: why now? Why do we happen to live at a time when the energy density of matter and dark energy are roughly comparable? 

Let us unpack the problem in a little more detail. The total age of the universe is estimated to be around 13.7 Gyrs \cite{arXiv:1807.06209}. The early phase of radiation domination represents only a small fraction of that history, giving way to matter after around 51 Kyrs, at a redshift of  $z_{eq} \approx 3400$. The transition to dark energy domination  occurred much later, after around 3.5 Gyrs, at a redshift of $z_{de} \approx 0.5$. This represents roughly a quarter of the universe's current lifetime \cite{arXiv:1410.2509}.  This means that the universe has been matter dominated for a significant fraction of its history, so much so that its density today is still comparable to that of dark energy. 

Why is this a problem? The issue is the rate at which dark energy dilutes in comparison to matter, if indeed it dilutes at all. If dark energy is a cosmological constant, as in the standard $\Lambda$CDM cosmology, its energy density remains constant while the energy density of matter falls off exponentially quickly once we enter the accelerating phase. If we take the standard scenario at face value, dark energy will dominate indefinitely. It seem implausible that we should find ourselves so close to the dawn of dark energy domination, within a single Hubble time. By way of comparison, we note that planetary orbits are expected to exist for another $10^5$ Hubble times \cite{arXiv:astro-ph/9701131}. 

In this sense, the coincidence problem can also be understood from the point of view of naturalness. Indeed, one could fix by hand that the matter-dark energy epoch happens at any {\it present time}. However, this requires a tremendous fine-tuning of the initial conditions during inflation due to the exponential dilution of matter during the dark energy era. 

There have been several proposals that touch upon aspects of this problem (see \cite{arXiv:1410.2509}). These range from anthropic considerations \cite{astro-ph/9906210, astro-ph/0005265, astro-ph/9907168, arXiv:1103.2401} to k-essence scenarios where dark energy is triggered by matter-radiation equality, coming to dominate within a few billion years  \cite{astro-ph/0004134, arXiv:astro-ph/0006373, arXiv:astro-ph/0304277}. Although the latter go some way to explaining the (limited) duration of the matter dominated epoch, they do not address the main question of why we happen to find ourselves so close to the dawn of dark energy domination. In \chpref{chap:coincidence}, we will tackle this problem from the point of view of stringy compactifications and in the remainder of this section we dedicate some time to review these proposed solutions and discuss to what extent these proposal solve the coincidence problem. 

We should note that the question of the coincidence problem can also be addressed in apocalyptic fashion by bringing the universe to a rapid conclusion. The basic idea is that dark energy domination begins with acceleration before triggering cosmic Armageddon, within a few efolds. The solution to the coincidence problem follows because the accelerated epoch is cut short - it does not go on indefinitely and endures for a similar time to matter domination. This was shown to occur in phantom cosmologies \cite{astro-ph/0410508, arXiv:1103.2401}, linear quintessence \cite{arXiv:astro-ph/0411033, arXiv:1103.2401} and sequestering scenarios \cite{Kaloper:2013zca, Kaloper:2014dqa, Kaloper:2014fca, Kaloper:2015jra, Kaloper:2016yfa, Kaloper:2016jsd, DAmico:2017ngr, Niedermann:2017cel, Coltman:2019mql}.  

\subsection{Anthropic considerations}
Following arguments that are similar in spirit to the ones in the last section, in a series of papers \cite{astro-ph/9906210, astro-ph/0005265, astro-ph/9907168}, the authors propose an anthropic resolution to the coincidence problem. Below, we provide a summarised version of the argument flow that led the authors to their anthropic solution:
\begin{enumerate}
\item We begin by demanding that the advent of dark energy domination, $t_\Lambda$, must happen late enough that it does not destroy structure formation.

\item Assume that most of the carbon-based life appears in our Universe around the time of peak carbon production, $t_{\text{carbon}}$.

\item The main contributors to carbon in the interstellar medium are stars in the few solar mass range \cite{Iben:1983ts}. Simulations in \cite{Livio:1998mq}, showed that peak carbon production happens roughly around the time of star formation and galaxy formation, $t_{\text{carbon}}\sim t_{\text{SFR}}\sim t_G$.

\item Further assume that the appearance of intelligent life is not delayed by more than a fraction of the lifetime of main sequence stars ($\sim$ 5--20 Gyrs), $t_G\sim t_{\text{IL}}$.

\item We take our civilisation as a typical example of carbon-based life and define the ``present time'' $t_0$ as the time necessary for a civilisation to become advanced enough to ask about the coincidence problem. On a cosmological timescale, given that our civilisation has been around for only some $10^4$ years, so we find $t_G\sim t_{\text{IL}}\sim t_0$.

\item Finally, we have to relate the time of dark energy domination to the time of galaxy formation. In \cite{Weinberg:1987dv}, Weinberg took values for the redshift such that galaxy formation is barely possible, giving a lower bound to the advent of dark energy domination. However, in \cite{astro-ph/9906210}, it is argued that if we are truly an average observer we would expect not to observe the limiting value found by Weinberg but rather one where a sizeable portion of the galaxies have already formed. This is found to be $t_G\sim t_\Lambda$.     
\end{enumerate}
Putting everything together we find $t_0\sim t_G\sim t_\Lambda$ and the coincidence problem seems to be resolved. It is clear that some assumptions must be made about the origin and evolution of intelligent life and these have drawn some criticism \cite{hep-th/0407174, astro-ph/0106143}. It is typically argued that, although anthropic constraints can select values of the cosmological constant which are close to the observed value, these values necessarily carry large uncertainties due to the lack of understanding of the conditions required for intelligent life to appear.

\subsection{k-essence scenarios}
Another proposed resolution to the coincidence problem are the k-essence scenarios originating in \cite{astro-ph/0004134, arXiv:astro-ph/0006373}. Let us follow the analysis of \cite{arXiv:astro-ph/0304277}. We restrict ourselves to models with Lagrangian
\be
\mathcal{L} = K(\phi) \tilde{p}(X)\ ,
\ee
where $K(\phi)>0$ and $X:={1\over 2} \nabla_\mu \phi \nabla^\mu \phi$. Fixing $K(\phi):=1/\phi^2$, loosely motivated by the expected perturbative supergravity form of the \kah metric, and defining the variable $y:=X^{-1/2}$ and the $\tilde{p}(X(y)):=g(y)/y$, the equation of state parameter and sound speed are given by
\be
w(y) = -{g(y)\over yg'(y)} \ , \qquad c_s^2 = {g(y)-yg'(y)\over y^2 g''(y)}\ .
\ee
Asking that $w<-1$ and $c_s^2>0$ implies that $g(y)$ is a convex and decreasing function of $y$, but is otherwise left undetermined. We would like the field $\phi$ to have a stable tracker solution (R) during radiation domination and then reach an attractor solution (K) during k-essence domination with $w\simeq -1$. In principle this would render the k-essence models mostly independent of initial conditions during inflation while attaining a observationally viable behaviour at late times. The question is whether these early tracker to late attractor is a generic behaviour for some family of $\tilde{p}(X)$.

\begin{figure}
\centering
\includegraphics[width=0.6\linewidth]{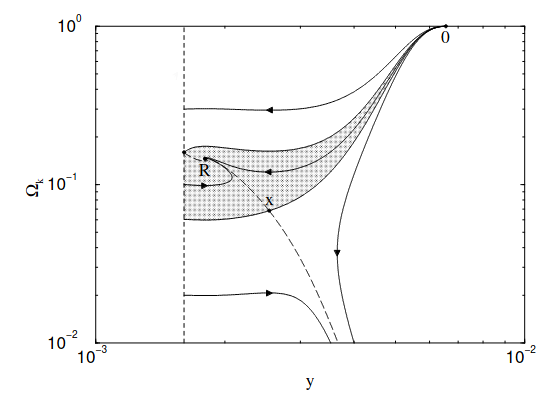}
\caption{The shaded region represents the basin of attraction for $\tilde{p}_2(X)$ during radiation domination. (R) is the tracker solution, (0) is the origin of the trajectories and (x) is the saddle point defined in \cite{astro-ph/0006373}. The vertical dashed line describes the point $y_c$, where $g''(y_c)=0$ and the sound speed diverges. Figure taken from \cite{arXiv:astro-ph/0304277}.}
\label{fig:model_2}
\end{figure}

In \cite{arXiv:astro-ph/0304277}, a dynamical analysis of the models of \cite{astro-ph/0004134, arXiv:astro-ph/0006373} was conducted. The models themselves are given by
\begin{gather}
\tilde{p}_1(X) := -2.01 + 2\sqrt{1+X} +3\cdot 10^{-17} X^3 - 10^{-24} X^4\ , \label{eq:model1} \\
\tilde{p}_2(X) := -2.05 + 2\sqrt{1+f(x)} \ ,\label{eq:model2}\\
f(X) = X - 10^{-8} X^2 + 10^{-12} X^3 -10^{-16} X^4 + 10^{-20} X^5 - {10^{-24}\over2^6 } X^6\ . \nonumber
\end{gather}
Since we are most interested in studying whether the fine-tuning of initial conditions is alleviated in any way, we concentrate on the behaviour of these models during radiation domination and whether the tracker solution exists for a wide range of initial conditions. The dynamical analysis showed that although some basin of attraction exists towards the radiation tracker (R), see for example the phase diagram in \figref{fig:model_2}, the basin is quite limited for these models. For most initial conditions, the model does not reach the tracker solution but rather rapidly evolves towards a period of early k-essence domination or a singularity given by a divergence of the sound speed in the theory. 
 
On top of this, it is clear that the models given in \eqref{eq:model1} and \eqref{eq:model2} have been engineered by hand to contain these tracker solution in the first place. So we are exchanging some slightly less fine-tuned initial conditions for highly contrived models. If the subset of initial conditions turned out to be physically important, then this exchange might be beneficial. However, without any prior or reason to believe that these regions of the parameter space would be more desirable, we find no clear benefits to using the k-essence models in \eqref{eq:model1} and \eqref{eq:model2}. Therefore, although an interestingly physical mechanism, k-essence does not constitute a resolution of the coincidence problem insofar as a large tuning of initial conditions seems to still be required.

In this chapter, we have reviewed some open questions of cosmology that will become important in the remainder of the thesis. The question on naturalness was the driving force behind our idea of developing a dark energy model involving the clockwork mechanism, which we describe in \chpref{chap:clockwork}. The coincidence problem will be the centrepiece of the discussion in \chpref{chap:coincidence}, where we propose how string compactifications might play a role in the resolution to the problem. Finally, the hierarchy problem and the microscopic nature of dark energy will play a key role in the goals and motivations for the second part of the thesis.
\blankpage
 \chapter{A naturally light dark energy model\label{chap:clockwork}}
 In \cite{1912.04905}, we proposed a model of dark energy driven by a pseudo-scalar field whose super-light mass emerges naturally from a simple microscopic theory with uniquely high scale couplings.  The model is a dark energy avatar of the  pion, its low mass emerging from the spontaneous breaking of a weakly broken symmetry.  It consists of \order{100} pseudo scalar fields (axions) with non-trivial mass mixing, one of which has a bilinear mixing with a four-form field strength.  All mass scales in the theory are assumed to lie at, or close to, the Planck scale. The motivation for this model arises as the marriage between two axion EFTs and we will begin by introducing these models: the clockwork mechanism \cite{1511.00132, 1511.01827} in \secref{sec:cw_mec} and the Dvali-Kaloper-Sorbo model \cite{hep-th/0410286, hep-th/0507215, 0811.1989, 1101.0026} in \secref{sec:dks_4}. \par
 
In \secref{sec:paper_nat} we will present our model in full and further motivate its UV completion in terms of a type IIA supergravity motivated toy model in \secref{sec:string_nat}. We will then proceed to scrutinise the type IIA embedding in more detail, being able to write down a no-go theorem for the clockwork mechanism at the level of the perturbative type IIA action with fluxes in \secref{sec:no_go}. Finally, in \secref{sec:braneworld} we perform the dimensional deconstruction of the discrete model, and study a family of braneworld configurations showing how the braneworld obsverver measures a low scale of dark energy on account of the warping of the extra dimension.
 
\section{Clockwork mechanism\label{sec:cw_mec}}
The underlying theory that we present in this section is based on the works \cite{1511.00132, 1511.01827}, where the authors present a mechanism to obtain trans-Planckian axion decay constants from a theory with $\mathcal{O}(\mp)$ mass parameters. These constructions are intimately related to the problem of obtaining viable natural inflation in string theory. In natural inflation, the large field excursions of the inflaton are protected from quantum corrections by virtue of the inflaton being an axion with a discrete shift symmetry. In this way, even if the inflaton range is super-Planckian, the axion range remains sub-Planckian as long as its decay constant is super-Planckian. However, it was pointed out that trans-Planckian decay constant are not easy to engineer in supergravity theories whilst keeping perturbative control \cite{hep-th/0303252}. The Kim-Nilles-Peloso (KNP) mechanism \cite{Kim:2004rp} provided a way to circumvent this issue by considering two-axion models with sub-Planckian decay constants, in such a way that there exists a linear combination of the axions with super-Planckian decay constant. From the point of view of the EFT, there exists a field with an effectively super-Planckian decay constant that could prove to be a good candidate for the inflaton. However, considerations related to strong forms of the weak gravity conjecture \cite{hep-th/0601001, 1503.04783, 1402.2287, 1503.00795} put some stringent bounds on the KNP mechanism. 

In \cite{1610.07962}, the authors go one step further and generalise the works \cite{1511.00132, 1511.01827} to show how to generate parametrically low energy scales from a theory with only Planckian-scale parameters. To do so, they will require the introduction of a technically natural parameter (in the sense of 't Hooft \cite{tHooft:1979rat}) $\sigma:=M^2/f^2\ll 1$, that represents the scale separation between spontaneously broken physics at a scale $f$ and softly broken symmetries at a scale $M$. In \cite{1610.07962}, the clockwork mechanism is generalised to scalars, vectors and tensor fields. For our discussion, we will concentrate on the scalar clockwork.

Let us consider a theory with a global symmetry $\mathcal{G}=U(1)^{N+1}$ spontaneously broken at a scale $f$. We will take $f\sim\mathcal{O}(\mp)$ but the mechanism generalises straightforwardly to any other scale, where $f$ would represent the cut-off of the would-be effective field theory. Below the symmetry breaking scale, the theory contains $N+1$ Goldstone bosons $\phi_j$ which can be packaged conveniently as 
\be
U_j = e^{i\phi_j/ f}\ , \qquad j=0,...,N\ ,
\ee
which transform by a phase under the corresponding $U(1)_j$ generator. We also softly break $\mathcal{G}$ by introducing $N$ mass parameters $M_i$, with $i=0,...,N-1$. These can be regarded as N spurion fields taking non-zero vacuum expectations values due to their charges under the $U(1)_i$
\be
Q_i[M_k^2]=\delta_{i\,k}-q\delta_{i\,k+1}\ .
\ee
Taking $q>1$, the symmetry breaking will be soft as long as $\sigma=M/f\ll1$.

The leading terms of the effective Lagrangian in terms of the true scalar and the N pseudo-scalar are
\be
\mathcal{L}=-\dfrac12\left[\sum_{i=0}^N(\partial\phi_i)^2+M^2\sum_{i=0}^{N-1}(\phi_i-q\phi_{i+1})^2 + \mathcal{O}(\phi^4) \right]\ ,
\ee
where we can rewrite the potential in terms of a mass matrix
\be
V(\phi)={1\over2}M^2\sum_{i=0}^{N-1}(\phi_i-q\phi_{i+1})^2 + \mathcal{O}(\phi^4) = {1\over2}\sum_{i,j=0}^{N} \phi_i{\cal M}_{ij}^2\phi_j + \mathcal{O}(\phi^4)  \ , 
\ee
{\small
\be
{\cal M}^2_{ij}=M^2\begin{bmatrix}
1 & -q & 0 & & \cdots & & 0 \\
-q & 1+q^2 & -q & & & & \\
0 & -q & 1+q^2 & \ddots & \ddots & & \vdots \\
 & & \ddots & \ddots & \ddots & & \\
\vdots & & \ddots & \ddots & 1+q^2 & -q & 0 \\
 & & & & -q & 1+q^2 & -q \\
0 & & \cdots & & 0 & -q & q^2
\end{bmatrix}\ .
\ee}
Diagonalising the mass matrix, we find for some orthogonal matrix $\mathcal{O}$
\begin{gather}
\tilde{\mathcal{M}}^2_{ij} = \mathcal{O}^T_{ik} \mathcal{M}^2_{kl} \mathcal{O}_{lj} = \text{diag}\left(m^2_0, ... , m^2_N\right) \ , \\
m^2_0 = 0 \ , \quad m^2_k = \left[q^2+1-2q\cos\left({k\pi\over N+1}\right)\right] M^2\ .
\end{gather}
The soft symmetry breaking has broken the axion shift symmetry, providing a mass for N pseudo-scalars. The mass spectrum of the theory contains a massless mode, related to the leftover spontaneously broken $U(1)$, and N massive modes. In the large $N$ limit, the massive modes span masses
\be
m_1\simeq (q-1)M \ , \qquad m_N \simeq (q+1)M=m_1 \left(1 + {2\over q-1}\right)\ ,
\ee
and the mass splitting $\delta m_k = m_{k+1} - m_k$ is given by
\be
{\delta m_k\over m_k}\simeq {q\pi\over N \left[q^2+1-2q\cos\left({k\pi\over N+1}\right)\right]} \sin \left({k\pi\over N+1}\right)\ .
\ee
Therefore, for a large number of fields (or ``gears'' in the language of \cite{1610.07962}), the $N$ massive states enter the theory at a mass scale $m_i\sim\mathcal{O}(M)$, with gaps $\delta m_i\sim \mathcal{O}(M/N)$. Furthermore, rewriting the theory in term of the mass eigenstates in the diagonal mass matrix basis $\pi_i = \sum_{j=0}^N \mathcal{O}_{ij}^T\phi_i$, we crucially find that $\mathcal{O}_{0j} \propto q^{-j}$. This means that if an operator couples to the {\it N-th} gear, the massless mode will feel its interaction suppressed by $q^N$ and for large $N$ this can become an efficient way to ``protect'' the massless mode against radiative corrections. We will see this explicitly in \secref{sec:paper_nat} once we couple the Dvali-Kaloper-Sorbo model to the {\it N-th} site of the clockwork chain.

\section{Dvali-Kaloper-Sorbo four-form model\label{sec:dks_4}}
Inspired by a long history of research on applications of membranes to gravity (see for example \cite{Abbott:1984qf, Brown:1987dd, Brown:1988kg}) arises the second element of our model. The Dvali-Kaloper-Sorbo model (DKS) \cite{hep-th/0410286, hep-th/0507215, 0811.1989, 1101.0026} has been proposed as an effective theory of axion monodromy in string theory \cite{0803.3085, 0808.0706}. However, the true string embedding of the DKS model remains unclear \cite{1611.00394}. Nonetheless, DKS, as an EFT, can provide an interesting description of single-field axion inflation since it avoids some of the issues related with super-Planckian excursions, in a similar vein to the KNP mechanism. 

The idea is to introduce a bilinear mixing between the axion and a four-form field strength. The theory admits a dual description in terms of a massive pseudoscalar and the magnetic flux of the four-form. The latter is locally constant in spacetime although it can jump between quantised values across a three dimensional membrane.  The effective inflaton is then a gauge invariant combination of the axion and the magnetic flux --- its large field values are obtained through the flux, which in turn may be identified with macroscopic quantities characterizing the system rather than high scale excitations of the inflaton field \cite{1607.06105}.  Small deformations of these models can also give rise to an emergent mechanism for screening the effects of vacuum energy at large scales \cite{1806.04740, 1903.07612, 1806.03308}.

The starting point for the DKS model is a massive $U(1)$ gauge theory given by \cite{1101.0026}
\be
\mathcal{L}=-\frac{1}{48}F^2-\frac{\mu^2}{12}\left(A_{\mu\nu\alpha}-h_{\mu\nu\alpha}\right)^2+\frac{\mu}{6}\phi \frac{\epsilon^{\mu\nu\alpha\beta} }{\sqrt{-g}} \partial_\mu h_{\nu\alpha\beta}\ , 
\ee
here $F_{\mu\nu\alpha\beta}=4 \del_{[\mu} A_{\nu\alpha\beta]}$ is the four-form field strength, $\epsilon^{\mu\nu\rho\sigma}$ is the totally antisymmetric Levi-Civita symbol defined such that $\epsilon^{0123}=1$, with indices raised and lowered with respect to the metric $g_{\mu\nu}$, and $\phi$ is an axion with periodicity $\phi:=\phi+2\pi f$, with $f$ left to be determined. The equation of motion for $\phi$ yields $h=db$, with $b_{\mu\nu}$ the Stueckelberg field invariant under $b\rightarrow b+d\lambda$. Integrating out the Stueckelberg through its equation of motion yields
\be
\mathcal{L}=-\frac{1}{48}F^2-\frac{1}{2}\left(\partial \phi\right)^2 + \frac{\mu}{24}\frac{\epsilon^{\mu\nu\alpha\beta} }{\sqrt{-g}}\phi F_{\mu\nu\alpha\beta}\ . \label{eq:L_U1}
\ee
We can eliminate the four-form by introducing a Lagrange multiplier $Q$ that fixes the Bianchi identity for the four-form on-shell. This allows us to rewrite the theory in terms of the pseudo-scalar
\be
\mathcal{L}=-\frac{1}{2}(\partial\phi)^2-\frac{\mu^2}{2}\left(\phi+\frac{Q}{\mu}\right)^2-\frac{Q}{6}\frac{\epsilon^{\mu\nu\alpha\beta} }{\sqrt{-g}}\partial_\mu A_{\nu\alpha\beta}\ , \label{eq:L_KS}
\ee
which is invariant under
\be
\phi\ \rightarrow\ \phi+2\pi f\ , \qquad Q\ \rightarrow \ Q- 2\pi q\ , \label{eq:sym}
\ee
where $f=q/\mu$. 

In the DKS model, one identifies the effective inflaton with $\varphi = \phi+{Q}/{\mu}$. We can see that, the inflaton gets a mass from its mixing with the four-form field strength, and the stronger the coupling $\mu$ the larger the effective mass. It is this mechanism for generating a mass that we exploit in our model of dark energy. By coupling one end of our clockwork to the four-form we guarantee that the mixing with the zero mode of the clockwork is exponentially small and as a result, a very small mass is generated in the low energy effective theory. 

\section{Obtaining a naturally light dark energy scale\label{sec:paper_nat}}
We begin with a chain of $N+1$ pseudo-scalar fields $\phi_0$, ..., $\phi_N$, all characterized by a single  ultraviolet mass scale $M$ and a nearest neighbour interaction with strength $q$  \cite{1511.00132, 1511.01827}.  The mass $M$ is assumed to lie at or close to the Planck scale. We further assume that one end of the chain is coupled to a four-form field strength as in the Dvali-Kaloper-Sorbo model. The combined set up is described by the following  Lagrangian density:
%commentstart
\begin{multline}
\mathcal{L}=-\dfrac12\left[\sum_{i=0}^N(\partial\phi_i)^2+M^2\sum_{i=0}^{N-1}(\phi_i-q\phi_{i+1})^2\right]\\
+ \dfrac{\mu}{24}\phi_N \frac{\epsilon^{\mu\nu\alpha\beta} }{\sqrt{-g}}F_{\mu\nu\alpha\beta}  -\dfrac{1}{48} F_{\mu\nu\alpha\beta} F^{\mu\nu\alpha\beta} \ .
\label{eq:L0}
\end{multline}
%commentend

In the first line we recognise the clockwork model of \cite{1511.00132, 1511.01827}.  In principle we could allow for site dependent masses $M_i \sim M$ and mixing strengths $q_i \sim q$, although for simplicity we take them all to be equal.  The coupling $q$ is dimensionless and assumed to be greater than 1, but it remains of order unity. The second line of \eqref{eq:L0} contains the DKS model for the $N$-th site in the chain, in the form of \eqref{eq:L_U1} and where we have identified $\phi_N=\phi$. The bilinear mixing between the axion and the four-form reveals another mass scale, $\mu \sim M$, which we also assume to be given by the characteristic ultraviolet scale of the theory. The gravitational sector of the theory is assumed to be described by Einstein gravity, although we will not need to include that in our discussion. 

The original clockwork Lagrangian gives rise to a massless Goldstone pseudo-scalar \cite{1511.00132, 1511.01827,1610.07962}, a consequence of the non-linearly realised shift  symmetry $\phi_i \to \phi_i + cq^{-i}$ for arbitrary $c$. This symmetry remains perturbatively unbroken through the mixing with the four-form, although in the presence of charged membranes, non-perturbative effects break the continuous symmetry down to a discrete subgroup \cite{hep-th/9510227, hep-th/0004134}. However, the symmetry is only {\it mildly} broken because $\phi_N$ has exponentially suppressed overlap with the zero mode. We thus expect that the zero mode acquires a mass, but that the latter remains tiny. In order to see this explicitly  it is convenient  to integrate out the four-form and pass to a dual description in which the four-form mixing generates a new mass term for the last axion in the chain.  This can be done in a straightforward manner by adding a Lagrange multiplier term of the form $\frac{1}{24}Q \frac{\epsilon^{\mu\nu\alpha\beta}}{\sqrt{-g}} (F_{\mu\nu\alpha\beta}-4\del_{[\mu}A_{\nu\alpha\beta]} )$ then eliminating $F$ using its algebraic equation of motion\footnote{Note that since the theory is quadratic in $F$ this amounts to performing the Gaussian  in the path integral, which can,  of course, be done exactly.} \cite{0811.1989}.  This yields a  dual theory described by the following Lagrangian:
%commentstart
\begin{multline}
\mathcal{L}=-\dfrac12\left[\sum_{i=0}^N(\partial\phi_i)^2+M^2\sum_{i=0}^{N-1}(\phi_i-q\phi_{i+1})^2\right]\\
-\dfrac12 (Q+\mu \phi_N)^2- \frac{Q}{6} \frac{\epsilon^{\mu\nu\alpha\beta} }{\sqrt{-g}}\del_{[\mu}A_{\nu\alpha\beta]}  \ .
\label{eq:Ldual}
\end{multline}
%commentend
The Lagrange multiplier $Q$ is fixed to be constant by the variation of the three-form.  If the latter is coupled to membrane charges, then $Q$ is quantised in units of the membrane charge, $e$, as in  $\langle Q\rangle=2\pi N e$ for integer values of $N$  \cite{hep-th/9510227, hep-th/0004134}.  This quantisation condition is compatible with the unbroken symmetry transformations, which take the form
$
\phi_i\to \phi_i+2\pi n \frac{e}{\mu}q^{N-i}, \ Q \to Q+2\pi ne
$
for integer values of  $n$. The mass matrix in the dual description is given by
%commentstart
{\small
\begin{equation}
{\cal M}_{ij}=M^2\begin{bmatrix}
1 & -q & 0 & & \cdots & & 0 \\
-q & 1+q^2 & -q & & & & \\
0 & -q & 1+q^2 & \ddots & \ddots & & \vdots \\
 & & \ddots & \ddots & \ddots & & \\
\vdots & & \ddots & \ddots & 1+q^2 & -q & 0 \\
 & & & & -q & 1+q^2 & -q \\
0 & & \cdots & & 0 & -q & r+q^2
\label{Eq:matrixG}
\end{bmatrix},
\end{equation}}
%commentend
where $r=(\mu/M)^2$ is the square of the ratio between the two mass scales. The eigenmasses of this matrix are given by  the roots of an $(N+1)$-th order polynomial. Again, we find that there is a tower of $N$ massive modes whose masses go with the ultraviolet scale $M$.  The remaining mode is massless in the limit where $r \to 0$ and is ultralight in general. We can find its mass by linearising the above eigenvalue problem; the resulting ultralight mass scale is given by
%commentstart
\begin{equation}
\begin{split}
m_0^2&\simeq(q^2-1)^2 r M^2\Big/\left[q^{2 (N+1)}(q^2+r -1)\right.
\\
&\left.\vphantom{q^{2 (N+1)}}\quad+(N+1) r(1-q^2) -q^2-r +1\right].
\label{eq:m0}
\end{split}
\end{equation}
%commentend
In the limit of large $N$ (and for $q$ larger than unity), $m_0$ is well approximated by
%commentstart
\begin{equation}
m_0^2\simeq\dfrac{1}{q^{2 (N+1)}}\dfrac{(q^2-1)^2 r M^2}{(q^2+r -1)}.
\label{eq:discrm0largeN}
\end{equation}
%commentend
Thus, $m_0$ acquires an exponential suppression with respect to $M$, the argument in the exponential being the total number of clockwork gears, $N+1$. If we take $M=\mu=m_\mathrm{P}$ ($m_\mathrm{P}=1.22\times10^{19}$~GeV being the Planck mass), $q=2$ and $N=200$, we get $m_0\simeq5.7\times10^{-33}$~eV,  which is the energy scale associated with the Hubble rate. At large scales, the dynamics will be equivalent to quintessence driven by a quadratic potential. However, the mass scale of  potential has not been tuned to the tiny value demanded by nature, rather it has arisen naturally on account of the clockwork mechanism and the coupling to the four-form.  The underlying theory is made up uniquely of high scale couplings. We emphasize the fact that the mass mixings need not be identical but can have site dependence. As long as they are  greater than unity, the clockwork mechanism will kick in as usual, even for order one couplings, and the suppression of the mass of the ultralight mode will occur as desired. 

Finally we note that if we allowed for a non-vanishing flux, the potential would actually go as 
\be
V_0=\frac12m_0^2  \left(\frac{2 \pi \mathcal{N}_Q e}{m_0}+ \pi_0\right)^2\ .
\ee
For natural values of $\pi_0 \sim m_P$, the flux term dominates the potential  for non-vanishing $\mathcal{N}_Q$ and there is too much dark energy. This is why we assumed a vanishing flux $\mathcal{N}_Q =0$, which is a robust condition provided the nucleation rate for bubbles of non-vanishing flux is suppressed\footnote{Although a detailed analysis of this interesting question is beyond the scope of our work, we  expect that there is indeed suppression for the following reason. For $e \sim m_P^2$ a unit of flux generates a large positive Planckian potential. Therefore, we can crudely model the transition from vanishing to non-vanishing flux in terms of a bubble of Planckian de Sitter curvature nucleating in a quasi de Sitter background with a very small curvature.  Such processes, tunnelling from true to false vacuum, are possible but are known to be suppressed relative to those going from false to true, becoming infinitely suppressed in the limit that the low scale curvature approaches zero (see e.g. \cite{Lee:1987qc}) .}.

\section{Warm-up: A string motivated toy model\label{sec:string_nat}}
In this section and the next one, we would like to study the possibility of a UV embedding in string theory for the model \eqref{eq:Ldual}. Although the question regarding the viability of a full embedding is complicated, we begin by considering the following: {\it can clockwork be embedded in fluxed type IIA perturbative supergravity?} The benefit of this setup is that, by turning all possible fluxes, one is allowed to fix all moduli at tree-level, with a number of complex structure moduli requiring non-geometric fluxes. The introduction of non-geometric fluxes does not come for free. Indeed, the fluxes backreact on the geometry leading to a half-flat non-Calabi Yau metric \cite{hep-th/0607223}. This creates a major caveat with regards to the trustability of these solutions. Nonetheless, we would like to move on and see if we can learn anything regarding a possible UV completion. Thus, in this section, we will see how the clockwork mechanism constraints the geometry of the internal space for a string motivated toy model. In the next section, we will study the scalar potential of fluxed type IIA and we will be able to provide a no-go for a supersymmetric type IIA clockwork, any embedding will require non-perturbative effects. Other considerations regarding the difficulty on clockwork embeddings can be found in \cite{Ibanez:2017vfl}.

Our model can be motivated from the point of view of a higher dimensional theory. Indeed, in a theory with extra dimensions, a large number of scalar fields in the four-dimensional EFT is often associated with the periods of differential p-forms living in the higher dimensional theory. In \cite{1507.06793, 1606.00508}, the authors showed that one may rewrite the democratic type IIA supergravity formulation \cite{hep-th/0103233} in terms of a pseudo-action containing Minkowski 4-forms and its dual fields, which is equivalent to the democratic action at the level of the equations of motion. Further details on type IIA supergravity can be found in \appref{app:iia_details}.

The democratic action is usually written in terms of a polyform gauge invariant field strength, $G=G_0+G_2+\ldots +G_{10}$ where  $\mathbf{G}=d\mathbf{C}+B\wedge\mathbf{C}+\mathbfcal{F}\wedge e^B$  and $B$ is the Kalb-Ramond two-form,  $\mathbf{C}=C_1+C_3+\ldots +C_9$ the polyform gauge field and $\mathbfcal{F}= \mathcal{F}_0+\mathcal{F}_2+\ldots +\mathcal{F}_{10}$ a formal sum of internal fluxes only. Let us take the following ansatz for the non-vanishing supergravity fields
 \begin{gather}
 B_2 = b^a(x) \omega_a, \quad C_3= c_3^0(x), \quad C_5=c_3^a(x)\omega_a\ , \nonumber\\
 C_7=\tilde{c}_{3a}(x)\tilde{\omega}^a, \quad \mathcal{F}_2=q^a \omega_a,\quad \mathcal{F}_4=e_a\tilde{\omega}^a,
 \end{gather}
where we note that this corresponds to the massless limit of type IIA supergravity.  We have introduced the cohomology basis of \{\text{two},\text{ four}\}-forms in the internal manifold $\mathcal{M}_6$ as \{$\omega_a$, $\tilde{\omega}^a$\}, respectively, and $\omega_6$ will denote the volume form of $\mathcal{M}_6$. With this ansatz, $\mathbf{G}$ reads\footnote{We have chosen to use the compact notation
\begin{gather}
F_4^0=dc_3^0, \quad F_4^a=dc_3^a+b^a F_4^0,\quad \tilde{F}_{4a}=d\tilde{c}_{3a}+\mathcal{K}_{abc}b^b dc_3^c. \nonumber
\end{gather}} 
%commentstart
\begin{gather}
G_2=q^a \omega_a,\qquad G_4=F_4^0+\left(e_a+\mathcal{K}_{abc}b^bq^c\right)\tilde{\omega}^a, \nonumber\\
G_6=F_4^a\omega_a+\left(e_a b^a+\frac{1}{2}\mathcal{K}_{abc}b^ab^bq^c\right)\omega_6,\quad G_8=\tilde{F}_{4a}\tilde{\omega}^a,
\end{gather}
%commentend
with $\mathcal{K}_{abc}=\int_{\mathcal{M}_6}\omega_a\wedge\omega_b\wedge\omega_c$, the triple intersection numbers of $\mathcal{M}_6$.

Below we describe the effective four-dimensional action. In doing so, we will assume that the volume moduli, including the dilaton, are stabilised by some other ingredient of the theory (see e.g \cite{hep-th/0411271}). Our main interest is to derive a plausible clockwork mechanism after compactifying the theory down to four dimensions. 

These fields enter the four-dimensional potential as follows (see for details \cite{1507.06793, 1606.00508})
%commentstart
\begin{gather}
V_{4D}\propto F_4^0\wedge\star F_4^0 + g_{ab}F_4^a\wedge\star F_4^b+{g}^{ab}\tilde{F}_{4a}\wedge\star \tilde{F}_{4b}\nonumber \\
 + F_4^0\rho_0 + F_4^a\rho_a + \tilde{F}_{4a}\tilde{\rho}^a\ ,
\end{gather}
%commentend
where $g_{ab}=\int_{\mathcal{M}_6}\omega_a\wedge\star\omega_b$, $g^{ab}=\int_{\mathcal{M}_6}\tilde{\omega}^a\wedge\star\tilde{\omega}^b$ and we have introduced the dual scalars to the Minkowski 4-forms
%commentstart
\begin{gather}
\rho_0=e_a b^a + \frac{1}{2} \mathcal{K}_{abc}q^ab^bb^c, \nonumber \\
-\rho_a= e_a + \mathcal{K}_{abc}q^bb^c, \quad \tilde{\rho}^a= q^a\ . \nonumber
\end{gather}
%commentend
Eliminating $F_4^a$ and $\bar{F}_{4a}$ through their equations of motion in favour of their dual scalars, the four-dimensional Lagrangian density with only one Minkowski four-form is given by
%commentstart
\begin{multline}
\mathcal{L}_{4D}=-\frac{1}{\kappa_4^2}\left[{e^{-2\phi}\over 4} g_{ab}db^a\wedge \star db^b+\frac{\mathcal{V}^3 e^{\frac{\phi}{2}}}{32}F_4^0\wedge\star F_4^0 \right. \\
\left. - \frac{1}{4}F_4^0\rho_0 + \frac{ e^{\frac{\phi}{2}}}{2 \mathcal{V}^2 }\left(g^{ab}\rho_a\rho_b+e^\phi g_{ab}\tilde{\rho}^a\tilde{\rho}^b\right)\right]\ .
\label{eq:L4D}
\end{multline}
%commentend

As this stands, the vacuum expectation value for the axions is not necessarily vanishing, owing to the presence of a tadpole in the potential for $b_a$.  We can fix this by shifting $b_a \to \beta_a+ b_a$ for some constant flux $\beta_a$ chosen so that it eliminates the tadpole. By comparison to \eqref{eq:L0}, one finds a Lagrangian density of the form\footnote{In writing \eqref{eq:four-dimensional_L} we have dropped  a next-to-leading order correction term to the coupling ($\sim F_4^0b^2$)  as well as a cosmological constant term  and a total derivative that do not contribute to our discussion.}
 \begin{gather}
\mathcal{L}=-\frac12 \left[\gamma_{ab} \del b^a \del b^b+M_{ab}b^a b^b +\dfrac{\alpha}{48} \left({F}_4^0\right)^2\right]+\frac{1}{24} \chi_a b^a F_4^0  \ ,
\label{eq:four-dimensional_L}
 \end{gather}
with
\be
\gamma_{ab} =\frac{e^{-2\phi}}{2 \kappa_4^2}\int_{\x}\omega_a\wedge\star\omega_b \ ,\quad \alpha=\frac{2\vol^3 e^{\frac{\phi}{2}}}{3\kappa_4^2} \nonumber \ ,
\ee
\begin{eqnarray}
M_{ab} &=&  \frac{e^{\frac{\phi}{2}}}{\vol^2 \kappa_4^2}  \mathcal{K}_{ade}\mathcal{K}_{bcf} q^d q^c\int_{\x}\tilde{\omega}^e\wedge\star\tilde{\omega}^f\ , \nonumber \\
 \chi_a &=&\frac{6 }{\kappa_4^2} \left( e_a+\mathcal{K}_{abc} \beta^b q^c\right)\ \nonumber \ .
\end{eqnarray}
To realise the clockwork dark energy we need two ingredients: a mass matrix with vanishing eigenvalue and a zero mode that overlaps with the remaining four-form. The former is the clockwork condition while the latter allows the zero mode to acquire a small mass.  If $s^a$ is an eigenvector of $M_{ab}$ with null eigenvalue, then the first condition requires $M_{ab}s^a\sim\mathcal{K}_{abc}q^cs^a=0$ while the second gives $s^a \chi_a \neq 0$. To obtain the clockwork structure, we must  therefore impose the following condition on the internal geometry
\be
\int_{\x}\left(\omega_a s^ a\right)\wedge\omega_b\wedge\left(\omega_c q^c \right) = 0\ .
\label{eq:condition}
\ee
A simple solution to the above constraint is provided by $s^a=q^a$ and the corresponding two-form given by a product of 1-forms, i.e. $\omega_cq^c=u \wedge v$. In that case, the constraint is immediately satisfied due to the antisymmetry of the wedge product on odd forms. As long as we require $q^a e_a \neq 0$, the second condition can also be satisfied. 

The proposed solution could be heuristically realised if we consider our internal manifold to be a product $\mathcal{M}_6\simeq T^2\times\Sigma_g^4$, where $g$ indicates the genus of the two-dimensional complex surface, which is assumed to be K\"ahler. Indeed, with this setup one naturally selects the zero mode, $q^i$, to point along the toroidal directions $T^2$. This is because the top form along $T^2$ is automatically a two-form corresponding to  a product of the 1-forms of the torus.  As long as we take the $e_i$ to overlap slightly with the $T^2$, but not completely,  we also satisfy the second condition, allowing the zero mode to gain a small mass through its four-form mixing. We  have also shown that for the clockwork mechanism to bring the mass of the lightest mode down to the dark energy scale one requires $\mathcal{O}\left(200\right)$ scalar fields. For simplicity of exposition, let us think of these fields as descending only from the Kalb-Ramond two-form. Then, the previous requirement translates into a constraint on the cohomology structure of $\Sigma_g^4$ after fixing the $T^2$ geometry, namely $g=b_1-b_2/2$ where $b_p$ is the {\it p-th} Betti number of $\Sigma_g^4$. Using, $b_2\sim\mathcal{O}\left(200\right)$ we find that $g\sim b_1-100$.  

\section{A more in-depth look into achieving the embedding\label{sec:no_go}}

Although the string motivated scenario of the last section seems exciting, we can now show that the perturbative embedding is not viable. Indeed, consider a potential of the form \cite{1507.06793}
\be
V = {1\over s^4}\left[{\rho_0^2\over2}\mathcal{V}+{g^{ab}\rho_a\rho_b\over 8\mathcal{V}}+2\mathcal{V}g_{ab}\tilde{\rho}^a\tilde{\rho}^b+{\mathcal{V}\over2}\rho_m^2-{1\over6\mathcal{V}} \left(u^\lambda h_\lambda\right)^2\right] + {h_0^2\over 2\mathcal{V}}{1\over s^2}\ . \label{eq:pot_iia_fluxed}
\ee
This scalar potential is obtained from the usual dimensional reduction of the fluxed type IIA supergravity action\footnote{Up to some overall constant $R=16 e^{2K^{\text{cs}}}\Re (Z_0)^4$ that we have renormalised to unity since it will not affect the discussion of this section.} (further details are in \appref{app:fluxes} and references therein). We have also allowed for a non-zero Romans mass term $m$ in the type IIA action \cite{Romans:1985tz} such that the dual scalar are now defined as
\begin{gather}
\rho_0=\varphi + e_a b^a + \frac{1}{2} \mathcal{K}_{abc}q^ab^bb^c -{m\over 6}\mathcal{K}_{abc}b^ab^bb^c, \label{eq:dual_massive_scalars} \\
-\rho_a= e_a + \mathcal{K}_{abc}q^bb^c -{m\over 2}\mathcal{K}_{abc}b^bb^c, \quad \tilde{\rho}^a= q^a-mb^a\ , \quad \rho_m=m \ ,\nonumber
\end{gather}
let us also introduce the short hand notation $\hat{e}_a:=e_a + \mathcal{K}_{abc}q^bb^c$
The scalar potential \eqref{eq:pot_iia_fluxed} contains $s$, the real part of the axio-dilaton, $u^\lambda$ the $h^{1,1}$ complex structure saxions, $\varphi$ a linear combination of axions, and $b^a$ the $h^{2,1}$ \kah axions. We will assume that $\mathcal{V}$ is a free parameter since the volume moduli can be fixed through fluxes to {\it a priori} large values (see \appref{app:fluxes}, and also \cite{hep-th/0505160}).

We concentrate on supersymmetric vacua where the \kah F-terms are given in \eqref{eq:b_stab} and \eqref{eq:v_stab} and fix
\begin{gather}
\tilde{\rho}^a=\rho_0=0 \ , \qquad 10\rho_a + 3m \mathcal{K}_{abc}v^bv^c=0 \ , \qquad b^a:={q^a\over 2m}\ , 
\end{gather}
Let
\be
h_{ab}:=-{3m\over 10}\mathcal{K}_{abc}v^c \ ,
\ee
so that
\be
\rho_a=h_{ab}v^b\ , \quad \mathcal{V}=-{5\over 9m}\rho_av^a\ .
\ee
The kinetic metric for the \kah sector can be written in terms of the \kah potential in \eqref{eq:kah_1} as
\be
g_{ab}={1\over 4} \partial_{v^a}\partial_{v^b} K^{\text{ks}} = {\vol_a\vol_b\over 4\vol^2}-{\vol_{ab}\over 4\vol}={9\over 4}{\rho_a\rho_b\over (\rho\cdot v)^2} - {3\over 2} {h_{ab}\over \rho\cdot v}\ ,
\ee
here $\rho\cdot v:=\rho_a v^a$. Its inverse is given by
\be
g^{ab} = -{2\over 3} (\rho\cdot v) h^{ab} -{2} v^a v^b\ . 
\ee 
The mass matrix for the $b^a$ fields takes the form
\be
M^a_b = {2\mathcal{V}\over s^4}\left(\delta^a_b-{g^{ac}g^{de}\over16m\vol^2}\mathcal{K}_{bcd}\rho_e\right)\ ,
\ee
with the clockwork conditions requiring the existence of an eigenvector $n^a$ with vanishing eigenvalues, {\it i.e.}
\be
M^a_b n^b \propto \left(1-{3\over 5}\right)n^a + {9\over 5} v^a {n\cdot \rho\over v\cdot \rho} = 0\ .
\ee
The only option to satisfy the eigenvalue condition is to choose $n^a = k v^a$ for some constant $k$, but then $M^a_b n^b=0 \Rightarrow k=0$ and there is no eigenvector satisfying this condition. Therefore, the clockwork conditions cannot be satisfied at the level of fluxed type IIA supergravity. It is then clear that, if clockwork is to be embedded in string theory, we will require the use of non-perturbative effects. 

\section{Deconstructing clockwork dark energy\label{sec:braneworld}}

In the final section of this chapter, we provide a description of a braneworld construction of our clockwork dark energy model. If the number of pseudo-scalar fields is very high, $N\to \infty$, the clockwork mechanism can arise as a deconstruction of an extra compact dimension \cite{1610.07962}. In this limit, the clockwork gears merge into a single field $\Phi$, and the gear $\phi_i$ becomes the value of $\Phi$ at site $i$; the interaction of the four-form with the last site of the discrete clockwork corresponds to the localisation of the four-form on a brane at the boundary of the compactified extra dimension. Our model can be obtained from the following five-dimensional theory, defined on a fixed geometrical background, 
\be \label{eq:S05D}
S_{5\mathrm{D}}=S_\text{bulk} +S_0+S_{\pi R}\ ,
\ee
where we have a canonical scalar in the bulk 
\be
S_\text{bulk} =\int d^4x \int_0^{\pi R}dy \  \sqrt{-g} \left[-\dfrac{m}{2}\ g^{IJ} (\partial_I \Phi) (\partial_J \Phi) \right]\ ,
\ee
and two branes, one containing the dark energy sector,
\be
S_0=
\int_{y=0} d^4 x \sqrt{-\gamma_0}\Big(\dfrac{\mu}{24}\Phi \frac{\epsilon^{\mu\nu\alpha\beta} }{\sqrt{-\gamma_0} }F_{\mu\nu\alpha\beta} -\dfrac{1}{48} F^{\mu\nu\alpha\beta}F_{\mu\nu\alpha\beta}\Big)\Big]\ ,
\ee
and the other containing the matter sector
\be
S_{\pi R }=\int_{y={\pi R}} d^4 x \sqrt{-\gamma_R} \mathcal{L}_m (\gamma_R^{\mu\nu}, \Psi)\ .
\ee
We define $g_{IJ}$  to be the five-dimensional metric, with indices $I, J$ running over the four spacetime dimensions $x^\mu$, as well as the additional fifth dimension $y$.  For its part, the $y$ dimension extends from 0 to $\pi R$, with a reflection symmetry at $y=0$ and $y=\pi R$. These two boundary surfaces represent the location of the branes with the induced metric on the brane at $y=0, \pi R$ given by $\gamma^0_{\mu\nu}, ~\gamma^R_{\mu\nu}$ respectively.  Upon dividing the $y$ dimension into $N+1$ sites and discretising the action \eqref{eq:S05D} accordingly,
one can recover the Lagrangian density \eqref{eq:L0}, provided we input the following geometry:
%commentstart
\be \label{eq:clockwork_geometry}
ds^2=e^\frac{4ky}{3} (dy^2+dx^2),
\ee
%commentend
where $dx^2$ entails the four-dimensional metric on a brane, and identifying $q^N=\mathrm{e}^{\pi kR}$ with the mass scale $M = N/(\pi R)$.  We imagine all such scales - $M, m, \mu$ and  $k$ - to be of the same order, corresponding to the UV scale of the underlying braneworld theory.  For this particular geometry, the latter corresponds to a five dimensional linear dilaton model with boundary terms living on the two branes \cite{1705.10162}.

However, the mechanism that suppresses the mass of the lightest mode is very general and works also for metrics that differ from \eqref{eq:clockwork_geometry}. 
For instance, it works also in the following family of metrics:
%commentstart
\be
ds^2 = e^{\frac{4ky}{3}} \left(e^{-4 \ell ky} \, dy^2 + dx^2 \right) \,.
\label{gen_metric}
\ee
%commentend
The clockwork geometry is recovered by choosing $\ell = 0$, while for $\ell = 1/3$ one gets Randall-Sundrum \cite{hep-ph/9905221}. 
The equation of motion for $\Phi$, once $F$ has been integrated out, is
%commentstart
\be \label{eq:eqPhi}
e^{2(\ell-\frac13)ky} \left[\Phi'' + 2( 1+ \ell) k \Phi' + \frac{\Box  \Phi}{e^{4k\ell y}} \right] = \delta(y) \frac{\mu(\mu \Phi + Q)} {m},
\ee
%commentend
where a prime stands for a $y$ derivative, and $\Box$ is the four-dimensional d'Alembertian. Setting $\Phi=-Q/\mu+\delta\Phi$, we can easily solve for $\delta\Phi$ in the bulk by taking the four-dimensional Fourier transform of eq.~(\ref{eq:eqPhi})\footnote{In general, one should embed the action \eqref{eq:S05D} in some geometric theory and study the perturbations of the geometric quantities as well as the fluctuations of the clockwork field. However, these fluctuations decouple at linear level. We can thus consistently examine the fluctuations of $\Phi$ on their own.}.
 The solution is then given by
 %commentstart
\be \label{eq:solution_general_metric}
\begin{split}
\delta\Phi(y, x^\mu) &= \int d^4p \, \Big[ A(p^2) \ J_{\frac{1+\ell}{2\ell}} \Big(\frac{\sqrt{-p^2}}{2k\ell e^{2k\ell y}}\Big) \\
&\quad+B(p^2) \ J_{-\frac{1+\ell}{2\ell}} \Big(\frac{\sqrt{-p^2}}{2k\ell e^{2k\ell y}}\Big) \Big]e^{i p_\mu x^\mu - k(1+\ell)y} ,
\end{split}
\ee
%commentend
where the $J_\alpha(z)$ are Bessel functions of the first kind, $A$ and $B$ being free functions. Equation \eqref{eq:eqPhi} further imposes some boundary conditions at $y=0$ and $\pi R$. As a result, $A$ and $B$ are linearly related, while $p^2$ can only take some values among a quantised set, $p^2=-m_n^2$, $n\in\mathbb{N}$. The masses $m_n$ are found as solutions of the following equation:
%commentstart
\begin{equation}
\begin{split}
&J_{\frac{\ell-1}{2 \ell}}\left(\frac{m_n}{2 k \ell}e^{-2 k \ell \pi  R}\right) \left\{J_{\frac{\ell+1}{2 \ell}}\left(\frac{m_n}{2 k \ell}\right) \left[4 k (\ell+1)m+\mu ^2\right]\right.
\\
&\left.-2 m_n m J_{\frac{3\ell+1}{2\ell}}\left(\frac{m_n}{2 k \ell}\right)\right\}+J_{\frac{1-\ell}{2\ell}}\left(\frac{m_n}{2 k \ell}e^{-2 k \ell \pi  R}\right)
\\
&\left[\mu ^2 J_{-\frac{\ell+1}{2 \ell}}\left(\frac{m_n}{2 k \ell}\right)-2 m_n m J_{\frac{\ell-1}{2 \ell}}\left(\frac{m_n}{2 k \ell}\right)\right]=0\ .
\end{split}
\end{equation}
%commentend
They are all of order $k$, except for the first which is very light, being suppressed by $e^{k \pi R} \gg 1$. It is possible to find an approximate expression for this light mass:
%commentstart
\be\label{eq:lightest_mode}
\begin{split}
m_0^2 (\ell) \simeq &\frac{\mu^2}{m} {8 k^2 \ell (1-\ell^2)} \Big/ 
\left\{2e^{2k(1-\ell)\pi R}\ell\left[2k(1+\ell)+\frac{\mu^2}{m}\right] \right.
\\
&\left.-(1+\ell)\left(4k\ell +\frac{\mu^2}{m}\right)-e^{-4k\ell\pi R}(\ell-1) \frac{\mu^2}{m}\right\}\,.
\end{split}
\ee
%commentend
The above approximation is based on an expansion of the Bessel functions assuming $m_0\ll k\ell$. This is no longer true when $\ell \to 0$. To compute the lightest mass in the clockwork geometry, one can solve directly 
$$m_0^2 (0) =\frac{4 k^2 \mu ^2}{e^{2 \pi  k R} \left(4 k m+\mu ^2\right)-4 k m-2 \pi  k \mu ^2 R+\mu ^2} \ .$$
\newline
Equation \eqref{eq:lightest_mode} shows that the suppression mechanism works for the whole family of metrics \eqref{gen_metric}, even if it gets less efficient as we depart from the clockwork geometry.  Note that \eqref{eq:lightest_mode} should not be trusted physically when $l\gtrsim1$, because the length of the $y$ dimension drops below the UV scale of the five dimensional theory.

As it stands, the mass of the lightest dark energy mode is suppressed relative to UV scales set by matter resident on {\it either} of the two branes. Why, then, have we placed matter on the brane at  $y=\pi R$?   This is because the energy density of the dark energy field in slow roll is enhanced by the effective four dimensional Planck scale. This enhancement exactly compensates for the suppression in the mass of the field, and  in the end the energy density during slow roll scales like $k^4$. In any event, it turns out that the energy density of dark energy will  only be suppressed if we calibrate our scales relative to the $\pi R$ brane, where the warp factor is exponentially large. This is why we put the visible sector of our theory on the right hand brane.  

\section{Conclusions}
In this chapter, we have studied the possibility of generating naturally small masses from Planckian scales. In particular, we have considered how a parametrically small dark energy field could arise in a (technically) natural way. By considering the interplay between clockwork gravity and the coupling between its pseudo-scalar fields and Minkowski top-forms we were able to generate a super-light dark energy field that arises solely from the characteristics of the high energy theory.

Furthermore, we have explored how this mechanism could be achieved from the point of view of a prospective UV completion of the theory. We have studied a type IIA toy model and translated the existence of a vanishing eigenvalue in the clockwork mass matrix as a constraint in the internal geometry. Even more so, we have been able to show that, after considering the fluxed perturbative type IIA scalar potential, a no-go arises. Our hopes of generating a stringy embedding of clockwork will then have to pass through the introduction of corrections to the tree-level action. Since non-perturbative effects are better understood in the type IIB dual \cite{1808.08967}, we will further only concern ourselves with type IIB supergravity in the remainder of the thesis. Finally, we were also able to write down a generalised braneworld scenario, motivated by the work of \cite{1610.07962}, where the DKS mode lives on a brane.
\blankpage
\chapter{A stringy perspective on the coincidence problem \label{chap:coincidence}}
In the last chapter, we tried to tackle the idea of how to obtain naturally small couplings in a theory with only $\mathcal{O}(\mp)$ parameters. In this one, we turn our attention to the coincidence problem discussed in \chpref{chap:openqs}. In \cite{2105.03426}, we argued that, for string compactifications broadly consistent with swampland constraints, dark energy is likely to signal the beginning of the end of our universe as we know it, perhaps even through decompactification,  with possible  implications for  the cosmological coincidence problem. Thanks to the scarcity (absence?) of stable de Sitter vacua, dark energy in string theory  is assumed to take the form of a quintessence field in slow roll.  As it rolls, a tower of heavy states will generically descend, triggering an apocalyptic phase transition in  the low energy cosmological dynamics after at most a few hundred Hubble times. As a result, dark energy domination cannot continue indefinitely and there is at least  a percentage chance that we find  ourselves in the first Hubble epoch. We begin by describing the some ideas borrowed from \secref{sec:swamp} that make the core of our discussion and, in \secref{sec:toy_model}, we use a toy model of quintessence coupled to a tower of heavy states to explicitly demonstrate the breakdown in the cosmological dynamics as the tower becomes light.

\section{Generic idea} \label{sec:idea}
We begin with the refined Swampland distance conjecture (rSDC) \cite{hep-th/0605264, 1610.00010, 1802.08264, 1803.04989}, which is one of the most well studied  and least controversial of the swampland conjectures (see also \cite{1602.06517, 1708.06761, 1801.05434, 1703.05776, 1802.08698, 1806.01874, 1808.05958, 1804.10504}). As we described in \chpref{chap:pert_II}, the rSDC states the following: consider two points in field space, $\phi_0$ and $\phi_0+\Delta \phi$,  separated by a geodesic distance $\Delta \phi$. As $\Delta \phi \to \infty$, there exists an infinite tower of states whose mass become exponentially light, 
\be
m(\phi_0+\Delta \phi) \sim m(\phi_0)e^{-\beta \frac{|\Delta \phi| }{M_{Pl}}} \label{distcon}
\ee
for some positive constant,  $\beta$, that we typically expect to be $\mathcal{O}(1)$. The offending tower of states is often associated with Kaluza-Klein modes or winding modes, depending on the direction of motion in moduli space ---although in some cases the tower can originate from localised sources that exist in the theory, see for example \cite{Font:2019cxq}.  For this reason, we take the initial mass $m(\phi_0)$ to be given by the scale of compactification, $1/R$, which  could be as low as a few meV in a braneworld setting where Standard Model fields are confined to a $3$-brane, although generically we expect it to be much larger, perhaps even just short of the Planck scale, $M_{Pl} \sim 10^{18}$ GeV. 

In its refined form, the de Sitter swampland conjecture (rdSC) \cite{1806.08362, 1807.05193, 1810.05506} concerns the form of  the potential $V(\phi)$ for scalar fields  in a low energy effective theory.  Assuming the effective theory  is obtained from a consistent theory of quantum gravity, the potential must satisfy either
\be
|\nabla V | \geq\frac{c}{M_{Pl}} V  \label{ds1}
\ee
or
\be
\text{min} \left( \nabla_i \nabla_j V\right) \leq -\frac{c'}{M_{Pl}^2} V \label{ds2}
\ee
where $c, c'$ are universal positive constants of $\mathcal{O}(1)$ and $\text{min} \left( \nabla_i \nabla_j V\right) $ is the minimum eigenvalue of the corresponding Hessian. Note that the conjecture forbids the existence of stable de Sitter vacua in string theory, for which we would have to have $V>0$. It does, however, allow for de Sitter vacua with a tachyonic instability of order the corresponding Hubble time $H^{-1} \sim M_{Pl}/\sqrt{V}$. Some constraints on the scale of the tachyon were derived in the context of 10D supergravity \cite{1807.09698}. 

We remark that we will not make use of the rdSC {\it per se}. Motivated by the difficulty to achieve de Sitter vacua in string theory, of which the rdSC is a symptom, we would like to ask the question of what does a dynamical model of dark energy look like for a generic string compactification. By generic we mean those string compactifications to which the rSDC applies. This is generic in so far that the rSDC applies to models in the {\it Dine-Seiberg region}, as discussed in \chpref{chap:pert_II}. This leaves out models that require perturbative and non-perturbative corrections to the tree-level action, those living in the {\it pheno region} depicted in \figref{fig:pheno_swamp_1}. An analysis of those models will be conducted in \chpref{chap:model_building}.

Of course, building a reliable model of quintessence within string theory is not without its own challenges \cite{Hebecker:2019csg, Cicoli:2021skd}, and we will also explore these in detail in \chpref{chap:model_building}. Nevertheless, we begin with a model of quintessence as a canonical scalar field, $\phi$, moving under the influence of a potential $V$.  Here we imagine that quintessence is generically described by the saxions of string theory with a non-compact field range. The energy density and pressure stored in the field are given respectively by $\rho_\phi=\frac12 \dot \phi^2+V$ and $p_\phi =\frac12 \dot \phi^2-V$. The dynamics of the scalar is governed by the following field equation
\be
\ddot \phi +3H\dot \phi+V'(\phi)=0
\ee
where $H(t)$ is the Hubble parameter and ``dot'' denotes differentiation with respect to cosmological time.  Our goal is to argue that dark energy domination is relatively short lived on account of the motion of the field  towards the infinite points in moduli space. Any motion of the field prior to dark energy domination will only reduce the amount it is allowed to move afterwards, bringing dark energy to an even quicker conclusion. Therefore, the most conservative scenario is to assume negligible motion of the moduli fields  until dark energy finally begins to dominate.  This is, in any event, likely as the field will be held up by Hubble friction. 

Once the dark energy field has come to dominate, to be compatible with the observed equation of state, it must be in slow roll, $\frac12 \dot \phi^2 \ll V, ~|\ddot \phi | \ll 3H|\dot \phi| $. Furthermore, we may assume that $H \approx H_0 \sim 10^{-33}$ eV $\sim 10^{-60} M_{Pl}$.  With these approximations, consider the field excursion in a short time $\delta t$. This is given by 
\be
\delta \phi \approx -\frac{V'}{3H_0} \delta t
\ee
If we accept the refined de Sitter conjectures, then one of \eqref{ds1} or \eqref{ds2} must hold. Let's assume we satisfy the condition on the gradient, given by \eqref{ds1}. It now follows that 
\be
| \delta \phi | \gtrsim \frac{c}{M_{Pl}}   \frac{V}{3H_0} \delta t \sim \mathcal{O}(1)  M_{Pl} H_0 \delta t
\ee
where we have used the Friedmann equation during dark energy domination  $H_0^2 \approx \frac{\rho_\phi}{ 3 M_{Pl}^2} \approx  \frac{V}{ 3 M_{Pl}^2} $ and the fact that $c$ is assumed to be $\mathcal{O}(1)$. We immediately see that the dark energy field rolls roughly a Planck unit in a Hubble time.  We now consider the implications for the distance conjecture \eqref{distcon}, assuming the tower of new states are initially very heavy, with masses $m(\phi_0) \sim M_{Pl}$ close to the Planck scale. After a single Hubble time, the dark energy field will move by around one Planck unit.  This only corresponds to a fractional change in the masses in the tower, and certainly not enough to contaminate the low energy physics. 

How far is the field allowed to roll before we have to start worrying about it?  In the local neighbourhood of the Earth the field should not be displaced from $\phi_0$ by more than around 30 Planck units. Anything more than that would bring the mass of the tower down from Planck scale to the scale of collider physics, opening up the possibility of producing these states at the LHC.  Of course, the details of this depend on the nature of the coupling between the tower and the Standard Model fields. Furthermore,  none of these considerations are relevant on cosmological scales, where we can certainly allow the field to move much further.  To ensure that the tower remains decoupled from the low scale cosmological dynamics, we conservatively impose a maximum displacement of  $|\Delta \phi|_\text{max} \sim \frac{M_{Pl}}{\beta} \ln(M_{Pl}/H_0)$. For $\beta \sim \mathcal{O}(1)$ this corresponds to a displacement of around 140 Planck units. Assuming the field continues to roll a Planck unit in every Hubble time, we see that the tower of states  will trigger a transition in the cosmological dynamics after no more than $\mathcal{O}(100)$ Hubble times.  The coincidence problem isn't solved but  it is significantly ameliorated. If a generic model of dark energy is destined to last at most $\mathcal{O}(100)$ Hubble epochs, and all epochs are equally probable,  we might expect there to be at least a percentage chance that we find ourselves in the first epoch\footnote{By way of comparison, we note that when Leicester City won the premier league in 2016, they started the season as 5000-1 outsiders.}.  

What if we assume that we satisfy the second of the two de Sitter swampland criteria in \eqref{ds2}, rather than \eqref{ds1}? If the condition on the gradient of \eqref{ds1} is violated, but the condition on the Hessian in \eqref{ds2} holds the dark energy field may move considerably less than a Planck unit in a Hubble time.  In fact, if we imagined the field sitting in a region where the gradient of the potential is negligible, we might even imagine it staying there indefinitely, giving a neverending period of dark energy.  However, this conclusion is too quick.  Quantum fluctuations will guarantee a displacement in the field of at least $\mathcal{O}(H_0)$ in the first Hubble time.   This initial displacement will grow thanks to the fact that the Hessian condition implies a tachyonic mass for the fluctuations in the dark energy field,  $\mu^2 = V''(\phi)<0$ with $|\mu^2| \gtrsim  \frac{c'}{M_{Pl}^2} V  \approx  3c' H_0^2$.   For $c' \sim \mathcal{O}(1)$, the corresponding instability can be as slow as a Hubble time but even so, its effect is amplified over several Hubble times by exponential growth.  There are two possibilities: the first is that the instability triggers a rapid transition  which brings the acceleration to a premature end  (as desired for the coincidence problem).  This could occur, for example,  by the potential changing sign so that  we no longer have a quasi de Sitter expansion. Alternatively, the background cosmology could remain roughly unchanged, at least beyond a few Hubble times. If this is the case, the tachyonic instability  amplifies the initial displacement to a value\footnote{To leading order, the scalar satisfies an equation $\ddot \phi +3H_0 \dot \phi-|\mu^2| \phi=0$, where we recall that $\mu^2 = V''(\phi)<0$ with $|\mu^2| \gtrsim  \frac{c'}{M_{Pl}^2} V  \approx  3c' H_0^2$. The general solution is then given by a sum of exponentials $e^{\lambda_\pm t}$ where $\lambda_\pm=\frac{-3H_0 \pm \sqrt{9H_0^2+4 |\mu^2| }}{2}$, which includes a growing mode, with $\lambda_+ \gtrsim H_0$,  thanks to the tachyonic instability with $|\mu^2|  \gtrsim H_0^2$.}
\be \label{tach}
| \delta \phi | \gtrsim  H_0 e^{\mathcal{O}(1) H_0 \delta t}
\ee

In the latter scenario, the tower of massive states would remain decoupled from the cosmological dynamics until $| \delta \phi |  \approx |\Delta \phi|_\text{max} \sim \frac{M_{Pl}}{\beta} \ln(M_{Pl}/H_0)$, at which point a transition is inevitable. For $\beta \sim \mathcal{O}(1)$, this will occur within at most  $\ln[M_{Pl}/H_0 \ln(M_{Pl}/H_0))] \sim 143$ Hubble times, so our conclusions are unchanged, and the coincidence problem isn't as serious as we previously thought. 

Of course, the de Sitter conjecture is less well established than the distance conjecture, and, in the simplest scenarios, may even be at odds with local measurements of $H_0$ \cite{2006.00244}. With this in mind, what can we say if  we  deny the validity of both criteria  \eqref{ds1} and \eqref{ds2} and abandon the de Sitter conjecture altogether? In this instance, we cannot rule out the possibility that the current phase of acceleration is approaching a stable de Sitter configuration in which  dark energy continues for an exponentially large number of Hubble epochs. If this is the case, then the coincidence problem is as problematic as ever.  However, we might tentatively speculate that the de Sitter conjecture is really a statement about what is {\it generic} within consistent models of dark energy within string theory.  Of course, it is much too early to make any definitive statement in this regard. Nevertheless, if it happens to be true that the generic scenarios are those for which one of the criteria \eqref{ds1} or \eqref{ds2} hold,  our results go through and the coincidence problem is tamed.  We might also worry about the fact we have assumed dark energy to be a single canonical scalar.   However, we  expect this to capture the generic dynamics of fields moving through moduli space,  with our canonical scalar tangential to the trajectory and all the orthogonal directions stabilised.

\section{A toy model} \label{sec:toy_model}
To better understand how an accumulation of light states can impact the cosmological evolution at late times, we consider a toy model of dark energy described by the following Lagrangian 
\be
\mathcal{L}=-\frac12 (\partial \phi)^2-\mu^4e^{- \frac{\alpha \phi}{M_{Pl}}} 
+\sum_{n=1}^{\infty}-\frac12 (\del \varphi_n)^2
-\frac12 n^2 M_{KK}^2 e^{-2  \frac{\beta \phi}{M_{Pl}}} \varphi_n^2 
\ee
Here $\phi$ is the quintessence field, driving dark energy, taken to be in slow roll on a runaway potential of the form $\mu^4e^{-\frac{\alpha \phi}{M_{Pl}}}$ where $\mu^4 = 3M_{pl }^2H_0^2$ is the  scale  of dark energy and $\alpha$ is some order one positive number.  To be compatible with observations we require\footnote{We should note that in \cite{1806.09718} the authors claim a more stringent bound of $\alpha\lesssim0.6$ at $3\sigma$ level. However, the more detailed analysis in \cite{1808.09440} gives the more relaxed bound $\alpha\lesssim1.02$ at $3\sigma$, which is the one that we use in this section. For a detailed analysis of the two approaches we point the interested reader to \cite{1808.09440}.} $\alpha\lesssim1.02$ \cite{1808.09440}, which is not in conflict with the swampland constraints on the potential \eqref{ds1} and \eqref{ds2}. In addition we have a tower of heavy states, $\varphi_n$, whose masses are originally set by some high scale $M_{KK}$ (imagined to be the Kaluza-Klein scale associated with the compact  internal manifold), becoming exponentially light as the dark energy field, $\phi$, rolls off to infinity. The rate at which the tower becomes light is set by the coupling $\beta$, which is also assumed to be order one and positive. 

At the  dawn of dark energy domination, at time $t_0$, we assume that the quintessence field is far away from the  tails of the exponentials, $\phi \sim \phi_0 \ll M_{Pl}/\alpha, M_{Pl}/\beta$.  Since the effective mass of the Kaluza-Klein tower, $M_{KK} e^{-  \frac{\beta \phi_0}{M_{Pl}}} \sim M_{KK} \gg H_0$ is high in this regime, far above the scale of the cosmological evolution, the Kaluza-Klein states are decoupled from the dynamics.  As a result, the quintessence field satisfies the following classical equation of motion on a homogenous and isotropic background
\be
\ddot \phi+3H\dot \phi-\frac{\alpha}{M_{Pl}}\mu^4e^{- \frac{\alpha \phi}{M_{Pl}}}=0
\ee
For a quasi-de Sitter expansion, with $H\approx H_0$, we find that $\phi \approx \phi_0+ \lambda (t-t_0)$ where 
\be
\lambda= \frac{\alpha \mu^4}{3 M_{Pl}H_0} =  \alpha M_{Pl} H_0< M_{Pl} H_0
\ee
This approximation works well as long as $\Delta \phi < M_{Pl}/\alpha$, or equivalently, $\Delta t <H_0^{-1}/\alpha^2$.  Of course, acceleration will continue beyond this time, only at a lower scale. There is a wealth of literature on the dynamics of similar quintessence models (for a review, see \cite{hep-th/0603057}). In this chapter, we wish to briefly explore another effect that is far less well studied - the time dependence on the mass of the Kaluza-Klein tower as the field begins to slowly roll. On the quasi-de Sitter background with constant curvature, this is given by
\be
M_{KK}^\text{eff}(t)=M_{KK} e^{-  \frac{\beta s}{M_{Pl}}} \approx M_{KK}  e^{-  \epsilon (t-t_0)} 
\ee
where $\epsilon=\beta \lambda/M_{Pl} =\alpha\beta H_0$. This will drive particle production in the Kaluza-Klein sector, kicking in as soon as the states stop being decoupled,  $M_{KK}^\text{eff}(t) \lesssim H_0$.  A complete analysis of this phenomena requires a detailed numerical study of particle creation on the dynamical background, taking into account the effect of the time varying mass and the de Sitter cosmology.

To get some immediate insight into what might happen we neglect the  curvature of the background spacetime and focus on the particle production due  to a mass varying exponentially with time on a Minkowski geometry.  Crucially,  these approximations allow us to make {\it preliminary} analytic estimates but we should also acknowledge their limitations. In particular, we are implicitly assuming that the dark energy field, which feeds into the effective mass of the Kaluza Klein tower, continues to evolve linearly in time beyond the first Hubble epoch. In truth, the Hubble scale changes and the dark energy field picks up additional temporal dependence which may affect some of the details.  Also, we are neglecting the effect of spacetime curvature. This is less of an issue, as our interest here is  on the effect of the  changing mass of the Kaluza-Klein tower, as opposed to the effect of quantum fields propagating on de Sitter.  We also expect this approximation to accurately capture the physics on sub-horizon scales. 

With these caveats in mind, let us take a closer look at the $n$th state in the Kaluza-Klein tower on a Minkowski background, whose dynamics described by the following Lagrangian
\be
\mathcal{L}=-\frac12 (\del \varphi_n)^2
-\frac12 n^2 M_{KK}^2 e^{- 2 \epsilon t}  \varphi_n^2 
\ee
where we have also set $t_0=0$ (without loss of generality). Our analysis follows the standard techniques reviewed in detail in \cite{Mukhanov:2007zz}, whose conventions we also follow.  As explained  in \cite{Mukhanov:2007zz},  the state operator for the quantum field  can be expanded in terms of creation and annihilation operators, $\hat a_{\bf k}^\dagger$ and  $\hat a_{\bf k}$, in the usual way,
\be
\hat \varphi_n(t, {\bf x})=\frac{1}{\sqrt{2} }\int \frac{d^3 \bf  k}{(2\pi)^\frac32} \left[ e^{i\bf{k \cdot x}} \bar u_k(t) \hat a_{\bf k}+ e^{-i\bf{k \cdot x}  }u_k(t) \hat a^\dagger_{\bf k} \right]
\ee
where ``bar'' denote the complex conjugate.  The mode functions are governed by  the equation for a time dependent  harmonic oscillator 
\be
\ddot u_{k}+\omega^2_k(t) u_k=0, \qquad \omega^2_k(t)=k^2+n^2 [M_{KK}^\text{eff}(t)]^2
\ee
and have  solutions that are conveniently expressed in terms of Hankel functions (of the first kind)
\be
u_k(t)=A H^{(1)}_{i\mu}(z)+B \bar H^{(1)}_{i\mu}(z) \label{uk}
\ee
where $z=\frac{n M_{KK} }{\epsilon} e^{-  \epsilon  t}$ and $\mu=\frac{k}{\epsilon}$.  Since $[M_{KK}^\text{eff}(t)]^2>0$, at any given time $t_*$ we can define the instantaneous vacuum state $|0 \rangle_*$ as the lowest energy state of the Hamiltonian at that time.  The mode functions that determine this state  satisfy the boundary condition \cite{Mukhanov:2007zz}
\be
u_k(t_*)=\frac{1}{\sqrt{\omega_k(t_*)}}, \quad \dot u_k(t_*)=i \sqrt{\omega_k(t_*)}
\ee
This fixes the constants in \eqref{uk} so that
\begin{eqnarray}
A&=&\frac{\pi i e^{-\pi\mu} }{4 \sqrt{\omega_k(t_*)}}\left[z_*\bar H^{(1)}_{i\mu}{}'(z_*)+ \frac{\omega_k(t_*)}{\epsilon} i \bar H^{(1)}_{i\mu}(z_*)\right] \\
B&=&-\frac{\pi i e^{-\pi\mu} }{4 \sqrt{\omega_k(t_*)}}\left[z_* H^{(1)}_{i\mu}{}'(z_*)+ \frac{\omega_k(t_*)}{\epsilon} i  H^{(1)}_{i\mu}(z_*)\right]~
\end{eqnarray}
The mode functions that determine the ``in'' vacuum at the beginning of the dark energy era, $u_k^\text{in}(t)$,  are given by these expressions for $A, ~B$ with the choice $t_*=0$. The mode functions  $u_k^\text{out}(t)$, that determine the ``out'' vacuum at some later time, $T$,  are given by the same formulae but with $t_*=T$. As usual, the two can be related by  a Bogoliubov transformation, of the form $u_k^\text{in}(t)=\alpha_k (T) u_k^\text{out}(t)+\beta_k(T) \bar u_k^\text{out}(t)$.  At time $T>0$, the true vacuum state differs from the initial vacuum state, and so the latter contains particles. As explained in \cite{Mukhanov:2007zz}, the mean particle number density for modes of momentum $\bf k$ is $N_k(T)=|\beta_k(T)|^2$ with corresponding energy density  $E_k(T)=\omega_k(T)N_k(T)$.  When we calculate this explicitly, it turns out that 
\be
N_k(T)=\frac{\pi^2}{16} e^{-2\pi\mu} \epsilon^2\left[X^2+Y^2\right]
\ee
where 
\be
X= \text{Im} \left[ \frac{ z_\text{in} H^{(1)}_{i\mu}{}'(z_\text{in} )}{\sqrt{\omega_k(0)}}\frac{ z_\text{out} \bar H^{(1)}_{i\mu}{}'(z_\text{out} )}{\sqrt{\omega_k(T)}} \right.  \left. - \frac{ \sqrt{\omega_k(0)} H^{(1)}_{i\mu}(z_\text{in} )}{\epsilon}\frac{\sqrt{\omega_k(T)} \bar H^{(1)}_{i\mu}(z_\text{out} )}{\epsilon} \right]
\ee
and
\be
Y= \text{Im} \left[ \frac{ z_\text{in} H^{(1)}_{i\mu}{}'(z_\text{in} )}{\sqrt{\omega_k(0)}} \frac{\sqrt{\omega_k(T)} \bar H^{(1)}_{i\mu}(z_\text{out} )}{\epsilon} \right.  \left. - \frac{ \sqrt{\omega_k(0)} \bar H^{(1)}_{i\mu}(z_\text{in} )}{\epsilon} \frac{ z_\text{out} \bar H^{(1)}_{i\mu}{}'(z_\text{out} )}{\sqrt{\omega_k(T)}}\right] 
\ee
Here $z_\text{in}=\frac{nM_{KK}}{\epsilon}$ and  $z_\text{out}=\frac{nM_{KK}}{\epsilon}e^{-\epsilon T}$.  We can obtain estimates for the energies stored in different momentum modes by using the approximations for Bessel and Hankel functions given in \cite{AS}. In particular,  we note that for large $ z \gg \mu^2+1$,  we have the following asymptotic expansion 
\be \label{approx1}
 H^{(1)}_{i\mu}(z) \approx
 \sqrt{\frac{2}{\pi z}} e^{\frac{\mu\pi}{2}+i\left(z-\frac{\pi}{4}\right)}\left[1+\frac{4 \mu^2+1}{8iz}+\ldots \right]
\ee
To obtain an approximation in the opposite limit,  for small $z$, we recall that $H^{(1)}_{i\mu}(z)=\frac{e^{\mu \pi} J_{i\mu} (z)-J_{-i\mu}(z)}{\sinh\mu\pi}$ and use the fact that 
\be \label{approx2}
J_{i\mu} (z) \approx \frac{e^{i\mu \ln \frac{z}{2} }}{ \Gamma(1+i\mu)}\left[1-\frac{z^2}{4(1+i\mu)}+\frac{z^4}{32(1+i\mu)i\mu}+\ldots\right]
\ee
whenever   $0<z \ll \sqrt{|\mu|+1} $.

\begin{figure}[h]
\includegraphics[width=\textwidth]{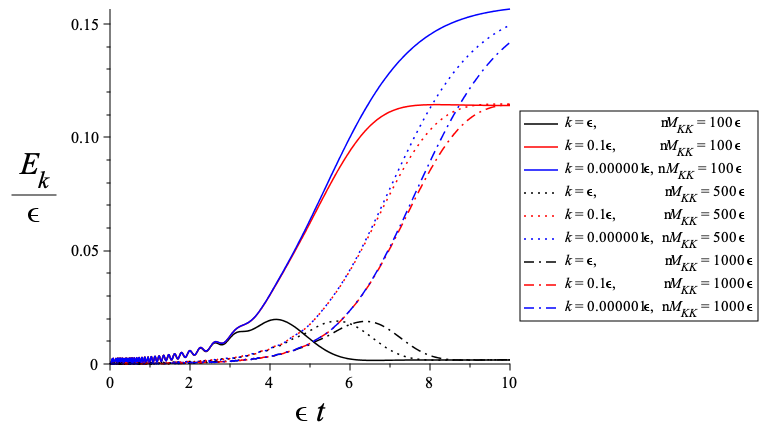}
\caption{Plot of energy versus time for  a range of scalar modes of different momentum  and  different values for the initial mass. These particular  plots were   produced using Maple 2020 and a numerical value of $\epsilon=0.01$. }\label{fig1}
\end{figure}

For high momentum modes, with $k \to \infty$, we find that
 $E_k \sim \frac{n^4 M_{KK}^4  \epsilon^2 }{4k^5}\left[\cosh^2\epsilon t-\cos^2kt\right]e^{-2\epsilon t}$. 
% $E_k \sim \frac{n^4 M_{KK}^4  \epsilon^2 }{k^5}\left[(1+e^{\epsilon t})^2-4\cos^2kt\right]e^{2\epsilon t}$. 
As expected, these modes are suppressed, being insensitive to the change in the mass of the field.  For modes of lower momentum, it is instructive to display the changes in the energy stored in each mode in a characteristic plot, such as the one shown in FIG. \ref{fig1}.  
The plot  shows the energy profile for modes running over a range of different momenta. We also vary over the initial mass of the scalar, $nM_{KK}$, or equivalently, over the level in the Kaluza-Klein tower.  For the $\varphi_n$ particles, at level $n$,  we see that the total energy density at late times is dominated by modes of momentum $k\lesssim \epsilon$, with significant particle production kicking in at a time $t_n \sim \frac{1}{\epsilon}\ln \left(\frac{nM_{KK}}{\epsilon}\right)$.  Using \eqref{approx1} and \eqref{approx2}, we can estimate  the energy stored in the low energy modes analytically, for $t \gg t_n$.  Since $nM_{KK} e^{-\epsilon t}  \ll \epsilon \ll nM_{KK}$ and $k \lesssim \epsilon$, we have $z_\text{out} \ll 1 \ll z_\text{in}$ and  $\mu \lesssim 1$, so that the energies approximate as
\be
E_{k} \approx 
\frac{k e^{-\frac{k\pi}{\epsilon}}}{2 \sinh\left(\frac{k\pi}{\epsilon}\right)}
\ee
In deriving this, we have used the relation $|\Gamma(i\mu)|^2 =\pi/\mu \sinh \mu\pi$, which follows from Euler's reflection formula and the fact that $\Gamma(-i\mu)=\overline{\Gamma(i\mu)}$. To obtain an estimate for the total  energy density stored in $\varphi_n$ particles at late times ($t \gg t_n$), we simply  integrate this result over all momentum. The result is
\be
\rho_n =\int d^3 {\bf k} E_k\approx \frac{\pi}{60}\epsilon^4
\ee
By itself, this would only have a tiny effect on the cosmological background, since $\epsilon^4 =(\alpha\beta)^4 H_0^4$, far below the scale of the critical density of the universe during the dark energy era, $M_{Pl}^2 H_0^2$. Of course, the formula will receive additional  corrections from the fact that the fields are actually propagating on a dynamical cosmological background, as opposed to Minkowski, although these are likely to be similarly suppressed, especially if $\alpha\beta \gtrsim 1$. Nevertheless,  by the time we have reached $t_N \sim \frac{1}{\epsilon}\ln \left(\frac{NM_{KK}}{\epsilon}\right)$ for some large $N$, we have started to produce particles for each of the first $N$ levels in the Kaluza-Klein tower.  The {\it total} energy density of all these particles, in this approximation,  is given by
\be
\rho_\text{total} (t \sim t_N)=\sum_{n=1}^N \rho_n \approx \frac{N\pi}{60}\epsilon^4
\ee
This will start to affect the cosmological dynamics for $N\sim N_\text{crit}$ where $ N_\text{crit} \sim \frac{M_{Pl}^2}{\epsilon^2}$, at a time $t_\text{crit} \sim \frac{1}{\epsilon}\ln \left(\frac{N_\text{crit}M_{KK}}{\epsilon}\right)$.  If we assume $\alpha\beta \sim \mathcal{O}(1)$, then $\epsilon \sim H_0$ and so $N_\text{crit}\sim 10^{120}$. Assuming $M_{KK} \lesssim M_{Pl}$, so that $N_\text{crit}  \gg\sqrt{N_\text{crit}}  \gtrsim \frac{M_{KK}}{\epsilon}$, we see that  $t_\text{crit}\lesssim \mathcal{O}(400)/H_0$, confirming our expectation that the dark energy era will not last beyond a few hundred or so Hubble times, before the space begins to decompactify.   If we lower the underlying Kaluza-Klein scale, $M_{KK}$, or consider models with $\alpha\beta>1$, the dark energy era is cut short even earlier. 

By the time dark energy succumbs to the extra dimensions, a huge number of particles species have already entered the low energy effective theory. We might be worried that this creates a ``species problem" at some earlier time, lowering the scale at which gravity becomes strongly coupled. However, for $N<N_\text{crit}$ species, the scale of strong coupling scales as $\Lambda_{QG} \sim M_{Pl}/\sqrt{N}>H_0$. In  other words,  four-dimensional gravity does not become strongly coupled on cosmological scales prior to $t_\text{crit}$, at which point the effective four-dimensional description breaks down anyway. 

As emphasized earlier, everything we are saying is only relevant to cosmological dynamics. On shorter scales, for example in the lab or in the solar system, the  quintessence field may be displaced from its cosmological value, so much so that the Kaluza-Klein tower remains heavy and decoupled from the low energy physics.  Indeed, if quintessence is coupled to matter with gravitational strength, such a displacement is likely to be necessary in order to avoid fifth force constraints \cite{astro-ph/0309300, astro-ph/0309411, 1511.05761}.  It would certainly be interesting to investigate the implications of this in more detail. 

\section{Further discussion} \label{sec:disc}
 
In this chapter, we have argued that string theory compactifications consistent with swampland constraints automatically prevent the dark energy era from extending beyond a few hundred Hubble times. This alleviates the coincidence  problem to some degree, inasmuch as we now have as much as a percentage chance of finding ourselves in the first e-fold of acceleration.  Our conclusions draw on two key features of string compactifications: (i) the scarcity (absence?) of stable de Sitter vacua, suggesting that most dark energy models will be driven by a quintessence field in slow roll \cite{1806.08362, 1807.05193, 1810.05506}; (ii) the accumulation of a large number light states as the field rolls off towards infinity  \cite{hep-th/0605264, 1610.00010, 1802.08264, 1803.04989}. Generically, the dynamics is such that the descending  tower of light states induces a cosmological phase transition, bringing the dark energy era to a conclusion.  We were able to demonstrate this explicitly with a toy model, whereby particle creation in the Kaluza-Klein sector starts to overwhelm the cosmological background within a few hundred Hubble times\footnote{We note that \cite{1909.11063} mentions a similar idea regarding the application of Swampland constraints to the coincidence problem. There, the authors claim that the trans-Planckian censorship conjecture (TCC) fixes the critical time to $\mathcal{O}(H_0)$. However, the analysis of \cite{2110.07300} seems to point towards either the TCC being incorrect or not being able to bound de Sitter lifetimes as claimed in \cite{1909.11063}.}.

From this stringy perspective, it seems that the  cosmological coincidence is readily reduced to one part in $\mathcal{O}(100)$ but can we do any better? Can we exploit the distance conjecture to bring the odds even further in? One speculative possibility is to consider dark energy a consequence of the clockwork dynamics of \chpref{chap:clockwork}. The clockwork set-up is brittle. After the dark energy field has moved just a single Planck unit, we could imagine a tower of states contaminating the low scale dynamics and spoiling the clockwork symmetry. Without the clockwork, the dark energy dynamics should also be spoilt. Whilst this idea is appealing, to be able to say anything with any degree of certainty, we would require a clockwork embedding in supergravity, which we have not been able to achieve so far.

Finally, it is natural to ask what would happen if we applied the same analysis to early universe inflation at a much higher scale than quintessence.  Because the Kaluza-Klein tower does not need to descend as far to contaminate the inflationary background, the inflaton excursion is limited to just a few Planck units. As a result, we would conclude that, generically, inflation should not extend beyond a few efolds. This seems problematic since inflation must last for at least 50 efolds in order  to address the horizon problem. Of course, these concerns aren't new - it is well known that early universe inflation is in some tension with the swampland constraints.

\blankpage
\chapter{Intermezzo: Supergravity and the Swampland\label{chap:intermezzo}}

In the first half of this thesis, we have discussed stringy motivated models and EFTs without making explicit a UV completion. For a consistent EFT treatment, this has the benefit of allowing for concrete and analytical treatment of the four-dimensional low-energy effective field theory by decoupling the microphysics of gravity.

However, this decoupling comes at a price. Without any intuition for what the microphysics are doing, it is hard to argue that the EFT is always well-posed. For example, consider the cosmological constant, a super-IR observable that we might expect to be UV independent. We have seen that the value of the cosmological constant, given by the minimum of the four-dimensional scalar potential, in a string compactification is directly affected by the stabilisation of the moduli fields, which are UV degrees of freedom. Arbitrarily exploring the parameter space of the EFT could, in principle, displace the moduli fields and change the value of the cosmological constant. In extreme cases, this displacement could lead to full destabilisation of the theory and to a break down of the validity of the EFT. Strikingly, this destabilisation might not look pathological on the EFT side and only become a clear problem when considering the full ten-dimensional theory. This kind of heuristic argument about computational control is what motivates, at least in part, the Swampland programme. If true, some of the Swampland arguments, like those in \ssecref{ssec:dsc}, can become truly problematic. Indeed, in \secref{sec:no_go_quint} we show that we are unable to produce a slow-roll regime in the asymptotics of moduli space, implying that there is a disconnect between string theory and cosmology.

In this chapter, we attempt to make more concrete statements regarding lack of computational control and to what extent we can extract phenomenological information from the Swampland programme, that are most relevant to the content of this thesis. A number of authors have explored ideas in the same vein, for example \cite{2008.12399} studies the obstacles to generating de Sitter vacua in string theory and its relations with the Swampland and \cite{1807.09698} tries to study the formal validity of some common Swampland related lore.

\section{A simple example of the backreaction on moduli space}
 Let us exemplify this situation by studying one of the models in \cite{1703.05776}. The type IIB toy model of interest consists of only a \kah modulus $T=\tau + i \theta$ and the axio-dilaton $S=s+i\sigma$ with a \kah potential and superpotential given by
\be
K = -\log(S+\bar{S}) - 3\log(T+\bar{T})\ , \qquad W= - if + i h S + i q T\ ,
\ee
where we note that $f$ is a non-geometric flux which, in \cite{1703.05776}, is needed for moduli stabilisation purposes. Although there are a vast number of models out there with more complicated compactifications that we may argue are under better control, we choose to concentrate on this simple example since it contains the minimal set of ingredients necessary to see the backreaction of large field excursion onto the theory in an analytical manner. 
 
The scalar potential for the four-dimensional effective theory is calculated from \eqref{eq:scalar_pot_1} and given by \cite{1703.05776} 
\be
V= {(hs+f)^2\over 16s\tau^3} - {6hqs-2qf\over 16s\tau^2}  - {5q^2\over 48s\tau}  +{\phi^2\over 16s\tau^3} \ ,
\ee
where $\phi:=q\theta+h\sigma$. We are interested in a non-supersymmetric minimum for the scalar potential. The non-SUSY minimum is fixed by
\be
\partial_{I}V=0\ , \label{eq:eq_pot}
\ee
where $I$ runs over the different moduli fields. Solving this set of equations, we find a non-SUSY AdS minimum\footnote{Note that the orthogonal direction to $\phi$ remains as a flat axionic direction as it does not appear in the potential.}
\be
s_0=\frac{f}{h}\ , \qquad \tau_0=\frac{6f}{5q}\ , \qquad \phi_0:=q\theta_0+h\sigma_0 = 0 \ , \qquad V_0=-\frac{25hq^3}{216f^2}\ , \label{eq:minimum}
\ee
with a cosmological constant $V_0$. We would like to consider $\phi$ as our candidate inflaton field and see what is the effect of the back-reaction of the rolling inflaton field. We are interested on the moduli dependence with $\phi$ when we displace it from its minimum $\phi_0$. To find $s(\phi)$ and $\tau(\phi)$, we relax \eqref{eq:eq_pot} and solve
\begin{gather}
\begin{cases}
  \partial_s V = 3h^2s^2 + 5q^2\tau^2 - 6fq\tau - 3f^2 - 3\phi^2 = 0  \\
  \partial_\tau V = 5q^2\tau^2 + 12q(3hs-f)\tau - 9\left(hs+ f\right)^2 - 9\phi^2 = 0
\end{cases}\ ,
\end{gather}
while allowing $\partial_\phi V\neq0$. From here, we find 
\be
s(\phi)=\frac{f}{h}\sqrt{1+\frac{5}{8}\left(\frac{\phi}{f}\right)^2} \ , \qquad \tau(\phi)= \frac{6f}{10q}\left(1+\sqrt{1+\frac{5}{8}\left(\frac{\phi}{f}\right)^2}\right)\ .
\ee
When $\phi$ becomes of order $\phi_c=\sqrt{\frac{8}{5}}f$, the back-reaction of the inflaton field is substantial and cannot be ignored. This supports the idea that naive displacements away from the minimum have to be bounded in order to maintain the validity of the EFT description. To derive a physically meaningful quantity, we concentrate on finding the critical value for the canonically normalised inflaton field.

We note that the canonically normalised inflaton is given by
\be
\Phi=\int \sqrt{K_{IJ} \frac{\partial m^I}{\partial \phi}\frac{\partial m^J}{\partial \phi}}d\phi=\int \left(n_I K^{IJ} n_J\right)^{-\frac{1}{2}}d\phi\ ,
\ee
where $m^I$ stands for the moduli fields with fluxes $n_I$
\be
\phi=\sum_I n_I m^I\ . 
\ee
Plugging back our scenario we find for the subcritical regime $\left(\phi<\phi_c\right)$
\be
\Phi(s,v)=\sqrt{\frac{3}{3h^2s^2+q^2v^2}}\,\phi \ , \qquad \Phi(s_0,v_0)=\frac{5}{\sqrt{37}}\frac{\phi}{f}\ .
\ee
For this simple model, we can see that there is a critical value for the canonical inflaton $\Phi_c=\sqrt{\frac{40}{37}}\simeq1.03$ which is flux independent and its value is around one Planck unit, {i.e.} it cannot be tuned by fluxes. 

These results are in agreement with the refined Swampland distance conjecture in \secref{sec:swamp}. Finally, to make contact with the light tower of states we note that
\be
M_{\text{KK}}\sim\frac{1}{\tau}\sim\frac{q}{f}e^{-\frac{\gamma}{2}\Phi}\ ,
\ee
which drop off exponentially fast as claimed by the refined conjecture. 

In this example, we have considered a type IIB toy model without complex structure moduli, with a flat direction $\sigma$ and a non-geometric flux $f$. Although one might be tempted to criticise this particular model for its simplicity, we find it more enlightening to take it as an example of a larger class of models that belong to the swampland which share some common ideas behind them. We discuss some of these ideas below.

\section{Calculations at the boundary of moduli space\label{sec:no_go_quint}}
The strict limit to the boundary of moduli space, thought of as the limit of the dilaton or the volume modulus going to infinity, is a limit to ten-dimensional Minkowski. Given that our Universe is not Minkowski, or ten-dimensional for that matter, this makes drawing conclusions for phenomenology in this limit somewhat useless. 

In the following, we will prove  that quintessence is not possible in the runaway region arbitrarily close to the boundary of moduli space, because one cannot satisfy the slow roll constraints. Related conclusions were drawn in \cite{0711.2512, 1810.09406, 2005.10168}, here we build on these works by analysing simultaneously the K\um{a}hler-dilaton sectors, by highlighting the role of supersymmetry and by considering corrections to the moduli potentials. Further difficulties with quintessence model  building in string theory were also discussed in \cite{1909.08625}. In the following we discuss the absence of a slow roll region in the runaway of type II and heterotic supergravities. For a comprehensive review of the pros-and-cons of the different theories regarding other model building aspects see \cite{1808.08967}.

To proceed with our proof, let us approach the boundary of moduli space in a string theory compactification on some Calabi-Yau threefold, $\x$.  In the strict boundary limit, we should think of the dilaton or the volume moduli going to infinity.  As is well known, such a limit is expected to bring down an infinite tower of light states,  spoiling the effective field theory description \cite{hep-th/0605264, 1610.00010, 1802.08264, 1803.04989}. In particular, the asymptotic limits of the volume moduli bring down either Kaluza-Klein modes, consistent with a decompactification of the internal space, or winding modes, consistent with decompactification of the dual. For the asymptotic limits of the dilaton, the tower of light states is less clear --- it has been suggested that this can introduce a tower of tensionless strings and domain walls  \cite{1904.05379}.

Since the physics at the boundary is clearly not phenomenologically viable, let us relax the limit and consider what happens when the dilaton or some \kah modulus becomes arbitrarily large but finite. In this case, the effective theory corresponds to tree-level supergravity close to the boundary, where the moduli fields are in a runaway regime. We are interested in theories that might admit no four-dimensional vacuum, or potentially a supersymmetric Minkowski or AdS vacuum somewhere in the bulk of moduli space. The supersymmetric requirement serves to align our discussion with the usual swampland lore of non-supersymmetric vacua being unstable \cite{1610.01533}. It follows from the arguments given in \cite{1810.05506} that de Sitter vacua cannot arise in any parametrically controlled regime of the moduli space, and our only hope to achieve phenomenologically viable models of dark energy is to consider runaway quintessence. For example, type IIB flux compactifications feature at tree-level supersymmetric Minkowski vacua where the K\"ahler moduli are flat directions due to the underlying no-scale structure \cite{Burgess:2020qsc}. Leading order $\alpha'$ corrections, or a non-supersymmetric stabilisation of the dilaton and the complex structure moduli, can then generate this type of de Sitter runaway for large volume.
 
{In the following we will assume that there exists some mechanism to fix the complex structure moduli at tree-level in a controlled manner}\footnote{Or the \kah moduli in case of the type IIA discussion.}, while the remaining moduli fields correspond to the axio-dilaton $S=s+i\alpha$ and the \kah moduli $T^a=\tau^a+i \theta^a$, which are identified with either two-cycles or four-cycles, depending on which is most convenient. In the runaway regime, close to the boundary,  the  dynamics of these moduli is governed by the following \kah potential in its tree-level sum separable form
\be 
K = -p\ln(\mathcal{V}) - \ln (S+\bar{S}) + K_0\ , \label{eq:kah_pot}
\ee
and a superpotential {$W$ to be specified for each supergravity}. We have denoted the internal volume  by $\vol(\tau^a)$. For any Calabi-Yau threefold $\x$, the volume is a homogeneous function of degree $3/2$ in the \kah moduli for four-cycles, or equivalently, of degree $3$ in the \kah moduli for two-cycles. Furthermore, for type IIB supergravity, we have  $p=2$, and work with four-cycles, while for type IIA and heterotic supergravities, we have $p=1$, and work with two-cycles. As already stated, the complex structure contribution to the \kah potential, $K_0$, is considered to be fixed. For further details, see \cite{hep-th/0507153, Cicoli:2013rwa}.

The Lagrangian for the scalar moduli is given  by 
\be
\mathcal{L} = K_{S\bar{S}} dS\wedge \star d\bar{S} +K_{a\bar{b}} dT^a\wedge \star dT^{\bar{b}} - V\ , \label{eq:lagrangian_gen}
\ee
where  $K_{I\bar J}=\del_I \del_{\bar J} K$ is the \kah metric for moduli space, which given equation \eqref{eq:kah_pot} is block diagonal. The scalar potential is obtained by computing 
\begin{equation} \label{eq:Fterm_V}
V=e^K (K^{I \bar J} D_I W D_{\bar J} \bar W-3 |W|^2)\ ,
\end{equation}
where $D_I W=\del_I W+W\del_I K$ is the \kah covariant derivative, and $I$ runs over all the moduli of the theory.

At the perturbative level, the axions, $\theta^a=\Im (T^a)$ and $\alpha$, do not contribute to the scalar potential and cannot play the r\^ole of runaway quintessence. Focusing instead on the saxions/moduli, $\tau^a=\Re (T^a)$ and $s$, we find that the relevant part of the Lagrangian is now given by
\be
\mathcal{L} = \frac{1}{4s^2} (\del s)^2 +K_{a\bar{b}} \del \tau^a \del \tau^b - V(s, \tau^a)\ , \label{eag2}
\ee
A necessary condition for phenomenologically viable quintessence is the existence of a slow roll regime, defined by the condition $\epsilon :=-\dot H/H^2< 1$ where $H$ is the Hubble parameter and  dot denotes differentiation with respect to proper cosmological time. For a generic multiscalar theory described by a Lagrangian  $\mathcal{L}=\frac12 Z_{IJ} \del \phi^I \del \phi^J- V(\phi^I)$ one can show in the limit of vanishing acceleration, $\ddot{\phi}^I=0$, the dynamics of the system is described by  
\be 
H^2\approx  \frac13 V \qquad\text{and}\qquad  3H \dot \phi^I +\Gamma^{I}_{J K} \dot{\phi}^J \dot{\phi}^K \approx  - Z^{IJ}\partial_{\phi^J} V, 
\ee
where $Z^{IJ}$ is the inverse of the field space metric and  $\Gamma^{I}_{J K}$ is the corresponding metric connection. These equations admit two distinct classes of slow roll solutions, whose existence and stability has been analysed in the context of dark energy in \cite{Cicoli:2020cfj}.
In this note we will focus on the cases when $3H \dot \phi^I \gg \Gamma^{I}_{J K} \dot{\phi}^J \dot{\phi}^K\rightarrow 0$, for which
\be
\dot H \approx \frac{\left(\del_{\phi^I} V \right) \dot \phi^I}{6H} \approx -\frac{(\del_{\phi^I} V) Z^{IJ} (\del_{\phi^J} V)}{18H^2}
\,,
\ee
and so
\be \epsilon =-\frac{\dot H}{H^2} \approx \frac12  \left(\del_{\phi^I} \ln V\right) {Z^{IJ}} \left( \del_{\phi^J} \ln V\right)\,.
\ee
In this regime, the requirement of quasi-de Sitter expansion implies the need for flat (multifield) scalar potentials. The other class of accelerating solutions, that we will not analyse here, forego the flatness of the  potential in exchange for large field space curvature, see \cite{Cicoli:2020cfj,Cicoli:2020noz}.\footnote{Strictly speaking the requirement is that $\Gamma^I_{JK} \dot{\phi}^J \dot{\phi}^K \gg 3 H \dot{\phi}^I$ for one of the scalars,  which can be achieved even in flat field space if, for instance, one uses polar coordinates as in \cite{Boyle:2001du}.} While promising, this avenue is not without challenges from a string model building perspective \cite{Brinkman:2022}.  

For the supergravity compactifications under consideration  \eqref{eag2}, the first slow roll parameter is given by $\epsilon=\epsilon_s+\epsilon_\vol$, where the dilaton contribution is
\begin{equation}
\epsilon_s=  s^2 \left(\del_s \ln V \right)^2\ ,
\end{equation}
and the \kah contribution is
\be
\epsilon_\vol=\frac14 (\del_{\tau^a} \ln V)   K^{a \bar b} (\del_{\tau^b} \ln V)\ ,
\ee
Throughout this section, we will make regular use of Euler's theorem and its corollary: if $P_n$ is a homogeneous function of degree $n$ in $\tau^a$, then $$\tau^a \del_{\tau^a} \ln P_n=n\ , \qquad \tau^a \del_{\tau^a} \del_{\tau^b}\ln P_n=-\del_{\tau^b} \ln P_n\ .$$ In particular, this allows us to infer the following conditions on the derivatives of  the \kah potential  
\be
K^{a \bar b} \del_{\bar T^b} K=-2 \tau^a\ , \qquad 
 (\del_{T^a} K)   K^{a \bar b} (\del_{\bar T^b} K) =3\ ,
 \ee
where the indices run over the \kah moduli only.
The latter result holds as long as we work with four-cycle \kah moduli in type IIB supergravity, and two-cycle \kah moduli in type IIA and heterotic.
 
Let us now derive the detailed form of the tree-level scalar potential, and the corresponding slow roll parameter, for the type II and heterotic supergravities, in turn. We will see that the slow roll condition $\epsilon<1$ can never be satisfied in the runaway regions close to the boundary of moduli space. 

\subsection{The type IIB runaway}
\label{sec:no_go_quint}

Let us begin with type IIB supergravity (for further details, see \cite{hep-th/0507153}). In this instance, at tree-level, the superpotential is linear in the dilaton $W=h_0 S + f_0$, with $h_0$ and $f_0$ set by the three-form fluxes, respectively $H_3$ and $F_3$, that stabilise the complex structure moduli supersymmetrically. Furthermore, the \kah moduli $T^a=\tau^a +i\theta^a$ that enter the \kah potential through the volume will be identified with four-cycles, so that the volume is a homogeneous function of degree $\frac32$ in the corresponding saxions, $\tau^a$. Using the form of the F-term scalar potential given by equation \eqref{eq:Fterm_V} and the \kah metric \eqref{eq:kah_pot} with $p=2$, we obtain a scalar potential 
\be
V= {e^{K_0}\over 2s\vol^2} |h_0 \bar S-f_0|^2\ ,
\ee
where $\bar S=s-i\alpha$. If $h_0\neq 0$, the imaginary part of the axio-dilaton can be stabilised at the point where $\del_\alpha V=0$, or equivalently, $\alpha=-\Im \left( \frac{f_0}{h_0}\right)$, so that the scalar potential becomes
\be \label{potIIB}
V(s,\vol) = {e^{K_0}\over 2s\vol^2}|h_0|^2 \left(s- \Re \left(\frac{f_0}{h_0}\right)\right)^2 \ .
\ee
It follows that
\be
\epsilon_s=\left( \frac{ s+ \Re \left(\frac{f_0}{h_0}\right) }{  s - \Re \left(\frac{f_0}{h_0}\right)} \right)^2
, \qquad \epsilon_\vol=3\ ,
\ee
defined whenever $V \neq 0 $. We immediately infer that $\epsilon=\epsilon_s+\epsilon_\vol \geq 3$, ruling out slow roll in the runaway regime. Notice that vacua with $h_0=0$ feature exactly $\epsilon_s=1\,\Rightarrow\,\epsilon=4$.

For type IIB supergravity, if $h_0\neq 0$, it is possible to further stabilise the dilaton at the location of the supersymmetric minimum, $s=\Re \left(\frac{f_0}{h_0}\right)$,  where the leading order potential for the \kah moduli vanishes. When this happens, we should consider additional perturbative and non-perturbative corrections to the potential. In order to be compatible with a robust large volume expansion, these go as \cite{Burgess:2020qsc, AbdusSalam:2020ywo, Cicoli:2021rub}
\be \label{pot}
V = {\mathcal{A}\over \vol^{2+p}} + {\mathcal{B}\, e^{-f}\over \vol^{2+q}}+\frac{\mathcal{C}}{\vol^{2+r} g^n}\ldots\ ,
\ee
where $\mathcal{A}$, $\mathcal{B}$ and $\mathcal{C}$ will in general depend on the stabilised dilaton and complex structure moduli, while  $f>0$ and $g>0$ are homogeneous functions  of degree one in the $\tau^a$. In particular, $f$  corresponds to the dominant saxion in the non-perturbative expansion. In order for us to trust the large volume expansion, we require each of these terms to scale away more quickly than the leading order $1/\vol^2$ term, which happened to vanish once the dilaton was stabilised {at its supersymmetric minimum}. This implies that   $p> 0$, $ q \geq 0$, and $r+\frac{2n}{3}>0$, suggesting that the volume direction is made even steeper and the volume modulus will not slow roll in this case either. To see this explicitly, we compute the slow roll parameter in the asymptotic regime where only one of the terms dominates, in accordance with our definition of the boundary of moduli space. If $V \sim  {\mathcal{A}\over \vol^{2+p}}$, we find that $\epsilon \sim 3 \left(1+\frac{p}{2}\right)^2>3$.  In contrast, if $V \sim  {\mathcal{B}\, e^{-f}\over \vol^{2+q}}$, we   find  that 
\be
\epsilon=3 \left(1+\frac{q}{2}\right)^2 +f   (2+q)+\frac14 K^{a \bar b} f_a f_b  \geq  3\ ,
\ee
where the inequality follows from the fact that $f, q \geq 0$ and $K^{a \bar b}$ being a symmetric matrix with all positive eigenvalues.  Finally, if $V \sim \frac{\mathcal{C}}{\vol^{2+r} g^n}$ we find that 
\be
\epsilon=3 \left(1+\frac{r}{2}\right)^2 +n  (2+r)+\frac{n^2}{4} K^{a \bar b} (\ln g)_a (\ln g)_b  \geq  3\ .
\ee
In each  case we see that slow roll is impossible in this asymptotic region, where a single one of the corrections in equation \eqref{pot} dominates. It is interesting to note that leading order supersymmetry, imposed for the sake of the stability of the compactification, only aggravates the problem, by cancelling off the $1/\vol^2$ term that would otherwise dominate the potential at the boundary of moduli space and replacing it with a steeper term.

Alternatively, we could consider a non-supersymmetric stabilisation of the complex structure sector. This would induce a correction to \eqref{potIIB} of the form $\lambda/s\vol^2$, where $\lambda $ is a positive constant proportional to the F-terms of the complex structure moduli. The vacuum expectation value of the dilaton is then shifted to a non-supersymmetric value leaving a $1/\vol^2$ runaway for the volume mode, which is too steep to give rise to slow roll. Note further that if we move into the bulk of moduli space, we might hope to stabilise the volume at some fixed value and achieve slow roll along some other saxion direction, through the last term in \eqref{pot}. This is indeed possible, although such a scenario runs into further difficulties associated with a light gravitino \cite{Cicoli:2021skd}. It is also possible that interference between two terms in equation \eqref{pot} can give rise to a (short) field range for the volume where $\epsilon< 1$. This avenue is likely to involve considerable tuning and, by definition, lies in the bulk of moduli space and therefore will not be analysed in this chapter.

\subsection{The type IIA runaway}

We now turn our attention to the type IIA runaway  (for further details, see \cite{hep-th/0507153}). The superpotential now contains RR fluxes $\left(e_0, e_a, q^a\right)$, the (3,0)-component of the $H_3$ flux, $h_0$, and the Romans mass, $m$ \cite{Romans:1985tz}:
\be
W= e_0 + e_a T^a +{1\over 2}\mathcal{K}_{abc}q^a T^b T^c + {m\over 6}\mathcal{K}_{abc} T^a T^b T^c -i h_0 S\ .
\ee
Here the \kah moduli $T^a=v^a+i\theta^a $ are identified with two-cycles and the volume is a homogeneous function of degree $3$ in the corresponding saxions, $v^a$. In particular,  $\vol=\frac16\mathcal{K}_{abc} v^a v^b v^c$, where $\mathcal{K}_{abc}$ are the triple intersection numbers for the Calabi-Yau. Without additional fluxes or corrections to the tree-level action, the complex structure sector remains flat. To continue the discussion on the slow roll regime, we must therefore assume that some other mechanism under computational control exists to stabilise the complex structure sector, so that, at tree-level, it only enters through some constant in the \kah potential, $K_0=\int_\x \Omega\wedge\bar{\Omega}$.

The tree-level scalar potential for the type IIA theory is then given by
\be
V = {e^{K_0}\over 2s \vol}\left[4|h_0|^2s^2 + K^{a\bar b}\rho_a \bar \rho_b +4s  \Im (W \bar h_0 ) - 4\tau^a\Re (W \bar \rho_a ) +|W|^2 \right]\ ,
\ee
where 
\be
\rho_a=\del_{T^a}W=e_a +\mathcal{K}_{abc}q^b T^c+\frac{m}{2}\mathcal{K}_{abc} T^b T^c\ .
\ee
We can always absorb the vacuum expectation value of the axions into a redefinition of the fluxes. Therefore, without loss of generality, we set the axions to vanish.  With the axions gone, the superpotential can be written as
\be
W=e_0+e_a v^a+\vol (q^a \omega_a +m) -ih_0s\ ,
\ee
and
\be
\rho_a=e_a+\vol \left[ q^b ( \omega_{ab}+\omega_a \omega_b )+m\omega_a\right]\ ,
\ee
where $\omega_a= \del_{v^a} \ln \vol$ and $\omega_{ab}= \del_{v^a}\del_{v^b}  \ln \vol$. Keeping only the leading order  \kah terms at large volume, we obtain a scalar potential that goes as 
\be
V ={e^{K_0}\over 2s \vol} \left[ |h_0|^2s^2  +14 \Im (m \bar h_0) \vol s +|m|^2 \vol^2 \right] +\ldots
\ee
We now compute the slow roll parameter. Since the scalar potential is a function of $\vol/s$ to leading order in this asymptotic regime, we have that $\epsilon_\vol=3 \epsilon_s$. Furthermore, 
\be
\epsilon_s=	\left[ \frac{|h_0|^2-|m|^2 \frac{\vol^2}{s^2}}{ |h_0|^2  +14 \Im (m \bar h_0) \frac{\vol}{ s} +|m|^2 \frac{\vol^2}{s^2}  } \right]^2\ .
\ee
We remark that, in order to keep $\alpha'$ corrections from becoming important, we require\footnote{In units of $\alpha'=1$.}
\be
\vol_{\text{S}}\gg 1\ ,\label{eq:alpha_con}
\ee 
where $\vol_{\text{S}}$ is the volume of \x in the string frame. After expressing the Einstein frame volume in string frame, $\vol=s^{3/2}\vol_{\text{S}}$, we see that $\vol/s = \sqrt{s} \vol_{\text{S}}\gg 1$ to be consistent with \eqref{eq:alpha_con} and weak coupling. Thus, in the runaway regime, the ratio $\vol/s$ has to be very large and it follows that $\epsilon_s  \approx 1$, and so $\epsilon=\epsilon_s+\epsilon_\vol\approx 4$, ruling out slow roll. 

\subsection{The heterotic runaway\label{ssec:heterotic_runaway}}

We finish with the heterotic runaway, where the tree-level superpotential does not depend on the dilaton or the \kah moduli  (for further details see \cite{Cicoli:2013rwa}), so that $W=W_{\rm cs}$. The complex structure moduli are assumed to lie in their supersymmetric vacuum by the vanishing of the corresponding F-terms. As a result, the scalar potential becomes  a runaway in the dilaton and the volume modulus
\be
V=e^{K_0} {\left| W_{\text{cs}}\right|^2\over 2s\vol}\ .
\ee
Computing the slow roll parameters, we find that  $\epsilon_s=1$ and $\epsilon_\vol=3$, and so $\epsilon=4$.  Clearly slow roll cannot be achieved in the heterotic runaway.

We have found that the tree-level type II and heterotic supergravities cannot contain a slow roll region in a parametrically controlled regime. The result is proven for any number of \kah moduli. It is clear that obtaining a slow roll region requires breaking the form of the \kah potential in \eqref{eq:kah_pot}. This can be done by introducing corrections to the tree-level action. 
In \chpref{chap:model_building}, we will step away from the runaway, by considering perturbative and non-perturbative corrections to this leading order behaviour in detail. There we find that phenomenologically viable quintessence requires  non-supersymmetric vacua. This suggests that proponents of the swampland should now object to quintessence  as vigorously as they object to de Sitter, placing them on a collision course with observations.

\section{Divergence of string diagrams and perturbative control or lack thereof}
A statement against the computational control of theory away from the boundary space is that, for finite couplings, the string diagrams can be written as an expansion in these couplings which is formally divergent when summed over. We do not seem to be able to find the origin of this statement for the Swampland programme, but a qualitatively similar argument exist for QED\footnote{We thank Tony Padilla for pointing out this analogy.}. 

Indeed, in \cite{Dyson:1952tj} it was already pointed out that the power series in $e^2$ that one obtains as a results of the perturbative treatment of QED are divergent, even though after renormalisation each individual coefficient in the series is finite. The argument given by Dyson is beautiful in its simplicity and we review it below. Let 
\be
F(e^2) = a_0 + a_2 e^2 + e_4 e^4 +... \label{eq:dyson_pert}
\ee 
be a physical quantity that admits a power expansion in $e^2$. Let $F(e^2)$ converge for some positive value of $e^2$, then the series must be an analytic function around $e=0$. By analytical continuation, one could then consider small enough values of $-e^2$ such that the series is also a well-behaved analytic function with a convergent expansion. Dyson thinks of this as a fictitious world where elementary particles of equal charge attract. 

In this fictitious world, the Hamiltonian is not bounded from below, and there is no good definition of a vacuum state. Indeed, the state of minimal energy will be occupied by an infinite number of these $-e^2$-charged states. Even if only a finite number of these states is initially present in the theory, spontaneous polarization of the would-be lowest-energy state would rapidly swamp the theory again. In this precarious situation, it is not possible to start with any given state in the fictitious world and, after integrating the equations of motion over some finite (or infinite) time, end up with well-defined analytic functions. Thus $F(-e^2)$ cannot be analytic and the series \eqref{eq:dyson_pert} must diverge. This is true for any $e^2\neq 0$; of course if the electron charge were to vanish the sum would be trivially convergent. 

However, nearly 70 years after the publication of \cite{Dyson:1952tj} we know that QED is one of the most precise theories available to physicist. One of its many successes is the theoretical calculation of the anomalous magnetic moment of the electron $a_e$. The QED contribution to $a_e$ can be written as an expansion in Schwinger factors to n-loops as
\be
a_e^{\text{(QED)}}=\sum_{n=1}^{\infty} \left({\alpha\over \pi}\right)^n a_e^{(2n)}\ , \label{eq:schwinger_sum}
\ee
where $a_e^{(2n)}$ is some finite quantity. This sum has been calculated to 5-loops \cite{1205.5368} and, in comparison with the experimental value \cite{Hanneke:2008tm}, one finds that 
\be
\left| a_e^{\text{(QED)}}-a_e^{\text{(exp)}}\right| \sim 10^{-12}\ .
\ee
The agreement of QED with Nature is hard to argue against, even though \eqref{eq:schwinger_sum} is formally divergent\footnote{In two different ways, in fact. One by Dyson's argument and two from the multiplicity of Feynman diagrams scaling as $n!$ so that at some point in loop expansion, $n! \alpha^n\sim 1$.}. Already in \cite{Dyson:1952tj}, the author provided an explanation to this uncomfortable problem. Either the theory is complete and we do not have the appropriate mathematical tools to make sense of the perturbative expansion or the theory is incomplete and the perturbative expansion is merely a workable approximation to a complete physical theory. 

An example that exemplifies these arguments by Dyson is given by the quantum anharmonic oscillator of \cite{Bender:1969si, Banks:1972xa},
\be
\left({d^2\over dx^2} + {x^2\over 4} + {\lambda x^4\over 4} - E(\lambda)\right)\psi(x) = 0\ .
\ee 
This theory can be solved, without making use of any approximation, by numerical techniques. An analytic solution requires making use of a perturbative expansion in the     quartic coupling $\lambda$ such that the energy eigenstates take the form
\be
E^{k} \sim k + {1\over 2} + \sum_{n=1}^{\infty} A_{n}^K \lambda^n\ , \qquad K=0,1,2,...
\ee
The perturbative approach suffers from the same issues as QED. The energy eigenstates have a zero radius of convergence in the $\lambda$ coupling. However, perturbation theory is able to reproduce the numerical results accurately \cite{Bender:1973rz}. Furthermore, the perturbative expansion has been shown to be Borel summable \cite{Graffi:1970erh}, which renders the perturbative theory convergent\footnote{We thank Oliver Gould for pointing out this nice example.}. We view this as an example of a theory that is closed, has perturbative infinities, and yet remains accurate with respect to the exact numerical results, and where new mathematical techniques give us a better understanding of the perturbative expansion.

Coming back to supergravity, we take Dyson's point of view. Even if there are formal infinities in the theory, as long as the perturbative expansion is well-behaved order by order, we should treat it as an approximation to the true physics of string theory. After all, supergravity only aims to capture the massless sector of string theory, so it was never a complete theory to begin with. 

\subsection{Perturbative and non-perturbative contributions spoiling the vacuum}

A number of Swampland conjectures exist to argue that, away from the boundary of moduli space, a number of corrections are bound to appear that spoil the original vacuum. For example, in \cite{2003.10452}, the authors argue that solutions that admit a leading order supersymmetric Minkowski vacuum, only broken at subleading order by instanton corrections, will exhibit non-perturbative corrections that will destabilise the leading order solution. A number of toroidal examples are given to justify this conjecture. 

In a more heuristic sense, this type of Swampland conjectures aim to draw non-model specific statements about string compactifications from a number of examples. These behaviours seem general enough that finding counterexamples is hard, if at all possible. At the same time, no formal and general proof for these conjectures exist. With regards to the topics in this thesis, conjectures about the existence and stability of de Sitter vacua are most relevant. These are not new or limited to the Swampland programme. We have already discussed these to some extent in \ssecref{ssec:dsc}. A very complete rundown of de Sitter and the Swampland can be found in \cite{1808.09440, 2008.10625}, and we will not dwell on this any further. 

We would like to finally remark that the question of systematically and consistently expanding supergravity order by order in the perturbative expansion of its coupling is an ongoing area of research. As we have alluded before, showing that the effective theory is not spoilt by obscure corrections is highly non-trivial due to supergravity having two expansion parameters. Some work in this direction has been done in \cite{2106.04592}, where the authors were able to show that the leading no-scale breaking terms appear at $\mathcal{O}(\alpha'^3)$ in the four-dimensional effective theory, modulo potential logarithmic corrections. Another work \cite{1908.04788}, showed that the ten-dimensional equations of motion and the four-dimensional equations of motion for KKLT \cite{hep-th/0301240} are equivalent while also providing insight into the cancellation of divergences at the ten-dimensional level.

We conclude that, Swampland programme aside, a number of interesting questions remain open. One is to complete the work of  \cite{2106.04592} in showing that no-scale can be consistently broken at $\mathcal{O}(\alpha'^3)$, {\it i.e.} the leading order vacuum generated by instanton and quartic corrections to the ten-dimensional Ricci scalar $\mathcal{R}^4_{(10)}$ is safe from instability woes. A second one is to better understand these expansions in terms of ten-dimensional physics, extending the work of \cite{1908.04788} to non-KKLT models.

We have seen that studying cosmology at the boundary of moduli space seems highly implausible, at best. In most of the remainder of our thesis we will shift our attention to challenges of model building our Universe in the bulk of moduli space. We will see that a hierarchy of scales between different orders in perturbation theory is necessary to stabilise the moduli fields. We argue that this scale hierarchy is precisely the one that protects the perturbative expansion against unwanted instabilities. Of course, this philosophy is not new but rather one of the usual sanity checks that one employs when constructing supergravity models at large (but finite) volume and small (but finite) coupling.

In the next chapter, we will introduce the necessary perturbative and non-perturbative tools to break the tree-level no-scale structure and generate a minimum for the \kah sector, thereby stabilising all the moduli fields in the theory. In \chpref{chap:model_building} we explore the challenges of building quintessence from string theory and provide a ``series of ingredients'' necessary that any phenomenologically viable quintessence model must contain. We will find that an understanding of next-to-leading order contributions can become important for model builders. Finally, in \chpref{chap:gbds} we begin to think in this direction and study whether higher curvature corrections can help produce de Sitter vacua.
\blankpage
\part{String models}
\blankpage
\chapter{Corrections to the supergravity action in type IIB\label{chap:non_pert}}

In this chapter we describe the most common and best understood ways to break the no-scale structure for the tree-level scalar potential described in \chpref{chap:pert_II}. Moduli stabilisation and the effects of perturbative and non-perturbative corrections to the tree-level action are best understood in type IIB supergravity. Therefore, we will limit ourselves to the discussion of type IIB constructions. For a review on the pros and cons of type IIB models see \cite{1808.08967}.

We begin by reviewing perturbative and non-perturbative corrections to the tree-level type IIB effective field theory in \secref{sec:review_iib}. In the subsequent sections, we dedicate some time to motivating the origin of the corrections that we consider from the point of view of the ten-dimensional type IIB theory. In \secref{sec:consistency_checks}, we discuss the regime of validity of the four-dimensional description. Finally, we introduce de Sitter model building within the context of: the KKLT scenario in \secref{sec:KKLT_fundamentals}, the racetrack scenario in \secref{sec:racetrack_fundamentals} and the large volume scenario in \secref{sec:LVS_fundamentals}.  

\section{Review of type IIB effective field theory\label{sec:review_iib}}
We begin with a brief review of the main techniques for deriving the form of the underlying scalar potential of string compactifications, with a view to building a robust model of dynamical dark energy with all moduli suitably stabilised. More detailed reviews can be found in \cite{Cicoli:2018kdo, AbdusSalam:2020ywo}. We assume that the potential is given by the F-term expression 
\be
V=e^K \left[ K^{i \bar j} D_i WD_{\bar j}\bar W-3 |W|^2 \right],
\ee
where $D_i W=(\partial_i +\partial_i K)W$ is the K\"ahler covariant derivative  and $K^{i \bar j}$ is the inverse of the K\"ahler metric $K_{i \bar j}=\partial_i \partial_{\bar j} K$. Our focus will be on type IIB string compactifications in which the complex structure moduli and the dilaton are fixed at semiclassical level, and so $\partial_i$ denotes partial differentiation with respect to the K\"ahler moduli $T_i=\tau_i+{\rm i} \theta_i$. Even though our focus here is on the effective action of type IIB string theory, our final phenomenological considerations on  quintessence also apply more generally to type IIA and heterotic setups. 

At tree-level, we have a K\"ahler potential, $K=K_0-2 \ln \mathcal{V}$ and a superpotential $W=W_0$, where $\mathcal{V}$ is the volume of the internal Calabi-Yau. $K_0$ and $W_0$ include the complex structure moduli and the dilaton that have already been stabilised, and are therefore assumed to be constant. Because of the `no-scale structure', the corresponding scalar potential vanishes identically. Therefore, to  generate the appropriate masses for the K\"ahler moduli, we must include at least one of the following: ($i$) perturbative corrections to the K\"ahler potential, $K\to K+\delta K_\text{p}$; ($ii$) non-perturbative corrections to the superpotential, $W \to W+\delta W_{\text{np}}$; ($iii$) higher derivative corrections to the scalar potential, $V \to V+\delta V_\text{hd}$.

A general formula giving the K\"ahler moduli dependence of perturbative and higher derivative corrections at all orders in $\alpha'$ and $g_s$ has been provided in \cite{Cicoli:2021rub} exploiting a combination of higher dimensional symmetries such as supersymmetry, scale invariance and shift symmetry, together with techniques from F-theory. This formula reproduces several known explicit computations of quantum corrections. Here we focus on those which have been used for cosmological applications:
\begin{itemize}
\item \textit{$\alpha'^3$ corrections}

These are perturbative corrections in $\ell_s = 2\pi\sqrt{\alpha'}$ to the K\"ahler potential. The leading one arises from $\mathcal{O}(\alpha'^3) \mathcal{R}^4$ terms in the ten-dimensional action and looks like \cite{Becker:2002nn}
\be
K\rightarrow K_0-2\ln\left(\mathcal{V}+{\hat \xi \over 2 }\right)\ , \label{eq:alpha_corrections}
\ee
with $\hat \xi=-\frac{\chi(\x)\zeta(3)}{2 g_s^{3/2}(2\pi)^3}$ where $\chi(\x)$ is the Euler number of the Calabi-Yau $\x$. 

\item \textit{Open string 1-loop corrections}

These are corrections in $g_s$ to the K\"ahler potential, $K\rightarrow K+\delta K_{g_s}$, and are conjectured to take the form \cite{Berg:2007wt, Cicoli:2007xp}
\be
 \delta K_{g_s}=\sum_i g_s {C_i(U,\bar{U}) t_i^{\perp} \over \mathcal{V}}+ \sum_i {\tilde{C}_i(U,\bar{U}) \over t_i^{\cap}\, \mathcal{V}}\ . \label{eq:loop_corrections}
\ee
Here there are two contributions: those of $\mathcal{O}(g_s^2 \alpha'^2)$ coming from the tree-level exchange of Kaluza-Klein closed strings, with $t_i^\perp$ denoting the 2-cycles perpendicular to the branes; and those of $\mathcal{O}(g_s^2 \alpha'^4)$ coming from winding strings, with $t_i^\cap$ denoting the 2-cycles of the intersection among branes. $C$ and $\tilde{C}$ are unknown functions of the complex structure moduli $U$, although, as the complex structure sector is fixed at tree-level, one can consider them to be constants. 

\item \textit{Higher derivative corrections}

These are also $\alpha'^3$ corrections to the scalar potential arising from the dimensional reduction of ten-dimensional higher derivative terms of the form $\mathcal{R}^2 G_3^4$, that yield $V \to V+\delta V_\text{hd}$ with \cite{Ciupke:2015msa}
\be
\delta V_\text{hd}=- {g_s^{-3/2}} {3^4 \lambda W_0^4\over \mathcal{V}^4}\,\Pi_i t^i \,,
\label{Vhd}
\ee
where $\lambda$ is an undetermined combinatorial number and $t^i$ are the 2-cycle volume moduli. $\Pi_i$ are topological quantities defined in terms of the $(1,1)$ forms $\hat{D}_i$ as 
\be
\Pi_i=\int_\mathcal{X} c_2\wedge \hat{D}_i\,, \label{eq:pi_is}
\ee
with $c_2$ the second Chern class of $\mathcal{X}$. Although these effects enter at higher F-term order, they can become important and comparable to string loop corrections.

\item \textit{Non-perturbative corrections}

These are corrections to the superpotential \cite{Blumenhagen:2009qh}
\be
 W \rightarrow W_0+\sum_i A_i\, e^{-a_i T_i}\,, \label{eq:inst_corrections}
\ee
related to the existence of E3-brane instantons ($a_i=2\pi$) or gaugino condensates on D7-branes ($a_i=2\pi/N$, where $N$ is the rank of the condensing gauge group). Similar non-perturbative corrections to the K\"ahler potential are subleading when compared to perturbative terms, which can arise from expansions in $\alpha'$ or $g_s$, and will thus be unimportant for our discussion here.
\end{itemize}
Together these corrections yield a scalar potential:
\be
V=\delta V_{\alpha'} +\delta V_{g_s} +\delta V_\text{hd} +\delta V_\text{np}
\ee
where $\delta V_\text{hd}$ is given by \eqref{Vhd} and (setting $e^{K_0}=1$)
\begin{eqnarray}
\delta V_{\alpha'} &=& \frac{3 \hat \xi W_0^2 }{4 \mathcal{V}^3} \label{dValpha} \\
\delta V_{g_s} &=&  \frac{W_0^2}{\mathcal{V}^2} \sum_i\left( g^2_s C_i^2 K^{\text{tree}}_{ii}-2 \frac{\tilde{C}_i}{\vo  t_i^{\cap}}\right) \label{eq:dVgs} \\
\delta V_\text{np} &=& \sum_{i, j} K_{\text{tree}}^{ij}a_ia_jA_iA_j \frac{e^{-a_iT_i-a_j\bar{T}_j}}{\vo^2}+\frac{2 W_0}{\vo^2}\sum_i A_i a_i\tau_i \left(e^{-a_iT_i} +e^{-a_i \bar T_i}\right) 
\end{eqnarray}
$K^{\text{tree}}_{ij}$ is the tree-level contribution to the K\"ahler metric, and $K_{\text{tree}}^{ij}$ its inverse. Notice how the Kaluza-Klein contribution to \eqref{eq:dVgs} enters at second order thanks to the `extended no-scale structure' \cite{Cicoli:2007xp}. All these corrections are under control when the overall volume $\mathcal{V}$ is large. In the regime where all 2-cycles scale as $t \sim \sqrt{\tau}\sim \mathcal{V}^{1/3}$, we have the scaling
\be
\delta V_{\alpha'} \sim {W_0^2\over \mathcal{V}^3}\ , \quad \delta V_{g_s} \sim {W_0^2\over\mathcal{V}^{10/3}}\ ,  \quad \delta V_{\text{np}} \sim {\mathcal{V}^{4/3} e^{-2 a \tau} + W_0 \mathcal{V}^{2/3} e^{-a \tau}\over \mathcal{V}^2}\ , \quad
 \delta V_\text{hd} \sim {W_0^4\over \mathcal{V}^{11/3}}\,, 
\label{eq:scaling_dV1} 
\ee
where we have taken $K^{\text{tree}}_{ij} \sim 1/\mathcal{V}^{4/3}$. To generate stable vacua, one has to find a balance between different terms in the potential. For example, in KKLT models \cite{Kachru:2003aw}, $W_0$ is tuned to exponentially small values, $W_0\sim (a\tau) e^{-a \tau}\ll 1$, so that the two contributions to $\delta V_{\text{np}}$ are comparable in size. This typically yields a supersymmetric AdS vacuum whose depth is parametrised by $-W_0^2/\mathcal{V}^2 $. Upon uplift the same scale controls the height of the barrier separating the vacuum from the decompactification limit \cite{Kachru:2003aw}. A notable exception to this rule is the racetrack setup which we describe in Sec. \ref{sec:kl_rev} \cite{Blanco-Pillado:2004aap, Kallosh:2004yh}, where the scale of the vacuum can be made arbitrarily small thanks to two instanton contributions that are aligned relative to one another.  

Another possible approach is to balance perturbative against non-perturbative corrections. Generically we expect the latter to be suppressed, except in the presence of small cycles, as this raises the size of the instanton correction. This is precisely what happens in the LVS scenario \cite{Balasubramanian:2005zx, Conlon:2005ki, Cicoli:2008va}, where $\delta V_{\alpha'}\sim{W_0^2\over \mathcal{V}^3}\sim \delta V_{\text{np}}$. This sets the scale of the potential, controlling both the depth of the non-supersymmetric AdS vacuum and the height of the barrier to infinity which develops after uplifting.   

Before moving onto the model building aspects, we would like to review in a bit more detail each of these corrections.

\section{Non-renormalisation theorems}
A first consideration might be to ask why are there no effects at $\mathcal{O}(\alpha')$ or $\mathcal{O}(\alpha'^2)$ in the type IIB supergravity theory. In turns out that the linear and quadratic order corrections in $\alpha'$ are proportional to the Ricci tensor of the internal manifold. For a Calabi-Yau compactification, the internal geometry is Ricci-flat and this corrections vanish as long as the internal manifold remains Calabi-Yau. This result is due to the sigma model calculations of \cite{Grisaru:1986dk}, to four-loops in the worldsheet action, and the non-renormalisation theorems of \cite{Witten:1985bz}. An explicit description of the non-renormalisation theorems for type IIB supergravity can be found in \cite{hep-th/0510213}. 

Below we concern ourselves with a simpler but qualitatively identical question. We consider the four-dimensional effective theory for the heterotic superstring as described in \ssecref{ssec:heterotic_runaway}. The supergravity approximation is controlled by the string coupling, which maps to the vev of the real part of the axio-dilaton field $e^{\phi} = Re(S)$, and the $\alpha'$ parameter, which maps to the vev of an overall \kah volume modulus $\vol$. The low-energy dependence of the compactified four-dimensional effective theory on the internal geometry is encapsulated by three functions: the \kah potential $K(S,T,U)$, the superpotential, which for heterotic supergravity has a no-scale structure in the axio-dilaton and \kah directions $\Phi:=\{S,T_\alpha\}$; and the gauge-kinetic coupling matrix $\mathcal{M}_{\kappa\lambda}$ depending only on the complex structure sector by virtue of being its kinetic metric\footnote{This is guaranteed to hold as long as the moduli space is a direct product between the complex structure and the quaternionic sector.}. The scalar potential, in the absence of D-terms, is given by the \kah potential and the superpotential only, $V=V(K,W)$. 

We will argue that the superpotential is protected by a number of shift symmetries in the moduli fields and becomes exact at any order in the perturbative expansion, obtaining only non-perturbative corrections. In the case of heterotic supergravity, the discussion is quite simple. We note that the axionic partners to the dilaton and the overall volume modulus benefit from an unbroken shift symmetry meaning that the supergravity action is invariant under transformations of the form $\Phi\to \Phi + i\lambda$ where $\lambda \in \mathbb{R}$. Supersymmetry implies that the superpotential $W$ and the matrix $\mathcal{M}_{\kappa\lambda}$ are holomorphic functions of the moduli fields. The axion shift invariance then forbids $W$ from depending on the imaginary part of $\Phi$ and thus, by virtue of these functions being holomorphic, it cannot depend on the real parts either at any order in the perturbative expansion. Then, any perturbative corrections to the scalar potential of the tree-level theory must arise from the \kah potential itself\footnote{As long as supersymmetry is not explicitly broken.}, and these enter the theory at $\alpha'^3$ as we will see in the next section. A similar argument can be made in more generality to show that the gauge-kinetic coupling matrix only receives corrections at first order in the loop expansion from \kah moduli dependent contributions, in agreement with the calculations of \cite{hep-th/0404087}.

The key point in supporting the previous arguments, and also in the cases of type II theories, is the survival of the axionic shift invariance in the presence of quantum corrections. This axionic symmetry can be understood as a subgroup of the larger accidental symmetry $SL(2,\mathbb{R})$ of the full string theory. In particular, the ten-dimensional type IIB supergravity action in \eqref{eq:iib_n2} can be made explicitly invariant under $SL(2,\mathbb{R})$ by exchanging the Kalb-Ramond and RR three-form field strength by their gauge invariant counterpart in $\hat{G}_3:=\hat{F}_3-iS\hat{H}_3$. With this choice of variables, we lose an explicit description of the \kah and complex structure sectors in terms of chiral multiplets, in contrast to \eqref{eq:iib_dual}, but we can see the symmetry group for the axio-dilaton explicitly. Indeed, the contents of the ten-dimensional theory in this form are given by $\hat{g}_{MN}$, $\hat{Q}:=i\hat{S}=\hat{C}_0+ie^{-\hat{\phi}}$, $\hat{G}_3$ and the imaginary self-dual five-form $\hat{F}_5$. Under an $SL(2,\mathbb{R})$ transformation \cite{Polchinski:1998rr}, the metric and the imaginary self-dual five-form remain invariant while
\be
\hat{Q}\to {a\hat{Q}+b\over c\hat{Q}+d}\ , \qquad \hat{G}_3\to {\hat{G}_3\over c\hat{\tau}+d}\ ,
\ee
for $ad-bc=1$ and $a,b,c,d\in \mathbb{R}$. The axion shift symmetry corresponds to a subset $SL(2,\mathbb{Z})\subset SL(2,\mathbb{R})$ with $c=0$ and $a=d=1$ for which $\hat{Q}\to \hat{Q}+b$. Critically, $SL(2,\mathbb{Z})$ is expected to be an invariance of string theory even in presence of non-perturbative effects and the perturbative non-renormalisation of the type IIB supergravity will be preserved. We remark that this does not prevent the superpotential from acquiring non-perturbative corrections, cf. \secref{sec:non_pert_corrections}, but these will be exponentially suppressed in a large volume/weak coupling limit.
 
\section{$\alpha'^3$ corrections\label{sec:alpha_corrections}}
We begin by discussing the corrections to the \kah potential to $\mathcal{O}(\alpha'^3)$ as calculated in \cite{hep-th/0204254, Candelas:1990rm}. The goal of this section would be to analise the effects of the higher order terms appearing in the M-theory effective action on the scalar potential of type IIB. However, only a few results regarding these terms are known and the best approach to obtain the leading order corrections to the type IIB action when compactifying on a CY is to use mirror symmetry and the c-map of \cite{Ferrara:1989ik}.

In the seminal work \cite{Candelas:1990rm}, the authors provide an exact result for the leading corrections to the \kah potential for the quintic and its mirror. In particular, the \kah prepotential for the \kah deformations with the leading order perturbative contributions is given by
\be
f(t) = -{1\over 6} \mathcal{K}_{ABC} {t^{\hat{A}}t^{\hat{B}}t^{\hat{C}}\over t^0} - (t^0)^2 {\hat{\xi}\over 4}\ , \label{eq:kah_pre_quintic}
\ee
where $t^{\hat{A}}=\{t^0, t^A\}$ are the adapted coordinates described in \secref{sec:special_geo}. Using the definition of the \kah potential in terms of its prepotential we find \eqref{eq:alpha_corrections}. Although the precise calculations are quite involved, we can provide a minimal qualitative derivation of the formula above. 

As our target manifold $\mathcal{M}$, we take the quintic $\mathbb{P}_4(5)$ which has two neat properties. First, its mirror, which we will take as the base manifold and denote by $\mathcal{W}$, is known thanks to the construction of \cite{Greene:1990ud}. Second, the Betti numbers for this threefold are $b_{1,1}=1$ and $b_{2,1}=101$. This allows for simple calculations in the \kah sector of the target manifold. The mirror of the quintic will have $b_{1,1}=101$ and $b_{2,1}=1$, which will simplify the calculations in the complex structure sector of the base manifold. We will denote the unique \kah (complex structure) coordinate of the target (base) manifold by $\psi$. 

We consider a one-parameter family of quintic hypersurfaces as a set in $\mathbb{C}\text{P}^4$ given by the roots of the polynomial
\be
p = \sum_{k=1}^5 x_k^5 - 5\psi \prod_{k=1}^5 x_k\ , \label{eq:quintic_metric}
\ee
which are invariant under the symmetry group generated by
\begin{gather}
g_0=(1,0,0,0,4)\ , \qquad g_1=(0,1,0,0,4)\ , \qquad g_2=(0,0,1,0,4) \ ,\nonumber \\
g_3=(0,0,0,1,4)\ , \label{eq:gen_quintic}
\end{gather}
where, for example, $g_3$ represents the $\mathbb{Z}_5$ action
\be
(x_1,x_2,x_3,x_4,x_5)\to (x_1,x_2,x_3,\alpha x_4, \alpha^4 x_5)\ ,
\ee
on the set of coordinates $x_k$ on the quintic and $\alpha=e^{2i \pi/5}$. The geometry of the complex structure in $\mathcal{W}$ is determined by the coordinate $\psi$. It is most natural to consider $\psi^5$ as the coordinate of the complex structure sector. This can be seen by noting that a scaling of $\psi$ to $\alpha \psi$ is equivalent to the coordinate transformation\footnote{This is equivalent to the coordinate transformations $(x_1, \alpha^{-1} x_2,x_3, x_4, x_5)=...=(x_1,x_2,x_3, \alpha^{-1}x_4, x_5)$ by the actions of the generators \eqref{eq:gen_quintic}. This freedom is due to the polynomial $p$ defining a covering space of $\mathcal{W}$, rather than $\mathcal{W}$ itself.} 
\be
(x_1,x_2,x_3,x_4,x_5)\to (\alpha^{-1} x_1,x_2,x_3, x_4, x_5)\ ,
\ee
which leads to the sample complex structure for the quintic in \eqref{eq:quintic_metric}. Furthemore, the complex structure geometry of $\mathcal{W}$ is singular for some values of $\psi$. Since we are interested in describing a Calabi-Yau we will have to smooth out the metric around the singularities. It is, therefore, instructive to describe the geometry in a neighbourhood of these singular points. To do so we note that on can identify the set of singular points of $\mathcal{W}$ with the points where the quintic fails to be transverse \cite{Candelas:1990rm}. This is the case when the equations
\be
\partial_{x_k} p = 0 \ ,
\ee
are satisfied simultaneously and we find
\be
\prod_{k=1}^5 x_k^5 = \psi^5 \prod_{k=1}^5 x_k^5\ ,
\ee
which will only be satisfied for $\psi^5=1$ and this in turn implies $x_k=\alpha^{n_k}$ with $\sum n_k=0$. All these points can be identified through the generators of \eqref{eq:gen_quintic}, and $\mathcal{W}$ at $\psi=1$ only has one singular point. In \cite{Candelas:1989ug}, it was shown that this singularity corresponds to a conifold for $\mathcal{W}$, where the singular point of the conifold is identified with the $\psi=1$ point of $\mathcal{W}$. Locally, the geometry is given by a cone with base $S^2\times S^3$ where, at the ``tip'' of the cone, the $S^3$ shrinks to zero volume with radius $r_{S^3}\sim \mathcal{O}(\sqrt{\psi-1})$. 

Now we can define the complex structure sector of the base manifold. We choose a symplectic basis $\{A^1,A^2,B_1,B_2\}$ for $H^3(\mathcal{W})$ such that the only non-vanishing intersections are $A^{\hat{K}}\cap B_{\hat{L}}=\delta^{\hat{K}}_{\hat{L}}$, and the dual basis $(\alpha_{\hat{K}}, \beta^{\hat{K}})$ such that
\be
\int_{A^{\hat{L}}} \alpha_{\hat{K}}= \delta_{\hat{K}}^{\hat{L}} \ , \qquad \int_{B_{\hat{L}}} \beta^{\hat{K}}= \delta^{\hat{K}}_{\hat{L}} \ .
\ee
Finally, the holomorphic three-forms can be expanded as
\be
\Omega = \mathcal{Z}^{\hat{K}} \alpha_{\hat{K}} - \mathcal{F}_{\hat{K}}\beta^{\hat{K}}\ .
\ee
We remark that we have 4 periods $(\mathcal{Z}^1,\mathcal{Z}^2,\mathcal{F}_1,\mathcal{F}_2)$ and only one paramter $\psi$. In the usual way discussed in \ssecref{ssec:cs_kahler}, we can fix half of the periods in terms of the other half and we choose $\mathcal{F}_{\hat{K}}:=\mathcal{F}_{\hat{K}} (\mathcal{Z}^{\hat{L}})$. Furthermore, \eqref{eq:gauge_omega} implies that the last free period can be fixed by absorbing the gauge freedom of complex rescaling of the holomorphic three-form.

We now discuss the monodromy structure around the singularity. Following the conventions of \cite{Candelas:1990rm}, we take $A^2$ to be the cycle related to the vanishing $S^3$ at $\psi=1$. By our choice of symplectic basis, the cycles $(A^1,B_1)$ do not intersect with the $S^3$ and can be taken to lie outside the neighbourhood of the singularity. The last cycle $B_2$ intersects $A_2$ at a point, in the neighbourhood of the conifold singularity. Under transport around $\psi=1$, $B_2$ can pick a multiple of $A_2$ while the other cycles are unambiguously defined. Then,
\be
\Pi:=
\begin{pmatrix}
  \mathcal{Z}^1 \\
   \mathcal{Z}^2\\ 
  \mathcal{F}_1 \\
   \mathcal{F}_2
\end{pmatrix} \to
\begin{pmatrix}
  1 && 0 && 0 && 0 \\
   0 && 1 && 0 && 0 \\
  0 && 0 && 1 && 0 \\
   0 && n && 0 && 1 \\
\end{pmatrix}
\begin{pmatrix}
  \mathcal{Z}^1 \\
   \mathcal{Z}^2\\ 
  \mathcal{F}_1 \\
   \mathcal{F}_2
\end{pmatrix}\ ,
\ee
where we have defined the period vector $\Pi$. The \kah potential is given by
\be
e^{-K} =\bar{\mathcal{Z}}^{\hat{K}}\mathcal{F}_{\hat{K}} - {\mathcal{Z}}^{\hat{K}}\bar{\mathcal{F}}_{\hat{K}}
\ee
and by making use of the monodromy structure, the asymptotic form of the \kah metric as $\psi \to \infty$ was calculated in \cite{Candelas:1990rm} to be
\be
g_{\psi\bar{\psi}} = {3\over 4\left|\psi\right|^2 (\ln \left|5\psi\right|)^2}\left(1- {48\zeta(3)\over 25 (\ln\left|5\psi\right|)^3}+ ...\right)\ , \label{eq:metric_psi}
\ee
where the ellipsis denotes further subleading terms in $1/\ln\left|\psi\right|$. The first term takes the usual inverse quadratic form upon the identification 
\be
t \sim {\sqrt{5}\over 2\pi} \ln(5\psi)\ , \qquad g_{\psi\bar{\psi}} \sim {3\over 4 v^2} dt\wedge \star dt\ , \label{eq:dual_metric}
\ee
where we have chosen the evocative notation of $v:=\text{Re}(t)$ as the $\psi$ in this regime will match the large volume limit on the dual side. Turning to the target quintic, the prepotential for the \kah sector prior to any corrections is given by the first term in \eqref{eq:kah_pre_quintic}, the results can be matched to the ones obtained in the dual side for the asymptotic metric in \eqref{eq:dual_metric}. The full details of this matching can be found in \cite{Candelas:1990rm} and references therein. The second term in \eqref{eq:metric_psi} will be related with the \kah corrections that we are looking for. 

Turning once again to the target quintic, the non-renormalisation theorem of \cite{Witten:1985bz} imply that the are no $\sigma$-model $\alpha'$ corrections to the prepotential couplings 
\ben
\partial^3 f(t) / \partial t^a \partial t^b \partial t^c\ .
\een
Furthermore, the prepotential being a homogeneous function of degree two in the $t^a$ coordinates is determined by the third derivative up to a quadratic term. This translates into the \kah potential being fixed by \cite{hep-th/0507153, Candelas:1990rm}
\be
e^{-K} = -2 t^{\hat{A}} \Re(\mathcal{N}_{\hat{A}\hat{B}}) t^{\hat{B}}\ ,
\ee 
where the matrix $\mathcal{N}_{\hat{A}\hat{B}}$ is the gauge-coupling matrix given in \eqref{eq:gc}, which is a matrix constructed from the second derivatives of the prepotential $f(t)$. We note that the imaginary part of $\mathcal{N}_{\hat{A}\hat{B}}$ drops out of the formula above. This can be understood by virtue of the complex rescaling invariance of $\Omega$, an imaginary shift in the matrix  $\mathcal{N}_{\hat{A}\hat{B}}$  can be reabsorbed into a redefinition of the $t^0$ coordinate. A shift to the real part, however, will map to a change in the volume of the quintic $\mathcal{V}(\mathbb{P}_4(5))\sim \Re(t)^3\sim v^3$ which, based on the mapping in \eqref{eq:dual_metric}, will affect the metric of the \kah sector and its complex structure dual. This will induce a correction to the prepotential function $f(t)$.

The first non-trivial contribution is unique at $\mathcal{O}(\alpha'^3)$ and corresponds to the combination 
\be
\mathbb{S} = {1\over 12(2\pi)^3} R_{IJ}^{\ \ \ KL}R_{KL}^{\ \ \ MN}R_{MN}^{\ \ \ \ \ IJ}-2R_{I\ J}^{\ K\ L}R_{K\ L}^{\ M\ N}R_{M\ N}^{\ \ \ I\ \ \,J}\ , \label{eq:S_alpha}
\ee
which for a six-dimensional manifold corresponds to its Euler density, 
\be
\int_\x d^6x\sqrt{g}\, \mathbb{S} = \chi\ .
\ee 
This type of contribution leads to corrections to the gauge kinetic coupling $\delta \mathcal{N}_{\hat{A}\hat{B}}$ which translate into a constant shift to the prepotential of the desired form upon dimensional reduction to the four-dimensional theory \cite{hep-th/0204254}, {\it i.e.} 
\be
\delta f(t) = - e^{-1/2 \phi_0} {\hat{\xi}\over 2}\ .
\ee
This expression matches \eqref{eq:kah_pre_quintic} up to an overall constant scale $\phi_0$, which can be absorbed in the definition of $t^0$.

Although the calculations of \cite{Candelas:1990rm, hep-th/0204254} are done on a pair of quintics, the results are independent of this choice as long as mirror symmetry is a true symmetry of string theory. The choice of the quintic allows for explicit calculations and it is argued in \cite{Candelas:1990rm, hep-th/0204254} that the form of the corrections to the prepotential will generalise to other smooth Calabi-Yaus.

\section{One-loop corrections}
In this section we concentrate on corrections due to the exchange of KK or winding modes. In \cite{hep-th/0508043}, the authors provide the one-loop string corrections to type IIB compactified on toroidal orientifolds by evaluating string scattering amplitudes. In \cite{0704.0737}, the authors study the consistency of the perturbative expansion of the string effective action in the presence of D-branes and O-planes. In particular, the authors make an educated guess for the generalisation of the toroidal corrections to smooth Calabi-Yaus and are able to obtain the characteristic scaling of such corrections and show that these are subleading to the $\alpha'$-corrections discussed in the previous section. This also served as a consistency check for the LVS scenario of \cite{hep-th/0502058} and which we discuss in \secref{sec:LVS_fundamentals}. Below, we provide a short overview of these arguments.

We begin by summarising the results of \cite{hep-th/0508043} for the toroidal orientifold. There, the \kah potential takes the form\footnote{Where we have adapted the notation to be consistent with the rest of this chapter.}
\begin{gather}
K = -2\ln(S+\bar{S}) - 2\ln (\vol) + K_{cs}(U,\bar{U}) - {\hat{\xi}\over 2\vol} \label{eq:orient_loop} \\ 
+\sum_i g_s {C_i(U,\bar{U}) \over 4\tau_i}+ \sum_i {\tilde{C}_i(U,\bar{U}) \over 4\tau_i\tau_j}\ . \nonumber
\end{gather}
The first line contains the tree-level contributions to the \kah potential and the $\alpha'$-corrections previously discussed. The second line are the corrections to the \kah potential in the presence of localised sources. In the following we exemplify the mechanisms with D-branes, but the statements would hold true for O-planes as well. 

The first term is due to parallel sources (D7/D3-branes or O7/O3-planes) exchanging KK modes between them. The D7-brane is wrapped around some internal four-cycle $\tau_i$ where the point-like component of the D3-brane also lies. The two branes can then interchange closed strings along some two-cycle $t_i$.

The second term is due to the intersection of D7-branes (O7-planes) wrapped around four-cycles $\tau_i\neq \tau_j$. The four-cycles will intersect in some two-cycle $t_i$. If, locally, this $t_i$ contains non-trivial one-cycles the winding modes can wound around them. It could seem that these modes would not appear in the CY generalisation. However, we note that from the point of view of the open string, these mechanism is based on Dirichlet strings that have their endpoints glued to each D7-brane. Therefore, the condition on the existence of these one-cycles is not a condition on the global geometry of the Calabi-Yau, but rather a condition on the topology of specific cycles within cycles. If the particular model to be considered contains no intersecting D7-branes, these corrections will vanish.
\begin{figure}[h!] 
	\centering
	\begin{minipage}{0.45\textwidth}
		\centering
		\includegraphics[width=75 mm]{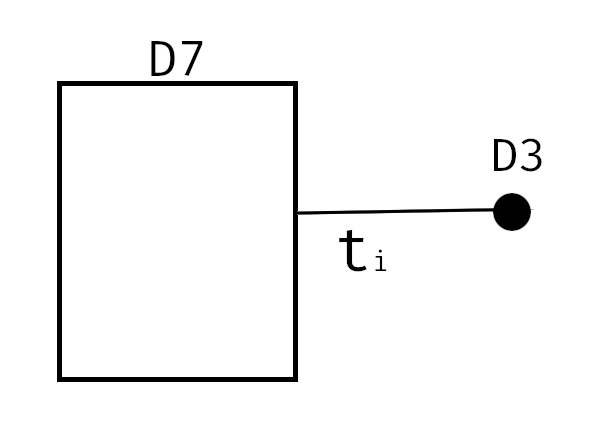}
	\end{minipage}\hfill
	\begin{minipage}{0.45\textwidth}
		\centering
			\includegraphics[width=45 mm]{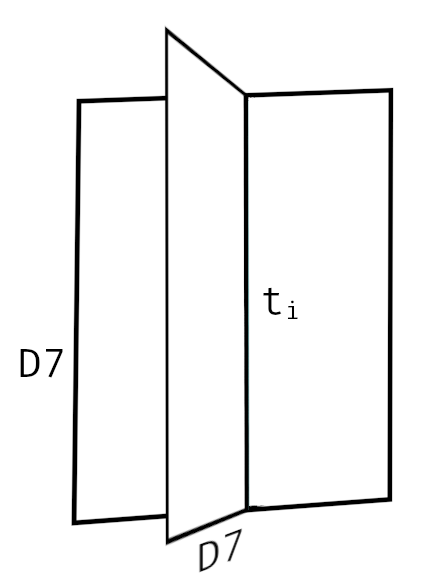}
	\end{minipage}\hfill
\caption{Left: Pictorical representation of the KK mode exchange mechanism for a coincident D7/D3-branes which can be separated by some distance parametrised by a two-cycle $t_i$. The plane of the physical page represents the four-cycle where the internal legs of D7, the two-cycle $t_i$ and the point-like internal component of the D3-brane lie. Right: Representation of the mechanism that allows the interchange of winding modes between two D7-branes. These modes are wrapped on a two-dimensional submanifold characterised by $t_i$ at the intersection of the D7-branes.}\label{fig:pict_exchange}
\end{figure}

In \cite{0704.0737}, it is then shown that the scaling in smooth Calabi-Yaus is given in terms of the internal volume \vol rather than particular four-cycles. In particular, these corrections take the form given in \eqref{eq:loop_corrections}
\be
 \delta K_{g_s}=\sum_i g_s {C_i(U,\bar{U}) t_i^{\perp} \over \mathcal{V}}+ \sum_i {\tilde{C}_i(U,\bar{U}) \over t_i^{\cap}\, \mathcal{V}}\ ,
\ee
where we can see that in the case of a toroidal geometry the volume dependence reduces to $\vol_{T^6}\sim t^3 \sim t \tau$ and the corrections take the form of \eqref{eq:orient_loop}.

We would like to remark that the coefficients $C_i,\tilde{C}_i$ are, in general, unknown. However, since these depend only on the complex structure sector, we can take them to be constant at the perturbative level since the complex structure sector is fixed by tree-level fluxes. Similarly, we have written some dependence on the string coupling $g_s$ which, a priori, depends on the vev of the axio-dilaton $S$. Taking the axio-dilaton fixed at tree-level by fluxes, we treat $g_s$ as a constant at this level.

The scaling with the volume is a good educated guess of \cite{0704.0737}. It is unclear how terms that scale with individual four-cycles (like in the case of the toroidal orientifold) could appear for smooth Calabi-Yaus. Although this idea cannot be completely disregarded, in all well-known examples where this computations can be carried out the volume dependence is the only one to appear. For example, in \cite{0704.0737} the exact one-loop corrections are calculated for the quintic $\mathbb{P}_4^{[1,1,1,6,9]}$ and are found to agree with \eqref{eq:loop_corrections}.

Recently, an important advancement was made to better understand the origin of loop corrections in generic Calabi-Yaus \cite{Gao:2022uop}. These results differ from those of \cite{hep-th/0508043} in the relative scaling of the winding and KK modes with the string coupling $g_s$. More importantly, the authors also flag the possible existence of log-enhanced corrections that could dominate the loop contributions. These log-corrections could arise from $\mathcal{R}^4$ operators localised on D7/O7 stacks or $\mathcal{R}^3$ localised on a 6D submanifold. A string amplitude calculation to confirm the existence of such terms is yet to be done. Models that rely on loop corrections to lift flat directions, such as large volume scenario models with multiple \kah moduli, could be in trouble if the existence of these operators were to be confirmed.

\section{Higher derivative corrections}
The last set of perturbative corrections that we would like to discuss are the higher derivative corrections of \cite{1505.03092}. These corrections are intimately related to the ones in \secref{sec:alpha_corrections}. Indeed, in that section we found that the \kah potential gains a perturbative corrections such that
\be
K\rightarrow K_0-2\ln\left(\mathcal{V}+{\hat \xi \over 2 }\right)\ .
\ee
These corrections will then contribute to the four-derivative expansion of the scalar potential as follows. In \cite{Freeman:1986br}, it was shown that the Ricci tensor along the internal components in the presence of $\alpha'$-corrections is given by
\be
R_{i\barj}\sim (\alpha')^3 \partial_i \partial_{\barj} \mathbb{S}\ , \label{eq:R_alpha}
\ee
where $\mathbb{S}$ was defined in \eqref{eq:S_alpha}. A priori, the metric is not Ricci-flat anymore and can be expanded as
\be
g_{i\barj} = g_{i\barj}^{(0)} + (\alpha')^3 g_{i\barj}^{(1)}\ ,
\ee
where $ g_{i\barj}^{(0)}$ is the original Ricci-flat metric in the absence of corrections and $g_{i\barj}^{(1)}$ solves the Einstein equations to $\mathcal{O}(\alpha'^3)$. At leading order in $\alpha'$, the only contribution from $g_{i\barj}^{(1)}$ enters through the dimensional reduction of the ten-dimensional Ricci scalar. However, this turns out to be a total derivative meaning that the first contribution from $g_{i\barj}^{(1)}$ comes at $\mathcal{O}(\alpha'^6)$. Therefore, at $\mathcal{O}(\alpha'^3)$ we can self-consistently ignore $g_{i\barj}^{(1)}$ and consider $g_{i\barj}= g_{i\barj}^{(0)}$ to be Ricci-flat. 

This implies that, at the four-dimensional effective level and at $\mathcal{O}(\alpha'^3)$, there are no curvature contributions coming from the internal piece. In the large volume limit and up to the four-derivative expansion, the corrections were calculated in \cite{1505.03092} to be given at leading order in the $\mathcal{O}(1/\vol)$ expansion by \eqref{Vhd}
\be
\delta V = - {g_s^{-3/2}} {3^4 \lambda W_0^4\over \mathcal{V}^4}\,\Pi_i t^i \,,
\ee
where $\Pi_i$ are defined as in \eqref{eq:pi_is}

\section{Non-perturbative corrections\label{sec:non_pert_corrections}}
Finally, we will consider non-perturbative corrections due to instantons in the theory. Below, we introduce the origin of instanton corrections in relation with type IIB model building. A complete overview of instanton effects on type II theories can be found at \cite{0902.3251}.

To obtain an $\mathcal{N}=1$ supersymmetric low energy effective theory from type IIB supergravity, we have seen that we need to introduce $O_p$-planes to project out half of supersymmetry generators. In particular, we introduce a holomorphic involution ($\sigma$) that acts on the \kah form and holomorphic 3-form as
\be
\sigma^* J \rightarrow J\ , \qquad \sigma^* \Omega \rightarrow \Omega\ .
\ee
The dimensionality of the orientifold is fixed by the dimensionality of the fix point set induced by $\sigma$. Taking $\Omega\propto dz^1\wedge dz^2\wedge dz^3$, for some set of $\{z^i, \bar{z}^i\}$ complex coordinates spanning the Calabi-Yau, we see that the condition $\sigma^* \Omega \rightarrow \Omega$ leads to fixed points\footnote{Submanifolds of a CY with one complex dimensions have to be given by a combination of holomorphic and antiholomorphic coordinates, as $H^2(\x)=H^{1,\,1}(\x)$, and thus these are not invariant under $\sigma$.} with complex dimensions 0 or 2, {\it i.e.} either points or complex planes. Given that the orientifold planes have to fill the external space to preserve four-dimensional Poincar\'e invariance, this particular choice of involution leads to $O3/O7$-planes.

To preserve the orientifolding, the instantons will have to be related to objects wrapping points or four-cycles. This gives rise to two possible contributions: one from Euclidean D3-branes and another one from gaugino condensation in D7-branes \cite{1001.5028}. In \cite{0902.3251}, it was shown that, for the Euclidean D3-brane instantons, the non-perturbative corrections to the superpotential takes the schematic form
\be
W(S,U) = W_0(S,U) \quad \rightarrow \quad W(S,U,T) = W_0(S,U, e^{-T}) + \sum_{E3_i} A_i(U) e^{-a_i T} + A_S(U) e^{-S}\ ,
\ee
where $E3_i$ denotes a stack of Euclidean D3-branes and $(S,U,T)$ are the axio-dilaton, complex structure and \kah chiral superfields, respectively. Noting that $(S,U)$ are fixed at tree-level by fluxes, we find an expansion in non-perturbative corrections of the form in \eqref{eq:inst_corrections}
\be
W(S,U,T) \simeq W_0(\langle S\rangle, \langle U\rangle) + \sum_{E3_i} A_i(\langle U \rangle) e^{-a_i T}\ ,
\ee
where $(\langle S\rangle, \langle U\rangle)$ are the vevs of the axio-dilaton and complex moduli sector, and we take $W_0$ and $A_i$ constant at the level of the low-energy supergravity description. The discussion for the D7-brane contribution is quantitatively similar and the coefficient $a_i$ in the exponential is related to the rank $N_i$ of the condensing gauge group on each stack of D7-branes, {\it i.e.} $a_i=2\pi/N_i$. 

We note that a third instanton contribution to the non-perturbative superpotential could arise from Euclidean D(-1)-branes. These effects would be the string theoretical analogue to classical localised solutions to the Euclidean equations of motion in QFT. We expect their contribution to the superpotential to be exponentially suppressed at weak coupling $W_{\rm ED(-1)}\sim\mathcal{O}(e^{-1/g_s})$ \cite{2201.04634} and thus will not consider them for model building purposes.

\section{Validity of the effective description\label{sec:consistency_checks}}
Since we will be treating the supergravity models at an effective four-dimensional level to study their phenomenology, it is good to keep in mind the consistency requirements to truncate the effective theory to the set of corrections given above. 

\noindent{\bf Tree-level hierarchy with respect to corrections.} Most of the work in the subsequent sections and chapters will assume that the axio-dilaton and complex structure moduli have been stabilised at tree-level by fluxes. This allows us to concentrate on lifting the \kah sector without having to worry about destabilising the other directions. To ensure that the other sectors are not destabilised we have to demand that the corresponding saxions are fixed at large vacuum expectation values by the fluxes. 

For the dilaton, this is of two-fold importance as it also guarantees that the loop expansion is well-posed, as we will remark below. For the complex structure moduli, this allows us to truncate the prepotential to the {\it large complex structure term}, the equivalent to the leading term of \eqref{eq:kah_pre_quintic} for the complex sector.

\noindent{\bf For the $\alpha'$ expansion.} We already made use of a consistency check in \chpref{chap:intermezzo}, namely that the $\alpha'$ expansion is under control when the string frame volume $\vol_S\gg 1$, this implied $\vol/s\gg 1$ for the volume in Einstein frame. A second, but related, check is that consistency in truncating the $\alpha'$ expansion to the leading term requires
\be
{\hat \xi \over g_s^{3/2} \vol} = {\hat \xi \over \vol_S} \ll 1\ . \label{eq:neglect_xi}
\ee
Indeed, further subleading terms carry extra derivatives with respect to the overall volume modulus. We expect these to be further suppressed in a $1/\vol$ expansion as compared to the leading order $\alpha'^3$-corrections. Thus, requiring that the $\alpha'^3$ term satisfies \eqref{eq:neglect_xi} guarantees the safety of the truncation to leading order $\alpha'$-corrections. 

\noindent{\bf For the loop expansion.} Given that the loop corrections are given to leading order in an expansion on the string coupling $g_s$, we require that the vev of the dilaton is fixed at large values
\be
{1\over g_s} := e^{-\langle \phi \rangle} = \langle \text{Re}( S) \rangle \gg 1\ .
\ee
We can see that this consistency check comes {\it for free} by requiring that there exists a hierarchy between the tree-level actions and its corrections.

\noindent{\bf For the expansion in non-perturbative effects.} For the instanton expansion to be well-posed we require that 
\be
e^{-a_i^a\, \text{Re}(T_a)} \ll 1\ , \qquad a_i^a\, \text{Re}(T_a) \gg 1
\ee
for all $(i,a)$, where $i$ labels the stack of Euclideanised D3-branes and $a=1,...,h^{1,\, 1}$. In the large volume scenario described in \secref{sec:LVS_fundamentals}, we will often work with only the leading term in this expansion, {\it i.e.} the smallest four-cycle $\text{Re}(T_s)$, which implies a further hierarchy
\be
1 \ll a_i^s\, \text{Re}(T_s) \ll a_i^b\, \text{Re}(T_b)\ ,
\ee
where the label ``$b$'' indicates any four-cycle other than $\text{Re}(T_s)$.

\noindent{\bf Supergravity scales.} Together with the moduli fields a number of Kaluza-Klein degrees of freedom will formally appear in the effective four-dimensional description. We are interested in integrating these out so that the complete hierarchy of scales appearing in the effective description has to look like
\be
 M_p > M_s > M_{KK}^{(i)} \gg m_i , m_{3/2}\ , 
\ee
where $M_s$ is the string scale mass, $M_{KK}^{(i)}$ are the Kaluza-Klein masses that can appear from bulk modes or associated to D7-branes wrapping 4-cycles, $m_i$ denotes the masses of the moduli fields and $m_{3/2}$ is the gravitino mass. It can be shown that these hierarchy can be made consistent when the other previous perturbative and non-perturbative checks are satisfied \cite{2005.11329}. Indeed, this is so due to the volume scaling for the different masses
\be
M_s \sim {1\over \sqrt{\vol}}\ , \qquad M_{KK} \sim {1\over \vol^{2/3}}\ , \qquad m_{3/2}\sim {1\over \vol}\ .
\ee
There will also be a number of phenomenological conditions on these scales (and others) due to observations (or lack thereof), but we will discuss them in the next chapter when they become more relevant. 

With all this in mind we move on to discuss particular realisations of de Sitter vacua within supergravity. In general, the scale of the scalar potential will be set by some combination of the form $\left|W_0\right|^p / \vol^q$, where $p,q>0$. Obtaining a small enough scale to fit cosmological observations allows a natural categorisation of the model building efforts into two classes: the ones that require $\left|W_0\right|\ll 1$, like KKLT \cite{Kachru:2003aw} and racetrack type scenarios \cite{Blanco-Pillado:2004aap, Blanco-Pillado:2006dgl}, and the ones requiring $\vol\gg1$, like the aptly named large volume scenario \cite{hep-th/0502058}.  

\section{KKLT model building\label{sec:KKLT_fundamentals}}
To exemplify the KKLT scenario, we will consider a single \kah modulus geometry where the volume of the Calabi-Yau is of the form $\vol = (T+\bar{T})^{3/2}$, with $T:=\tau + i \theta$ the single 4-cycle volume modulus. The key ingredient to generate a minimum for the \kah direction in the KKLT scenario is the use of the non-perturbative corrections in \eqref{eq:inst_corrections} and the leading $\alpha'$ corrections in \eqref{eq:alpha_corrections}. In this case, the superpotential and \kah potential are given by
\be
W=W_0+A\,e^{-a T}\,, \qquad K=K_0 -2\ln\left(\mathcal{V}+\frac{\hat{\xi}}{2}\right).
\ee 
We note that neglecting loop and higher derivative contributions to the potential can be done consistently due to the volume hierarchy in \eqref{eq:scaling_dV1} and that we will still require $\vol\gg1$. The dynamics of the moduli, to leading order in $\hat{\xi}$, is given by the Lagrangian
\begin{equation}
    \mathcal{L}=K_{T\bar{T}} \,\partial T \partial \bar{T} - V(\tau,\theta) = \frac{3}{4\tau^{2}}\left(1-\frac{5\hat{\xi}}{2^{7/2}\tau^{3/2}}\right)\left[(\partial\tau)^2+(\partial\theta)^2\right]-V(\tau,\theta)\ ,
\end{equation}
where the F-term potential is
\be \label{Vkklt}
V(\tau, \theta)=\frac{a^2A^2e^{-2a\tau}}{6\tau}\left(1+\frac{3}{a\tau}\right)-\frac{aA|W_0|e^{-a\tau}}{2\tau^2}\cos (a \theta)+\frac{3 W_0^2\hat{\xi}}{64 \sqrt{2}\tau^{9/2}}\,,
\ee
and, without loss of generality, we have assumed $W_0$ to be real and negative, $W_0 = -|W_0|$. It then follows that a SUSY minimum exists at zeroth order in $\hat\xi$, located at $\theta=0$ and  $\tau=\tau_{\text{min}}$,  where the latter satisfies the following relation
\be \label{W0kklt}
|W_0|=A\,e^{-a\tau_{\text{min}}}\left( 1 +\frac{2}{3}a\tau_{\text{min}}\right)\simeq \frac23 A a \tau_{\rm min} \,e^{-a\tau_{\text{min}}}\,.
\ee
We work in a regime where the $\alpha'^3$ corrections can be consistently neglected in the vicinity of the minimum thanks to \eqref{eq:neglect_xi}. At large volumes these corrections will induce a maximum in the potential as stressed in \cite{Conlon:2018eyr}, which will be relevant to the discussion in \chpref{chap:model_building}. In the past these very same corrections have been used for uplifting the SUSY AdS minimum to Minkowski \cite{Westphal:2006tn}, a more extreme regime which we will not consider here. For the moment, however, we concentrate on the dynamics close to the minimum, neglecting the subleading effects of $\hat\xi$. The potential at the minimum is AdS, and given by 
\be
V_{\rm AdS} \equiv V(\tau_{\text{min}}, 0)=-\frac{a^2A^2e^{-2a\tau_{\text{min}}}}{6\tau_{\text{min}}}= -3\left(\frac{ |W_0|}{\vol_{\rm min}}\right)^2\ .
\ee
When the axion is at its minimum, the corresponding KKLT potential for the saxion is given by $V_\text{KKLT}(\tau) =V(\tau, 0) +V_{\text{up}}(\tau)$. The uplift is tuned so that the new dS minimum, located at $\tau_\text{dS}$, is compatible with current bounds on the cosmological constant, that is $V_\text{KKLT}(\tau_\text{dS}) =V(\tau_\text{dS}, 0) +V_{\text{up}}(\tau_\text{dS}) \lesssim 10^{-120}\,\mp^4$. For the instanton expansion to be under control at the minimum, it must be placed at some large value of $\tau$. As a consequence, the uplift does not have a huge effect on its position, and we can take  $\tau_{\text{dS}}\simeq\tau_{\text{min}}$. It follows that the scale of the uplift is simply given by the scale of the AdS vacuum, as one might already have expected,  $V_{\text{up}}(\tau_\text{dS}) \approx |V_\text{AdS}|$. Furthermore, since the original scalar potential decays exponentially quickly in comparison to the uplift term at large $\tau$, the metastable vacuum is separated from the runaway region by a barrier whose height is fixed  at the same scale, $V_{*}\sim V_{\text{up}}(\tau_\text{dS})\approx \left|V_{\text{AdS}}\right|$. The generic shape of the potential with and without uplift is shown in \figref{kklt.fig}.
\begin{figure}[h!] 
		\centering
		\includegraphics[width=75 mm]{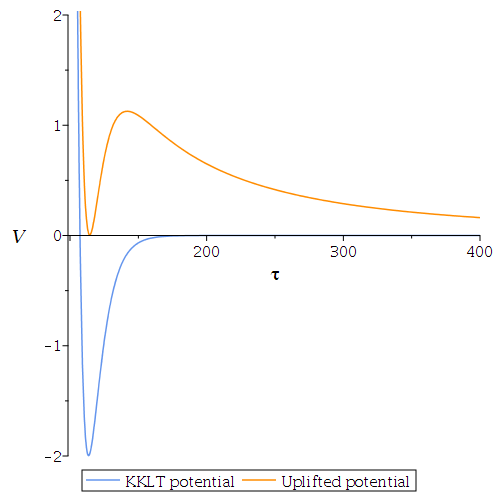}
\caption{KKLT scalar potential with and without uplift.}	\label{kklt.fig}
\end{figure}

It is clear now that, if the scale of the de Sitter vacuum is fixed by $V_{\rm AdS}=3\left(\frac{ |W_0|}{\vol_{\rm min}}\right)^2$, to match the current observational bounds on the cosmological constant we require $W_0\ll 1$, $\vol \gg 1$ or a combination of small tree-level fluxed superpotential and large volume at the minimum. We also note that in this particular case with a single \kah modulus, the volume and the fluxed superpotential are not independent. Indeed, from \eqref{W0kklt}
\be
|W_0|\simeq \tau_{\rm min} \,e^{-\tau_{\text{min}}}\sim \vol_{\rm min}^{2/3} e^{-\vol_{\text{min}}^{2/3}}\ , \qquad V_{\text{dS}} \simeq \left|V_{\text{AdS}}\right| \sim {e^{-\vol_{\text{min}}^{2/3}}\over \vol_{\text{min}}}\ ,
\ee
and with $V_{\text{dS}}\sim 10^{-120}$ we find $\vol_{\text{min}}\sim \mathcal{O}(10^3)$ and $W_0\sim\mathcal{O}(10^{-114})$. As noted before, this value of the volume at its minimum allows us to pass the consistency checks of \secref{sec:consistency_checks} and provide a hierarchy in \eqref{eq:scaling_dV} so that subleading corrections can be safely ignored.

The usual lore goes that given that the string landscape could have as many as $\mathcal{O}(10^{272000})$ flux vacua\footnote{This was originally computed using the techniques from the works of Ashok, Denef and Douglas \cite{hep-th/0307049, hep-th/0404116, hep-th/0411183}.} \cite{1511.03209}, albeit with a large degeneracy in the fluxes, finding such a small value for the fluxed superpotential should not be difficult. Recently, a mechanism to obtain a naturally small $W_0$ has been proposed in \cite{2107.09064}, where flux configurations where found so that $W_0\lesssim\mathcal{O}(10^{-123})$.

Finally, we note that with supersymmetry now broken by the anti D3-brane, the gravitino acquires a mass at the uplifted minimum\footnote{We remark that the precise uplifting mechanism is unimportant at this stage as supersymmetry must be broken by any uplift and the quantitative discussion about the gravitino mass follows through.}, given by 
\be
m^2_{3/2}=\left[e^K|W|^2\right]_{\tau=\tau_{\text{dS}},\theta=0}\approx \left[e^K|W|^2\right]_{\tau=\tau_{\text{min}},\theta=0} = \left(\frac{ |W_0|}{\vol_{\rm min}}\right)^2 = \frac{|V_\text{AdS}|}{3}
\ee
where we have used the fact that $D_i W=0$ and so $V_\text{AdS}=-3e^K|W|^2$ for the supersymmetric AdS vacuum. 

As was noted in \cite{hep-th/0411011}, the KKLT scenario suffers from the so-called Kallosh-Linde (KL) problem and a solution in the form of a racetrack type superpotential was offered. We will discuss this problem in detail in the following chapter and we continue by  introducing the racetrack construction. At this point, suffice it to say that the issue is related to fact that the gravitino mass and the potential barrier are of the same order, since both are generated at the same order in the corrections.

\section{Racetrack model building\label{sec:racetrack_fundamentals}}
\begin{figure}[h!]
		\centering
		\includegraphics[width=75 mm]{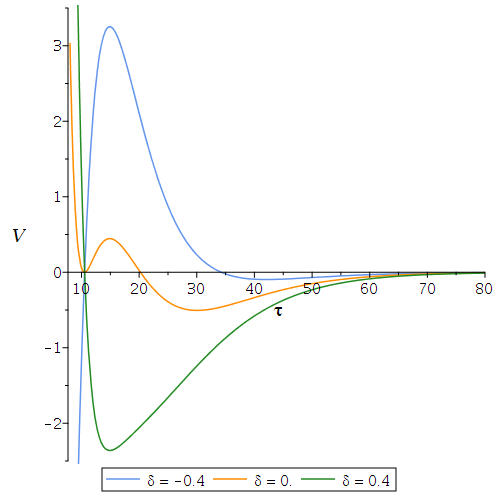}
	\caption{Racetrack potential at $\theta=0$ for parameter choice $
    A = 1, B = -1, a = 0.1, b = 0.09$
and different values of $W_0=W_0^{\text{crit}}(1+\delta$).} \label{racetrack.fig}
\end{figure}

The  racetrack superpotential \cite{Blanco-Pillado:2004aap, Blanco-Pillado:2006dgl} receives a second instanton contribution
\be
W=W_0+A\,e^{-a T}+B\,e^{-b T}\,,
\ee
as one would expect from gaugino condensation in a theory with a product gauge group\footnote{This is equivalent to considering instanton corrections generated by two stacks of $N$ and $M$ Euclidean D3-branes.}. In particular, for $SU(M) \times SU(N)$ we expect $a=2\pi/M$ and $b=2\pi/N$. The corresponding F-term potential is given by
\begin{eqnarray}
V(\tau, \theta)&=&\frac{a^2A^2e^{-2a\tau}}{6\tau}\left(1+\frac{3}{a\tau}\right)+\frac{b^2B^2e^{-2b\tau}}{6\tau}\left(1+\frac{3}{b\tau}\right)  \nonumber \\
&+&\frac{abAB e^{-(a+b)\tau}}{3\tau} \left(1+\frac{3}{2a\tau}+\frac{3}{2b\tau}\right)\cos[(a-b)\theta]
 \nonumber\\
&-& \frac{aA|W_0|e^{-a\tau}}{2\tau^2}\cos (a \theta)-\frac{bB|W_0|e^{-b\tau}}{2\tau^2}\cos (b \theta) \,. 
\label{raceV}
\end{eqnarray}
In contrast with KKLT, the racetrack model admits a SUSY Minkowski vacuum for a critical value of $W_0$, given by 
\be
|W_0|^{\text{crit}}=A\,\mathcal{R}^\frac{a}{b-a}+B\,\mathcal{R}^\frac{b}{b-a}
\ee
where we assume $\mathcal{R}=-\frac{a A}{b B}>1$ and $a>b$ for definiteness. The minimum is located at $\theta=0$ and  $\tau_{\rm min}=\frac{1}{(a-b)} \ln \mathcal{R}$. The gravitino mass vanishes at the Minkowski vacuum since supersymmetry remains unbroken. The shape of the potential for different values of $W_0$ is shown in Fig. \ref{racetrack.fig}.

In this case, the tuning to exponentially small $W_0$ is alleviated in exchange for some extra tuning between the values of the coefficients $(a,b)$ in the exponentials of the superpotential. Indeed, taking as an example the numbers in \cite{hep-th/0411011}
\be
A=1\ ,\qquad B=-1.03\ , \qquad a={2\pi\over 100}\ ,\qquad b={2\pi\over 99}\ , 
\ee
yields $W_0^{\text{crit}}\sim -10^{-4}$ and $\vol \sim \mathcal{O}(1000)$. Obtaining a de Sitter vacuum now follows analogously to KKLT. Introducing an uplift term $\sim {C\over \tau^2}$ will lead to a de Sitter vacuum and it is assumed that the $C$ coefficient can be fine-tuned to obtain a viable scale for the cosmological constant. Critically, the mass of the gravitino is now generated by the small uplift whereas the potential barrier is generated by the leading order non-perturbative corrections. This allows the racetrack model to side-step the KL problem.

\section{LVS model building\label{sec:LVS_fundamentals}}
A fundamentally different approach to constructing de Sitter in type IIB effective theories goes by the name of the large volume scenario (LVS) \cite{hep-th/0502058}. In this case, the required smallness of the cosmological constant scale is obtained through an exponentially large internal volume with an $\mathcal{O}(1)$ value for $W_0$. The key ingredients for the LVS are an internal geometry containing large and small four-cycles and the introduction of loop and higher derivatives corrections to the theory. 

In \cite{hep-th/0502058}, it was shown that an AdS minimum can be generated in the large volume limit of type IIB supergravity in the presence of $\alpha'$ corrections and non-perturbative effects. We review the discussion below \cite{0907.0665}. Consider a Calabi-Yau \x whose \kah sector is given by a number of moduli fields
\begin{equation}
\begin{cases}
\quad \tau_i , \qquad \text{which remain small } \ \forall i=1,..., N_{s} \ , \\
\vol\to\infty \qquad \text{for } \tau_j\to\infty\ , \qquad \forall j=N_{s}+1,...,h^{1,\,1}(\x)\ , 
\end{cases}\label{eq:small_big_cycle}
\end{equation}
for a type IIB $\mathcal{N}=1$ four dimensional effective theory whose \kah potential and superpotential are given by
\begin{equation}
\begin{cases}
\quad K=K_0-2\ln\left(\vol + {\hat{\xi}\over 2}\right) \ , \\
\quad W=W_0 + \sum_{i\leq N_{s}} A_i e^{a_i T_i} \ .
\end{cases} \label{eq:init_data}
\end{equation}
Then the four-dimensional scalar potential admits non-SUSY AdS vacua at exponentially large volume $\vol\sim e^{a\tau_i}$ ($\forall i\leq N_{s}$) along $h^{1,\,1}(\x)-N_s-1$ directions corresponding to the $\tau_j$ blow-up modes if and only if $\hat{\xi}>0$ and $\tau_j$ resolves a point-like singularity, {\it i.e.} $K_{jj}^{-1}\sim\vol\sqrt{\tau_j}$. 

Let us consider a minimal example in which the volume of the Calabi-Yau is given by $\vol=\tau_b^{3/2}-\tau_s^{3/2}$ where $\tau_s$ is a small cycle and $\tau_b$ is a blow-up mode, as defined in \eqref{eq:small_big_cycle}. The four-dimensional scalar potential when considering this geometry and \eqref{eq:init_data} is given at leading order by
\be
V= {8 a^2 A^2 \sqrt{\tau_s} e^{-2 a \tau_s}\over 3\vol} {-}{4\tau_s\over \vol^2}aAe^{-a\tau_s} |W_0| \cos (a\theta ) +{3W_0^2\hat{\xi}\over 4\vol^3}  \, \label{eq:pot_lvs_0}
\ee
where we have set $K_0=1$ and taken $W_0<0$ without loss of generality. The axion is stabilised at $\theta_*=0$ and we have dropped subleading $e^{-b\tau_b}$ contributions in anticipation that the big cycle will be stabilised at $1\ll a\tau_{s, *} \ll b\tau_{b,*}$. Extremising with respect to the volume and the small cycle modulus we find the well-known LVS results
\begin{gather}
\vol = {3\sqrt{\tau_{s,*}}\left|W_0\right| \over 4 aA} e^{a\tau_{s,*}} f\left(\tau_{s,*}\right)\ , \qquad f\left(\tau_{s,*}\right) := {1-{1\over a\tau_{s,*}}\over 1-{1\over 4a\tau_{s,*}}}\ , \\
{\hat \xi \over 2} = \tau_{s,*}^{3/2} {f\left(\tau_{s,*}\right)[4-f\left(\tau_{s,*}\right)]\over 3} =\left[{1\over b}\ln \left(\vol\over \left|W_0\right|\right)\right]^{3/2}\ .
\end{gather}
The minimum of the potential is a non-SUSY AdS vacuum
\be
\langle V\rangle= -{3\hat{\xi} W_0^2 \over 2 \mathcal{V}^3 }\left[\frac{1-f(\tau_{s,*})}{4-f(\tau_{s,*})}\right] \simeq -{3\hat{\xi} W_0^2 \over 8 a \tau_{s,*} \mathcal{V}^3 }\ .
\ee
Once again, the introduction of an uplift mechanism to obtain de Sitter or Minkowski vacua is necessary.

Here we have seen the LVS in action. Even though the volume becomes exponentially large, the presence of the small cycle $\tau_s$ means that the non-perturbative contributions can compete against the leading $\alpha'$-corrections generating a minimum along the $\tau_s$ direction. We remark that, in the presence of $h^{1,\,1}$ \kah moduli the leading order scalar potential would only fix $N_s$ directions and overall the volume modulus, leaving $h^{1,\,1}-N_s-1$ directions flat that could be lifted by subleading perturbative effects, like the loop corrections or higher derivative corrections introduced in the earlier sections. 

Now that we have a handle on how model building in the bulk of moduli space works, in the next chapter we would like to answer the question of what are the necessary ingredients of a successful quintessence model.
\blankpage
 \chapter{Obstructions to quintessence model building \label{chap:model_building}}
 
In this chapter, we consider the challenges in establishing the microscopic origin of both inflation and dark energy. In \cite{Obied:2018sgi} it has been conjectured that scalar potentials that can be derived from putative quantum gravity theories obey the bound
\be
V_\phi\ge \frac{c}{M_p}\,V\,,
\ee
where $c$ is a positive and dimensionless order one constant. If true, this conjecture has serious implications for inflation in the early universe and dark energy at present times. The most obvious consequence is that de Sitter (dS) vacua are forbidden, ruling out the cosmological constant as the source of dark energy. However, the bound is also in some tension with the requirement of slow roll in two derivative scalar actions, both for inflation and dark energy. While this tension is stronger in the context of inflation, it may be acceptable for dark energy models given that current bounds on $\omega_{\rm DE}$ \cite{Planck:2018vyg} are more relaxed that those derived from the scalar spectral tilt, $n_s$, for inflationary models \cite{Planck:2018jri,Amendola:2016saw}. It was later realised that this bound would rule out the experimentally tested Higgs potential, and would preclude electroweak symmetry breaking which requires $V_\phi=0$ for $V_{\phi\phi}<0$ and $V>0$ \cite{Denef:2018etk}. Moreover, it would also rule out supersymmetric AdS vacua that are accompanied by dS maxima at large field values \cite{Conlon:2018eyr}. This unsatisfactory state of affairs prompted the proposal of a refined conjecture that took the form \cite{Garg:2018reu,Ooguri:2018wrx}
\be
V_\phi\ge \frac{c}{M_p}\,V \qquad\text{or}\qquad V_{\phi\phi}\leq -\frac{c'}{M_p^2}\,V\,, 
\label{dscon}
\ee
where $c$ and $c'$ are positive and dimensionless order one constants. These conjectures are not based on rigorous proofs and several counterexamples have been proposed \cite{Louis:2012nb, Cicoli:2012vw, Cicoli:2012fh, Cicoli:2013cha, Blaback:2013ht, Braun:2015pza, Cribiori:2019hrb, Antoniadis:2019rkh, Crino:2020qwk, Cicoli:2021dhg}. Rather, the logic behind their formulation is the theoretical difficulty in  establishing the existence of a dS vacuum in a fully convincing manner, mainly due to the need to break supersymmetry. Strong evidence in favour of the refined version of the conjecture has been given in \cite{Ooguri:2018wrx} for any parametrically controlled regime of string theory using a combination of the distance conjecture and entropy considerations. This is the regime where the semiclassical approximation can be made arbitrarily good by sending  the parameters that control the string loop and the $\alpha'$ expansions to zero. These are, respectively, the real part of the axio-dilaton $S$, which sets the string coupling $g_s = 1/{\rm Re}(S)$, and the extra-dimensional volume in string units $\vo$ which controls the $\alpha'$ expansion since $1/\vo^{1/3} = \alpha' /{\rm Vol}^{1/3}$ (where ${\rm Vol}$ is the dimensionful volume). The asymptotic limit where ${\rm Re}(S)\to \infty$ and $\vo\to \infty$ corresponds to the semiclassical approximation with no dS vacua. 

However, dS vacua could still exist in the bulk of moduli space where the quality of the approximations should be carefully checked. In particular, a necessary condition to have control over the effective field theory is the existence of small expansion parameters such as the flux-generated superpotential $W_0\ll 1$ in KKLT models \cite{Kachru:2003aw} and the inverse of the internal volume $1/\vo\ll 1$ in LVS vacua \cite{Balasubramanian:2005zx, Conlon:2005ki, Cicoli:2008va}. Much progress has been made in this direction by determining perturbative \cite{Becker:2002nn, Berg:2005ja, vonGersdorff:2005ce, Berg:2007wt, Cicoli:2007xp, Ciupke:2015msa, Antoniadis:2018hqy} and non-perturbative corrections \cite{Blumenhagen:2009qh}, or by estimating their moduli dependence using higher dimensional arguments based on symmetries \cite{Burgess:2020qsc} and geometry \cite{Cicoli:2021rub}. However, it is fair to say that the existence of dS vacua in the interior of the moduli space has still to be established in a fully convincing manner and there are a growing number of no-go theorems explicitly demonstrating their absence in particular compactifications of string inspired effective theories \cite{Maldacena:2000mw, Green:2011cn, Gautason:2012tb,Dasgupta:2014pma,Kutasov:2015eba, Quigley:2015jia, Dine:2020vmr, Montero:2020rpl,Cunillera:2021fbc}. Even if they \emph{did} exist, dS vacua in string theory might well be short-lived, as suggested by the TCC conjecture \cite{Bedroya:2019snp}. Of course, this is not a problem for dark energy as observations only require it to be dominant for a single efolding of accelerated expansion. This may even be desirable in the context of the cosmological coincidence problem, as discussed in \chpref{chap:coincidence} \cite{Zlatev:1998tr, Velten:2014nra, Cunillera:2021izz}.

All these considerations show that the existence of dS vacua in string theory is still an open problem which requires further scrutiny. It is, therefore, interesting to investigate if the alternative to a cosmological constant --- namely, quintessence --- shares the same technical difficulties. To this end, we shall focus on the microscopic origin of dark energy as a dynamically evolving scalar field emerging from a compactification of string theory.

First of all, let us mention that, similar to dS constructions, quintessence cannot be realised in any parametrically controlled regime of string theory since dilaton or volume mode runaways in the asymptotics of the moduli space are too steep to drive an epoch of accelerated expansion \cite{Cicoli:2021fsd}, see \chpref{chap:intermezzo}.  The obstruction echoes some of the obstructions to dS vacua \cite{Ooguri:2018wrx}, with related results for quintessence also being derived in  \cite{Hertzberg:2007wc, Garg:2018zdg, ValeixoBento:2020ujr}. Note that the situation does not improve if one performs a multifield evolution including their corresponding axionic fields. In fact, even if non-geodesic trajectories on curved field manifolds could, in principle, yield a period of accelerated expansion for steep potentials \cite{Cicoli:2020cfj, Cicoli:2020noz}, this is never the case for either ${\rm Re}(S)$ or $\vo$ \cite{Brinkman:2022}. 

As a result, quintessence can only be realised in the bulk of moduli space where it generically shares the same control issues as dS model building \cite{Cicoli:2018kdo}. On top of the technical difficulties in trusting the effective field theory, quintessence is known to feature some phenomenological challenges including the `light volume problem' and the `F-term problem' \cite{Hebecker:2019csg}. The `light volume problem' relates to quintessence driven by a saxion, typically a volume modulus. To be compatible with the acceleration we see today, this modulus needs to be extremely light, with its mass bounded above by the current Hubble scale. As it also couples to matter with gravitational strength, this would yield an additional long range  scalar force, in violation of fifth force constraints \cite{Will:2014kxa}. The `F-term problem' is  associated with radiative corrections involving supersymmetric particles running in loops, producing contributions to the scalar potential that are much larger than the dark energy scale. Traditional quintessence, at least in  a perturbative regime,  also has some observational problems, having been shown to enhance the so-called Hubble tension \cite{Colgain:2019joh,Banerjee:2020xcn} which is already at $5\sigma$ for $\Lambda$CDM \cite{Riess:2021jrx}. 

Here we add to the challenges facing quintessence in string theory. In particular, we show how a version of the so called `Kallosh-Linde (KL) problem' \cite{Kallosh:2004yh} drastically constrains the spectrum of possibilities. The KL problem is one of runaway behaviour in the volume mode during inflation. It is normally used to constrain the scale of inflation against the gravitino mass. We use it to constrain the form of the underlying scalar potential responsible for dark energy, exploiting the huge hierarchy of scales between the acceleration today and in the early universe. This hierarchy makes it extremely difficult to have a scalar potential that is compatible with current observations and is protected from the KL runaway during inflation. 

Let us briefly run through the logic. We begin with $V_0(\vo)$, the potential that fixes the volume mode. However, the volume mode also couples to any source of energy-momentum thanks to the Weyl rescaling to four-dimensional Einstein frame. As a result, in Einstein frame,  there is a direct coupling between $\vo$ and the potentials for both the inflaton $\sigma$ and the quintessence field $\phi$. The total scalar potential describing the dynamics of all three fields is given by $V_{\rm tot} = V_0(\vo) + V_1(\sigma, \vo) + V_2(\phi, \vo)$ where $V_1(\sigma, \vo)$ is generated from the inflaton potential and  $V_2(\phi, \vo)$ from the quintessence potential. Recall that there exists an enormous hierarchy between the energy scales of inflation and dark energy:  $V_{\rm inf} \gtrsim (1\,{\rm MeV})^4 \gg (1\,{\rm meV})^4 \sim V_{\rm DE}$.

During inflation, with $\sigma$ in slow roll, it follows that the quintessence field $\phi$ should be frozen, with $V_1(\sigma,\vo) \gg V_2(\phi, \vo)$. Furthermore, in order to avoid destabilising the volume direction \cite{Kallosh:2004yh}, we need to impose the condition $ |V_0(\vo_*)| \gtrsim V_1(\sigma,\vo) \gg V_2(\phi,\vo)$, where $\vo_*$ is the value of $\vo$ controlling the barrier against decompactification. For Minkowski vacua $\vo_*=\vo_{\rm max}$, the value of $\vo$ at the top of the barrier, while for AdS vacua $\vo_*=\vo_{\rm min}$, the value of the $\vo$ at the minimum\footnote{For dS vacua the story is slightly different: we  need to impose $V_1(\sigma, \vo)\lesssim \left(V_0(\vo_{\rm max})-V_0(\vo_{\rm min})\right)$, although we shall ignore this case since quintessence model building is less well motivated in the presence of a dS vacuum.}. Of course, for AdS vacua inflation is possible only if $V_1(\sigma,\vo)$ acts as an uplifting term such that $V_{\rm inf}  \simeq V_0(\vo_{\rm min}) + V_1(\sigma, \vo_{\rm min})>0$. (Here we are assuming that the location of the minimum, $\vo_\text{min}$, does not change significantly in the presence of the uplift.)

After the end of inflation $V_1(\sigma, \vo)$ goes to zero and so $V_{\rm tot}\simeq V_0(\vo_{\rm min}) + V_2(\phi,\vo_{\rm min})$. For the case where  $V_0(\vo)$ admits a (near) Minkowski vacuum with $V_0(\vo_{\rm min})\simeq 0$,  it follows that $V_{\rm tot}\simeq V_2(\phi, \vo_{\rm min})\simeq V_{DE} \lesssim 10^{-36} V_{\rm inf}\lesssim 10^{-36} V_0(\vo_{\rm max})$, implying a \emph{huge} hierarchy between the energy scales associated with the potential that stabilises $\vo$ and the one which drives quintessence.  For the case where $V_0(\vo)$ admits an AdS vacuum, the hierarchy of scales ensures that, after inflation, $V_{\rm tot}\simeq V_0(\vo_{\rm min}) + V_2(\phi,\vo_{\rm min}) \simeq V_0(\vo_{\rm min}) <0$, implying that quintessence model building is not possible.

Notice that similar considerations would apply if the volume also plays the r\^ole of the inflaton ($\sigma=\vo$), notwithstanding that explicit constructions of volume inflation look rather contrived \cite{Cicoli:2015wja}. Alternatively, if the  volume plays the r\^ole of quintessence ($\phi=\vo$) its potential would, again, be destabilised by the inflationary energy density. Finally if the volume is everything ($\sigma=\phi=\vo$), we would require the presence of two slow roll regions at hierarchically different field values. Given that plateau-like regions can be obtained only by balancing competing terms, if the quintessence epoch at large field values is under control, the inflationary era would lie in a region where perturbation theory would tend to break down. Reheating after the end of inflation and fifth force constraints would also present additional problems in this particular case.

These considerations can be combined with implications of the refined dS conjecture \cite{Ooguri:2018wrx}. The refined dS conjecture rules out quintessence models with a very shallow potential, as in \cite{Cicoli:2012tz}, but allows for quintessence rolling near a hilltop at positive energy (perhaps in the presence of a global AdS or supersymmetric Minkowski vacuum) or down an exponential potential of the form $V=V_0 \ e^{-\lambda \phi/M_p}$. In the latter case, it has been shown that agreement with data requires $\lambda\le 1.02$ at $3 \sigma$ \cite{Akrami:2018ylq}\footnote{A stronger bound of $\lambda \le 0.6$ was obtained in \cite{Agrawal:2018own}. We refer the reader to \cite{Akrami:2018ylq} for a discussion of the two approaches.}. However, our analysis suggests  that these two scenarios are not under better control than dS vacua.

Exponential potentials arise from no-scale breaking perturbative effects for saxions and are typically not small enough to produce the required hierarchy in scales between inflation and dark energy. Therefore, these models are expected to be destabilised by the inflationary dynamics, as well as suffering from problems with the light volume and the F-term. The KL problem also applies to hilltop quintessence near a maximum at positive energy, with a global AdS minimum.
%since this scenario would be plagued by the KL problem. 
We shall present explicit examples of these scenarios and  elucidate their problems in Sec. \ref{sec:KL}. 

To avoid the KL problem, we could consider hilltop quintessence models with a supersymmetric (near) Minkowski vacuum. However, in these models the gravitino mass would be of order the dark energy scale resulting in violation of current bounds \cite{Ferrer:1997yz,Brignole:1997sk, Brignole:1998me,Kawasaki:2008qe}. Moreover, supersymmetric Minkowski solutions are highly  constrained, requiring a very precise form for the superpotential $W$. Therefore, even if the presence of supersymmetry might seem a powerful tool to keep computational control over these solutions, proving their existence in the interior of the moduli space might still be a challenge. As an illustrative example, consider the well-understood type IIB compactifications with $H_3$ and $F_3$ flux, and a  tree-level $W$ that does not depend on the complex volume mode $T=\tau + {\rm i}\theta$. In this set-up, any supersymmetric Minkowski solution at tree-level would necessarily feature a complex flat direction, given by $T$. The existence of a global Minkowski solution with all moduli stabilised would, therefore,  have to rely on the existence of non-perturbative corrections, which lack a full systematic understanding. They would also lift $\tau$ and $\theta$ at the same level of approximation, without generating the right hierarchy between the would-be quintessence field $\theta$ and the volume mode $\tau$.

In the end, we arrive at a generic picture for building a viable quintessence model in string phenomenology. Let us summarize the main points:
\begin{itemize}
\item At leading order (in either perturbative or non-perturbative expansions), the scalar potential $V_0(\vo)$ should feature a (near) Minkowski vacuum with a stabilised volume mode. Notice that non-supersymmetric Minkowski vacua typically require the inclusion of uplifting sectors, and so look qualitatively similar to dS vacua. Although supersymmetric Minkowski solutions could give better computational control, the subdominant effects which generate dark energy would also be responsible for supersymmetry breaking. The gravitino mass (and the soft terms) would not be decoupled from the dark energy scale, in strong tension with both particle physics \cite{Ferrer:1997yz,Brignole:1997sk, Brignole:1998me} and cosmological observations \cite{Kawasaki:2008qe}. Thus the leading order Minkowski vacuum should be \emph{non}-supersymmetric.

\item At the leading order of approximation, the quintessence field should remain flat in order to be able to create the required hierarchy between $V_0(\vo_{\rm max})$ and $V_{\rm DE}$,  with the latter generated by subdominant contributions. The presence of a flat direction can be guaranteed by shift symmetries which fall into two categories: ($i$) non-compact rescaling symmetries for saxions arising from the underlying no-scale structure \cite{Burgess:2014tja, Burgess:2016owb}. However, these are broken by perturbative effects, and so are not  generally efficient enough to provide the required hierarchy; and ($ii$) compact shift symmetries for axions which can potentially generate huge hierarchies, being broken only by tiny non-perturbative effects. Moreover, the smallness of these non-perturbative corrections ensures that the energy density associated with the quintessence potential does not destabilise the volume minimum.

\item Axion quintessence automatically avoids the fifth-force problem (being driven by a pseudo-scalar) and ensures radiative stability thanks to the fact that the axionic shift symmetry is exact at the perturbative level.

\item The main problem with axion quintessence is that its potential is flat enough to drive a period of accelerated expansion only if the axion decay constant is trans-Planckian. However, this situation is very difficult to realise since explicit string constructions with control over the effective field theory tend to have axions with sub-Planckian decay constants \cite{Cicoli:2012sz}, as also implied by the weak gravity conjecture \cite{Arkani-Hamed:2006emk}. There could be counter-examples based on alignment mechanisms \cite{Kim:2004rp, Dimopoulos:2005ac}, although their trustability requires further scrutiny. 

\item For generic axion potentials with sub-Planckian decay constants, we might seek quintessence from a hilltop model. Even if this possibility looks attractive from a model building perspective, we shall see in Sec. \ref{sec:hills} that, when combined with quantum diffusion during the inflationary epoch \cite{Kaloper:2005aj}, it relies on two conditions: ($i$) \emph{very} finely tuned initial conditions; and ($ii$) an extremely low inflationary scale ($H_{\rm inf}\lesssim 1$ MeV), at least for axion decay constants in the regime where the effective field theory is under control.
\end{itemize}
\vskip 0.5cm
\noindent \emph{In other words, from the point of view of theoretical and phenomenological control,  quintessence model building in string theory  is at least as challenging as the search for dS vacua.} 
\vskip 0.5cm
This conclusion raises doubts over the validity of the swampland dS conjecture. Taken alongside the challenges to quintessence, it would imply strong tension between quantum gravity and observation. This might be an indication that phenomenologically relevant solutions to string theory, like dS vacua, lie in the bulk of the moduli space. In this case, it might still be true that perturbation theory is a valid approximation but to be confident of this, we need to refine our technical ability to compute quantum corrections. In the end Nature has already shown an affinity for couplings (as in standard gauge theories and cosmological perturbation theory) that are weak enough to allow us to describe it to a good approximation, even if they cannot be made arbitrarily small.

Finally, if data were to prefer dynamical dark energy, our analysis shows that quintessence models are very unlikely to be axion hilltops since they require highly tuned initial conditions and a very low Hubble scale during inflation. In this regard, axion quintessence models based on alignment mechanisms look more promising even if they need further studies to be convincingly established  in fully fledged string compactifications with moduli stabilisation.

\section{Old challenges for quintessence in string theory} 
\label{sec:modelbuild}

We begin by recalling some relevant results from \chpref{chap:non_pert}. In particular, the scalar potential in the presence of the leading $\alpha'$, $g_s$, higher derivative and non-perturbative corrections is given by
\be
V=\delta V_{\alpha'} +\delta V_{g_s} +\delta V_\text{hd} +\delta V_\text{np}
\ee
where $\delta V_\text{hd}$ is given by \eqref{Vhd} and (setting $e^{K_0}=1$)
\begin{eqnarray}
\delta V_{\alpha'} &=& \frac{3 \hat \xi W_0^2 }{4 \mathcal{V}^3} \label{dValpha} \ ,\\
\delta V_{g_s} &=&  \frac{W_0^2}{\mathcal{V}^2} \sum_i\left( g^2_s C_i^2 K^{\text{tree}}_{ii}-2 \frac{\tilde{C}_i}{\vo  t_i^{\cap}}\right) \label{dVgs}\ , \\
\delta V_\text{np} &=& \sum_{i, j} K_{\text{tree}}^{ij}a_ia_jA_iA_j \frac{e^{-a_iT_i-a_j\bar{T}_j}}{\vo^2}+\frac{2 W_0}{\vo^2}\sum_i A_i a_i\tau_i \left(e^{-a_iT_i} +e^{-a_i \bar T_i}\right) \ .
\end{eqnarray}
In the regime where all 2-cycles scale as $t \sim \sqrt{\tau}\sim \mathcal{V}^{1/3}$, we have the scaling
\be
\delta V_{\alpha'} \sim {W_0^2\over \mathcal{V}^3}\ , \quad \delta V_{g_s} \sim {W_0^2\over\mathcal{V}^{10/3}}\ ,  \quad \delta V_{\text{np}} \sim {\mathcal{V}^{4/3} e^{-2 a \tau} + W_0 \mathcal{V}^{2/3} e^{-a \tau}\over \mathcal{V}^2}\ , \quad
 \delta V_\text{hd} \sim {W_0^4\over \mathcal{V}^{11/3}}\,, 
\label{eq:scaling_dV} 
\ee
where we have taken $K^{\text{tree}}_{ij} \sim 1/\mathcal{V}^{4/3}$. To generate stable vacua, one has to find a balance between different terms in the potential. For example, in KKLT models \cite{Kachru:2003aw}, $W_0$ is tuned to exponentially small values, $W_0\sim (a\tau) e^{-a \tau}\ll 1$, so that the two contributions to $\delta V_{\text{np}}$ are comparable in size. This typically yields a supersymmetric AdS vacuum whose depth is parametrised by $-W_0^2/\mathcal{V}^2 $. Upon uplift the same scale controls the height of the barrier separating the vacuum from the decompactification limit \cite{Kachru:2003aw}. A notable exception to this rule is the racetrack setup which we describe in Sec. \ref{sec:kl_rev} \cite{Blanco-Pillado:2004aap, Kallosh:2004yh}, where the scale of the vacuum can be made arbitrarily small thanks to two instanton contributions that are aligned relative to one another.  

Another possible approach is to balance perturbative against non-perturbative corrections. Generically we expect the latter to be suppressed, except in the presence of small cycles, as this raises the size of the instanton correction. This is precisely what happens in the LVS scenario \cite{Balasubramanian:2005zx, Conlon:2005ki, Cicoli:2008va}, where $\delta V_{\alpha'}\sim{W_0^2\over \mathcal{V}^3}\sim \delta V_{\text{np}}$. This sets the scale of the potential, controlling both the depth of the non-supersymmetric AdS vacuum and the height of the barrier to infinity which develops after uplifting.   

\subsection{Fifth forces and radiative instability}

Some dynamical dark energy models have already been built within the framework of string compactifications \cite{Choi:1999xn, Kaloper:2008qs, Panda:2010uq, Cicoli:2012tz, Kamionkowski:2014zda, Olguin-Trejo:2018zun}. Typically the quintessence field corresponds to the lightest mode and the other moduli are stabilised at tree-level and by leading order corrections. In this way dynamical dark energy appears as a next-to-leading order effect, allowing us to retain perturbative control. It also guarantees that the slow roll of the quintessence field away from the minimum only displaces the volume mode from its original vacuum expectation value by a small amount. 

However none of the existing quintessence models in the literature is really satisfactory due to several challenges which were already highlighted in \cite{Cicoli:2018kdo, Hebecker:2019csg}. These challenges are related to the phenomenological requirements that a prospective stringy quintessence field would have to satisfy, namely:
\begin{enumerate}
\item A \textit{light quintessence modulus} $\phi$ with $m_\phi\lesssim H_0 \sim 10^{-60}\,M_p$. This follow directly from requiring that the scalar field $\phi$ is in slow roll at the current epoch.

\item \textit{Heavy superpartners} with masses $M_{\rm soft}\gtrsim10^{-15}\,M_p$. Supersymmetric partners must be above the threshold set by the LHC \cite{Zyla:2020zbs}. This, in turn, yields large perturbative corrections from loops of visible sector supersymmetric particles.

\item \textit{Heavy Kaluza-Klein scale} with $M_{\rm KK}\gtrsim10^{-30}\, M_p$. Sub-millimetre scale tests of Newtonian gravity put a bound on the Kaluza-Klein scale \cite{Kapner:2006si}. 

\item \textit{Heavy volume modulus} with $m_{\mathcal{V}}\gtrsim 10^{-30}\,M_p$. Upon compactification, the four-dimensional Ricci scalar gets a prefactor which depends on the volume modulus which couples to matter fields after Weyl rescaling to Einstein frame. There are stringent bounds on such fifth force effects given by sub-millimetre experiments \cite{Kapner:2006si, Damour:2010rp, Acharya:2018deu}.
\end{enumerate}

The authors of \cite{Hebecker:2019csg} discuss the implications of these requirements for string models of dark energy, with a focus on LVS-motivated scenarios for concreteness. Two main issues arise.

\paragraph{The light volume problem:} 

The Kaluza-Klein mass is given by
\begin{equation}
    M_{\rm KK}={M_s\over R}\sim \frac{M_p}{\mathcal{V}^{2/3}}\gtrsim 10^{-30}\,M_p\qquad \Rightarrow \qquad \mathcal{V}\lesssim 10^{45}\ , \label{eq:mkk_bound}
\end{equation}
where we have used $M_s\simeq M_p\,\mathcal{V}^{-1/2}$, the fact that the radius of the compact space $R\sim \mathcal{V}^{1/6}$, and the bound on the Kaluza-Klein mass given above.

In the LVS scenario, the mass for the volume modulus is generated through leading $\alpha'$ corrections \eqref{dValpha}, while at subleading order loop corrections \eqref{dVgs} lift additional K\"ahler moduli which could play the r\^ole of the quintessence field $\phi$. Using \eqref{eq:scaling_dV} and \eqref{eq:mkk_bound}, one finds
\begin{equation}
    \frac{m_\phi}{m_\mathcal{V}}\sim \sqrt{\frac{\delta V_{g_s}}{\delta V_{\alpha'}}}\sim \frac{1}{\mathcal{V}^{1/6}} \gtrsim 10^{-7}\ .
    \label{eq:modVol}
\end{equation}
In \cite{Cicoli:2012tz} loop contributions are suppressed due to low energy supersymmetry in the bulk and an anisotropic shape of the extra dimensions. The quintessence field $\phi$ is instead lifted by poly-instanton effects which give
\begin{equation}
    \frac{m_\phi}{m_\mathcal{V}}\sim \sqrt{\frac{\delta V_{\rm poly}}{\delta V_{\alpha'}}}\sim \frac{1}{\sqrt{\vo}} \gtrsim 10^{-22}\ .
    \label{eq:modVol2}
\end{equation}
However, both (\ref{eq:modVol}) and (\ref{eq:modVol2}) are in contradiction with the phenomenological bound imposed by fifth force constraints and the value of $H_0$, i.e. ${m_\phi \over m_{\mathcal{V}}}\lesssim 10^{-30}$. A way to avoid this issue is to introduce subleading effects that modify the volume scaling of \eqref{eq:modVol} and (\ref{eq:modVol2}) \cite{Cicoli:2011yy}. For example, in the model of \cite{Cicoli:2012tz}, $\phi$ does not mediate any fifth force since its coupling to Standard Model fields is weaker than Planckian due to sequestering effects in the extra dimension (see also \cite{Acharya:2018deu} for estimates of moduli couplings in sequestered models with large extra dimensions). Nevertheless, the volume mode would lead to new long range interactions since, due to (\ref{eq:modVol2}), it is much lighter than 1 meV and it couples with ordinary matter with standard Planckian strength (however, see \cite{Burgess:2021qti} for a possible screening effect due to the kinetic coupling of $\vo$ to its associated axionic field). 

\paragraph{The F-term problem:}  

The mass of the superpartners, which we approximate by the gaugino mass, is of the order
\begin{equation}
    M_{\rm soft}\sim M_{1/2} = \frac{F^i \partial_i f}{{\rm Re}(f)}\,M_p\,,
\end{equation}
where $f$ is the gauge kinetic function. If we assume that supersymmetry breaking is mediated through some higher-dimensional operator at some scale $M_b$, for a simple toy model with a single spurion field $X$ and F-term $F_X$, the contribution to the scalar potential is 
\begin{equation}
    \delta V_X \sim F_X^2 \sim M_b^2 M_{\rm soft}^2\ ,
\end{equation}
where $M_{\rm soft}$ enters the scalar potential after canonical normalisation of the spurion term. If we require that supersymmetry breaking is mediated above the TeV scale, together with the phenomenological constraints on the superpartner masses, we find $\delta V_X\gtrsim 10^{-60} M_p\gg H_0^2$. This contribution would raise the scale of the potential well beyond the  dark energy scale. A loophole is to consider a new contribution to the scalar potential that would cancel supersymmetry breaking effects with some fine-tuning, as in \cite{Cicoli:2012tz}, where the additional effect is assumed to come from the backreaction of  non-supersymmetric visible sector branes (see also \cite{Burgess:2021obw} for recent developments of quintessence models in scenarios with non-linearly realised supersymmetry).

The challenges for quintessence outlined in \cite{Cicoli:2018kdo, Hebecker:2019csg} are just the tip of the iceberg. In the next section, we identify an even bigger problem: disruption of the energetic dynamics by the inflationary energy density, resulting in destabilisation of the volume mode and decompactification. 

\section{The KL problem for quintessence}
\label{sec:KL}

In \cite{Kallosh:2004yh} Kallosh and Linde argued that the scale of inflation is bounded from above by the gravitino mass in the standard KKLT scenario \cite{Kachru:2003aw}.  The constraint arises in order to avoid a runaway in the volume mode, leading to  decompactification at early times. Similar considerations were used to place limits on thermal corrections to the scalar potential, imposing a maximum temperature in the four-dimensional effective theory \cite{Buchmuller:2004tz, Anguelova:2009ht}. The KL problem extends beyond KKLT, and has also been shown to affect LVS models where the constraint turns out to be even stronger \cite{Conlon:2008cj}. We begin by reviewing the key aspects of the original argument of \cite{Kallosh:2004yh}. Later we will show that it has implications also for string models of dynamical dark energy. 

\subsection{Review of the KL problem}
\label{sec:kl_rev}

Consider the one-instanton KKLT model of \secref{sec:KKLT_fundamentals} with superpotential and $\alpha'$-corrected K\"ahler potential given by
\be
W=W_0+A\,e^{-a T}\,, \qquad K=K_0 -2\ln\left(\mathcal{V}+\frac{\hat{\xi}}{2}\right),
\ee
where $\mathcal{V}=(T+\bar T)^{3/2}$ is the volume of the internal Calabi-Yau manifold with $T=\tau+{\rm i} \theta$ the 4-cycle volume modulus. The dynamics of the moduli, to leading order in $\hat{\xi}$, is given by the Lagrangian
\begin{equation}
    \mathcal{L}=K_{T\bar{T}} \,\partial T \partial \bar{T} - V(\tau,\theta) = \frac{3}{4\tau^{2}}\left(1-\frac{5\hat{\xi}}{2^{7/2}\tau^{3/2}}\right)\left[(\partial\tau)^2+(\partial\theta)^2\right]-V(\tau,\theta)\ ,
\end{equation}
where the F-term potential is
\be \label{Vkklt}
V_\text{KKLT}(\tau, \theta)=\frac{a^2A^2e^{-2a\tau}}{6\tau}\left(1+\frac{3}{a\tau}\right)-\frac{aA|W_0|e^{-a\tau}}{2\tau^2}\cos (a \theta)+\frac{3 W_0^2\hat{\xi}}{64 \sqrt{2}\tau^{9/2}}\,.
\ee
and, without loss of generality, we have assumed $W_0$ to be real and negative, $W_0 = -|W_0|$. From this potential, we found SUSY AdS vacua located at
\be \label{Vads}
V_{\rm AdS} \equiv V_\text{KKLT}(\tau_{\text{min}}, 0)=-\frac{a^2A^2e^{-2a\tau_{\text{min}}}}{6\tau_{\text{min}}}=- 3\left(\frac{ |W_0|}{\vo_{\rm min}}\right)^2.
\ee
which was uplifted via
\be
V_{\text{up}} = \frac{C}{\tau^2} \,.
\ee
Furthermore, by demanding that after uplifting the minimum of this potential sat at $V_{\text{KKLT}'}(\tau_\text{dS}) =V_\text{KKLT}(\tau_\text{dS}, 0) +V_{\text{up}}(\tau_\text{dS}) \lesssim 10^{-120}\,M_p^4$, we obtained a relation between the mass of the gravitino $m_{3/2}$ and the height of the potential barrier, $V_*$, protecting the volume modulus from a runaway
\be
m^2_{3/2}= \frac{|V_\text{AdS}|}{3} \sim V_*\ .
\ee
\begin{figure}[h!] 
	\centering
	\begin{minipage}{0.45\textwidth}
		\centering
		\includegraphics[width=75 mm]{Figs/plot_potential.png}
	\end{minipage}\hfill
	\begin{minipage}{0.45\textwidth}
		\centering
			\includegraphics[width=75 mm]{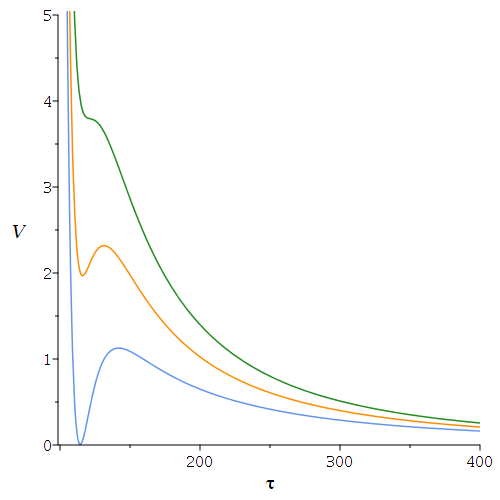}
	\end{minipage}\hfill
\caption{Left: KKLT scalar potential with and without uplift. Right: Uplifted potential with increasing inflationary corrections.}	\label{kklt.fig}
\end{figure}
The KL problem emerges when we consider inflation in this particular setup. To begin with, one could consider hilltop inflation from the top of the potential barrier $V_*$. In this case, the Hubble parameter during inflation is related to the gravitino mass at present through
\be
H_{\rm inf}^2 \approx  \frac{V_*}{3} \sim \frac{\left|V_{\text{AdS}}\right|}{3} \approx  m_{3/2}^2\ . \label{eq:H1}
\ee
Another mechanism for inflation could be due to the dynamics of branes in the compact space \cite{Hsu:2003cy}. In this case, the inflaton is some other modulus field $\sigma$ controlling the location of the D-branes in the internal space. The uplifted KKLT potential then receives a contribution from the inflaton due to the structure of the supergravity F-term potential. The inflationary potential will generically take the form
\be
V_{\text{inf}}(\tau)= V_{\text{KKLT}'}(\tau)+\frac{V(\sigma)}{\tau^3}\ . \label{eq:KL_pot}
\ee
As shown in \figref{kklt.fig}, the inflationary corrections raise the level of the minimum more than they raise the height of the barrier. Eventually, if the scale of inflation is high enough, the local minimum turns into an inflection point and the barrier disappears completely. 
For the volume modulus to stay stabilised during inflation, we require that $V_{\text{inf}}\lesssim V_* \sim 3 m_{3/2}^2 M_p^2$, and again we find a relation between the value of the Hubble parameter during inflation and the gravitino mass today
\be
H_{\rm inf}^2 \approx  \frac{V_\text{inf}}{3 M_p^2} \lesssim m_{3/2}^2\ . \label{eq:H2}
\ee
Equations \eqref{eq:H1} and \eqref{eq:H2} are the main results of \cite{Kallosh:2004yh} and show that, unless one can parametrically decouple the value of the gravitino mass from the height of the potential barrier, the scale of inflation is bounded from above by the gravitino mass, $H_{\rm inf} \lesssim m_{3/2}$. This presents a problem in that it sets the gravitino mass to be extremely large, which by proxy sets the scale of supersymmetry breaking to be much larger than the TeV scale. Notice that this is not necessarily a problem for models with a high scale of supersymmetry breaking. However, to obtain the observed value of the Higgs mass, these require severe fine tuning  or a sequestered visible sector on D3 branes at singularities where $M_{\rm soft}\ll m_{3/2}$ \cite{Aparicio:2014wxa}. The tension between low scale supersymmetry and inflation is even more acute in LVS models since the barrier is generically not as high, scaling as $V_* \sim m_{3/2}^3 M_p$. Using similar arguments as above, this constrains the scale of inflation to be $H_{\rm inf} \lesssim m_{3/2} \sqrt{m_{3/2}/M_p}$.

\subsection{Racetrack solution to the KL problem}

Kallosh and Linde \cite{Kallosh:2004yh} proposed a resolution to this problem within  a racetrack model \cite{Blanco-Pillado:2004aap}, see \secref{sec:racetrack_fundamentals} for the details of the model.  The key observation of Kallosh and Linde is that the racetrack potential admits a supersymmetric Minkowski vacuum for a critical value of $W_0$, 
\be
|W_0|^{\text{crit}}=A\,\mathcal{R}^\frac{a}{b-a}+B\,\mathcal{R}^\frac{b}{b-a}\ , \qquad \mathcal{R}=-\frac{a A}{b B}\ ,
\ee
while the gravitino mass vanishes at the Minkowski vacuum since supersymmetry remains unbroken. As such  there is no relation between the gravitino mass and the height of the potential barrier. The latter scales as $V_* \lesssim (a-b)^3 A^2 \mathcal{R}^\frac{2a}{b-a}$, and can be made arbitrarily high provided we assume $(a-b) \ll 1$, while the minimum for the volume modulus $\tau_{\rm min} = {1\over a-b} \ln \mathcal{R}$ is pushed to a large value of $\tau$. 

\subsection{General implications for quintessence}
\label{implications}

Although \cite{Kallosh:2004yh} were interested in constraining the scale of inflation and the form of the inflationary potential,  similar considerations can be applied to the low energy potential describing the dynamics of dark energy today.  We focus on the dynamics of up to three moduli: a quintessence field $\phi$ describing dark energy, an inflaton $\sigma$ and the volume modulus $\tau$. In principle, these could be three different moduli, or they could overlap - we consider all possibilities. 

We start by assuming they are all different. In general, the full scalar potential can be written as
\be
V_{\rm tot}(\phi, \sigma, \tau)=V_0(\tau)+V_1(\sigma, \tau)+V_2(\phi, \tau)
\ee
where $V_0$ is the potential that fixes the volume mode, $V_1$ the contribution of the inflaton and $V_2$ from quintessence.
In this case, dark energy is assumed to be described by a scalar potential $V_{\rm DE}(\phi, \tau)=V_0(\tau)+V_2(\phi, \tau)$ with $\phi$ in slow roll at some scale $\phi \sim \phi_0$ today and the volume stabilised at some large value $\tau =\tau_0$. 
However, during inflation, we generically expect the full scalar potential to receive an inflaton-dependent correction as described in the original KL scenario \cite{Kallosh:2004yh}. In other words, 
\be
V_{\rm inf}(\phi, \sigma, \tau)=V_{\rm DE}(\phi,\tau)+\frac{V(\sigma)}{\tau^3}\ .
\ee
where we have set $V_1(\sigma, \tau)=\frac{V(\sigma)}{\tau^3}$, as a result of Weyl rescaling. The inflaton field rolls slowly through at least $50$ efoldings of inflation, starting out at $\sigma_\text{inf}$ and ending at $\sigma_0$, with $V(\sigma_0) \approx 0$. The volume modulus and the quintessence field are assumed to be spectators during inflation, stabilised at $\tau \approx \tau_\text{inf}$ and $\phi \approx \phi_\text{inf}$ respectively. Once inflation has ended, the inflaton dumps energy into the Standard Model sector. During this phase of reheating, the volume and quintessence moduli are allowed to move, if necessary, towards their current values, $\tau \to \tau_0$, $\phi \to \phi_0$. However, in order to avoid potential problems with light element abundances \cite{Pitrou:2018cgg} and the spectrum of the cosmic microwave background radiation \cite{Planck:2018vyg}, all three moduli must remain stabilised from nucleosynthesis onwards, right up until the current epoch of dark energy domination, at which point $\phi$ starts to slow roll.  

The key observation is that 
$V_{\rm DE}(\phi_0,\tau_0)\simeq H_0^2 \ll H_{\text{inf}}^2 \simeq V(\sigma_{\text{inf}})/\tau_{\text{inf}}^3$, 
where $H_0$ is the current Hubble scale, and $H_{\rm inf}$ is the scale of inflation. The hierarchy is a considerable one: the scale of dark energy is $H_0 \sim 10^{-60} M_p$, whereas the scale of inflation is assumed to lie somewhere in the range $10^{-42} M_p \lesssim H_\text{inf} \lesssim 10^{-5} M_p$.\footnote{Here, the lower bound comes from the scale of BBN (around MeV), although the actual temperature of the primordial bath might be higher, of $\mathcal{O}(\text{GeV})$ \cite{Davidson:2000dw}. The upper bound comes from constraints on the tensor-to-scalar ratio \cite{Planck:2018vyg}.} The considerations of \cite{Kallosh:2004yh}, now suggest a parametric separation in the scale of the underlying quintessence potential in the early universe and at late times. In particular, we can constrain the scale at early times, given by $V_{\rm DE}(\phi_\text{inf},\tau_\text{inf})$ by demanding that the volume modulus remains stabilised during inflation.  Following the same logic as \cite{Kallosh:2004yh}, we note that to avoid the runaway in the volume, the corresponding minimum at $\tau_\text{inf}$ should be separated from the asymptotic region by a barrier, $V_*$, as high as the scale of inflation, $V_* \gtrsim H^2_\text{inf}$. Given that generically we expect $V_* \sim |V_{\rm DE}(\phi_\text{inf},\tau_\text{inf})|$, to avoid a runaway we require $|V_{\rm DE}(\phi_\text{inf},\tau_\text{inf})| \sim V_* \gtrsim  H^2_\text{inf} \gg H_0^2 \simeq V_{\rm DE}(\phi_0,\tau_0)$.  As we will see in a moment, it is hard to see how we can achieve this separation of scales in a controlled setup.

One of the lessons from Sec. \ref{sec:modelbuild} is that, at leading order and weak coupling, one scale typically controls the scale of the AdS vacuum and the height of the barrier, both going as $W_0^p/\mathcal{V}^q \ll 1$, for some $p, q>0$. If the volume, inflaton and quintessence fields, correspond to three different moduli, we have seen how the considerations of Kallosh and Linde \cite{Kallosh:2004yh} suggest that the barrier height should be at least as large as the scale of inflation to avoid a runaway. This fixes the scale of the underlying potential to be far in excess of the dark energy scale, $W_0^p/\mathcal{V}^q \gtrsim H_\text{inf}^2 \gg H_0^2$. The AdS vacuum, even if it could be uplifted to Minkowksi by the inflationary energy density, would now be too deep for any next to leading order correction to be a viable dark energy candidate, where the potential must be positive.

Although our arguments have focused on the case where the inflation, the volume and quintessence field are three different moduli, the situation is not improved when we relax this assumption. Let us consider each of the alternatives:
\begin{itemize}
\item \textit{The volume accounts for dark energy but not for inflation:} The inflaton is once again assumed to be some other (s)axion orthogonal to the volume mode. It is then required that the volume mode stays stabilised during inflation and finds itself in a gentle slope at late times, giving rise to dynamical dark energy. However, in order to avoid the runaway during inflation, the potential must have a large barrier, far in excess of the scale of the late time potential. This scenario is very similar to the one we have already described, and as such, suffers from the same difficulties. Volume driven quintessence  will also give rise to long range forces that violate fifth force constraints (see e.g. \cite{Will:2014kxa}).

\item \textit{The volume accounts for inflation but not for dark energy:} In this scenario, the potential for the volume contains a high scale plateau, allowing the volume to roll slowly during inflation. After inflation, the volume should settle into a low scale Minkowksi vacuum. This could then be stabilised at leading order, with some next-to-leading effect giving rise to dynamical dark energy through another modulus. In \cite{Conlon:2008cj,Cicoli:2015wja} volume inflation near an inflection point has been realised by considering different competing contributions: non-perturbative effects, string loops, higher derivative corrections, anti-branes and charged hidden matter fields. Besides looking very contrived and tuned, these constructions raise doubts as to the level of perturbative control since the value of the volume during inflation is relatively small. Moreover, one should make sure that the quintessence field away from the minimum does not result in the volume being destabilised.\footnote{Notice that the tension between $H_{\text{inf}}$ and $H_0$ could be relaxed by  also having $W_0$ evolve from large to small values during inflation, as in the toy model of \cite{He:2010uk}. However we are not aware of a robust model that realises this effect while remaining under computational control.}
 
\item \textit{The volume accounts for everything:} In this case, the volume modulus is responsible for both inflation and dark energy. The situation is similar to the previous case but now we require a flat enough plateau later on as well. This seems to require competing terms at both small volumes (during inflation) and large volumes (during the dark energy period), with a significant hierarchy built in. Such hierarchies would need to be generated by exponentials, which are generated non-perturbatively. This suggests the early time behaviour may not be under perturbative control. Furthermore, if inflation ends with the inflaton rolling in a steep potential, and not approaching a minimum, reheating would need to be non-standard. Crucially the late time behaviour would also fall foul of fifth force constraints.
\end{itemize}

We can try to get around these  problems by assuming that the stabilisation of the volume lies at some low scale, near Minkowski  vacuum generated at leading order, breaking the connection between the scale of the vacuum and the height of the barrier. (Recall that the barrier height should exceed the scale of inflation to avoid decompactification.) If this leading order stabilisation leaves, say,  an axionic flat direction which is lifted only at subdominant order by tiny non-perturbative effects, one could reproduce the required hierarchy between $H_{\rm inf}$ and $H_0$ without inducing any destabilisation of the volume mode. Notice, however, that generating a supersymmetric Minkowski minimum ($W=0$) by solving the F-terms equations ($D_i W=0$) requires a finely tuned cancellation between all contributions to the superpotential, both at tree and non-perturbative level, as in the  racetrack scenario \cite{Blanco-Pillado:2004aap}, which was already identified as a way to skirt around the original KL problem \cite{Kallosh:2004yh}. Moreover, axion quintessence in agreement with swampland bounds on the associated decay constant, requires dynamical dark energy to occur close to the maximum of the axion potential where the scale of supersymmetry breaking would be extremely low, set by the  scale of dark energy. This leads us to conclude that {\it non}-supersymmetric Minkowski vacua are actually more appealing, at least if we want to build a viable  model of quintessence in string theory.

\subsection{A closer look at quintessence models with a KL problem}
\label{DetailedImplications}

When we consider quintessence in string theory, commitment to the refined dS and weak gravity conjectures forbids a dynamical model of dark energy based on either of the following scenarios: 
\begin{enumerate}
\item Minkowski vacuum with saxion slow roll down a very shallow potential \cite{Cicoli:2012tz}
\item Minkowski vacuum with axion quintessence with trans-Planckian decay constant \cite{Arkani-Hamed:2006emk}
\end{enumerate}
Whilst this leaves some alternatives, our consideration of the KL problem in the previous section suggests that most of these are also ruled out. In particular, the following scenarios 
\begin{enumerate}
\item Saxion hilltop for a Minkowski or AdS vacuum
\item Axion hilltop for a Minkowski vacuum with no hierarchy
\item Saxion slow roll down a moderate slope, with a runaway or a Minkowski vacuum
\end{enumerate}
are all compatible with the dS conjecture. Two minor clarifications are in order here. By `hierarchy' we mean the existence of an exponential hierarchy of scales between the leading order potential for the volume and the axion potential responsible for quintessence. By a `moderate slope' we mean order one in Planck units, {\it i.e.} steep enough to satisfy the refined dS conjecture but shallow enough to allow for at least one efolding of slow roll. 

Each of these three alternatives suffers from the KL problem. They also suffer from a variety of other problems, not least that of an unacceptably light volume modulus and a light gravitino. In this section we study specific examples of each scenario, explicitly demonstrating how many of these problems emerge.

\subsubsection*{Saxion hilltop for a Minkowski or AdS vacuum}

The racetrack scenario has a  supersymmetric Minkowski vacuum, separated from the runaway regime by a maximum in the volume mode. We can therefore imagine a dynamical model of dark energy where  the volume mode is rolling close to the hilltop, and the axion is fixed at its minimum, at $\theta=0$.  As we saw previously, the racetrack scenario was proposed as a way around the original KL problem, since the height of the barrier can be taken to be higher than the scale of inflation without any consequence on the gravitino mass. However, the height of the barrier is the height of the maximum in the volume direction (or better, the height of saddle in the ($\tau, \theta$)-plane). For hilltop quintessence driven by the volume mode, this height is now set by the dark energy scale
\begin{equation}
H_0^2 \sim V_\text{race}(\tau_{\rm max}, 0)\,.   
\end{equation}
Clearly this barrier is too small to protects us from the KL problem. Indeed, the contribution from inflation,  driven by a different field $\sigma$,  couples to the volume mode due to Weyl rescaling.  As expected, it will induce  destabilisation of the volume towards decompactification since 
\begin{equation}
\frac{V(\sigma)}{\tau^3} \sim H_{\rm inf}^2  \gg H_0^2 \sim V_\text{race}(\tau_{\rm max}, 0)  
\end{equation}

Of course, similar considerations also apply  to saxion hilltops where the global minimum is supersymmetric AdS. As an example, consider  KKLT models where a hilltop in the volume modulus is generically present, even in the absence of an anti-D3 brane uplift, as a consequence of $\alpha'^3$ corrections to the K\"ahler potential, as already stressed in \cite{Conlon:2018eyr}.  This model is also tractable  enough to easily demonstrate other issues that can emerge beyond the KL problem, such as the light volume modulus and the light gravitino. Let us run through some of the details.

The KKLT potential was already given in \eqref{Vkklt}. If we assume that the axion $\theta$ is stabilised at its minimum at $\theta=0$, the dynamics of the volume modulus $\tau$ in a neighbourhood of the maximum is controlled by the last two terms, in other words
\be
V_\text{KKLT}(\tau, 0) \approx -\frac{aA|W_0|\,e^{-a\tau}}{2\tau^2}+\frac{3 W_0^2\hat{\xi}}{64 \sqrt{2}\tau^{9/2}}\ ,
\ee
with $|W_0|$ given by \eqref{W0kklt}. This simplification allows us to show that the maximum is located at $\tau_\text{max}$, defined  by the relation
\be
\hat \xi=\frac{64 \sqrt{2} a A \tau_\text{max}^{5/2} e^{-a \tau_\text{max}} (a \tau_\text{max}+2)}{27 |W_0|}.
\ee
Since $a\tau_\text{max} \gg 1$, the height of the potential at the maximum is 
\be \label{Vkkltmax}
V_\text{KKLT}(\tau_\text{max}, 0)\approx \frac{ a^2 A |W_0|\,e^{-a \tau_\text{max}}}{9 \tau_\text{max}} = \frac{4}{9} a \tau_\text{min}\left(\frac{\tau_\text{min}}{\tau_\text{max}}\right) e^{-a \Delta \tau}|V_\text{AdS}|
\ee
where we made use of \eqref{W0kklt} and the expression for the scale of the leading order AdS minimum \eqref{Vads}. The minimum at $\tau_\text{min}$ and the maximum at $\tau_\text{max}$ are separated by distance $\Delta \tau= \tau_\text{max}-\tau_\text{min}>0$. 

Clearly the height of the maximium should be  fixed by the current Hubble scale, $H_0^2 \sim V_\text{KKLT}(\tau_{\rm max}, 0) $. By the same reasoning as for the racetrack scenario, we run into a KL problem.  In this simple model, it is also instructive to demonstrate the smallness of the mass of the gravitino and the volume modulus explicitly.

Current observational bounds require $V_\text{KKLT}(\tau_\text{max}, 0)\sim 10^{-120}$ in Planck units. Such low values can be achieved either by having $\tau_{\rm max}$ large or 
$|W_0|$ exponentially small, two requirements that are not independent in KKLT, as can be seen from \eqref{W0kklt}. 
In order to estimate the choice of  parameters that leads to the correct value for $H_0$, we use the fact that the maximum and the minimum are not too far apart, and  compute the height of the maximum to zeroth order in $\Delta \tau$
\be
V_\text{KKLT}(\tau_\text{max}, 0)\approx  \frac{4}{9} a \tau_\text{min}|V_\text{AdS}|\approx\frac{2 a^3 A^2 }{27}e^{-2 a\tau_\text{min}}
\ee
where we made use of \eqref{Vads}. Assuming $\frac{2 a^3 A^2 }{27}=\mathcal{O}(1)$, matching the observed value of $H_0$ requires $a \tau_\text{min}\sim 140$, which through \eqref{W0kklt} translates as $|W_0|\sim 10^{-59}$ and a gravitino mass of $m_{3/2}\sim 10^{-33}$ eV. This is unacceptably light \cite{Ferrer:1997yz,Kawasaki:2008qe}. The fact that the gravitino mass is of order the dark energy scale can be traced back to the fact that the leading order vacuum is supersymmetric and very close to Minkowski. Indeed, from equation \eqref{Vkkltmax}, we see that the scale of the supersymmetric AdS vacuum is bounded above by the dark energy scale. This failure to decouple $m_{3/2}$ and $H_0$ is clearly typical of any model featuring a leading order  supersymmetric Minkowski, or near Minkowski,  vacuum. 

To compute the mass of the volume mode, one has first to switch to a canonical field via $\phi = \sqrt{\frac32} \ln \tau$ and then compute $m_\phi^2\simeq V_{\phi\phi}$ at the location of the maximum. This yields $m_\phi^2 \simeq -3 a \tau_\tmax V_0 \simeq -3 a \tau_\tmax m_{3/2}^2$. Since  $a\tau_\text{max} \gtrsim a\tau_\text{min}\sim 140$, this implies that the mass of the volume mode is only one order of magnitude above the gravitino mass, explicitly showing the existence of a light volume problem.

\subsubsection*{Axion hilltop for a Minkowski vacuum with no hierarchy}

Let us return to the racetrack model and consider using $\theta$, instead of $\tau$, to drive quintessence. Once again, since the dark energy scale now sets the scale of the potential; this will immediately run into a KL problem.  As it happens, this model suffers from another problem, closely related to the KL problem, but applied only to late time dynamics. Indeed, even if we ignore the contributions from inflation, the volume barrier disappears as soon as we move the axion sufficiently far away from its minimum. In other words, in attempting to move the axion to the hilltop, the volume itself is immediately destabilised. 

\begin{figure}[h!]
		\centering
		\includegraphics[width=75 mm]{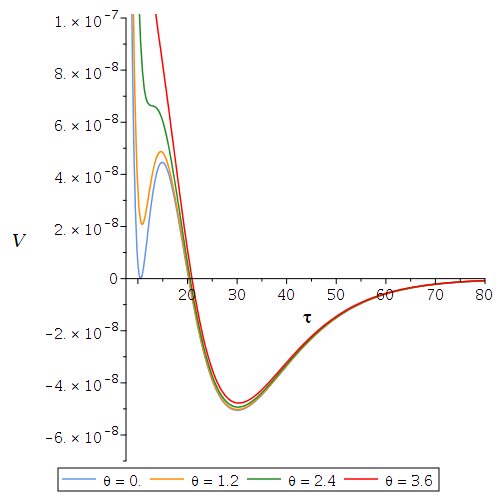}
	\caption{Racetrack potential at different values of the axion $\theta$ for parameter choice $
    A = 1, B = -1, a = 0.1, b = 0.09$
.} \label{racetrackbarrier.fig}
\end{figure}

This is demonstrated numerically in \figref{racetrackbarrier.fig}. Here we plot the form of the racetrack potential \eqref{raceV} as a function of the volume modulus $\tau$ for different values of the axion $\theta$. When the axion lies at its minimum at $\theta=0$, we see that the volume is stabilised at the Minkowski minimum. However, as we increase $\theta$ in units of the instanton coupling $a$, the volume barrier begins to shrink, and eventually disappears completely. At this point the volume will roll towards the AdS vacuum and any hope of exploiting the axion as a dynamical dark energy model is lost. 

These problems might have been anticipated in the racetrack scenario, as both the stabilisation of the volume and the dynamics of the corresponding axion rely on the {\it same} non-perturbative corrections to the superpotential. There was always a danger that the stabilisation would fail the moment the axion began to roll. As already pointed out, to proceed with a viable model of quintessence, we need to break the connection between the stabilisation of the volume and the dynamics of the would-be dark energy field, creating a hierarchy in mass between these two fields.

\subsection*{Saxion slow roll down a moderate slope, with a runaway or a Minkowski vacuum}

Let us now focus on a saxion runaway model, where the saxion is asymptotically rolling slowly down a moderate slope. 
At leading order, our example contains a non-supersymmetric Minkowski vacuum where one  of the saxion  directions is flat. The saxion runaway potential is then generated perturbatively, beyond leading order. However, since it is perturbative, it is not possible to generate a large enough hierarchy between the leading and subleading order terms to prevent the KL problem and destabilising the volume. Note that similar considerations would apply if additional subleading corrections were to generate a global non-supersymmetric Minkowski minimum, as opposed to a runaway. As shown in \cite{Hellerman:2001yi}, the case with a global supersymmetric Minkowski minimum is actually incompatible with slow roll down a moderate slope due to the stability condition on the form of the scalar potential.

Consider a fibred Calabi-Yau whose volume takes the form \cite{Cicoli:2011it,Cicoli:2018tcq}
\begin{equation}
\vo = \sqrt{\tau_1}\tau_2
\end{equation}
The saxion kinetic terms look like (we ignore the corresponding axions)
\begin{equation}
\mathcal{L}_{\rm kin}= \frac12 \left[(\partial \ln \tau_2)^2 +\frac12 (\partial \ln \tau_1)^2\right]
\end{equation}
and can be brought into canonical form by the following field redefinition 
\begin{equation}
\tau_1 = e^{\sqrt{\frac23} \chi + \frac{2}{\sqrt{3}}\phi}    \qquad\qquad \tau_2=e^{\sqrt{\frac23}\chi -\frac{1}{\sqrt{3}}\phi}
\end{equation}
Notice that $\chi$ corresponds to the volume mode $\vo$, and $\phi$ to the ratio $u=\tau_1/\tau_2$ since
\begin{equation}
\vo = \sqrt{\tau_1} \tau_2 = e^{\sqrt{\frac32}\chi} \qquad\qquad
u = \frac{\tau_1}{\tau_2} = e^{\sqrt{3}\phi}
\end{equation}
Let us consider an effective field theory defined by the following K\"ahler potential and superpotential
\begin{equation}
K = -2\ln\left(\vo+\frac{\xi}{2 g_s^{3/2}}-\gamma\sqrt{g_s} \ln \vo\right) - \frac{\tilde{C}}{\vo \sqrt{\tau_1}} \qquad\qquad W=W_0
\label{KandW}
\end{equation}
where $\xi$ controls $O(\alpha'^3)$ corrections, while $\gamma$ controls brane loop corrections at $O(\alpha'^3 g_s^2)$ \cite{Antoniadis:2018hqy}. For $\vo\gg 1$ these can naturally compete with the tree-level $O(\alpha'^3)$ term due to the $\ln\vo$ enhancement factor. The term proportional to $\tilde{C}$ represents $O(\alpha'^4 g_s^2)$ string loop corrections due to exchange of winding modes at the intersection of D7-branes. This contribution is subleading since it is suppressed by an additional power of $\alpha'$ with respect to the terms proportional to $\xi$ and $\gamma$. We do not include Kaluza-Klein loop corrections since they would be suppressed with respect to winding contributions by an additional power of $g_s^2$, and, moreover, they could be absent by construction if all branes intersect each other. We also neglect higher derivative $F^4$ contributions to the scalar potential since they would arise with additional volume suppression factors.

The K\"ahler potential and superpotential in (\ref{KandW}) generate the following scalar potential
\begin{equation}
V = V_{\rm lead}(\vo) + V_{\rm sub}(\vo, u)
\end{equation}
where
\begin{equation}
V_{\rm lead}(\vo) = \frac{C_{\rm up}}{\vo^{8/3}} + \frac{3W_0^2}{4\vo^3} \left(-2\gamma\sqrt{g_s}\ln\vo+\frac{\xi}{g_s^{3/2}}\right)
\end{equation}
and
\be
V_{\rm sub}(\vo ,u)=   \frac{2 \tilde{C} W_0^2}{\vo^{10/3}} \frac{1}{u^{1/3}} 
\ee
Notice that in $V_{\rm lead}$ we included also a term proportional to $C_{\rm up}$ representing the positive contribution of a T-brane background \cite{Cicoli:2015ylx} which is a generic feature of type IIB compactifications with 3-form fluxes and magnetised D7-branes. In the limit where the supergravity approximation is under control, {\it i.e.} for $\vo\gg1$ and $g_s\ll 1$, $V_{\rm sub}$ is indeed subdominant with respect to $V_{\rm lead}$ since
\begin{equation}
\frac{V_{\rm sub}}{V_{\rm lead}} \sim \frac{g_s^{3/2}}{\vo^{1/3}} = \frac{g_s^2}{\vo_s^{1/3}} \ll 1
\end{equation}
where $\vo_s = \vo g_s^{3/2}$ denotes the string frame volume. Thus at leading order the potential features a flat direction parametrised by $u$. At this level of approximation, by a suitable tuning of $C_{\rm up}$, $V_{\rm lead}$ features a non-supersymmetric Minkowski minimum where the volume scales as (for $k=\xi/\gamma$)
\begin{equation}
\vo_{\rm min} \sim e^{\frac{k}{g_s^2}}\gg 1\qquad\text{for}\qquad g_s\ll 1
\label{min}
\end{equation}
The subleading contribution $V_{\rm sub}$ determines just a small shift of the volume minimum and generates a runaway for $u$ which, when written in terms of the canonically normalised field $\phi$, looks like
\begin{equation}
V_{\rm sub} (\phi)=   V_0\,e^{-\lambda\phi}\quad\text{with}\quad \lambda=\frac{1}{\sqrt{3}}\quad\text{and}\quad V_0 = \frac{2 \tilde{C} W_0^2}{\vo_{\rm min}^{10/3}}
\label{saxionrunaway}
\end{equation}
This gentle runaway could provide an interesting model of quintessence in agreement with the refined dS conjecture since $\lambda\simeq 0.577$ is of order unity and it marginally satisfies the bound $\lambda \le 0.6$ obtained in \cite{Agrawal:2018own}. However the requirement to avoid volume destabilisation due to the inflationary energy implies (similar considerations would apply also to the case where (\ref{saxionrunaway}) describes a quintessence potential with a global Minkowski minimum)
\begin{equation}
V_{\rm lead} = \left(\frac{V_{\rm lead}}{V_{\rm sub}}\right) V_{\rm sub} \sim \frac{\vo^{1/3}}{g_s^{3/2}} H_0^2\gtrsim H_{\rm inf}^2\quad \Leftrightarrow\quad \frac{\vo^{1/3}}{g_s^{3/2}} \gtrsim \left(\frac{H_{\rm inf}}{H_0}\right)^2 \gtrsim 10^{-36}\,,
\end{equation}
where we have used $H_{\rm inf}\gtrsim 10^{-42}\,M_p$ as the extreme lower bound on the Hubble scale during inflation to be compatible with a BBN reheating temperature of at least 1 MeV.
Using (\ref{min}) which implies $g_s \sim \left(\ln\vo\right)^{-1/2}$, this bound becomes
\begin{equation}
\vo^{1/3} \left(\ln\vo\right)^{3/4}\gtrsim 10^{36}\quad\Leftrightarrow\quad \vo \gtrsim 10^{103}
\end{equation}
This would yield a string scale $M_s$ well below the TeV scale and a gravitino mass $m_{3/2}$ well below the meV scale since
\be
M_s \simeq g_s^{1/4} \frac{M_p}{\sqrt{\vo}} \sim  \frac{M_p}{\left(\ln\vo\right)^{1/8}\sqrt{\vo}} \lesssim 10^{-52}\,M_p
\ee
\begin{equation}
m_{3/2} \simeq g_s^{1/2} \frac{M_p}{\vo} \sim  \frac{M_p}{\left(\ln\vo\right)^{1/4}\vo} \lesssim 10^{-104}\,M_p
\end{equation}
Similar considerations imply that the mass of the volume mode is also very suppressed with respect to the meV scale. Hence the hierarchy between $V_{\rm lead}$ and $V_{\rm sup}$ is not big enough to prevent the KL and light volume problems. The reason is that the effective shift symmetry for $u$ is already broken at perturbative level.

\section{Axion hilltop quintessence and initial conditions} 
\label{sec:hills}

As explained in the previous section, a viable quintessence model has to feature a leading order non-supersymmetric Minkowski vacuum with hierarchy, {\it i.e.} where, at leading order, the axion is a flat direction while the saxion (in particular, the volume mode) is heavy. The axionic flat direction is then lifted by subdominant instanton effects which can lead to axion hilltop quintessence. In this section we therefore focus on this model, providing first an explicit realisation in LVS string models, and then studying the issue of initial conditions.

\subsection{LVS axion hilltop quintessence}
\label{LVSaxion}

The simplest way to realise an axion hilltop quintessence model in type IIB string theory is through the simplest LVS scenario, with two K\"ahler moduli $T_b=\tau_b+{\rm i}\theta_b$ and $T_s=\tau_s+{\rm i}\theta_s$, where the `big' modulus $\tau_b$ turns out to be much larger than the `small' modulus $\tau_s$. The important point is that the scalar potential only  depends on the volume axion, $\theta_b$, at next to leading order, without affecting the stabilisation of the volume mode which occurs at leading order. As a result, the volume axion can potentially play the r\^ole of quintessence when it is rolling near the top of its potential, without having any of the adverse consequences we saw for the racetrack scenario. We should, however, be mindful of the fact that the simplest LVS setup leads to a non-supersymmetric AdS vacuum which needs to be uplifted to Minkowski by the inclusion of additional sources of energy like T-branes \cite{Cicoli:2015ylx}. Notice that these positive contributions to the scalar potential are generic features of consistent type IIB compactifications \cite{Cicoli:2011qg, Cicoli:2012vw, Cicoli:2013mpa, Cicoli:2013cha, Cicoli:2017shd} due to the presence of hidden sector stacks of D7-branes (induced by D7 tadpole cancellation), 2-form gauge fluxes (induced by Freed-Witten anomaly cancellation) and 3-form background fluxes (used to freeze the dilaton and the complex structure moduli). For further details and a comprehensive discussion of dS model building in string theory, see \cite{Cicoli:2018kdo, Danielsson:2018ztv}. 

The main ingredients of this model are $\mc{O}(\alpha'^3)$ contributions to the K\"ahler potential and non-perturbative corrections to the superpotential of the form
\be
K = -2\ln\left(\vo+\frac{\hat\xi}{2}\right) \qquad\qquad 
W =W_0 + A_s\,e^{-a_s T_s} + A_b\,e^{-a_b T_b}
\ee
where the internal geometry corresponds to a simple Swiss-cheese scenario with the volume given by $\vo=\tau_b^{3/2}-\tau_s^{3/2}$. The resulting potential looks like
\begin{eqnarray}
V&=& {4 A_b^2 a_b^2 \over 3\tau_b}\,e^{-2 a_b \tau_b}+{8 A_s^2 a_s^2\sqrt{\tau_s} \over3\tau_b^{3/2}}\,e^{-2a_s\tau_s}+{3|W_0|^2\hat{\xi}\over 4\tau_b^{9/2}}  - {4 A_b a_b \over \tau_b^2}|W_0|\,e^{-a_b\tau_b}  \cos (a_b\theta_b )  \nonumber
\\ &-& {4 A_s a_s\tau_s\over \tau_b^3}|W_0|\,e^{-a_s\tau_s}\cos (a_s\theta_s)
+{8 A_b A_s a_b a_s \tau_s\over \tau_b^2 }\,e^{-(a_b\tau_b+a_s\tau_s)}\cos\left(a_b\theta_b-a_s\theta_s\right), \nonumber
\end{eqnarray}
where we have used $\tau_b\gg \tau_s$, $a_b\tau_b\gg 1$, $a_s\tau_s \gg 1$ and we have assumed $W_0<0$ so that the axions are minimised at $\theta_b=\theta_s=0$. With the axions settled at their minima, we consider the stabilisation of $\tau_s$ and $\tau_b$. Dropping all terms suppressed by $e^{-a_b \tau_b}$ or more, one finds the well-known LVS results from the variation with respect to $\tau_s$ and $\tau_b$
\be
\vo\simeq \langle \tau_b\rangle ^{3/2} \simeq {3|W_0| \sqrt{\la\tau_s\ra}\over 4 A_s a_s}\, e^{a_s\la\tau_s\ra} \qquad\qquad
{\hat\xi\over 2}\simeq \la\tau_s\ra^{3/2} \simeq \left[{1\over a_s} \ln\left(\vo\over |W_0|\right)\right]^{3/2}
 \label{eq:stab_lvs}
\ee
The minimum is AdS, breaks supersymmetry and, to leading order, is given by
\be
\langle V \rangle \simeq  -{3\hat\xi |W_0|^2 \over 8 a_s \la\tau_s\ra \vo^3 }\,.
\ee
There are several sources of uplifting to Minkowski which can be expressed as 
\be
V_{\rm up}={\kappa \over \mathcal{V}^\alpha}\ , \label{eq:uplift_term}
\ee
where $\kappa$ is a positive coefficient and $0<\alpha<3$. For example,  $\alpha=8/3$ for T-branes while $\alpha=4/3$ for an anti D3-brane at the tip of a warped throat, although the particulars of the uplifting mechanism are unimportant for the discussion that follows. The uplift term modifies the second relation in \eqref{eq:stab_lvs} as
\begin{equation}
    {\hat{\xi}\over 2}=\la\tau_s\ra ^{3/2} - {2 \alpha \kappa\over 9 |W_0|^2} \mathcal{V}^{3-\alpha}\ , \label{eq:eqxi}
\end{equation}
and we will fix 
$\kappa$, to zeroth order in $e^{-a_b\tau_b}$, by demanding that the uplifted LVS vacuum is Minkowski, {\it i.e.}
\begin{eqnarray}
\langle V\rangle &=& -\left({|W_0|^2\la\tau_s\ra^{3/2} \over  \vo^3} - {\kappa \left(3-\alpha\right)\over 3\vo^\alpha}\right)=0 \nonumber \\
&\Rightarrow&\quad \kappa = {3|W_0|^2 {\la\tau_s\ra}^{3/2}\over \left(3-\alpha\right)\vo^{3-\alpha}} \ . \label{eq:eq3}
\end{eqnarray}
The equations \eqref{eq:stab_lvs}, \eqref{eq:eqxi} and \eqref{eq:eq3} form a system fixing $\left(\la\tau_b\ra,\la\tau_s\ra,\kappa\right)$ for a particular choice of $\left(\xi, W_0, A_s, A_b, a_s, a_b, \alpha\right)$. Once the Minkowski vacuum is fixed in this way, we focus on $\theta_b$ as a dark energy candidate. The hierarchy of scales between $e^{a_b \tau_b}$ and $e^{a_s \tau_s}$ guarantees that shifts in the $\theta_b$ direction do not destabilise the Minkowski vacuum. 

The uplift term can be further adjusted at $\mc{O}\left(e^{-a_b \tau_b}\right)$  to guarantee a Minkowski vacuum at $\tau_b=\la\tau_b\ra, \tau_s=\la\tau_s\ra$ and $\theta_b=\theta_s=0$. Releasing the volume axion, $\theta_b$, its dynamics is then described by the following dark energy potential to leading order
\begin{eqnarray}
V_{\rm DE}&=& \left[
{4 A_b a_b \over \la\tau_b\ra^2}|W_0|\,e^{-a_b\la\tau_b\ra}     -{8 A_b A_s a_b a_s \la\tau_s\ra \over \la\tau_b\ra^2}\, e^{-(a\la\tau_b\ra+a_s \la\tau_s\ra)}\right]\left(1-\cos (a_b\theta_b)\right) \nonumber \\
&\simeq &
{4 A_b a_b \over \la\tau_b\ra^2}|W_0|\,e^{-a_b\la\tau_b\ra}  \left(1-\cos (a_b\theta_b)\right) ,
\label{VDE}
\end{eqnarray}
where we explicitly see the Minkowski minimum at $\theta_b=0$. The maximum is located at $\theta_b=\pi/a$. 
From the form of the K\"ahler metric, the canonically normalised axion and the corresponding decay constant turn out to be
\be
\phi \simeq \sqrt{3\over 2} {\theta_b\over \la\tau_b\ra}\qquad \qquad f_a = \sqrt{\frac32}\frac{M_p}{a_b\la\tau_b\ra}\,,
\label{LVSfa}
\ee
so that (\ref{VDE}) can be rewritten in a more standard way as
\be
V_{\rm DE} = V_0 \left(1-\cos \frac{\phi}{f_a}\right) \qquad\text{where}\qquad V_0 \equiv {4 A_b a_b \over \la\tau_b\ra^2}|W_0|\,e^{-a_b\la\tau_b\ra} 
\label{VDELVS}
\ee
The $\eta$ parameter at the maximum of the dark energy potential, where $\theta_b=\pi/a_b$, becomes  
\be
\eta_{\rm hilltop}= \left.{V_{\text{DE}, \, \phi\phi}\over V_{\rm DE}} \right|_{\phi_{\rm max}}  =\left.\frac23 \la\tau_b\ra^2  {V_{\text{DE}, \, \theta_b\theta_b }\over V_{\rm DE}}\right|_{\theta_b=\pi/a_b} = -\frac13 a_b^2 \la\tau_b\ra^2 
\ee
To estimate this, notice that the value of the potential at the hilltop should be $\mc{O}(10^{-120})$ in Planck units to be compatible with dark energy at late times. This suggests $a_b \la\tau_b\ra\sim \mathcal{O}(100)$, and so $\eta_\text{hilltop}\sim- \mathcal{O}(3000)$. Clearly the curvature at the hilltop is compatible with the improved swampland bound of \eqref{dscon}. However, the large absolute value of the $\eta$ parameter requires a high degree of fine tuning of the initial conditions for $\theta_b$ if it is to give rise to a viable quintessence model as it rolls away from the maximum, as we will illustrate in Sec. \ref{sec:constraining}. 

\subsubsection*{Other approaches to axion hilltop quintessence}

Before discussing the issue of initial conditions, let us mention other two possible approaches which can lead to a viable quintessence models via axion hilltop:
\begin{enumerate}
\item \emph{Uplifted KKLT with an orientifold-odd axion}

The standard uplifted KKLT scenario with a single K\"ahler modulus $T=\tau + i \theta$ features a non-supersymmetric Minkowski vacuum with no flat direction. 
This cannot be used to drive quintessence since both $\tau$ and $\theta$ are lifted by the same non-perturbative effect. However, in the presence of an extra orientifold-odd modulus $G= c+ S b$ (where $S$ is the axio-dilaton), $b$ would also be lifted by the non-perturbative superpotential $e^{-a T}$ (with a minimum at $b=0$), while the axionic mode $c$ would remain flat. This axionic direction can, instead, be lifted at subleading order by the inclusion of fluxed E3-instanton corrections to $W$ of the form $e^{- a (T+S+i G)} \sim e^{-a T} e^{-a/g_s}$ for $b=0$ \cite{Cicoli:2021tzt}. Therefore, for $g_s\ll 1$, the scale of the potential for $c$ is exponentially suppressed with respect to the potential for $T$, providing a promising candidate for a viable quintessence model with decay constant $f_a\simeq  \sqrt{g_s/\tau}\,M_p \ll M_p$ for $\tau \gg 1$ \cite{Cicoli:2021gss}.

\item \emph{Non-geometric fluxes}

A second possibility is to consider the effect of non-geometric fluxes which extend the GVW superpotential to \cite{Shelton:2005cf, Benmachiche:2006df, Blaback:2013ht, Plauschinn:2018wbo}
\begin{equation}
W=P_1(U) + S P_2(U) + \sum_i T_i P^{(i)}_3(U)\ ,
\end{equation}
where $P^{(i)}_3$ are cubic polynomials of the complex structure sector $U$. Combined with the tree-level expression of the K\"ahler potential, the dependence of $W$ on $T$ generates no-scale breaking contributions to the scalar potential. 
With regards to the previous discussion, if we are able to stabilise all but one (axionic) modulus  at a non-supersymmetric Minkowski vacuum (or even a near Minkowski AdS minimum) at tree-level through an appropriate choice of fluxes, then the hierarchy between non-perturbative effects and tree-level would guarantee that lifting the leftover flat direction would not displace the heavy moduli from the tree-level minimum. Furthermore,  the leftover axion could be made parametrically light and may be used to drive quintessence.\footnote{Assuming that backreaction effects on the K\"ahler potential can be kept under control and the internal volume can be made large enough to trust the perturbative expansion.}
\end{enumerate}

\subsection{Constraining hilltops} 
\label{sec:constraining}

Hilltop models are classically unstable. The rate of the instability is controlled by the $\eta$ parameter, describing the rate of change of the gradient close to the maximum of the potential. The larger the $\eta$ parameter, in absolute value, the closer the field needs to start near the maximum in order to obtain the required period of acceleration. Fortunately, for quintessence, we only require one efolding of accelerated expansion (this is in contrast to early universe inflation which requires at least $50$). Nevertheless, in string theory, the $\eta$ parameter can sometimes be quite large forcing the field to start very close to the top of the hill. Classically, this is not problematic if one accepts the inevitable tuning of initial conditions, although as we will see later, quantum diffusion at early times can push the field away from the sweet spot, spoiling any realistic chance of late time quintessence. 

In this section we will derive the constraints on the parameters and the initial conditions of a generic model of late time acceleration. As explained in the previous sections, our main interest will be in axion hilltop models, although we will also generalise our analysis also to saxion hilltop models which we approximate in a neighbourhood of the maximum as an inverted quadratic.

\subsubsection*{Axion hilltop quintessence}

In the context of late time acceleration, axions are the prototype of thawing quintessence models \cite{Scherrer:2007pu, Dutta:2008qn}: models where the field is frozen due to Hubble friction until the very recent past. These models are known to be sensitive to the choice of initial conditions and relatively insensitive to the particular form of the potential. A generic axion potential has the usual trigonometric form
\begin{comment}
\be
V=V_0-\tilde{V_0}\cos\frac{\phi}{f_a}
\ee
\end{comment}
\be
V=V_0\left(1-\cos\frac{\phi}{f_a}\right)
\ee
where we have (for simplicity) assumed that the vacuum expectation value of the axion lies at vanishing potential, consistent with a Minkowski vacuum. This can lead to accelerated expansion in two distinct regimes $f_a>M_p$ and $f_a<M_p$.  For $f_a>M_p$  acceleration takes place in the concave region of the potential, whereas if $f_a<M_p$ it happens in the vicinity of the maximum. While models with super-Planckian decay constants are less sensitive to initial conditions, getting these large values for $f_a$ has proven challenging from a UV point of view due to the tension with the weak gravity conjecture \cite{Arkani-Hamed:2006emk} and with explicit computations \cite{Cicoli:2012sz, Cicoli:2021gss}. For example, the LVS axion model presented in Sec. \ref{LVSaxion} features for example a sub-Planckian decay constant since (\ref{LVSfa}) gives $f_a\simeq M_p/(a_b \la\tau_b\ra)\sim 0.01\, M_p$. 

\begin{figure}[h!]
\begin{center}
\includegraphics[width=0.6\textwidth]{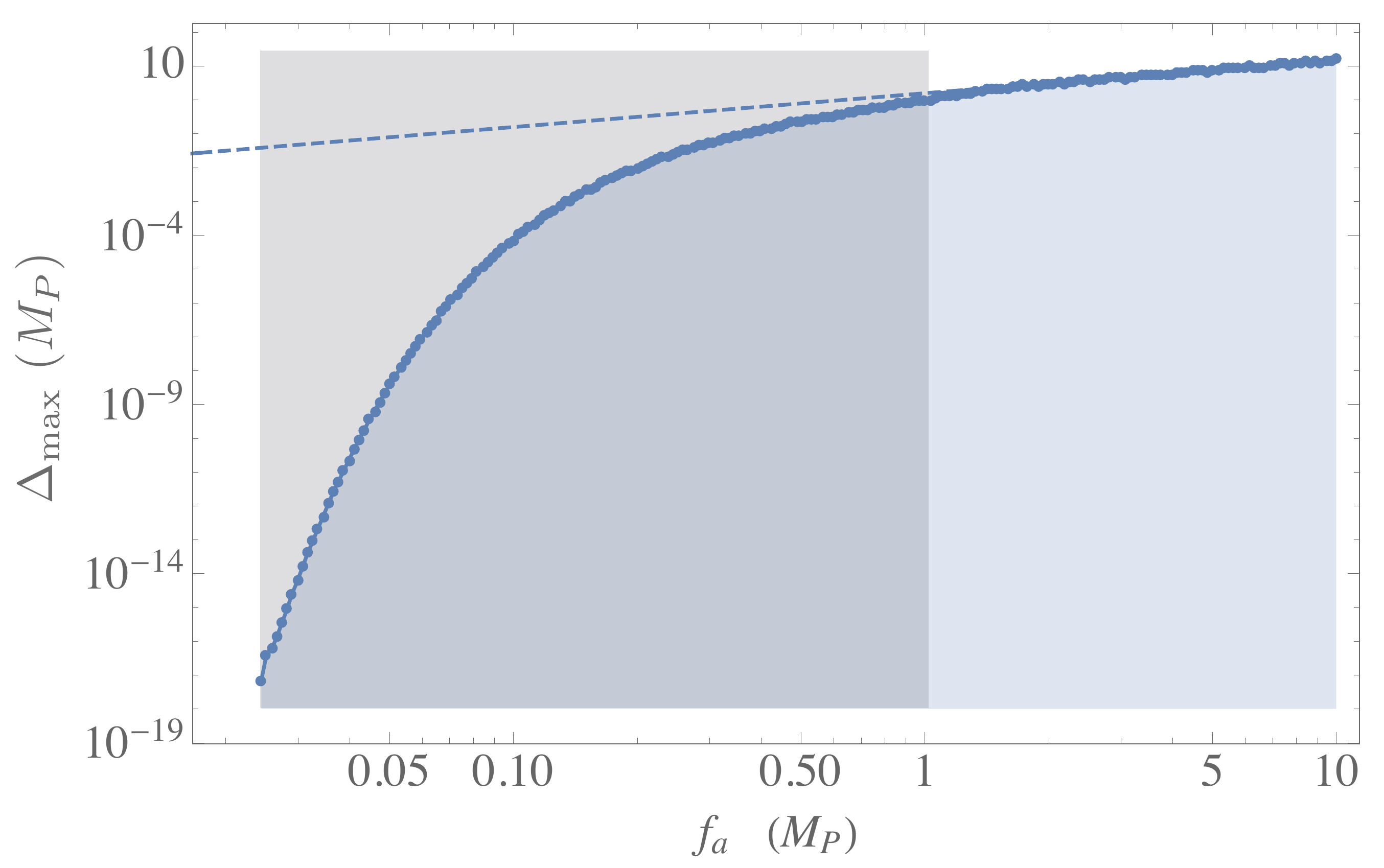}
\caption{Constraints on initial axion displacement from the maximum compatible with a viable quintessence hilltop model as a function of the decay constant $f_a$. For at least one e-fold of accelerated expansion, the initial value $\phi_\text{in}$ should  satisfy $|\phi_\text{in}-\phi_\text{max}|<\Delta_{\text{max}}$, where $\phi_\text{max}$ is the location of the maximum and $\Delta_{\text{max}}$ is given by the solid blue line. The dashed line shows the position of the inflection point $\phi=\pi f_a/2$. The blue shaded region corresponds to $f_a>M_p$ which is in tension with the weak gravity conjecture.}
\label{fig:f_vs_Phi0}
\end{center}
\end{figure}

Of course, axion hilltop quintessence can take place irrespective of the value of the decay constant, though it may require finely tuned initial conditions. In Fig. \ref{fig:f_vs_Phi0} we plot the deviation from the maximum as a function of the decay constant for a range of $f_a$ that is compatible with swampland constraints. We see that the range of $f_a$ that is more naturally achieved in UV constructions, $f_a<M_{\rm GUT}$, is also the one that suffers from an extreme sensitivity to the initial position of the field. In the region $f_a\in[0.02, 0.1]\,M_p$, the curve bounding the viable region can be approximated by
\be
\ln\Delta_{\text{max}} =c_0 +c_1 \ln f_a+c_2 (\ln f_a)^2\,,
\label{approximation}
\ee
where $\Delta_{\text{max}}$ denotes the maximum distance from the maximum compatible with late time acceleration, $c_0=-32.6$, $c_1=-28.977$ and $c_2=-8.2302$ .

\subsubsection*{Saxion hilltop quintessence}

For completeness, we now turn our attention to saxion models of quintessence, which, in the vicinity of the hilltop can approximated by an inverted quadratic
\be
V=V_0-\frac12\,m^2\, \phi^2.
\label{eq:V_HT}
\ee
It is useful to define the following parameter
\be
\eta_0=\frac{V_{\phi\phi}M_p^2}{V}\Big|_{\phi=0}=-\frac{\left(m M_p\right)^2}{V_0}
\ee
which describes the curvature of the scalar potential at the origin. Let us recall that the swampland conjecture (\ref{dscon}) requires $|\eta_0|\leq c'\sim\mc{O}(1)$. 

\begin{figure}[h!]
\begin{center}
    \includegraphics[width= 0.5\textwidth]{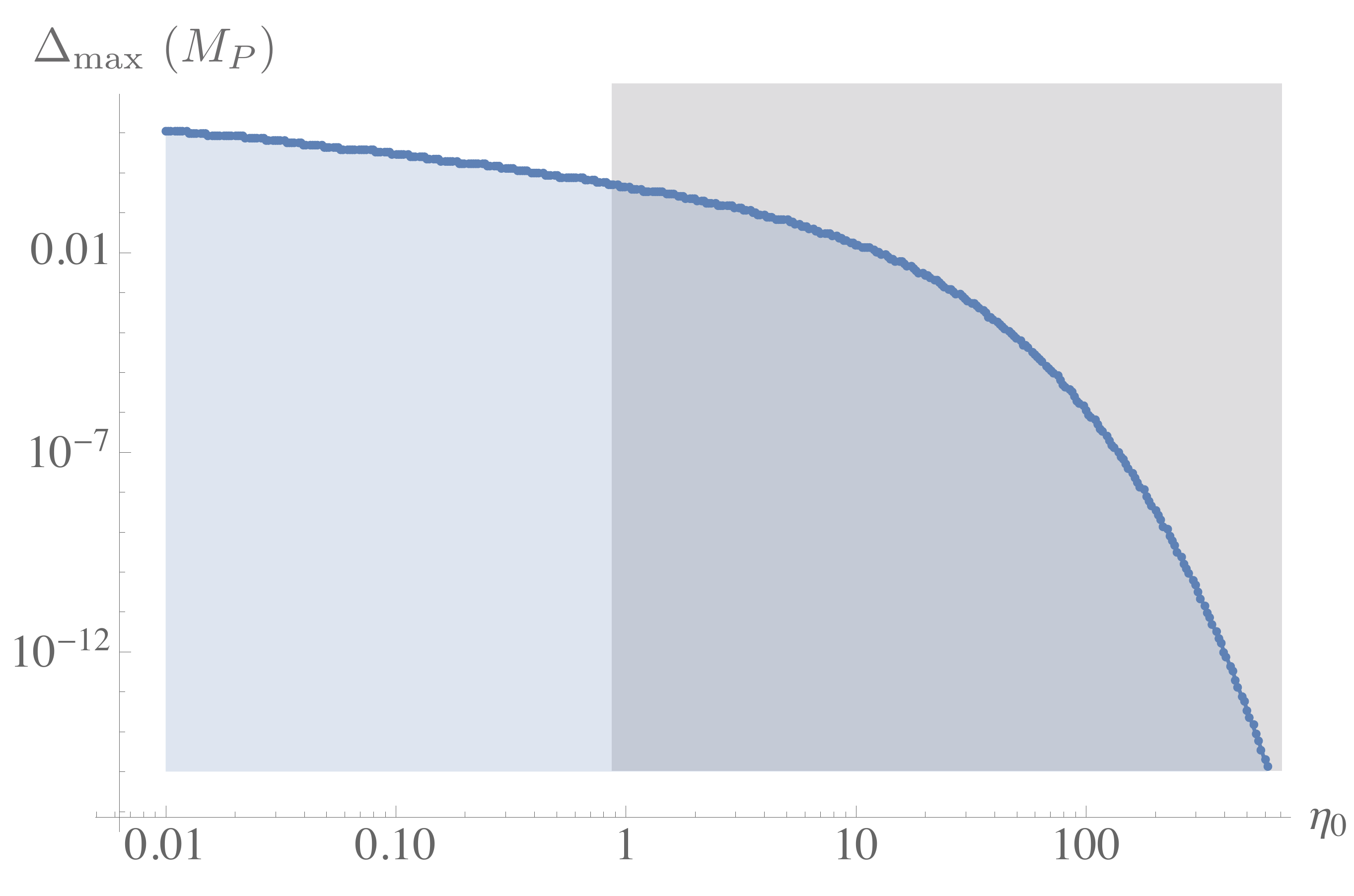}
\caption{Constraints on initial saxion displacement from the maximum compatible with a viable quintessence hilltop model as a function of $\eta_0$.  The grey region corresponds to $\eta_0>1$ which is in tension with the refined swampland conjecture. The steeper the maximum (larger $\eta_0$), the more fine tuned the initial conditions (smaller $\Delta_{\text{max}}$).}
\label{fig:VKS}
\end{center}
\end{figure}

Hilltop models, no matter how steep the potential  (or how large $\eta_0$), \emph{classically} always lead to extended periods of accelerated expansion, given that at the maximum $\epsilon_V=0$. This would allow for a description of the late time acceleration, regardless of the swampland limits on the slope and curvature of the potential that can be attained within a UV complete framework. On the flip-side, this comes at the price of tuning the initial position of the quintessence field, $|\phi_{\rm in}|<\Delta_{\text{max}}$ -  the steeper $V$, the closer $\phi_{\rm in}$ needs to be to the maximum in order to have an extended period of acceleration. One would think that this problem is slightly less of an issue in quintessence models which require only $\mc{O}(1)$, instead of $\mc{O}(50)$, efoldings as in inflation. However for large $\eta_0$ the level of tuning is similar.  

In Fig. \ref{fig:VKS} we show the maximal allowed initial displacement from the maximum, $\Delta_{\text{max}}$, compatible with late time acceleration as a function of $\eta_0$ for quadratic quintessence models. Each point corresponds to numerical solutions that start in matter domination with zero initial velocity and different values for $\phi_{\rm in}$. We see that the steeper the potential the more tuned is the initial value of $\phi$. For $\eta_0\ge1$ (grey region) $\phi_{\rm in}$ must lie within a fraction of $M_p$ from the top of the hill. In the absence of a dynamical mechanism, such initial conditions look rather unnatural. The curve bounding the viable region can be approximated by
\be
\ln\Delta_{\text{max}} =c_1+c_2 \ \eta_0^{-p}\ ,
\ee
where $c_1=1.7$, $c_2=-2.1$ and $p=0.44$. 

Even if one is willing to accept this level of tuning in order to  describe  the observed accelerated expansion and be in agreement with putative bounds from a UV theory, one must ponder if quantum effects will spoil the required tuning of the initial conditions. We address this issue in Sec. \ref{sec:stochastic}.

\subsection{Stochastic effects and initial conditions}
\label{sec:stochastic}

In this section we investigate whether the judicious choices of initial conditions described in the previous section survive the unavoidable stochastic fluctuations in the early universe. For our purposes we model inflation as an exact dS background, and therefore fix $H_{\rm inf}$ to be constant. Generally speaking this approximation is adequate when $H_{\rm inf}$ varies slowly, like in plateau models, but needs to be refined in the context of monomial inflationary models as shown in \cite{Hardwick:2017fjo}. In the cases of interest, where the stochastic processes are diffusion dominated and the equilibrium distribution (if it exists) is of little relevance, the exact dS approximation gives an adequate description of the system.

The quintessence field $\phi$, due to the large hierarchy $V(\phi)\ll H_{\rm inf}^2 M_p^2$ behaves as a spectator during inflation  and is, to leading order in the slow roll expansion, described by the Langevin equation
\be
\frac{\partial \phi}{\partial N}=-\frac{V_\phi}{3 H_{\rm inf}^2}+\frac{H_{\rm inf}}{2\pi}\,\xi\ ,
\label{eq:Langevin}
\ee
where $N$ denotes the number of efoldings and $\xi(N)$ is a stochastic variable with unit variance $\langle \xi(N_1) \xi(N_2)\rangle=\delta(N_1-N_2)$ and zero mean $\langle \xi(N)\rangle=0$. The last term in (\ref{eq:Langevin}) describes the backreaction of the short-wavelength modes of $\phi$ onto the homogeneous mode and turns the deterministic slow roll evolution of $\phi$ into a stochastic process. The stochastic nature of \eqref{eq:Langevin} implies that the system can equivalently be described in terms of the Fokker-Planck equation for the probability density function $P(\phi,\phi_{\rm in},N)$
\be
\frac{\partial P}{\partial N}=\frac{1}{3 H_{\rm inf}^2}\frac{\partial }{\partial \phi}\left(V_\phi P\right)+ \frac{H_{\rm inf}^2}{8 \pi^2 }\, P_{\phi\phi}\ ,
\label{eq:Focker_Planck}
\ee 
where we take $\phi_{\rm in}$ to be fixed at the onset of inflation. Once the solution to \eqref{eq:Focker_Planck} is known, all relevant moments of the distribution can be computed:
\be
\langle\phi^n\rangle(N)=\int d\phi\  \phi^n  P(\phi,\phi_{\rm in},N)\ .
\ee
In what follows we will be interested in the  first two moments: the mean, $\langle\phi\rangle$, and the mean square, $\langle\phi^2\rangle$, which allow us to determine the variance of the distribution $\sqrt{\langle\phi\rangle^2-\langle\phi^2\rangle}$.

Given that the energy scale of quintessence is hierarchically smaller than the scale of inflation,  the quintessence field is classically frozen during the inflationary epoch. It is only expected to thaw once the background Hubble parameter drops to around $H\sim m_\phi$, which should happen during the matter phase, after the Big-Bang. This should hold true regardless of the shape of the quintessential potential.

The existence of a vast hierarchy between the value of the Hubble parameter today and during inflation, $H_0^2\simeq V_0/M_p^2\ll H_{\rm inf}^2$, implies that the quintessence field is a spectator during inflation, and that it is undergoing pure Brownian motion with \eqref{eq:Langevin} well approximated by
\be
\phi'=\frac{H_{\rm inf}}{2\pi}\,\xi \ ; 
\ee
or equivalently \eqref{eq:Focker_Planck}, by the one-dimensional heat equation
\be
\frac{\partial P}{\partial N}= \frac{H_{\rm inf}^2}{8 \pi^2}\, P_{\phi\phi}\,.
\label{eq:heat}
\ee
This dominance of the stochastic effects over the classical evolution has severe consequences for the retention of memory of the initial conditions for the quintessence field.
Exact solutions to \eqref{eq:heat} take the form (see e.g. \cite{Uhlenbeck:1930zz})
\be
P=\sqrt{\frac{2\pi}{N H_{\rm inf}^2}}\exp \left(-\frac{2 \pi^2 }{N}\frac{(\phi-\phi_{\rm in})^2}{H_{\rm inf}^2}\right)
\label{eq:heatKernel}
\ee
from which one can show that 
\be
\langle\phi\rangle =\phi_{\rm in}\ ,
\ee
{\it i.e.} the ensemble average is frozen at the specified initial value for the classical field $\phi_{\rm in}$, in accordance with the fact that classically the field is frozen by Hubble friction. One can also show that 
\be
\langle\phi^2\rangle =\left( \frac{H_{\rm inf}}{2\pi} \right)^{2} N +\phi_{\rm in}^2\ ,
\ee
which implies 
\be
\sqrt{\langle\phi^2\rangle-\langle\phi\rangle^2} =\frac{H_{\rm inf}}{2\pi} \sqrt{N}\ .
\label{eq:var_ax}
\ee
Therefore in one efolding of inflationary expansion, the spectator field will be kicked on average by $H_{\rm inf}/(2\pi)$. Depending on the sensitivity of a given hilltop to the choice of initial conditions, and on the exact value of $H_{\rm inf}$, these stochastic effects can push the field away from the top of the hill, and into a region where it cannot account for the observed present day accelerated expansion.

\begin{figure}[h!]
\begin{center}
    \includegraphics[width= 0.5\textwidth]{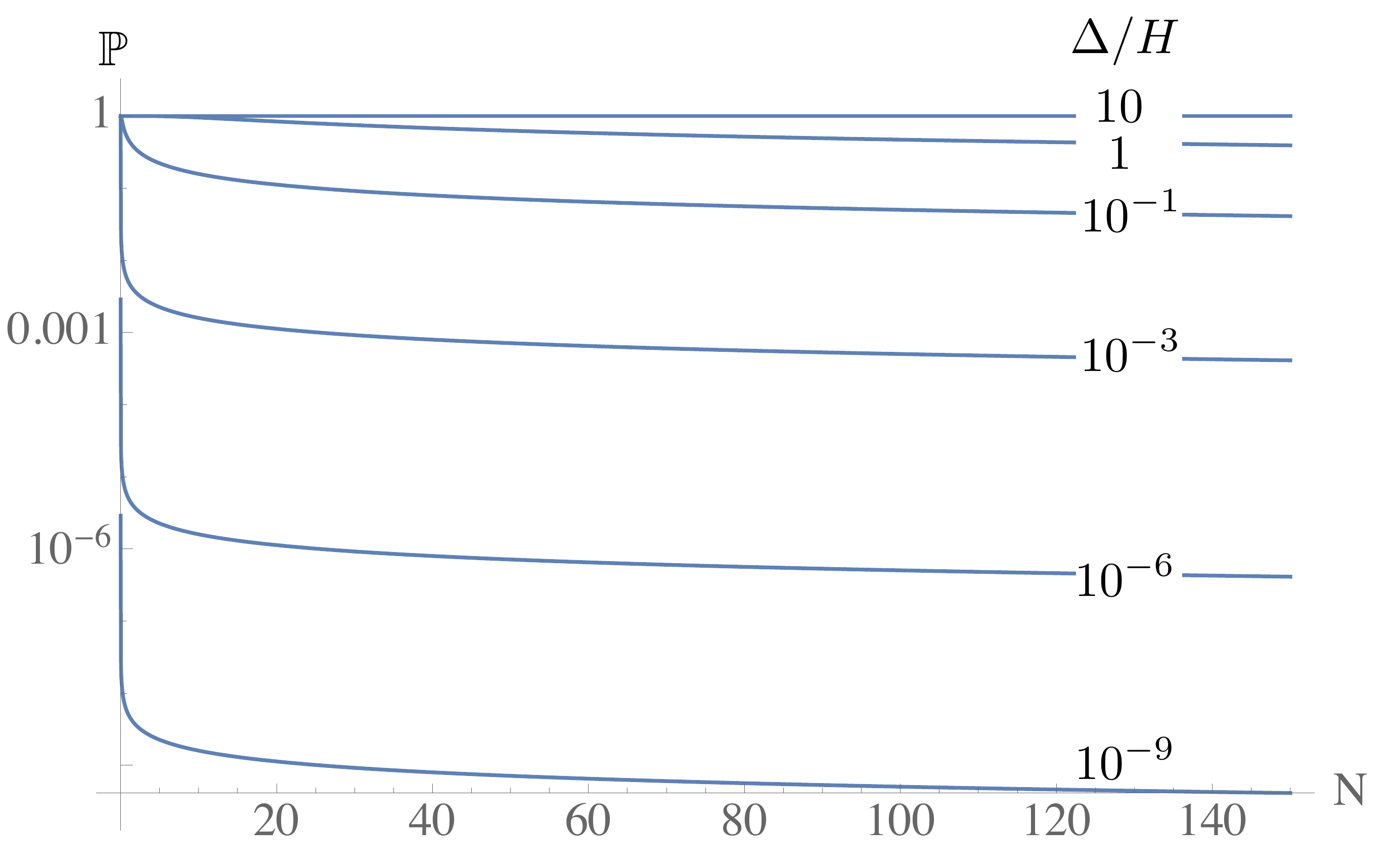}
\caption{Survival probability to remain within a distance $\Delta$ from the initial condition after $N$ efoldings of inflation, as a function of the ratio $\Delta/H_{\rm inf}$.}
\label{fig:Psurv}
\end{center}
\end{figure}

From \eqref{eq:heatKernel} we can compute the probability that a given choice of initial conditions survives the stochastic diffusion during a period of inflation. Setting $\phi_{\rm in}=0$ (assuming that this corresponds to the location of the maximum) and asking that after $N$ efoldings of inflation $\phi$ remains within a distance $\Delta$, we find that the survival probability is given by 
\be
\mathbb{P}(|\phi|\le \Delta)=\int_{-\Delta}^{\Delta} d\phi\,P= \text{erf}\left( \sqrt{\frac{2 \pi^2}{N}}\frac{\Delta}{H_{\rm inf}}\right)\ ,
\label{eq:survP}
\ee
where $\text{erf}$ is the complementary error function. For $N\gg 2\left(\frac{\pi \Delta}{H_{\rm inf}}\right)^2$ we can approximate \eqref{eq:survP} by
\be
\mathbb{P}(|\phi|\le \Delta)\simeq 2 \sqrt{\frac{2\pi}{N}}\frac{\Delta }{H_{\rm inf}}\,.
\label{important}
\ee
Once again we see that if $H_{\rm inf}\le \Delta$ the memory of the choice of initial conditions is preserved for a long period. In Fig. \ref{fig:Psurv} we plot the survival probability for various choices of $\Delta/H_{\rm inf}$.

\subsubsection*{Axion hilltop quintessence}

Stochastic effects on axionic hilltop quintessence models were studied in \cite{Kaloper:2005aj} and were pivotal in making the case for super-Planckian decay constants. Let us revisit these models in the light of the formalism reviewed above.

\begin{figure}[h!]
\begin{center}
\includegraphics[width=0.6\textwidth]{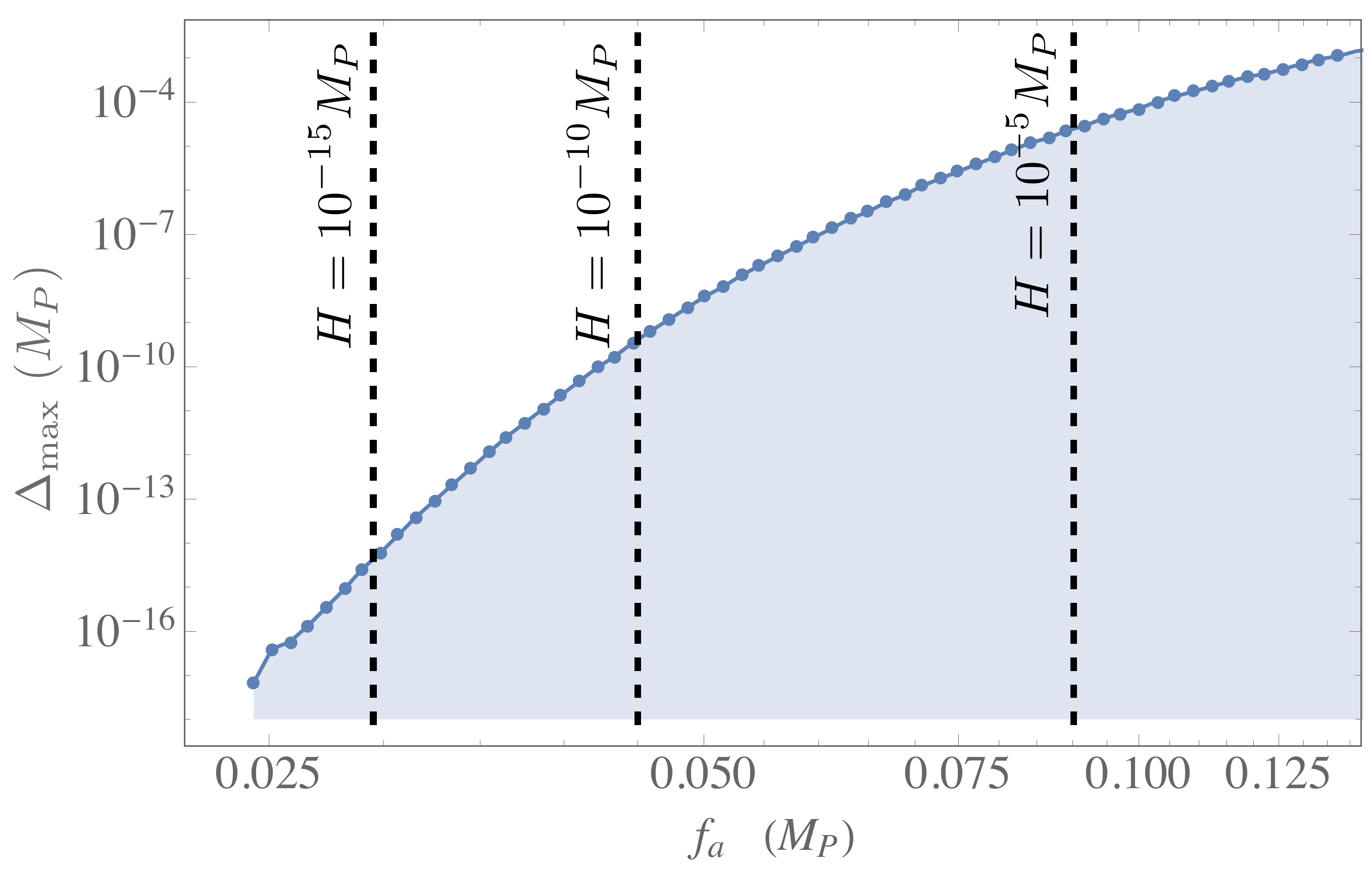}
\caption{Effect of quantum diffusion on the choice of initial conditions for axion hilltop quintessence models. Stochastic effects do not push the axion away from the hilltop region which yields late time acceleration only if $H_{\rm inf}\lesssim \Delta_{\text{max}}(f_a)$.}
\label{fig:f_vs_Phi0_ZOOM}
\end{center}
\end{figure}

Recall that current bounds on primordial tensor modes imply that, during inflation, $H_{\rm inf}\lesssim 10^{-5} M_p$ and that stochastic effects become relevant when the width of the region around the maximum of $V$ that gives rise to late time expansion is at, or below, the inflationary energy: $\Delta_{\text{max}}\lesssim H_{\rm inf}$. In Fig. \ref{fig:f_vs_Phi0_ZOOM} we zoom in on the low $f_a$ region of Fig. \ref{fig:f_vs_Phi0} and superimpose the constraints from $H_{\rm inf}$. We are led to the conclusion that axionic quintessence hilltop models with $f_a<0.1\,M_p$ are subject to stochastic fluctuations that (depending on the inflationary energy scale) may push $\phi>\Delta_{\text{max}}$ ruining the late time dynamics of those models. These estimates are in agreement with those of \cite{Kaloper:2005aj} and provide a worst case scenario. We note that a sharper statement can only be made with the knowledge of the exact energy scale of inflation.

\begin{figure}[!t]
\begin{center}
\includegraphics[width=0.47\textwidth]{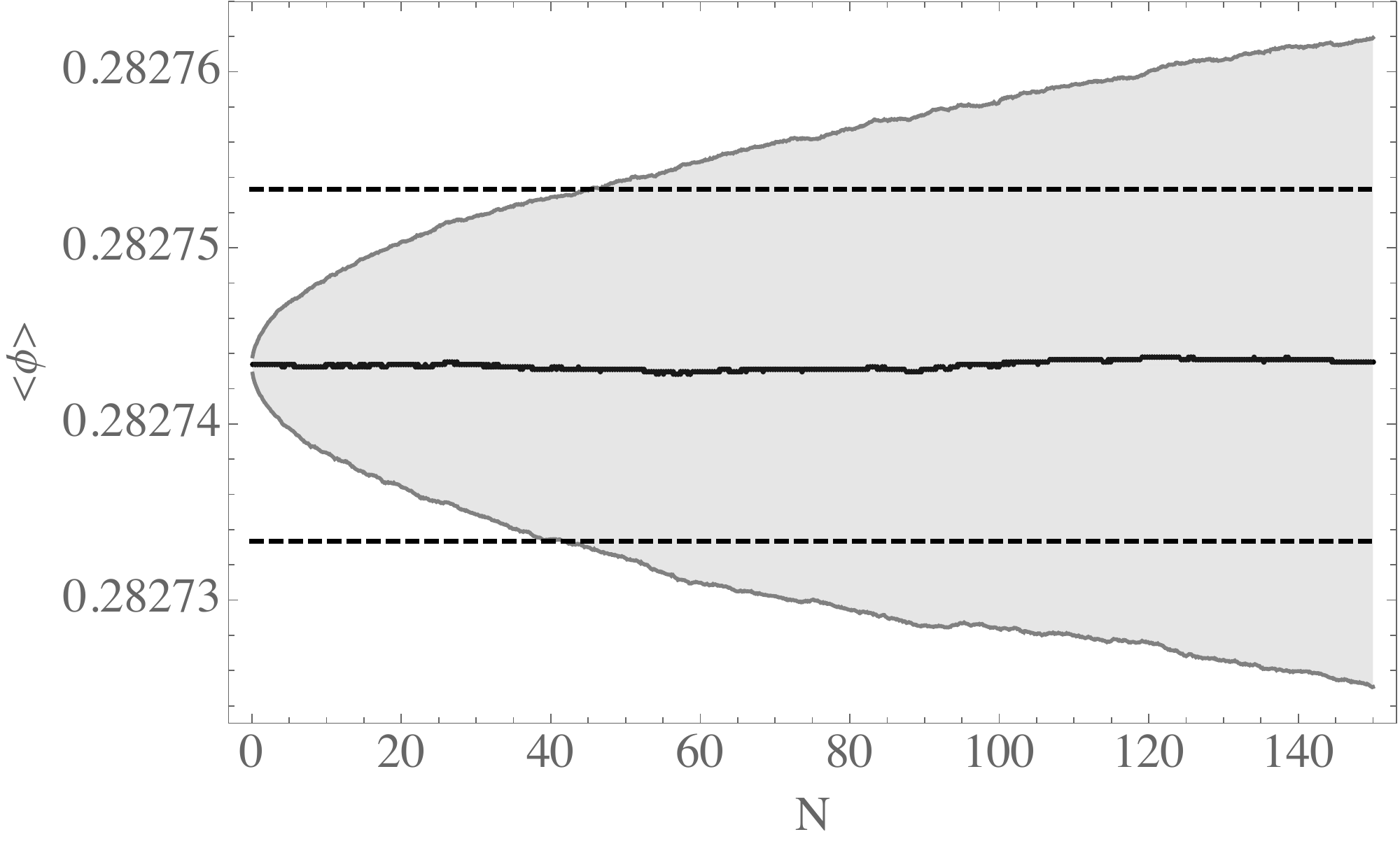}\hspace{0.5cm }\includegraphics[width=0.47\textwidth]{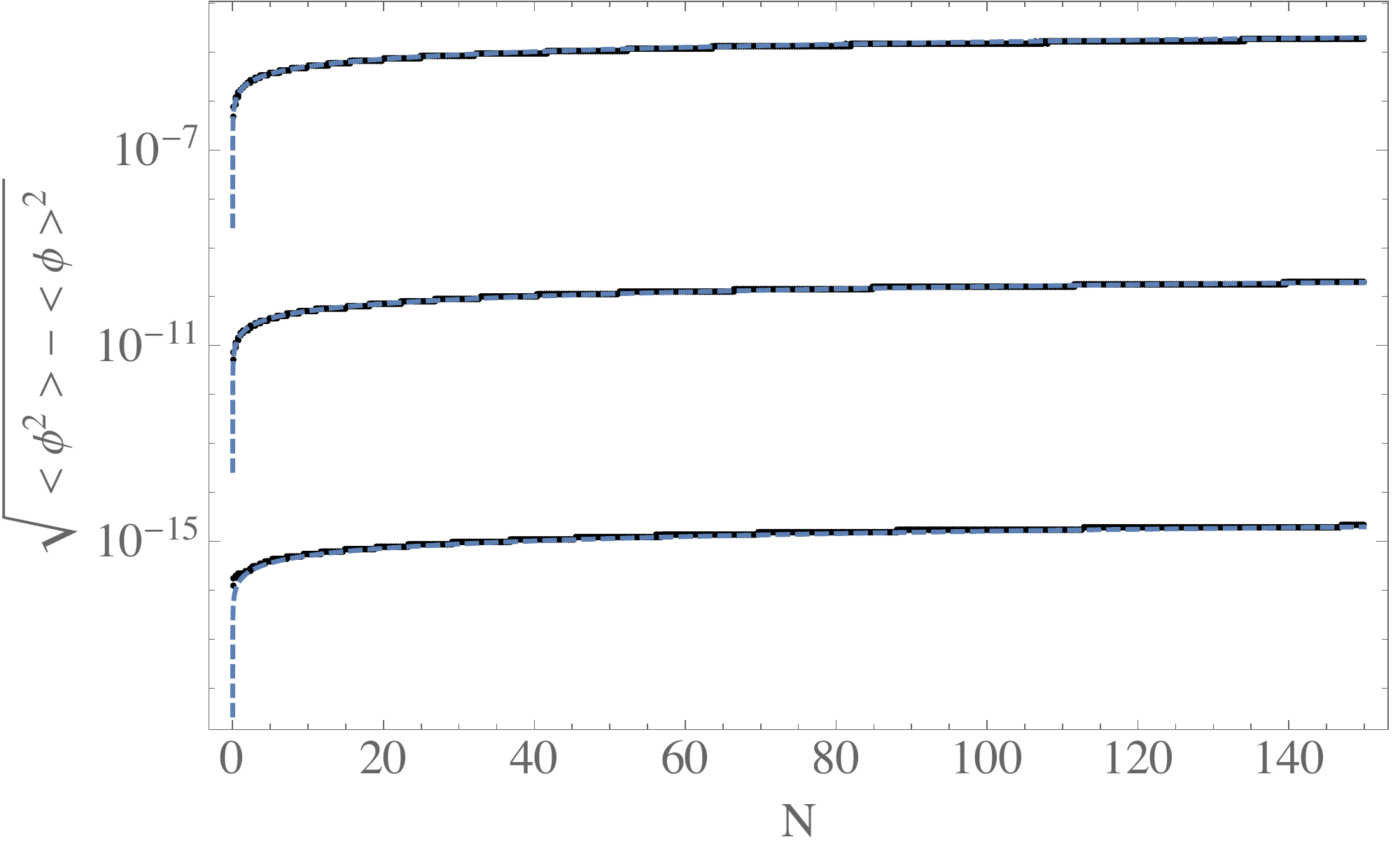}
\caption{Stochastic evolution for an axionic hilltop spectator with $f_a=0.09\, M_p$, $V_0=10^{-120} M_p^4$ and $\phi_{\rm in}=\pi f_a$ from $1000$ numerical solutions of the Langevin equation. Left panel: $H_{\rm inf}=10^{-5} M_p\simeq \Delta_{\text{max}}$, the grey region corresponds to the $1 \sigma$ band and the grey dashed lines denote the interval around the hilltop where the field should find itself in the matter phase to be able to drive quintessence. Right panel: standard deviation for (top to bottom) $H_{\rm inf}=\{ 10^{-5}, 10^{-10}, 10^{-15}\} \ M_p$, the black line corresponds to the $1000$ realisations of the Langevin equation while the dashed blue line corresponds to the analytical solution of \eqref{eq:var_ax}.}
\label{fig:P_FP_ax}
\end{center}
\end{figure}

In Fig. \ref{fig:P_FP_ax} we plot the solutions to the Langevin and Fokker-Planck equations for an axionic spectator with $H_{\rm inf}\simeq \Delta_{\text{max}}$, where it is evident that stochastic effects lead to a loss of memory of initial conditions after $N>\mc{O}(10)$ efoldings of inflation. 

The situation is even worse for $H_{\rm inf}>\Delta_{\text{max}}$ since the probability of lying within a distance $\Delta_{\text{max}}$ away from the maximum after $N\simeq 50-60$ efoldings of inflation turns out to be extremely small, as can be seen from (\ref{important}) with $\Delta = \Delta_{\text{max}}(f_a)$:
\be
\mathbb{P}(|\phi|\le \Delta_{\text{max}}(f_a))\simeq 2 \sqrt{\frac{2\pi}{N}}\frac{\Delta_{\text{max}}(f_a)}{H_{\rm inf}}\,,
\label{important2}
\ee
where $\Delta_{\text{max}}$ depends on the axion decay constant $f_a$ (for $f_a\in[0.02, 0.1]\,M_p$, $\Delta_{\text{max}}$ is very well approximated by (\ref{approximation})). For example, in the explicit LVS axion model of Sec. \ref{LVSaxion}, equations (\ref{LVSfa}) and (\ref{approximation}) give a decay constant $f_a \simeq 0.02\,M_p$ and a maximum displacement, $\Delta_{\text{max}}\simeq 2.4\times 10^{-20}\,M_p$. Choosing the largest value of $H_{\rm inf}$ compatible with the lack of observation of primordial gravity waves, $H_{\rm inf}\simeq 2\times 10^{-5}\,M_p$, and $N\simeq 50$ (\ref{important2}) would give $\mathbb{P}(|\phi|\le \Delta_{\text{max}})\simeq 10^{-15}$. Clearly smaller values of $H_{\rm inf}$ would give a smaller survival probability. Notice from Fig. \ref{fig:f_vs_Phi0} that $\Delta_{\text{max}}$ drops very quickly for smaller values of $f_a$, reducing the survival probability even further. 

We therefore conclude that a viable axion hilltop quintessence model requires two crucial conditions: ($i$) a tuning of initial conditions close to the maximum which becomes more severe for smaller values of $f_a$; and ($ii$) $H_{\rm inf}\lesssim \Delta_{\text{max}}$, else  stochastic effects will very quickly push the field away from the hilltop region compatible with a late time period of accelerated expansion. This second constraint turns out to be very strong since explicit computations of axion decay constants from string theory typically yield $f_a\lesssim 0.02\,M_p$ in the regime where the effective field theory is under control. For these small values of $f_a$, the maximum displacement is bounded as $\Delta_{\text{max}}\lesssim 10^{-20}\, M_p$\footnote{Numerically, we did not consider decay constants $f_a<0.02\,M_p$ since the high degree of tuning of the initial conditions rapidly brings about numerical precision issues. This prevent us from explicitly determining $\Delta_{\text{max}}$ for such low values of $f_a$, although the result can be obtained by extrapolating the validity of the formula \eqref{eq:heat}. In any event, the precise estimate is not really needed for drawing our general conclusions.}. Inserting this into the second constraint above, we see that we require a very low scale of inflation, $H_{\rm inf} \lesssim 10^{-20}\, M_p$.
When combined with the observed amplitude of scalar perturbations $A_s$, this bound on $H_{\rm inf}$ can then be translated into a severe upper bound on the inflationary slow roll parameter
\be
\epsilon_V = \frac{1}{8\pi^2 A_s}\left(\frac{H_{\rm inf}}{M_p}\right)^2 \lesssim 10^{-35}\,.
\ee
This is in strong tension with the dS swampland conjecture (\ref{dscon}) which requires $\mc{O}(1)$ values of $\epsilon_V$ (unless inflation is also realised extremely close to a maximum). Thus we conclude that axion hilltop quintessence would either be in the swampland or, if we ignore the refined dS swampland conjecture, would require a high tuning of initial conditions combined with a Hubble scale during inflation below $\mc{O}(1-10)$ MeV. Because these models are so contrived, we expect that dynamical dark energy, if supported by data, will have to be driven by a different mechanism, probably along the lines of axion alignment \cite{Kim:2004rp}.

\subsubsection*{Saxion hilltop quintessence}

The effects of diffusion in saxion hilltop models are qualitatively similar to those of axionic models. They will constrain the steeper hilltops, as these are the ones where the initial conditions are more severely tuned. From Fig. \ref{fig:VKS} we see that quadratic hilltops with $\eta_0\gtrsim 70$ require $\Delta_{\text{max}}\lesssim 10^{-5} M_p$ rendering them potentially vulnerable to diffusion effects during inflation, as CMB observations imply $H_{\rm inf}\lesssim 10^{-5} M_p$. Notice that the results for the effects of quantum diffusion obtained for the axion case can  also be used for saxion hilltop quintessence as long as we identify $|\eta_0| = \left(M_p/f_a\right)^2$.

\section{Discussion}
\label{Conclusions}

With compelling observational evidence for dark energy \cite{Perlmutter:1998np,SupernovaSearchTeam:1998fmf, Aghanim:2018eyx},  we cannot avoid the question of its microscopic origin. But should we be looking for a cosmological constant or quintessence? If the latter, then is it driven by a scalar or pseudo-scalar, on a shallow potential or at a hilltop? At present all options are observationally viable, but we can also ask which is easiest to build into a fundamental theory. 

In this chapter, we have outlined several challenges facing string theory models of quintessence focusing on effective field theories where perturbation theory is under numerical control - i.e. where the dilaton, ${\rm Re}(S)\gg 1$, and the volume mode, $\vo\gg 1$, are large enough to trust both the string loop and the $\alpha'$ expansion. This is arguably the most interesting region of moduli space since deep in the bulk, where ${\rm Re}(S)\sim \vo \sim \mc{O}(1)$, one would need a full knowledge of the whole quantum theory, while at boundary of the moduli space, where ${\rm Re}(S)\to\infty$ and $\vo \to \infty$, there is strong evidence indicating the absence of both dS vacua \cite{Ooguri:2018wrx} and a viable quintessence dynamics \cite{Cicoli:2021fsd, Hertzberg:2007wc, Garg:2018zdg, ValeixoBento:2020ujr}. 

Some of the problems of dynamical dark energy models are shared with the pure cosmological constant (like the smallness of $H_0$), while others are particular to quintessence (like constraints from fifth forces, tuning of initial conditions and radiative stability of the mass of the quintessence field). Here we were particularly concerned with  the destabilisation of the volume modulus  during inflation (the KL problem applied to quintessence) and generating the large hierarchy between the scale of the universe today and during inflation.  We have argued that the need to overcome all of these challenges has singled out a preferred model building scenario for dynamical dark energy. 
%We have argued that the need to avoid these problems - most notably,  the destabilisation of the volume during inflation (the KL problem applied to quintessence) and the large hierarchy between $H_0$ and $H_{\rm inf}$, singles out a preferred model building scenario for dynamical dark energy. 
The leading order contributions to the scalar potential should yield a vacuum with the following properties:
\begin{enumerate}
\item it should admit a flat direction in order to decouple the dark energy scale from the inflationary scale;
\item this flat direction should be axionic. This is because saxions are already lifted  at perturbative level without being able to generate the required hierarchy between $H_0$ and $H_{\rm inf}$. Axions, in contrast, develop a potential via highly-suppressed non-perturbative effects;
\item it should be (nearly) Minkowski since otherwise subleading corrections would not be able to push it up to the positive energies required to drive an epoch of accelerated expansion; 
\item it should break supersymmetry in order to decouple the gravitino mass from the dark energy scale. 
\end{enumerate}

It is interesting to combine these results with the swampland dS conjecture that would rule out dS vacua. From a model-building perspective, dS vacua look qualitatively the same as  non-supersymmetric Minkowski, leaving dynamical dark energy as the only explanation for the present acceleration of the universe. However we have found that it is extremely hard to realise a working model of quintessence in any scenario which would be in agreement with the swampland dS conjecture, like moderately sloped runaways, or supersymmetric AdS or Minkowski vacua. This tension raises some doubts on the validity of the swampland dS conjecture since it would imply that quantum gravity is in contradiction with observations. 

At this point it is tempting to favour the humble cosmological constant as the simplest empirical model of dark energy: it fits the available data and avoids the additional complications associated with quintessence. However, it is important to note that quintessence can open up opportunities to solve other cosmological problems.  For example, in \chpref{chap:coincidence}, it was shown how dynamical models of quintessence in string theory may shed new light on the cosmological coincidence problem \cite{Zlatev:1998tr,Velten:2014nra}. An evolving scalar on cosmological scales may also allow for self adjustment mechanisms to address the naturalness problems associated with vacuum energy (see \cite{Weinberg:1987dv, Niedermann:2017cel} for relevant no go theorems, and \cite{Copeland:2021czt} for a recent way around them). But perhaps most importantly,  future observations may rule out the cosmological constant as the driver of late time acceleration. 

If this were indeed the case, our analysis provides guidance for successful quintessence model building in string theory. In fact, we studied axion hilltop quintessence  in detail since vanilla string compactifications lead to axion decay constants at least two orders of magnitude below the Planck scale. We found that hilltop models are rather contrived since, even if the initial conditions are tuned very close to the maximum, quantum diffusion effects during inflation would kick the quintessence field away from the accelerating region close to the maximum, unless the Hubble scale during inflation is extremely low, $H_{\rm inf}\lesssim \mc{O}(1-10)$ MeV. Of course, one could envisage a scenario where a suitable coupling between the inflaton and the quintessence field makes the latter heavy during inflation, thereby suppressing any stochastic effect. However, after the end of inflation, the inflaton would typically settle down at the minimum of its potential, reaching its present day value. Quantum diffusion would then still play an important r\^ole in the reheating phase  and after, implying that the results of Sec. \ref{sec:stochastic} would still hold qualitatively after inflation with $H_\text{inf}$ replaced by the Hubble scale during a given epoch. We conclude that a more promising avenue to build a working model of dynamical dark energy is to rely on alignment mechanisms to obtain an effective axion decay constant which is trans-Planckian \cite{Kim:2004rp}. 

As stated earlier, for dynamical dark energy, we first seek a scenario where the volume is stabilised at leading order to a vacuum that is uplifted to Minkowski. Non-perturbative corrections can then be exploited to drive dark energy at the correct scale. Although it might seem a little uneconomical to uplift and then do quintessence, if dark energy turns out to be dynamical both steps may be necessary to explain the present state of the universe in the context of string compactifications. 

In truth, both the cosmological constant and quintessence face formidable challenges from the perspective of consistent model building in string theory, while remaining perfectly compatible with observational constraints. It behooves us to  better understand the limitations imposed by perturbative string theory in both cases. Indeed, does a microscopic understanding of dark energy require input from non-perturbative strings, through string field theory, or M-theory?  Since this may be a question of properly understanding the vacuum structure of the theory, this seems like a reasonable possibility.

\blankpage
\chapter{Gauss-Bonnet curvature corrections and the absence of de Sitter vacua \label{chap:gbds}}

In the previous chapter, we described a number of phenomenological impediments to quintessence model building. Many of the arguments we described rely on compactifications of string theory on Calabi-Yau manifolds, originally chosen to preserve some supersymmetry in the external space. Of course, to recover de Sitter space, supersymmetry must be broken completely\footnote{We note that supersymmetry may be realised non-linearly on de Sitter space whenever the symmetry is broken spontaneously, as in \cite{Dudas:2015eha, Bergshoeff:2015tra, Bandos:2015xnf, Nagy:2019ywi, Bansal:2020krz}}, so it is worth exploring string compactifications on manifolds that are not Calabi-Yau, at least in the rare cases where one can solve the equations and retain  some calculational control. To this end, in \cite{Montero:2020rpl}, the authors  took a generic class of string inspired models and studied compactifications on a $p$-sphere with internal $p$-form fluxes. They identified a parametric bound (henceforth known as the Montero-van Riet-Venken (MvRV) bound), that needed to hold for de Sitter solutions to exist. The MvRV bound corresponds to a remarkably simple inequality between the slope of the potential and the derivative of a $p$-form gauge coupling. A number of stringy models were studied where it could be explicitly shown that the inequality is not satisfied, ruling out de Sitter solutions in the most realistic set-ups. 

In this section, we expand on the work of \cite{Montero:2020rpl} by introducing higher order curvature corrections in the form of a Gauss-Bonnet contribution to the action, as well as generalising the set-up to contain internal or external fluxes with spherical or toroidal internal geometries. The goal is to establish whether the curvature corrections  strengthen or weaken the MvRV bound of \cite{Montero:2020rpl} for existence of de Sitter vacua. We will consider actions of the following form:
\begin{equation}
	S=\displaystyle \int{\mathrm{d}^Dx\sqrt{-g}\left[\dfrac{R}{2\kappa_D^2}-\dfrac12 (\partial\varphi)^2-V(\varphi)-{g(\varphi)}\left|H_k\right|^2+f(\varphi)\mathcal{G}\right]},
	\label{eq:action}
\end{equation}
where $\mathcal{G}$ is the Gauss-Bonnet term, cf. \eqref{eq:GB_def}, and take our ansatz manifold to be the warped product $\mathcal{M}=\Sigma_n\rtimes_\rho\Sigma_p$ with line element
\begin{equation}
\label{eq:metansatz}
	\mathrm{d}s^2=g_{AB}\mathrm{d}x^A\mathrm{d}x^B=e^{2\alpha \rho}\tilde{g}_{\mu\nu}\mathrm{d}\tilde{x}^\mu\mathrm{d}\tilde{x}^\nu+e^{2\beta \rho}\hat{g}_{ab}\mathrm{d}\hat{x}^a\mathrm{d}\hat{x}^b,
\end{equation}
where $\alpha$ and $\beta$ are constants, capital Latin indices refer to the $D$-dimensional spacetime, Greek indices refer to the external spacetime of dimension $n$ with metric $\tilde{g}_{\mu\nu}$, lower case Latin indices refer to the internal manifold of dimension $p$ with metric $\hat{g}_{ab}$, and $\rho$ is a scalar that varies over the external spacetime only. $H_k$ is a top-form that is either purely external ($k=n=D-p$) or purely internal ($k=p$) since there is no preference for one or the other at the level of the effective field theory.  We will consider $\Sigma_n$ to be either de Sitter $dS_n$ or Minkowski $\mathbb{R}^{1,n-1}$, while $\Sigma_p$ will be either  a sphere $S^p$ or a torus $T^p$.

We initially adopt a perturbative approach, working to leading order in the Gauss-Bonnet coupling, $f(\varphi)$.  In this limit, we find  that the MvRV bound obtained in \cite{Montero:2020rpl}  can be marginally violated and de Sitter solutions can still exist.  This is true as long as a second parametric bound holds on the gradient of $f(\varphi)$. Unfortunately, this second bound does {\bf not} hold for choices of $f(\varphi)$ and  $g(\varphi)$ best motivated from string theory. To reinforce this latter point, we extend our analysis beyond perturbation theory, this time specialising to the string motivated choices for $f(\varphi)$ and $g(\varphi)$.  Consistent with our perturbative analysis, we find that the MvRV bound is still a necessary condition for the existence of de Sitter solutions. However, it is no longer sufficient, suggesting that for string motivated potentials, the higher curvature corrections make it parametrically harder, not easier to find de Sitter vacua. 

The structure of this chapter is as follows. In \secref{sec:intro_het}, we review the results of \cite{Montero:2020rpl} and motivate the introduction of the Gauss-Bonnet term as a higher order curvature correction within the framework of heterotic strings. In \secref{sec:genpot}, we gain some insight into the effects of the Gauss-Bonnet correction by performing a small coupling expansion. Then, in \secref{sec:specifpot}, we focus on specific string-inspired potentials and discuss in detail what conditions are necessary and sufficient for the existence of solutions. We conclude with a discussion in \secref{sec:discussion}.

%%%%%%%%%%%%%%%%%%%%%

\section{Review of flux compactifications of the heterotic string action and stability conditions}
\label{sec:intro_het}

In \cite{Montero:2020rpl}, the authors considered the existence and stability of $dS_{n}\times S^p$ flux compactifications. The action of \cite{Montero:2020rpl} can be cast as \eqref{eq:action} with $f(\varphi)=0$ and with
$k=p$, so that $H_k$ is a top-form living in the internal space, providing the flux that aims to stabilise the potential for the scalar $\varphi$, and prevent the runaway. In this particular case, the coefficients $\alpha$ and $\beta$ which parametrise the compactification are chosen to be
\be \label{albe}
\alpha^2=\frac{p}{2(n-2)(n+p-2)}\ , \qquad \beta=-\frac{n-2}{p}\alpha\ ,
\ee
so that the $n$ dimensional effective action obtained after dimensional reduction can be expressed in Einstein frame, with a canonical kinetic term for the breathing mode. The effective potential for $\varphi$ and $\rho$ in the external spacetime is found to be\footnote{Up to an overall internal volume factor which can be reabsorbed in a Weyl redefinition of the four-dimensional metric.}
\be
V_\text{eff}=-\dfrac{p(p-1)}{2\kappa_D^2} e^{2(\alpha-\beta)\rho}+ {Q^2}e^{2(n-1)\alpha \rho}g(\varphi)+V(\varphi)e^{2\alpha\rho}\ .
\ee
As a result, for constant $\varphi$ and $\rho$, the metric field equations in the external $n$ dimensional space take the form $\tilde{G}_{\mu\nu}=-V_\text{eff}\,\tilde{g}_{\mu\nu}$; $Q$ is the magnetic charge associated with the internal top-form, fixed by the equations of motion to be
\be
Q^2=-e^{-2\alpha(n-1)\rho}\frac{V'}{g'}\ ,
\ee
where the primes denote derivative with respect to $\varphi$. Demanding that the potential is positive at its minimum leads to the  MvRV constraint
\be
(p-1)\left|\frac{V'}{V}\right|\leq\left|\frac{g'}{g}\right|\qquad \text{or~equivalently} \qquad  (p-1)\left|\frac{V}{g}\right|\geq e^{-2\alpha(D-p-1)\rho}\frac{1}{Q^2}. \label{eq:constraint}
\ee
To check the stability of such solutions, one must specify the form of the scalar potential $V$ and the coupling between the top-form and the potential $g$, while also stabilising the scalar. In \cite{Montero:2020rpl}, the authors consider a number of string-motivated scenarios with the key takeaway being the instability of the de Sitter minimum. For illustrative purposes, we present below the case where $V(\varphi)=V_0 e^{-\gamma \varphi}$ and $g(\varphi)=e^{\delta \varphi}$, for $\gamma, \delta>0$.  The Hessian of the potential along the $\varphi,\rho$ directions is given by 
\be
\mathcal{H} = {\partial^2 V_{\text{eff}}\over \partial \phi^i \partial \phi^j} \ , \qquad \phi^i=\{\varphi,\rho\}\ .
\ee
and its corresponding determinant by
\be
\det(\mathcal{H})=-\frac{(n-2)[\delta+\gamma(n-1)][\gamma(1-p)+\delta]\delta e^{2\delta \varphi + 2(n-1)\rho}}{4\gamma p}Q^4.
\ee
Demanding that \eqref{eq:constraint} holds at the minimum ---~or in other words $(p-1)\gamma \leq \delta$~--- fixes the determinant of the Hessian to be negative whenever $n>2$. This means all the corresponding solutions are  unstable. The only loophole to the above discussion is the case where the number of external dimensions $n=2$. In this case, a specific analysis is required
%the determinant of the Hessian vanishes 
and one may obtain a meta-stable de Sitter  solution while satisfying \eqref{eq:constraint}. 

As mentioned in the introduction, here we expand upon the above obstruction by considering the effects of a Gauss-Bonnet term in the action \eqref{eq:action}. Let us now review the origin of this term. Our starting point  is the effective action presented in  \cite{Campbell:1990fu}
\be
S=\int \text{d}^D x \sqrt{\left|g\right|}\left[{R\over 2\kappa_D^2}-\frac{1}{2}(\partial\varphi)^2-V(\varphi)-g(\varphi)\left|H_3\right|^2+f(\varphi)\mathcal{G}\right]\ , \label{eq:EA_camp}
\ee
with $H_3=\text{d}B_2+\omega_\text{L}-\omega_\text{Y}$ where $B_2$ is the Kalb-Ramond field, $\omega_\text{L}$ is the Lorentz spin connection, $\omega_\text{Y}$ is the Yang-Mills gauge connection, and with
\be
f(\varphi)= \dfrac{\alpha'}{16\kappa_D^2} e^{-\kappa_1\varphi}\ ,\quad g(\varphi)={e^{-2\kappa_1\varphi}\over {\kappa_D^2}} \ ,\label{eq:func_form}
\ee
with $\kappa_1=2{\kappa_D}/\sqrt{D-2}$ and $D=10$. The Gauss-Bonnet scalar is defined as
\be
\mathcal{G}= R_{ABCD}R^{ABCD}-4R_{AB}R^{AB}+R^2. \label{eq:GB_def}
\ee
This is  the combination of the quadratic curvature corrections that emerges in the effective theory for supersymmetric strings after super-completing the multiplet with the Lorentz-Chern-Simons terms, guaranteeing  the absence of ghosts \cite{Zwiebach:1985uq, Zumino:1985dp, Cecotti:1985nf}.  For $V=0$, the action \eqref{eq:EA_camp} describes the low-energy dynamics of the heterotic superstring \cite{Gross:1986mw}.  Alternatively, by orbifolding $E_8\times E_8$ strings one can obtain a non-supersymmetric theory whose low-energy limit is given by $SO(16)\times SO(16)$ strings, where the vacuum energy generates a potential \cite{Alvarez-Gaume:1986ghj},
\be \label{V}
V(\varphi)=\frac{0.037}{\alpha'^{5}}e^{5\kappa_1\varphi}\ .
\ee
The effective action  \eqref{eq:EA_camp}  is now a consistent truncation of this $SO(16)\times SO(16)$ action. 

In this chapter, we shall work with a slight generalisation of  \eqref{eq:EA_camp}, as given by  \eqref{eq:action}, allowing arbitrary spacetime dimension, $D$,  and  a $k$-form that is not necessarily a 3-form. We will also consider more general potentials, focussing on  exponentials with arbitrary coefficients and arbitrary slopes. Actually, these coefficients are not \emph{entirely} arbitrary ---~we will assume they have the same sign as in the known example from the heterotic string \eqref{eq:EA_camp}. This means that all the cases of interest will follow the philosophy of \cite{Montero:2020rpl}, a runaway potential to be lifted by flux contributions. Note that the relative sign of the slopes is all that really matters in the corresponding potentials. This because one can always perform a redefinition of the scalar $\varphi\to-\varphi$. With our conventions, the runaway is assumed to occur in the limit $\varphi\to-\infty$, in contrast with the conventions of \cite{Montero:2020rpl}, but consistent with \cite{Campbell:1990fu}. 

In the remaining sections, we consider compactified solutions of the action \eqref{eq:action} that are a direct product of $n$-dimensional de Sitter/Minkowski times a $p$-sphere/torus. In terms of the Riemann tensors associated with the external and internal metrics, this translates as
\begin{align}
	\tilde R_{\mu\nu\rho\sigma}&=\kappa\left(\tilde g_{\mu\rho} \tilde g_{\nu\sigma}-\tilde g_{\mu\sigma} \tilde g_{\nu\rho}\right),
	\label{eq:extR}
	\\	
	\hat R_{abcd}&=\lambda\left(\hat g_{ac} \hat g_{bd}-\hat g_{ad} \hat g_{bc}\right),
	\label{eq:intR}
\end{align}
where $\kappa$ is the curvature scale of the physical spacetime, while $\lambda$ is the curvature scale of the internal spacetime ($\lambda>0$ corresponds to a $p$-sphere and $\lambda=0$ to a $p$-torus).  We will also assume that $k$ is either $n$ or $p$, ensuring that the $k$-forms fills either the external or internal space with flux. We remark that in the standard  heterotic case, with $D=10$ and $k=3$ this requires compactification on a 3-sphere or a 3-torus, down to seven external dimensions, requiring three more compact directions to recover a four dimensional universe.  Of course, an explicit embedding of this kind with stable moduli is a non-trivial matter and further consistency considerations, such as tadpole cancellation  \cite{Becker:2005nb}, must be taken into account, but this is beyond the scope of this chapter.

Let us conclude by noting that the cases described in this chapter do not fall into the usual Maldacena-Nu\~nez no-go theorem \cite{Maldacena:2000mw} since the Gauss-Bonnet term represents higher curvature corrections to the action that are not considered in the original paper.

%%%%%%%%%%%%%%%%%%%%%

\section{Extending the Montero-van Riet-Venken bound in the small Gauss-Bonnet coupling limit}
\label{sec:genpot}

The goal of this chapter is to study the impact of the higher curvature corrections on the MvRV bound for the existence of de Sitter solutions. To develop some intuition, we begin by studying the  Gauss-Bonnet correction using a perturbative expansion in the dimensionless parameter $\varepsilon\equiv {\kappa_D^4\langle f(\varphi)/ V(\varphi)\rangle} \geq 0$, where $f(\varphi)$ and $V(\varphi)$ are evaluated at the vev of $\varphi$. As a result, we extend the MvRV bound \eqref{eq:constraint} to account for the higher curvature terms and show that the condition on the existence of de Sitter vacua can be recast in the form
\begin{equation}
    \left|\frac{g'}{g}\right|\geq (p-1)\left|\frac{V'}{V}\right| - \varepsilon\cdot h(\varphi,\rho) +\mathcal{O}\left(\varepsilon^2\right)\ , \label{eq:Montero_cond_GB}
\end{equation}
where the sign of $h(\varphi,\rho)$ determines whether, at linear order, the Gauss-Bonnet correction favours the existence of solutions or not. In particular, if $h>0$, the bound is more easily satisfied and the higher curvature corrections should make it easier to find de Sitter solutions; if $h<0$, the bound is harder to satisfy and de Sitter solutions are less forthcoming.  

Throughout this section we consider the top form to be purely internal and we keep the sign of the slopes for the potential and the couplings as given in \eqref{eq:func_form}, without specifying the functional form:
\be
\text{sgn}\left({V'\over V}\right) = 1 \ , \qquad \text{sgn}\left({g'\over g}\right) = -1 \ , \qquad \text{sgn}\left({f'\over f}\right) = -1 \ .
\label{eq:genslopes}
\ee
The Lagrangian for the $n$-dimensional theory, after compactifying the $D$-dimensional action \eqref{eq:action} in a $p$-dimensional internal space, is 
\begin{multline}
\mathcal{L} = \sqrt{-\tilde{g}}\left\{ \dfrac{\tilde{R}}{2{\kappa_D^2}} + \dfrac{\hat{R}}{2{\kappa_D^2}} e^{2\left(\alpha-\beta\right)\rho} - e^{2\alpha \rho} V(\varphi) - g(\varphi) Q^2 e^{2(n-1)\alpha\rho}\right.
\\
\left. + f(\varphi)\left[\tilde{\mathcal{G}} e^{-2\alpha\rho}  +2 \tilde{R}\hat{R}e^{-2\beta\rho}+\hat{\mathcal{G}}e^{2\left(\alpha-2\beta\right)\rho}\right] + ...\vphantom{\dfrac{\tilde{R}}{2{\kappa_D^2}}}\right\}\ ,
\label{eq:Vpot}
\end{multline}
where we have omitted the terms involving derivatives of the scalar fields, as we are interested in studying the system at the minimum in the $\left(\varphi,\rho\right)$ directions. Extremising the Lagrangian with respect to the metric and the two fields yields the conditions, 
\begin{align}
\begin{split}
Q^2&={{{e}^{-2n\alpha\rho}}\over g'(\varphi)} \left\{f'(\varphi) \left[ \lambda^2 p(p-1)\left( p-2 \right)  \left( p-3 \right) {{e}^{4
  \left( \alpha-\beta \right) \rho}}\right .\right.
 \\
 &\left.\left.\vphantom{e^{()}} +2\lambda\kappa n(n-1)p(p-1) {{e}^{2  \left( \alpha-\beta
 \right) \rho}}+\kappa^2 n(n-1)(n-2)(n-3)\right]  -{{e}^{4 \alpha \rho}}V' \left( \varphi\right)\right\}\ ,
 \label{eq:Qcond}
 \end{split}
 \\
\begin{split}
0&=\lambda\left\{{\dfrac{1}{\kappa_D^2}}\left(p-1\right)\left(p+n-2\right)e^{2\left(\alpha-\beta\right)\rho}+2\left(p-1\right)f(\varphi)e^{-2\beta\rho}\left[2\kappa\left(n-2\right) n\left(n-1\right)\vphantom{e^{2\left(\alpha-\beta\right)\rho}}\right.\right.
\\
&\quad\left.\left.+\lambda\left(p-2\right)\left(p-3\right)\left(p+2n-4\right)e^{2\left(\alpha-\beta\right)\rho}\right]\vphantom{\dfrac{1}{\kappa_D^2}}\right\} -2{Q^2} g(\varphi)\left(n-1\right)e^{2\left(n-1\right)\alpha\rho}
\\
&\quad -2\kappa^2 n\left(n-1\right)\left(n-2\right)\left(n-3\right)f(\varphi) e^{-2\alpha\rho}- 2V(\varphi) e^{2\alpha\rho} \vphantom{\left\{e^{()}\right\}}\ ,
\label{eq:kcond}
\end{split}
\\
\begin{split}
0&=\kappa \left\{\vphantom{\dfrac{1}{\kappa_D^2}}2(n-1)(n-2) f(\varphi) \left[\kappa (n-3)(n-4)e^{-2\alpha\rho}+2\lambda p(p-1)e^{-2\beta\rho}\right] \right.
\\
&\quad \left.+{\dfrac{1}{\kappa_D^2}}(n-1)(n-2)\vphantom{e^{-2\beta\rho}}\right\}+ \lambda p (p-1)\left[2\lambda(p-2)(p-3) f(\varphi) e^{2\left(\alpha-2\beta\right)\rho} + \dfrac{e^{2\left(\alpha-\beta\right)\rho}}{{\kappa_D^2}}\right]
\\
&\quad- 2 V(\varphi) e^{2\alpha \rho} - 2 g(\varphi) Q^2 e^{2(n-1)\alpha\rho} \vphantom{\dfrac{e^{2\left(\alpha-\beta\right)\rho}}{\kappa_D^2}}\ ,
\label{eq:kappacond}
\end{split}
\end{align}
It is not possible to solve the above system for arbitrary potentials and dimensions. However, we can solve the system order by order in a perturbative Gauss-Bonnet coupling expansion\footnote{Note that Gauss-Bonnet gravity, being quadratic in curvature, typically yields two distinct vacua, one which is perturbative in the coupling, $f$, and one which is non-perturbative. Our analysis here does not capture the latter, which one might have expected to be unstable anyway \cite{Charmousis:2008ce}.}. To do so, we take the ansatz
\begin{gather}
    Q^2= Q^2_0 + \varepsilon\, Q^2_1\ , \quad     \lambda= \lambda_0 + \varepsilon \,\lambda_1\ , \quad   \kappa= \kappa_0 + \varepsilon\, \kappa_1\ ,\label{eq:ansatz}
\end{gather}
where the zeroth order quantities solve the previous system of equations with $f(\phi)=0$ and are given by
\begin{align}
Q^2_0&={{e}^{-2\left(n-2\right)\alpha\rho}}\left|\frac{{
V'}}{g'  }\right|\ , 
 \label{eq:Q_exp}
\\
{\lambda}_0&={2{\kappa_D^2} V{{e}^{2\beta\rho}}\over (p-1){\left( 
n+p-2\right)}}\left|{g\over g'}\right|\left[\left|{g'\over g}\right|  +\left( n-1
 \right)  \left|{V'\over V}\right|  \right] \ ,
\label{eq:k_exp}
\\
\kappa_0 &= {2{\kappa_D^2}V e^{2\alpha \rho} \over (p+n-2) (n-1)}\left|{g\over g'}\right|\left[\left|{g'\over g}\right|-\left(p-1\right)\left|{V'\over V}\right|\right]\ .
\label{eq:kappa_exp}
\end{align}
Solving for the ansatz \eqref{eq:ansatz} to leading order in the Gauss-Bonnet coupling, we find the external curvature to be
\begin{equation}
\kappa = {2{\kappa_D^2} V e^{2\alpha\rho}\over (p+n-2)(n-1) }\left|{g\over g'}\right|\left[\left|{g'\over g}\right| - (p-1) \left|{V'\over V}\right|+\varepsilon h_{n,p}+\mathcal{O}\left(\varepsilon^2\right)\right]\ , \label{eq:vmin}
\end{equation}
where $(\varphi,\rho)$ are understood to be evaluated at their vevs. Demanding that the external curvature is positive leads to an extended condition of the form  \eqref{eq:Montero_cond_GB}, with  the leading order correction given by a polynomial in $|g/g'|$
\begin{equation}
\begin{split}
h_{n,p}&={{V^2}\over (p-1)(n-1)(p+n-2)}\left[A_{n,p}\left|{g'\over g}\right|+B_{n,p}\left|{f'\over f}\right|+C_{n,p}\left|{V'\over V}\right|\vphantom{\left|{g'\over g}\right|^2}\right.
\\
&\quad\left.+\left(D_{n,p}\left|{f'\over f}\right|+E_{n,p}\left|{V'\over V}\right|\right)\left|{V'\over V}\right|\left|{g\over g'}\right|+F_{n,p}\left|{f'\over f }\right|\left|{V'\over  V}\right|^2\left|{g\over g'}\right|^2\right]\ .
\label{eq:gcond_app}
\end{split}
\end{equation}
The coefficients of the polynomial depend on the dimensions $\left(n,p\right)$: 
\begin{align}
    A_{n,p}&= 4n(p-7)(1-p) + 4n^2(1-p) + 4p(p-11) + 48 \ ,
    \\
    B_{n,p}&= -4\left[(n - 1)p^2 + (n^2 - 4n + 3)p - n^2 + 3n\right](p - 1)\ ,
    \\
    C_{n,p}&= 8\left[(n -1 )p^3 + (n^2 -11n + 14)p^2 - (2n^2 - 19n + 27)p + n^2 - 7n + 12\right]\ ,
    \\
    D_{n,p}&= 8(p- n)(np - n - p + 3)(p - 1)\ ,
    \\
    E_{n,p}&=4\left[2(3n - 5)p^3 - (2n^2 + 15n - 31)p^2 + (3n^2 + 12n - 31)p + n^2 - 7n + 12\right],
    \\
    F_{n,p}&=4(p - 1)(4n^2p^2 - 7n^2p - 7np^2 + n^2 + 14np + p^2 - 3n - 3p)\ .
\end{align}
Although these  expressions for the coefficients are not very illuminating, we can remark on some interesting behaviour. 

For the particular stringy potentials  given in \eqref{eq:func_form} and with the well motivated  choice,  $n=4$, $p=6$ we find that $h_{n,p}>0$.  However, this does not point to the existence of new de Sitter solutions as the MvRV bound is already strongly violated at leading order.  For the Gauss-Bonnet corrections to yield something interesting in this perturbative set-up, our best hope is ask what happens when the MkRV bound is \emph{only just} violated at leading order.  In particular, we consider the case where
\begin{equation}
\left|\dfrac{g'}{g}\right|=(p-1) \left|\dfrac{V'}{V}\right|(1-\Delta)
\end{equation}
for some $0<\Delta \ll \varepsilon$.  In such a scenario, the existence of de Sitter solutions is only marginally ruled out by the original MvRV bound \eqref{eq:constraint}. Could the curvature corrections rule them back in? In this particular case we find that 
\begin{equation}
h_{n,p}=  {4p(p-2)(p-3)\over(p-1)^3}{V^2}\left[\left|\dfrac{g'}{g}\right|-(p-1)\left|\dfrac{f'}{f}\right|\right]  [1+\mathcal{O}(\Delta)] \ , \label{eq:comb}
\end{equation}
Since $0<\Delta \ll \varepsilon$, the external curvature is dominated by this form of $h_{n, p}$,  giving
\be
\kappa \approx  \frac{8\kappa_D^2 p(p-2)(p-3) V^3 e^{2\alpha\rho}}{(p+n-2)(n-1)(p-1)^3 }  \varepsilon  \left|{g\over g'}\right|
\left[\left|\dfrac{g'}{g}\right|-(p-1)\left|\dfrac{f'}{f}\right|\right] 
\ee
which is clearly positive whenever 
\be
\left|\dfrac{g'}{g}\right| > (p-1)\left|\dfrac{f'}{f}\right|\ . \label{eq:extended_bound}
\ee
This suggest that de Sitter solutions {\it might} be possible for these parametric choices,  although we are not aware of any stringy motivated  supergravity model that satisfies them. 

Another possibility is to consider the case where  $|g'/ g|\ll |V'/V|$.  When this hierarchy is big enough, the last term in \eqref{eq:gcond_app} could be the dominant contribution to the external curvature,  dominating over the zeroth order piece in the perturbative expansion in the Gauss-Bonnet coupling. Whilst this casts some doubt on the validity of the expansion, it would, if true,  suggest that \be
\kappa \approx {2{\kappa_D^2} V^3 e^{2\alpha\rho}\over (p+n-2)^2(n-1)^2(p-1) } \varepsilon F_{n,p}\left|{f'\over f }\right|\left|{V'\over  V}\right|^2\left|{g\over g'}\right|^3
\ee
This could point towards new de Sitter solutions  since we also have  that $F_{n,p}>0$ for $n,p\geq2$.   Of course, such an extreme hierarchy is not especially well motivated since generically we expect $|g'/g|$ and $|V'/V| $ to be $\mathcal{O}(1)$ as  both couplings are expected to be fixed by Weyl rescaling to Einstein frame.

\section{Non-perturbative conditions for the existence of de Sitter solutions with a Gauss-Bonnet term}
\label{sec:specifpot}

We shall now establish some exact results, seeking de Sitter and non-trivial Minkowski vacua beyond the perturbative approach of the previous section.  To this end, it is necessary  to fix some of the potentials in order to make concrete progress. We choose the Gauss-Bonnet  potential $f$ and the gauge coupling $g$ to coincide with the heterotic string, as per equation~\eqref{eq:func_form}, so that 
\be \label{fg}
f(\varphi)= \dfrac{\alpha'}{16\kappa_D^2} e^{-\kappa_1\varphi}\ ,\quad g(\varphi)={e^{-2\kappa_1\varphi}\over {\kappa_D^2}} 
\ee
while  the remaining potential is given more generally as  
\begin{equation}
    V(\varphi)=V_0e^{q\kappa_1\varphi},  \label{f}
\end{equation}
with $V_0>0$ and $q>0$ a free dimensionless parameter.  There is sufficient generality left in $V$ to mirror the analysis of \cite{Montero:2020rpl}, and investigate parametric constraints on the slope of $\ln V$. Note that for  the string motivated scenario of \cite{Alvarez-Gaume:1986ghj}, we have $q=5$ and $V_0={0.037\over\alpha'^{5}}$. 

In \cite{Montero:2020rpl}, the slope of $\ln V$ is constrained in the form of the  MvRV bound   \eqref{eq:constraint}. For these exponential potentials, this bound  takes a particularly simple form
\begin{equation}
    q\leq\dfrac{2}{p-1}.
    \label{eq:MvR}
\end{equation}
Our goal here is to ask if the parametric constraint on the slope of $\ln V$ is affected by the Gauss-Bonnet correction. We will also allow for both internal and external fluxes, further generalising \cite{Montero:2020rpl}. If  $H_k$ is allowed to carry external flux, we find that there are only trivial vacua:  a Minkowski external space, vanishing flux and a toroidal internal manifold. If, instead, $H_k$ is allowed to carry internal flux, the set-up is more interesting. The perturbative analysis of the previous section suggested that one could find de Sitter solutions even when the  MvRV  bound is violated, provided an extended condition  \eqref{eq:extended_bound} holds for the Gauss-Bonnet potentials. However, it turns out that this extended bound  is {\it not} satisfied for the string motivated potentials for $f$ and $g$  given by \eqref{fg}.  Indeed, we shall see that in this instance, the MvRV condition \eqref{eq:MvR} is necessary for the existence of de Sitter solutions, but it is not sufficient. This means the higher curvature correction makes de Sitter solutions no more forthcoming, constraining the parametric dependence  of the scalar potential $\ln V$ at least as much as in \cite{Montero:2020rpl}. 

Note that in this section, we shall not assume the particular choice of $\alpha$ and $\beta$ given in \eqref{albe}, but will keep things general. Furthermore,  we work directly with the $D$-dimensional field equations derived from the variation of the action \eqref{eq:action}, which are given by 
\begin{align}
	\begin{split}
		\label{eq:EinsteinD}
		0&=f(\varphi)\left(4R_A^{~C}R_{BC}+\dfrac12\mathcal{G}g_{AB}-2RR_{AB}+4R^{CD}R_{ACBD}-2R_A^{~CDE}R_{BCDE}\right)
		\\
		&\quad+\dfrac12\nabla_A\varphi\nabla_B\varphi-\dfrac14(\nabla\varphi)^2g_{AB}+2f'(\varphi)\left(R\,\nabla_A\nabla_B\varphi-4R_{(B}^{~C}\nabla_{A)}\nabla_C\varphi+2G_{AB}\Box\varphi\right.
		\\
		&\quad+2g_{AB}R^{CD}\nabla_C\nabla_D\varphi-2R_{ACBD}\nabla^C\nabla^D\varphi\left.\vphantom{R_{(B}^{~C}}\right)+2f''(\varphi)\left[R\,\nabla_A\varphi\nabla_B\varphi\right.\vphantom{\dfrac12}
		\\
		&\quad\left.-4R_{C(B}\nabla_{A)}\varphi\nabla^C\varphi+2G_{AB}(\nabla\varphi)^2+2g_{AB}R^{CD}\nabla_B\varphi\nabla_C\varphi-2R_{ACBD}\nabla^C\varphi\nabla^D\varphi\right] \vphantom{\dfrac12}
		\\
		&\quad-\dfrac{1}{2\kappa_D^2}G_{AB}-g(\varphi)\left[\dfrac{1}{2k!}H_{A_1...A_k}H^{A_1...A_k}g_{AB}-\dfrac{1}{(k-1)!}H_{A_1...A_{k-1}A}H^{A_1...A_{k-1}}_{~~~~~~~~~B}\right]
		\\
		&\quad-\dfrac12V(\varphi)g_{AB},
	\end{split}
	\\
	0&=\Box\varphi+f'(\varphi)\mathcal{G}-V'(\varphi)-\dfrac{g'(\varphi)}{k!}H_{A_1...A_k}H^{A_1...A_k},
	\label{eq:scalD}
	\\
	0&=\nabla_B\left[g(\varphi)H^{A_1...A_{k-1}B}\right].
	\label{eq:topform}
\end{align}

%%%%%%%%%%%
\subsection{Internal flux}
\label{ssec:int}
If we assume that $H$ is a top form along the internal manifold, then the solution to equation~\eqref{eq:topform} is given by
\begin{equation}
	H_{a_1...a_p}= Q \hat\epsilon_{a_1...a_p}
	\label{eq:intform}
\end{equation}
where $Q$ is a constant, all other components of $H$ vanish, and $\hat\epsilon_{a_1...a_p}$ is the volume form associated with the internal submanifold ($\hat\epsilon_{1...p}=\sqrt{\hat g}$, etc). We substitute this solution, together with \eqref{eq:extR} and \eqref{eq:intR} into eqs.~\eqref{eq:EinsteinD} and \eqref{eq:scalD}, projecting equation~\eqref{eq:EinsteinD} over $\tilde g_{\mu\nu}$ and $\hat g_{ab}$ respectively. Let us define the following master variables
\begin{equation}
	\label{eq:changevarint}
	\begin{split}
		K&=\kappa e^{2(\beta-\alpha)\rho},\qquad  \hat H=Q^2e^{-2\kappa_1\varphi-2(p-1)\beta\rho},\qquad U=V_0\kappa_D^2e^{q\kappa_1\varphi+2\beta\rho},
		\\
		X_1&=\dfrac{\alpha'}{16} e^{-\kappa_1\varphi-2\beta\rho},\qquad X_2=\dfrac{\alpha'}{16} \kappa e^{-\kappa_1\varphi-2\alpha\rho},\qquad X_3= \dfrac{\alpha'}{16}  \kappa^2 e^{-\kappa_1\varphi+2(\beta-2\alpha)\rho},
	\end{split}
\end{equation}
Given that $\alpha'>0$, for a consistent de Sitter (or Minkowski) vacuum, we necessarily have that all of the master variables, with the possible exception of $U$, are non-negative. This will be crucial to our subsequent proofs. The system of equations can now be cast in the following linear form, which is particularly useful for analysing the system
\begin{align}
	\begin{split}
		\label{eq:Einsteinnint2}
		0&=-2 \hat H - 2 U + (n-1)(n-2) [K + 2 (n-3)(n-4) X_3] 
		\\
		&\quad+ \lambda p(p-1) [1 + 4 (n-1)(n-2) X_2]  +  2\lambda^2 p(p-1)(p-2)(p-3) X_1,
	\end{split}
	\\
	\begin{split}
		\label{eq:Einsteinpint2}
		0&=2 \hat H- 2 U + n(n-1)[K + 2 (n-2)(n-3)X_3] 
		\\
		&\quad+\lambda  (p-1)(p-2)  [1 + 4 n(n-1) X_2] +  2 \lambda^2(p-1)(p-2)(p-3)(p-4) X_1 ,
	\end{split}
	\\
	\begin{split}
		\label{eq:scalint2}
		0&= 2 \hat H - q U - n(n-1)(n-2)(n-3) X_3 - 2\lambda n (n-1) p(p-1) X_2 
		\\
		&\quad- \lambda^2 p(p-1)(p-2)(p-3) X_1 .
	\end{split}
\end{align}
 Given that $n\geq2$ and $q>0$, we can always solve the above equations for $U$, $\hat H$ and $K$ in terms of $X_1$, $X_2$ and $X_3$. In particular, we have 
\begin{equation}
\begin{split}
K=K^\mathrm{(int)}_\mathrm{sol}(X_1,X_2,X_3)&\equiv-\dfrac{1}{(n-1) [2 + (n-1) q]}\left\{\lambda(p-1) [q(p-1)-2]\right.
\\
&\quad+2 \lambda^2(p-1)(p-2)(p-3) [p-4 + (p-2) q] X_1 \vphantom{\dfrac{1}{(n-1) [2 + (n-1) q]}}
\\
&\quad + 4\lambda(p-1) (n-1)\{-p (q-2) + n [p-2 + (p-1) q]\} X_2 \vphantom{\dfrac{1}{(n-1) [2 + (n-1) q]}}
\\
&\quad \left.+2 (n-1)(n-2)(n-3)[4 + n + (n-2) q] X_3\right\}\vphantom{\dfrac{1}{(n-1) [2 + (n-1) q]}}.
\end{split}
\end{equation}
The idea behind the proof is to show that $K$ is maximal at $(X_1,X_2,X_3)=(0,0,0)$ (restricting of course to non-negative $X_i$ values), but that even at its maximum, $K<0$ as soon as we choose $q>2/(p-1)$.  Since $K$ carries the same sign as the external curvature, this would prove the absence of de Sitter solutions whenever $q>2/(p-1)$. Said another way, the MvRV condition \eqref{eq:MvR} is a necessary condition for the existence of de Sitter vacua, even in the presence of  the Gauss-Bonnet correction. 

To this end, we  consider
\begin{align}
\dfrac{\partial K^\mathrm{(int)}_\mathrm{sol}}{\partial X_1}&=-\dfrac{2\lambda^2 (p-1)(p-2)(p-3)[p-4 + (p-2) q]}{(n-1) [2 + (n-1) q]},
\\
\dfrac{\partial K^\mathrm{(int)}_\mathrm{sol}}{\partial X_2}&=-\dfrac{4\lambda (p-1)\{-p (q-2) + n [p-2+ (p-1) q]\}}{2 + (n-1) q},
\\
\dfrac{\partial K^\mathrm{(int)}_\mathrm{sol}}{\partial X_3}&=-\dfrac{2(n-2)(n-3)[n+4+(n-2)q]}{2 + (n-1) q}.
\end{align}
$\partial K^\mathrm{(int)}_\mathrm{sol}/\partial X_3$ is obviously negative. For $p=1,2,3$, $\partial K^\mathrm{(int)}_\mathrm{sol}/\partial X_1=0$ while for $p\geq4$, $\partial K^\mathrm{(int)}_\mathrm{sol}/\partial X_1$ is also obviously negative. Finally, for $p=1$, $\partial K^\mathrm{(int)}_\mathrm{sol}/\partial X_2=0$, while for $p\geq2$ and $n\geq2$,
\begin{equation}
-p (q-2) + n [p-2+ (p-1) q]\geq(p-2)q+4(p-1)>0.
\end{equation}
Hence, $\partial K^\mathrm{(int)}_\mathrm{sol}/\partial X_2\leq0$ in all cases, and $K^\mathrm{(int)}_\mathrm{sol}$ is indeed maximal at $(X_1,X_2,X_3)=(0,0,0)$, where it takes the value
\begin{equation}
K^\mathrm{(int)}_\mathrm{max} \equiv K^\mathrm{(int)}_\mathrm{sol}(0,0,0)=\dfrac{\lambda(p-1) [2-(p-1)q]}{(n-1) [2 + (n-1) q]}.
\label{eq:Kmax}
\end{equation}
The above equation shows that $K$ can vanish for $\lambda=0$ ($p$-torus) or $p=1$ (1-sphere), i.e. when the curvature of the internal manifold vanishes. These solutions correspond to a Minkowski spacetime, and exist only for a vanishing flux $Q$ and a vanishing potential $V(\varphi)$, which makes them uninteresting. The existence of a de Sitter solution, on the other hand, requires that $K>0$. This is not possible if $q>2/(p-1)$, as shown by equation~\eqref{eq:Kmax}, which completes our proof. The condition \eqref{eq:MvR} is necessary for de Sitter solutions to exist.

In terms of the variables $K$, $\hat H$, $U$, $X_i$, \eqref{eq:MvR} is also sufficient; indeed, it is possible to show that, when $q\leq2/(p-1)$, $K$, $\hat H$ and $U$ are all positive in the neighbourhood of the point $(X_1,X_2,X_3)=(0,0,0)$. However, this is an artifact of the choice of variables. Sending all $X_i$ to 0 actually corresponds to sending the Gauss-Bonnet coupling to 0. Hence, it is not surprising to find that condition \eqref{eq:MvR} is necessary and sufficient in this case, as we are back to the action studied in \cite{Montero:2020rpl}. If instead, we think of the Gauss-Bonnet coupling as fixed, we can show that the condition \eqref{eq:MvR} is not sufficient, by an analysis that is very similar to the one we carried out in \secref{sec:genpot}. Indeed, let us saturate the bound \eqref{eq:MvR} by choosing $q=2/(p-1)$ ---~which corresponds to being on the edge of the existence region when the Gauss-Bonnet coupling is absent. Then, $K_\text{sol}^{(\text{int})}$ becomes
\begin{equation}
\begin{split}
K_\text{sol}^{(\text{int})} &=-\dfrac{1}{(n-1)(n+p-2)}\left\{(n-1) (n-2)(n-3) [n-8 + (4 + n) p] X_2\vphantom{(p-3)^2}\right.
\\
&\quad\left.+ 2p(p-1)(n-1) [(2 + n) p-n-4] \lambda X_3 + p(p-1)(p-2)(p-3)^2 \lambda^2 X_1\right\}.
        \label{eq:Ksolintsat}
        \end{split}
\end{equation}
It is easy to check that for $p\geq2$ (we already saw that the case $p=1$ was uninteresting), the coefficients of $X_1$, $X_2$ and $X_3$ in the above equation are all negative. The coefficient of $X_3$ is actually strictly negative. Hence, there can be no de Sitter (or Minkowski) solution for this value of $q$ ---~and neighbouring ones, by continuity. Therefore, \eqref{eq:MvR} is not sufficient. This reinforces the conclusion of \secref{sec:genpot}, as it is a non-perturbative statement. The presence of the Gauss-Bonnet term never enlarges the space of solutions, but instead reduces it, at least when the potentials, $f$ and $g$, take on the form best motivated by string theory.

\subsection{External flux}

So far, we have assumed $H$ to be an  internal flux, and,  as we have demonstrated, this leads to the absence of de Sitter solutions to the effective action~\eqref{eq:action} as long as $q>2/(p-1)$. Now let us consider the opposite scenario: that $H$ is a top form along the external directions. In this case, the solution to equation~\eqref{eq:topform} is given by
\begin{equation}
	H_{\mu_1...\mu_n}= \dfrac{Q}{g(\varphi)} e^{(n\alpha-p\beta)\rho} \tilde\epsilon_{\mu_1...\mu_n}
	\label{eq:extform}
\end{equation}
where $Q$ is a constant, all other components of $H$ vanish, and $\tilde\epsilon_{\mu_1...\mu_n}$ is the volume form associated with the internal submanifold ($\tilde\epsilon_{1...n}=\sqrt{-\tilde g}$, etc). We will prove that no de Sitter solutions at all exist in this case, for any value of $q$. The proof is very similar to the one presented in \secref{ssec:int}. Employing the same variables $U$, $K$, $X_i$ as defined in equation~\eqref{eq:changevarint}, and trading $\hat H$ for
\begin{equation}
\tilde H =Q^2e^{2\kappa_1\varphi-2(p-1)\beta\rho},
\end{equation}
the field equations are identical to eqs.~\eqref{eq:Einsteinnint2}--\eqref{eq:scalint2} upon the substitution $\hat H\to\tilde H$ in eqs.~\eqref{eq:Einsteinnint2}-\eqref{eq:Einsteinpint2} and $\hat H\to-\tilde H$ in equation~\eqref{eq:scalint2}. We need to distinguish a few more sub-cases with respect to the internal case, depending on how $q$ compares to $2/(n-1)$.

\subsubsection{$q>2/(n-1)$}

In this case, the proof is almost exactly identical to the internal case. We solve the field equations for $K$, $U$ and $\tilde H$ in terms of $X_1$, $X_2$ and $X_3$, to obtain
\begin{equation}
\begin{split}
K=K^\mathrm{(ext)}_\mathrm{sol}(X_1,X_2,X_3)&\equiv-\dfrac{1}{(n-1) [(n-1) q-2]}\left\{\lambda(p-1) [q(p-1)+2]\right.
\\
&\quad+2 \lambda^2(p-1)(p-2)(p-3) [p+4 + (p-2) q] X_1\vphantom{\dfrac{1}{(n-1) [2 + (n-1) q]}}
\\
&\quad+4 \lambda(n-1)(p-1)\{-p (q+2) + n [p+2 + (p-1) q]\} X_2 \vphantom{\dfrac{1}{(n-1) [2 + (n-1) q]}}
\\
&\quad \left.  +2 (n-1)(n-2)(n-3)[n-4 + (n-2) q] X_3\right\}\vphantom{\dfrac{1}{(n-1) [(n-1) q-2]}}.
\end{split}
\end{equation}
It immediately follows that
\begin{align}
\dfrac{\partial K^\mathrm{(ext)}_\mathrm{sol}}{\partial X_1}&=-\dfrac{2\lambda^2 (p-1)(p-2)(p-3)[p+4 + (p-2) q]}{(n-1) [(n-1) q-2]},
\\
\dfrac{\partial K^\mathrm{(ext)}_\mathrm{sol}}{\partial X_2}&=-\dfrac{4\lambda (p-1)\{-p (q+2) + n [p+2+ (p-1) q]\}}{(n-1) q-2},
\\
\dfrac{\partial K^\mathrm{(ext)}_\mathrm{sol}}{\partial X_3}&=-\dfrac{2(n-2)(n-3)[n-4+(n-2)q]}{(n-1) q-2}.
\end{align}
Since $q>2/(n-1)$, $\partial K^\mathrm{(ext)}_\mathrm{sol}/\partial X_1$ is obviously negative. For $n=2,3$, $\partial K^\mathrm{(ext)}_\mathrm{sol}/\partial X_3=0$ while for $n\geq4$, $\partial K^\mathrm{(ext)}_\mathrm{sol}/\partial X_3$ is also obviously negative. Finally, for $p=1$, $\partial K^\mathrm{(ext)}_\mathrm{sol}/\partial X_2=0$, while for $p\geq2$ and $n\geq2$,
\begin{equation}
-p (q+2) + n [p+2+ (p-1) q]\geq(p-2)q+4>0.
\end{equation}
Hence, $\partial K^\mathrm{(ext)}_\mathrm{sol}/\partial X_2\leq0$ in all cases, and $K^\mathrm{(ext)}_\mathrm{sol}$ is maximal at $(X_1,X_2,X_3)=(0,0,0)$, where it takes the value
\begin{equation}
K^\mathrm{(ext)}_\mathrm{max} \equiv K^\mathrm{(ext)}_\mathrm{sol}(0,0,0)=-\dfrac{\lambda(p-1) [2+(p-1)q]}{(n-1) [(n-1) q-2]}.
\end{equation}
This is always negative in the sub-case that we are considering, hence no de Sitter solutions exist. Again, trivial Minkowski solutions can exist for $p=1$ or $\lambda=0$.

\subsubsection{$q<2/(n-1)$}

Once again, the proof is very analogous to the previous ones, but now we consider the solution for $\tilde H$ rather than $K$. This is given by
\begin{equation}
\begin{split}
\tilde H=\tilde H_\mathrm{sol}(X_1,X_2,X_3)&\equiv\dfrac{1}{2[(n-1) q-2]}\left\{\lambda(p-1) q(n+p-2)\right.
\\
&\quad+2 \lambda^2(p-1)(p-2)(p-3) [p + (2n+p-4) q] X_1\vphantom{\dfrac{1}{(n-1) [2 + (n-1) q]}}
\\
&\quad+4 \lambda n(n-1)(p-1)[p + (n-2) q] X_2 \vphantom{\dfrac{1}{(n-1) [2 + (n-1) q]}}
\\
&\quad \left.  -2 n(n-1)(n-2)(n-3)(q-1) X_3\right\}\vphantom{\dfrac{1}{(n-1) [(n-1) q-2]}}.
\end{split}
\end{equation}
and therefore,
\begin{align}
\dfrac{\partial \tilde H_\mathrm{sol}}{\partial X_1}&=\dfrac{2\lambda^2 (p-1)(p-2)(p-3)[p+(p+2n-4)q]}{(n-1) [(n-1) q-2]},
\\
\dfrac{\partial \tilde H_\mathrm{sol}}{\partial X_2}&=\dfrac{4\lambda n(n-1)(p-1)[p+(n-2)q]}{(n-1) q-2},
\\
\dfrac{\partial \tilde H_\mathrm{sol}}{\partial X_3}&=\dfrac{n(n-1)(n-2)(n-3)(1-q)}{(n-1) q-2}.
\end{align}
$\partial \tilde H_\mathrm{sol}/\partial X_1$ and $\partial \tilde H_\mathrm{sol}/\partial X_2$ are obviously negative in the subcase we consider. For $n\leq3$, $\partial \tilde H_\mathrm{sol}/\partial X_3=0$. For $n\geq4$, since $q<2/(n-1)$, $1-q>0$ and therefore $\partial \tilde H_\mathrm{sol}/\partial X_3$ is negative. Hence,  $\tilde H_\mathrm{sol}$ is maximal at $(X_1,X_2,X_3)=(0,0,0)$, where it takes the value
\begin{equation}
\tilde H_\mathrm{max} \equiv \tilde H_\mathrm{sol}(0,0,0)=\dfrac{\lambda q(p-1) (n+p-2)}{2 [(n-1) q-2]}<0.
\end{equation}
Since $\tilde H$ should be non-negative for real values of the flux, we conclude that  the underlying assumptions are inconsistent and therefore no de Sitter or non-trivial Minkowski solutions can exist.

\subsubsection{$q=2/(n-1)$}

In the case where $q$ takes exactly the value $2/(n-1)$, we can no longer solve the system of equations for $K$, $U$ and $\tilde H$ simultaneously. However, we can solve it for $X_2$, $U$ and $\tilde H$ instead, provided that $\lambda\neq0$ and $p\geq2$ (if this is not the case, once again only trivial Minkowski solutions can exist). We obtain
\begin{equation}
\begin{split}
X_2=X_{2,\,\mathrm{sol}}(X_1,X_3)&\equiv-\dfrac{1}{2\lambda n(n-1) (p-1) [(n-1)p+2(n-2)]}
\\
&\quad\left\{ \lambda^2 (p-1)(p-2)(p-3) [p(n+1)+4(n-2)] X_1 \right.\vphantom{\dfrac{1}{(n-1) [2 + (n-1) q]}}
\\
&\quad \left. +n(n-1)(n-2)(n-3)^2 X_3 +\lambda(p-1) (n + p-2)\right\}\vphantom{\dfrac{1}{(n-1) [(n-1) q-2]}}.
\end{split}
\end{equation}
Hence,
\begin{align}
\dfrac{\partial \tilde X_{2,\,\mathrm{sol}}}{\partial X_1}&=-\dfrac{\lambda(p-2)(p-3)[p(n+1)+4(n-2)]}{2 n(n-1)[(n-1)p+2(n-2)]},
\\
\dfrac{\partial \tilde X_{2,\,\mathrm{sol}}}{\partial X_3}&=-\dfrac{(n-2)(n-3)^2}{2\lambda (p-1) [(n-1)p+2(n-2)]},
\end{align}
These quantities are obviously negative. Therefore,  $X_{2,\,\mathrm{sol}}$ is maximal at $(X_1,X_3)=(0,0)$, where it takes the value
\begin{equation}
X_{2,\,\mathrm{max}} \equiv \tilde X_{2,\,\mathrm{sol}}(0,0)=-\dfrac{n+p-2}{2 n(n-1) [(n-1)p+2(n-2)]}<0.
\end{equation}
Again, for consistent de Sitter or non-trivial Minkowski solultions, $X_2$ should be non-negative, in contradiction with the above inequality. This concludes the proof for the external case.

\section{Discussion}
\label{sec:discussion}
We have considered solutions to a general class of gravitational actions, motivated by quadratic curvature corrections to the effective action for the heterotic string.  The actions include a scalar field (identified with the dilaton), a $k$-form field strength, and gravity. We considered warped compactifications on spheres and tori, with $k$-form flux and investigated the existence of de Sitter solutions along the external directions.  A similar analysis was carried out in \cite{Montero:2020rpl} without the quadratic curvature corrections, which in our case corresponded to the Gauss-Bonnet operator coupled to the dilaton.  In \cite{Montero:2020rpl}, the MvRV bound \eqref{eq:constraint} was derived for the dilaton potentials, specifiying the condition for de Sitter solutions to exist. In the presence of the quadratic  curvature corrections, we saw that this bound could be violated  perturbatively and de Sitter solutions could still exist, as long as a second bound \eqref{eq:extended_bound} on the Gauss-Bonnet coupling  also held.  However, for potentials that are best motivated by string theory, the second bound does not hold, suggesting no violation of the MvRV bound is possible in this instance.  Indeed, for these well motivated set-ups, we were also able to show that the MvRV bound was a necessary, but not sufficient, condition for the existence of de Sitter solutions in the presence of quadratic curvature corrections. This leads to our main conclusion:  higher curvature corrections that are well motivated by string theory generically made it parametrically harder, not easier, to find de Sitter solutions. 

It would be very interesting to ask if this were a generic feature of higher order  corrections in all versions of string theory. Clearly this is difficult to establish and more evidence needs to be found.  For  type II strings, curvature corrections kick in at fourth order \cite{Gross:1986iv}, while for heterotic strings we may also consider higher order mixings between the gauge field and curvature \cite{Gross:1986mw}.  

We can make a crude and simplistic argument to support a claim that curvature  corrections are unlikely to improve the search for de Sitter vacua, at least at finite order in the curvature expansion.  In $D$ dimensional General Relativity without sources, the absence of de Sitter vacua can roughly be understood in terms of an energetic balance between the curvature of the external space and that of the compact space. If the compact space carries positive or vanishing curvature (as in a sphere or a torus), then the external space looks to counter that, delivering negative or vanishing curvature (as in AdS or Minkowski).  For higher curvature operators to help facilitate new de Sitter vacua, the details of this balance have to change. In particular, on one side of the energetic balance, a positive curvature space (be it a sphere along the compact directions, or de Sitter along the external directions)  must contribute negative energy through the higher order operators. This might indicate the presence of ghost-like instabilities and the associated negative energy excitations. Such instabilities should certainly be absent in effective actions derived from string theory.

 Of course, the situation becomes more complicated in the presence of additional fields  and sources, including branes, fluxes, Casimir energies, and with more exotic compact spaces.  The challenge here is to retain calculational control over the effective description, be it parametrically or even  just numerically.  Indeed, the level of control one is willing to sacrifice seems to be at the core of the ongoing debate over de Sitter space and the \emph{Swampland}.

\blankpage
\chapter{Conclusions and outlooks\label{chap:future}}

In this thesis, we began by developing the necessary mathematical machinery to understand Calabi-Yau compactifications and proceeded to study the perturbative tree-level type IIB supergravity action in \chpref{chap:pert_II}. We saw that the \kah sector remained flat and anticipated that to be able to generate a potential for the \kah moduli we would need to introduce perturbative and non-perturbative corrections to the tree-level action. This prompted the introduction of the Swampland programme, in particular with reference to the phenomenological side of the Swampland, and the questions of parametric control (or lack thereof).

In \chpref{chap:openqs}, we introduced a number of important open questions in modern cosmology thereby focusing the aim of the remainder of the thesis on obtaining explanations for some of these questions within the framework of string theory. Of particular interest were the naturalness and hierarchy problem, as these will plague any effective description of a ``theory of everything''\footnote{On this occasion, by ``theory of everything'' we refer to any theory encompassing many different physical scales, as would be expected from a UV complete description of our Universe arising from $\mathcal{O}(\mp)$ physics.}, and the nature of dark energy.   

The last two chapters of the first half of the thesis explore these aspects of modern cosmology from the point of view of effective descriptions motivated by supergravity. In \chpref{chap:clockwork} we introduce the clockwork mechanism which aims to naturally produce scale hierarchies in theories with many scalars. In the original formulation, a theory with $N$ scalars obtains a massless direction and $N-1$ heavy fields. Furthermore, we apply this mechanism to a cosmological setting by introducing the Dvali-Kaloper-Sorbo 4-form mixing. This mixing can softly break the clockwork symmetry, giving a small mass to the originally massless clockwork mode while preserving the hierarchy between this new light mode and the heavy modes. We anticipated that for some $\mathcal{O}(100)$ scalar fields the light clockwork mass would be of order the dark energy scale observed today and thus this mechanism could prove useful to naturally generate dark energy models. We then provided a no-go theorem for embedding this mechanism into the tree-level type IIA supergravity. 

In ongoing research, the question remains whether non-perturbative type effects can generate the necessary clockwork structure in type IIB. Heuristically, we expect non-perturbative effects to enter the theory through the superpotential as
\be
W_{np}= \sum_i A_i e^{a_i^j T_j}\ ,
\ee
where $i$ runs over the number of E3-branes in the theory, then by appropriately choosing a basis $a_i^j$ in instanton parameter space, one would hope to generate the right structure if the internal geometry is anisotropic, {\it i.e.} there are some small and large cycle such that a hierarchy $\tau_b \gg \tau_i$ for all $i\neq b$ exists. This would allow to single out a \kah direction with an associated axion much lighter than the other directions. Although this picture is attractive, obtaining a full ten-dimensional description is not trivial. For example writing down the required basis $a_i^j$ could, a priori, lead to inconsistencies on the ten-dimensional theory. An example of such inconsistencies is given by strong forms of the weak gravity conjecture \cite{1504.00659, 1611.06270}. 

In \chpref{chap:coincidence} we concentrated on the coincidence problem and its implications for dynamical models of dark energy obtained from generic properties of string compactifications close to the boundary of moduli space, where we assumed that the dynamics for the quintessence field are described by some runaway potential toward a Dine-Seiberg type vacuum at large vevs. We were able to show that, due to the mass of the heavy states depending on the runaway quintessence field, particle production is enhanced over time. This allows us to set an upper bound on the time it takes the particle production to swamp the effective description. We found that in some $\mathcal{O}(100) H_0$ time, the particle production becomes too much and the quintessence regime is destabilised. This ameliorated the coincidence problem to a 1-in-100 chance to find ourselves close to the dawn of the dark energy domination epoch. Finally, we argued that a more stringent bound might be obtained from a true Calabi-Yau compactification of the supergravity action, rather than a (well) motivated EFT.

Before moving on to the second half of the thesis, we spent some time in \chpref{chap:intermezzo} discussing the differences between supergravity computations at the boundary of moduli space and at the bulk of moduli space. We argued that, as long as the couplings are small enough (but finite) in the bulk of moduli space\footnote{This is to say that we are at large volume and weak coupling, rather than deep in the bulk of moduli space where the volume can be small or the physics being strongly coupled.} the perturbative expansion is well-posed and this allows us to work in a region of moduli space that is phenomenologically interesting. Furthermore, we were able to prove that on the runaway of moduli space one cannot achieve a slow-roll regime. This, together with the usual no-go of \cite{Maldacena:2000mw}, prevents us from obtaining a phenomenologically viable description of our Universe. We pointed out that this implies that there exists a disconnect from the phenomenological claims of the Swampland programme and cosmology.

Moving on to the second half of the thesis, in \chpref{chap:non_pert} we introduced a number of perturbative and non-perturbative corrections that are typically used to model build de Sitter vacua in type IIB supergravity. We reviewed each of these in some detail, as well as the best understood effective de Sitter constructions in type IIB, namely KKLT, the racetrack models and the large volume scenario. Armed with this knowledge, we tried to construct a viable model of quintessence in \chpref{chap:model_building}. However, we found when considering the cosmological evolution of our Universe, from an early period of inflation to late time dark energy, that a mixture between the Kallosh-Linde problem and quantum diffusion allowed only axionic hilltop models of quintessence with an extreme fine of the initial conditions. We argued that these constraints make quintessence less attractive to model builders than a cosmological constant. Therefore, without a clear input from cosmological data, {\it i.e.} finding with some degree of confidence an equation of state parameter for dark energy different than unity, string phenomenologists are better off focusing on de Sitter vacua.

Moving forward, there are two clear directions. One is to continue advancing our understanding of the de Sitter constructions that have already been developed by the community. This will require a better handle on perturbative and non-perturbative corrections, as well as, a clear ten-dimensional picture of all the necessary ingredients to generate such solution. A second option would be to introduce new ingredients that generate new solutions, like non-linearly realised supersymmetry which is the main idea behind the de Sitter supergravity constructions of \cite{Bergshoeff:2015tra, Bergshoeff:2016psz}. These new solutions might be the key to a more complete picture of supergravity.

In \chpref{chap:gbds} we follow the former approach. We consider runaway effective field theory that have some connection with fluxed supergravity and show the effects of higher curvature corrections to the leading order solution. In particular, we considered the Gauss-Bonnet corrections as motivated in the heterotic supergravity framework. We found that, on the boundary of parameter space where the leading order theory would yield an external four-dimensional space, the higher order corrections contributed with negative curvature thus generating anti-de Sitter solutions. \chpref{chap:gbds} served to point out the power of the effective theory approach with regards to calculating higher order corrections. This approach can be very helpful in flagging interesting model building terms. However, some notion of the form of the corrections is known. Of course, historically it has been very hard to obtain any specifics about the behaviour of these corrections in any generality. A recent and commendable effort in this direction is \cite{Cicoli:2021rub}, where the existence and volume scaling of $\alpha'$-corrections is explored.

Another interesting and qualitatively different approach is the one followed by \cite{Conlon:2020wmc, Conlon:2021cjk}. In this case, the authors take advantage of the AdS/CFT correspondence to study the holographic behaviour of the AdS vacua of a number of supergravity models, where the \kah moduli can be stabilised consistently (like KKLT or LVS). Interestingly, they find that each of these models presents a certain universality in the conformal dimensions of the CFT operators dual to the moduli fields in the AdS vacua. This leads the authors to speculate about the origin of this universality. If true for any AdS construction, it would seem to imply that the {\it a priori} vast string landscape maps to a limited set of conformal dimensions when translated into the CFT language, which are, for example, independent of choices of fluxes.

In this thesis, we have concerned ourselves with the study of the microscopic origin of our Universe with a focus on the nature of dark energy. In the first half of the thesis, we considered the consequence of some stringy motivated effective field theory on naturalness and the coincidence problem. In the second half, we got our hands dirty working with the effective field theory descending from type IIB supergravity. Although we were able to provide some very strong phenomenological results, sadly, we do not find ourselves any closer to a microscopic description of our Universe. We are able to say what doesn't work but lack an intuition of what might work from a string theory perspective. However, one should not be despondent. This might just be because a more complete picture of the non-perturbative aspects of string theory is needed, or maybe it requires new interpretations of supergravity. 

The microscopic origin of our Universe is a very challenging open problem which has provided a concrete framework to test and better understand string theory. To some extent, this could be seen as one of the last great challenges of theoretical physics. In a somewhat related position we would have found Michelson and Kelvin at the turn of the $20^{\rm th}$ century\footnote{I would like to thank Prof. Javier Tejada for bringing this anecdote to my attention some years ago while I was an undergraduate.}. In \cite{michelson}, Michelson claimed the following,

\say{The more important fundamental laws and facts of physical science have all been discovered, and these are now so firmly established that the possibility of their [sic.] ever being supplanted in consequence of new discoveries is exceedingly remote.}

which is, in hindsight, a {\it brave} claim. In \cite{kelvin}, Kelvin would go on to point out the \say{two clouds over the dynamical theory of heat and light}. These two clouds where related to the nature of ether and the partition energy as set out by Maxwell and Boltzmann and the specific heat of gases. The first cloud would be dispersed by special relativity and the second one by quantum physics, in the meantime generating storms of which this thesis is just a small part. It is worth noting that in \cite{michelson}, Michelson would go on to add the now famous quote \say{our future discoveries must be looked for in the sixth place of decimals}, which would turn out to be true, albeit a bit misdirected when taking into consideration his former quote on the laws of physics.

As some concluding words to this thesis, let us see what appears after the clouds covering the cosmological constant are dispersed and which fundamental laws need rewriting. It is certainly a very exciting time for Physics.

\blankpage

\bookmarksetup{startatroot}
\appendix
\chapter{Details and calculations for type IIB supergravity}
In this appendix, we write the details of some of the least enlightening and more annoying calculations for \chpref{chap:pert_II}. We will describe the details of the dimensional reduction of the $\mathcal{N}=2$ type IIB action and discuss the dualisation of the Kalb-Ramond and RR two-form axion.

\section{Calculations for the dimensional reduction of the $\mathcal{N}=2$ type IIB action\label{app:dim_redux_iib}}
We would like to find the dimensional reduction of \eqref{eq:iib_n2}. It is most convenient to work in string frame and do an overall Weyl rescaling at the end. Therefore, our starting point is the string frame action
\begin{gather}
\mathcal{S}_{IIB}=\int{e^{-2\hat{\phi}}\over2}\left(\hat{R} \star \mathbf{1} + 4d\hat{\phi}\wedge\star d\hat{\phi}- \hat{H}_3\wedge\star\hat{H}_3\right)\\
-\int\frac{1}{4}\left(\hat{F}_1\wedge\star\hat{F}_1+\hat{F}_3\wedge\star\hat{F}_3+{1\over 2}\hat{F}_5\wedge\star\hat{F}_5+\hat{C}_4\wedge\hat{H}_3\wedge \hat{F}_3\right)\ ,\nonumber
\label{eq:app_iib_n2}
\end{gather}
where the field strength are \eqref{eq:f_strength}
\begin{gather}
\hat{H}_3=d\hat{B}_2\ , \quad \hat{F}_1=d\hat{C}_0\ , \quad \hat{F}_3=d\hat{C}_2+ \hat{B}_2 \wedge d\hat{C}_0\ , \nonumber \\
\hat{F}_5=d\hat{C}_4 + \hat{B}_2\wedge d\hat{C}_2\ , \quad  
\label{eq:app_f_strength}
\end{gather}
and with the field content given in \eqref{eq:field_content_iib}
\begin{gather}
\hat{B}_2=B_2(x)+b^A(x)\omega_A\ , \quad \hat{C}_2 = C_2(x) + c_2^A\omega_A \ , \nonumber \\
\hat{C}_4= c_4^A(x)\omega_A + \xi^{\hat{K}}(x)\wedge\alpha_{\hat{K}}-\tilde{\xi}_{\hat{K}}(x)\wedge\beta^{\hat{K}}+\rho_A(x)\tilde{\omega}^A\ . \label{eq:app_field_content_iib}
\end{gather}

\subsection{The dimensional reduction of the NSNS sector \label{ssec:app_iib_dimredux}}
We take the ansatz
\be
ds^2 = g_{\mu\nu}(x)dx^\mu dx^\nu + g_{mn}(x,z)dz^md\bar{z}^{n}\ ,
\ee
where $g_{\mu\nu}$ (with $\mu,\nu=0,...,3$) is the external metric and $g_{mn}$ (with $m,n=1,...,6$) is the CY metric of \x.

Following 
\be
\hat{R} = g^{MN} \hat{R}^P_{MPN} = g^{\mu\nu}\left(\hat{R}^{\rho}_{\mu\rho\nu} + \hat{R}^p_{\mu p \nu}\right) +  g^{mn}\left(\hat{R}^{p}_{mpn} + \hat{R}^\rho_{m\rho n}\right) \ ,
\ee
we write the expansion for the ten-dimensional Ricci scalar 
\be
\int \hat{R}= \int R+ R_\x + \frac{1}{4}\left(g^{m n}\partial^\mu g_{mn}\right)\left(g^{pq}\partial_\mu g_{pq}\right) -\frac{1}{4}\left(g^{pq}\partial_\mu g_{q m}\right)\left(g^{m n}\partial^\mu g_{n p}\right)\ ,
\ee
where $\left(R,R_\x\right)$ are the Ricci scalar for the external and Calabi Yau three-fold, respectively. We are interested in introducing deformations that preserve the Ricci-flatness of the internal space
\be
g_{mn}\rightarrow g_{mn}^0(y)+\delta g_{mn}(x,y)\ , \qquad R_{m n}(g+\delta g)=0\ .
\ee
To discuss the deformations, it is convenient to write the indices in terms of a holomorphic/anti-holomorphic notation, like in the main text, such that $m=\left\{i,\bari\right\}$ with $i=1,2,3$. The deformations are given by the $h^{1,\,1}$ \kah moduli and $h^{2,1}$ complex structure moduli
\begin{gather}
\delta g_{i\barj} = -iv^A(\omega_A)_{i\barj}\ ,\\
\delta g_{i j} =\frac{i}{||\Omega||^2}\bar{z}^K(\bar{\chi}_K)_{i\bari\barj}\Omega^{\bari \barj}_j\ , 
\end{gather}
which follow from the results of \secref{sec:special_geo}. Thus, even after deforming the CY metric, $R_\x=0$. Expanding to quadratic order in $\delta g$ we find
\be
{1\over 4} \left(g^{m n}\partial_\mu g_{mn}\right)\left(g^{pq}\partial^\mu g_{pq}\right) = -g^{i\barj}_0 g^{k\bar{l}}_0 \partial^\mu \delta g_{i\barj} \partial_\mu \delta g_{\bar{l}k} = \partial_\mu v^A \partial^\mu v^B (\omega_A)^{i}_i(\omega_B)^{j}_j\ ,
\ee
and
\begin{gather}
 -\frac{1}{4}\left(g^{pq}\partial_\mu g_{q m}\right)\left(g^{m n}\partial^\mu g_{n p}\right) = -{1\over 2} \left(g^{i\barj}_0 g^{\bar{k} l}_0 \partial_\mu \delta g_{i \bar{k}} \partial^\mu \delta g_{\barj l}+ g^{i\barj}_0 g^{k \bar{l}}_0 \partial_\mu \delta g_{i k} \partial^\mu \delta g_{\barj \bar{l}}\right)= \nonumber \\
{1\over 2} \left[\partial_\mu v^A \partial^\mu v^B (\omega_A)^{i\barj}(\omega_B)_{i\barj}+ \partial_\mu z^K \partial^\mu \bar{z}^L {\chi_K\wedge \bar{\chi}_L \over \Omega \wedge \bar{\Omega}}\right]  = \\
2\mathcal{V} \left[G_{AB} dv^A\wedge \star_4 dv^B + G_{K\bar{L}} dz^K\wedge \star_4 d\bar{z}^L\right]\ ,
\end{gather}
where $\mathcal{V}$ is the volume of the CY, $\{G_{AB},G_{K\bar{L}}\}$ are the \kah structure and complex structure metrics defined in \eqref{eq:ks_metric} and \eqref{eq:cs_metric} respectively, and $\{v^A,z^K\}$ are the moduli fields spanning the \kah structure and complex structure manifolds. Putting everything together we find for the dimensional reduction of the Ricci term
\be
{1\over 2}\int e^{-2\hat{\phi}} \hat{R}\star \mathbf{1}  =\frac{1}{2} e^{2\phi}\left[R\star_4 \mathbf{1}  + 2 G_{AB} dv^A\wedge\star_4 dv^B + 2 G_{K\bar{L}}dz^K\wedge \star_4 d\bar{z}^L \right]\ .
\ee
where we have used the convenient definition for the four-dimensional dilaton in \eqref{eq:4dim_dilaton}
\be
e^\phi := {1\over \mathcal{V}^{1/2}}e^{\hat{\phi}} \ .
\ee
The dilaton term in the ten-dimensional action reduces straightforwardly
 \be
\frac{1}{2}\int e^{-2\hat{\phi}}\ 4 d\hat{\phi}\wedge\star d\hat{\phi}=2 \int e^{-2 \phi}\ d\phi\wedge\star_4 d\phi\ .
 \ee
Finally, the Kalb-Ramond form becomes
 \be
 -\frac{1}{4}\int e^{-2\hat{\phi}}\ \hat{H}_3\wedge\star \hat{H}_3=-\frac{1}{4} \int e^{-2\phi}\left(  4 G_{AB}db^A\wedge\star_4db^B + dB_2\wedge\star_4 dB_2\right)\ , 
 \ee
 Altogether, the NSNS sector can be written in the string frame as
 \be
\hspace{-5em} \mathcal{S}_{NSNS}^{(4)}=\int \frac{e^{-2 \phi}}{2}\left(R\star \mathbf{1} + 4 d\phi\wedge\star d\phi - 2 G_{AB} dt^A\wedge\star d\bar{t}^B - 2 G_{K\bar{L}}dz^K\wedge \star d\bar{z}^L - \frac{1}{2} dB_2\wedge\star dB_2 \right)\ , \qquad
\ee
where we have made use of the definition of the \kah complexified fields $t^A:=v^A+ib^A$ and we have dropped the subindex in the Hodge star operator to simplify the notation.

\subsection{The dimensional reduction of the RR sector}
Moving onto the RR sector we find the following dimensional reduction for the RR field strengths 
\be
-{1\over 4}\int \hat{F}_1\wedge \star \hat{F}_1 = -{\vol\over 4} \int dC_0\wedge \star_4 dC_0 \ ,
\ee
\begin{gather}
-{1\over 4} \int \hat{F}_3\wedge \star \hat{F}_3 =  -{1\over 4} \int\vol \left(dC_2+B_2\wedge dC_0\right)\wedge \star_4 \left(dC_2+B_2\wedge dC_0\right) \\
-{1\over 4} \int 4\vol G_{AB}\left(dc_2+bdC_0\right)^A\wedge \star_4\left(dc_2+bdC_0\right)^B \nonumber\ , 
\end{gather}
\begin{gather}
-{1\over 4} \int \hat{F}_5\wedge \star \hat{F}_5 = -{1\over 4} \int d\xi^{\hat{K}}\wedge\star_4 d\xi^{\hat{L}} \int_\x \alpha_{\hat{K}}\wedge\star_6 \alpha_{\hat{L}}\\
\hspace{-5em}-{1\over 4} \int d\tilde{\xi}_{\hat{K}} \wedge \star_4 d\tilde{\xi}_{\hat{L}} \int_\x \beta^{\hat{K}}\wedge\star_6 \beta^{\hat{L}} - {1\over 4} \int d{\xi}^{\hat{K}} \wedge \star_4 d\tilde{\xi}_{\hat{L}}\int_\x \alpha_{\hat{K}}\wedge\star_6 \beta^{\hat{L}}-  {1\over 4} \int d\tilde{\xi}_{\hat{K}} \wedge \star_4 d{\xi}^{\hat{L}} \int_\x  \beta^{\hat{K}}\wedge \star_6 \alpha_{\hat{L}}\nonumber \\
+ {1\over 4} \int {1\over 4\vol} G^{AB} d\rho_A\wedge \star_4 d\rho_B-{1\over 4} \int \left(dc_4+b dc_2\right)^A \wedge \star_4\left(dc_4+bdc_2\right)^B\int_\x \omega_A\wedge\star_6\omega_B \nonumber\\
= {1\over 4} \int (\text{Re}\ \mathcal{M})_{\hat{K}\hat{L}} F^{\hat{K}} \wedge  F^{\hat{L}} + (\text{Im}\ \mathcal{M})_{\hat{K}\hat{L}} F^{\hat{K}} \wedge\star_4  F^{\hat{L}}   \nonumber \\
-{1\over 16\vol} \int G^{AB}\left(d\rho_A-\mathcal{K}_{ACD}b^C dc^D_2\right) \wedge \star_4\left(d\rho_B-\mathcal{K}_{BEF}b^Edc^F_2\right)\nonumber
\end{gather}
where we have used the self-duality of $\hat{F}_5$ to write everything in terms of $\xi^K$ and $\rho_A$, the definition of $F^{\hat{K}}:=d\xi^{\hat{K}}$ from the main text, that ${\omega}_A=\mathcal{K}_{ABC}\tilde{\omega}^{B}\tilde{\omega}^{C}$ and also the period matrices 
\begin{gather}
\int \alpha_{\hat{K}}\wedge\star\alpha_{\hat{L}}=-\left[\text{Im}\ \mathcal{M} + (\text{Re}\ \mathcal{M}) (\text{Im}\ \mathcal{M})^{-1}(\text{Re}\ \mathcal{M})\right]_{\hat{K}\hat{L}} \ , \\
\int \beta^{\hat{K}}\wedge\star\beta^{\hat{L}}=-\left(\text{Im}\ \mathcal{M}\right)^{-1\ \hat{K}\hat{L}} \ , \\
\int \alpha_{\hat{K}}\wedge\star\beta^{\hat{L}}=-\left[(\text{Re}\ \mathcal{M})(\text{Im}\ \mathcal{M})^{-1}\right]_{\hat{K}}^{\hat{L}} \ . 
\end{gather}
Finally, for the topological form we find
\be
-{1\over 4}\int \hat{C}_4\wedge \hat{H}_3\wedge d\hat{F}_3 = -{1\over 2} dC_2 \wedge \left( \rho_A db^A - b^A d\rho_A\right) + {1\over2}dB_2\wedge c^A_2d\rho_A - {1\over 4} \mathcal{K}_{ABC} c^A_2 c^B_2 dB_2 \wedge db^c\ ,
\ee
Putting everything together we find\footnote{Where once again we drop the subindex in the four-dimensional Hodge star to avoid cluttering the notation.}
\begin{gather}
\mathcal{S}_{IIB}^{(4)}=\int -{e^{-2\phi}\over2} R\star{\mathbf 1}+{1\over 4} \text{Re}\ \mathcal{M}_{\hat{K}\hat{L}} F^{\hat{K}}\wedge F^{\hat{L}}+{1\over 4} \text{Im}\ \mathcal{M}_{\hat{K}\hat{L}} F^{\hat{K}}\wedge\star F^{\hat{L}} \\
+ 2e^{-2\phi} d\phi\wedge\star d\phi - e^{-2\phi} G_{AB} dt^A\wedge\star d\bar{t}^B - e^{-2\phi} G_{K\bar{L}}dz^K\wedge \star d\bar{z}^L - \frac{e^{-2\phi}}{4} dB_2\wedge\star dB_2  \nonumber \\
-\vol G_{AB}\left(dc_2^A+b^AdC_0\right)\wedge \star\left(dc_2^B+b^BdC_0\right) \nonumber \\
-{1\over 16\mathcal{V}} G^{AD}\left(d\rho_A-\mathcal{K}_{ABC}c_2^Bdb^C\right)\wedge\star\left(d\rho_D-\mathcal{K}_{DEF}c_2^Edb^F\right)\nonumber\\
-{1\over 2} dC_2 \wedge \left( \rho_A db^A - b^A d\rho_A\right) + {1\over2}dB_2\wedge c^A_2d\rho_A - {1\over 4} \mathcal{K}_{ABC} c^A_2 c^B_2 dB_2 \wedge db^c\ ,
\label{eq:app_iib_dim_redux}
\end{gather}
After Weyl rescaling to Einstein frame\footnote{Note that one can match the expressions on both frames through
\be
F_p^{(s)}\wedge \star^{(s)}_4 F_p^{(s)} \quad \rightarrow \quad e^{2\phi(2-p)}F_p^{(e)}\wedge \star^{(e)}_4 F_p^{(e)}\ ,
\ee
where the superindex $(s)$,$(e)$ indicate quantities in the string or Einstein frame, respectively, and that the rescaling of the Ricci scalar in four-dimensions introduces a shift to the kinetic dilaton term that goes like $-6d\phi\wedge\star_4 d\phi$.}, $g_{\mu\nu}\rightarrow e^{2\phi} g_{\mu\nu}$, and dualising the $C_2$ and $B_2$ axions (see next section for details) we find the full action as given in \eqref{eq:iib_dual}
\begin{gather}
\mathcal{S}^{(4)}_{IIB}=\int -{1\over2} R\star{\mathbf 1}+{1\over 4} \text{Re}\ \mathcal{M}_{\hat{K}\hat{L}} F^{\hat{K}}\wedge F^{\hat{L}}+{1\over 4} \text{Im}\ \mathcal{M}_{\hat{K}\hat{L}} F^{\hat{K}}\wedge\star F^{\hat{L}}\nonumber \\
- G_{K\bar{L}} dz^K\wedge\star d\bar{z}^{\bar{L}} - h_{pq} dq^p\wedge\star dq^q\ ,
\end{gather}
with
\begin{gather}
h_{pq} dq^p\wedge\star dq^q := (d\phi)^2 + G_{AB} dt^A \wedge\star dt^B + {1\over 4} e^{2\phi} \mathcal{V} (dC_0)^2 \nonumber\\ 
+e^{2\phi} \mathcal{V} G_{AB}\left(dc_2^A-C_0db^A\right)\wedge\star\left(dc_2^B-C_0db^B\right) \nonumber\\
+{1\over 16\mathcal{V}} e^{2\phi} G^{AD}\left(d\rho_A-\mathcal{K}_{ABC}c_2^Bdb^C\right)\wedge\star\left(d\rho_D-\mathcal{K}_{DEF}c_2^Edb^F\right)\nonumber\\
+{1\over 4\mathcal{V}} e^{2\phi} \left[dh-{1\over2}\left(\rho_A db^A - b^A d\rho_A\right)\right]^2\nonumber\\
+{1\over 2} e^{4\phi}\left[d\tilde{h}+C_0dh+c_2^Ad\rho_A+{1\over2}C_0\left(\rho_A db^A - b^A d\rho_A\right)-{1\over4}\mathcal{K}_{ABC}c_2^Ac_2^Bdb^C\right]^2\ . 
\end{gather}

\section{Dualising the $C_2$ and $B_2$ axions \label{app:dual_axions}}
To put the type IIB action into its gauged supergravity form, we have had to dualise the $B_2$ and $C_2$ axions. In this section we provide the brief calculation necessary to dualise them, for the sake of completeness. Since both objects we wish to dualise are two-form gauge fields, we consider a generic Lagrangian of the form
\be
\mathcal{L} = -{\alpha\over 4} dA_2 \wedge \star dA_2 - {1\over 4}dA_2 \wedge J_1\ , \label{eq:L_A2}
\ee
where $J_1$ is a generic one-form field. To enforce the Bianchi identity $F_3:=dA_2$, we introduce a Lagrange multiplier $da$ 
\be
\mathcal{L} = -{\alpha\over 4} F_3 \wedge \star F_3 - {1\over 4}F_3 \wedge J_1 + F_3\wedge da\ , \label{eq:L_a}
\ee
such that its equation of motion fixes $dF_3=0$. On the other hand, by completing the square for the three-form field strength, the Lagrangian can be written in the convenient form
\be
\mathcal{L} = -{\alpha\over4} \hat{F}_3\wedge \star \hat F_3 - {1\over 4\alpha} \hat J_1\wedge \star \hat J_1\ ,
\ee
where $\hat F_3:= F_3+{1\over \alpha} \star \hat J_1$ and $\hat J_1:=J_1 - 4 da$. Integrating out $\hat F_3$ yields the Lagrangian for the dual scalar $a$
\be
\mathcal{L}_a =  - {1\over 4\alpha} \hat J_1\wedge \star \hat J_1 =  - {4\over \alpha} \left(da - {1\over 4} J_1\right)^2\ .
\ee
To dualise the $C_2$ and $B_2$ axions into the $h$ and $\tilde{h}$ scalars, respectively, we find from \eqref{eq:app_iib_dim_redux}
\begin{gather}
\mathcal{L}_{C_2} = - {\mathcal{V}e^{-2\phi}\over 4}  dC_2 \wedge \star dC_2 -{1\over 4} dC_2 \wedge \left( 2\rho_A db^A -2 b^A d\rho_A  \right) \ , \\ 
\hspace{-3em}\mathcal{L}_{B_2} = - {e^{-2\phi}\left( e^{-2\phi}+\mathcal{V} C_0^2\right)\over 4}  dB_2 \wedge \star dB_2 -{1\over 4} dB_2 \wedge \left( \mathcal{K}_{ABC} c_2^A c_2^B db^C - 2 c_2^A d \rho_A  - \mathcal{V} C_0 e^{2\phi} \star dC_2 \right) \ ,
\end{gather}
and following the relations set out in this section the dual Lagrangian can be given as
\begin{gather}
\mathcal{L}_{h,\,\tilde{h}} = - {1\over 4\mathcal{V}} e^{2\phi} \left[dh-{1\over2}\left(\rho_A db^A - b^A d\rho_A\right)\right]^2\nonumber\\
- {1\over 2} e^{4\phi}\left[d\tilde{h}+C_0dh+c^Ad\rho_A+{1\over2}C_0\left(\rho_A db^A - b^A d\rho_A\right)-{1\over4}\mathcal{K}_{ABC}c_2^Ac_2^Bdb^C\right]^2\ . 
\end{gather}
\chapter{Type IIA perturbative supergravity \label{app:iia_details}}
In this appendix we provide details on how to obtain the $\mathcal{N}=1$ scalar potential for fluxed type IIA supergravity. We will begin by finding the analogue of the type IIB $\mathcal{N}=2$ scalar potential. Then, we will see how the orientifold projection works for type IIA theories and we will present the orientifolded scalar potential. Lastly, we will discuss moduli stabilisation in type IIA theories.

\section{$\mathcal{N}=2$ supergravity action}
In spirit, this section follows very closely the analogue type IIB calculation in \secref{sec:n2}. Again, it is convenient to work in the string frame and do an overall Weyl rescaling at the end. We begin by writing down the 10-dimensional type IIA supergravity action in string frame \cite{Polchinski:1998rr}
\begin{gather}
\mathcal{S}_{IIA}=\int {e^{-2\hat{\phi}} \over2} \left(\hat{R} \star \mathbf{1} + {4}d\hat{\phi}\wedge\star d\hat{\phi}-{1\over2}\hat{H}_3\wedge\star\hat{H}_3\right)\\
-\frac{1}{2}\int\left(\hat{F}_2\wedge\star\hat{F}_2+\hat{F}_4\wedge\star\hat{F}_4+\mathcal{L}_{\text{top}}\right)\ ,\nonumber
\label{eq:iia_n2}
\end{gather}
where the NSNS sector is the same as in the type IIB theory with the ten-dimensional dilaton $\hat{\phi}$ and the Kalb-Ramond two-form $\hat{B}_2$, whereas the RR sector now contains odd-form gauge fields whose field strengths we define as follows
\begin{gather}
\hat{F}_2=d\hat{C}_1\ , \quad \hat{F}_4=d\hat{C}_3-d\hat{C}_1\wedge \hat{B}_2 \ .
\label{eq:app_f_strength_iia}
\end{gather}
where $(\hat{B}_2)^p := \underbrace{\hat{B}_2\wedge...\wedge\hat{B}_2}_{p-\text{times}}$. The topological term is given by
\begin{gather}
\mathcal{L}_{\text{top}} = \hat{B}_2\wedge d\hat{C}_3\wedge d\hat{C}_3 - (\hat{B}_2)^2\wedge d\hat{C}_3\wedge d\hat{C}_1+{1\over 3}(\hat{B}_2)^3\wedge d\hat{C}_1\wedge d\hat{C}_1  \ .
\end{gather}
We can expand in the bases of harmonic (p,q)-forms that we found in \secref{sec:special_geo} and that we summarise below in \tabref{tab:app_bases} for convenience. The fields are given in these bases by
\begin{gather}
\hat{B}_2=B_2(x) + b^A\omega_A\ , \quad \hat{C}_1=A^0(x)\ , \quad \hat{C}_3=C_3(x)+A^A(x)\wedge\omega_A+\xi^{\hat{L}}\alpha_{\hat{L}}+\tilde{\xi}_{\hat{K}}\beta^{\hat{K}}\ . \label{eq:fc}
\end{gather}
Here, we have used notation that is reminiscent of the type IIB description, since the moduli will fill similar roles to their IIB counterparts. Let us analyse the dimensional reduction of the type IIA action \eqref{eq:iia_n2}. As in the type IIB case, we will take the ansatz for the ten-dimensional metric to be
\be
ds^2 = g_{\mu\nu}(x)dx^\mu dx^\nu + g_{i\barj}(x,z)dz^id\bar{z}^{\barj}\ ,
\ee
where $g_{\mu\nu}$ (with $\mu,\nu=0,...,3$) is the external metric and $g_{i\barj}$ (with $i,\barj=1,...,3$) is the CY metric of \x.
\begin{table}[h!]
\centering
\begin{tabular}{|c|c|c|}
\hline
Cohomology group & Dimension    & Basis                                           \\ \hline
$H^{1,1}$        & $h^{1,1}$    & $\omega_A$                                      \\ \hline
$H^{2,2}$        & $h^{1,1}$    & $\tilde{\omega}^A$                              \\ \hline
$H^3$            & $2+2h^{2,1}$ & $\left(\alpha_{\hat{K}},\beta^{\hat{L}}\right)$ \\ \hline
$H^{2,1}$        & $h^{2,1}$    & $\chi_K$                                        \\ \hline
$H^{3,3}$        & $1$          & $\text{dvol}(\x)$                                       \\ \hline
\end{tabular}
\caption{Bases for a generic Calabi-Yau threefold.} 
\label{tab:app_bases}
\end{table}
The reduction of the NSNS sector is identical to the type IIB case, and we refer the reader to \ssecref{ssec:app_iib_dimredux}.

\subsection{The RR sector and the topological term of the type IIA action}
Below, we plug the field content \eqref{eq:fc} into the different terms of the R-R sector. For the 2-form we find
\be
 -\frac{1}{2}\int_\x \hat{F}_2\wedge\star \hat{F}_2=-\frac{\mathcal{V}}{2} dA^0\wedge\star_4dA^0  \ , \label{eq:f2}
\ee
where \vol is the volume of \x and $G_{AB}$ is given by \eqref{eq:ks_metric}. The 4-form dimensional reduction reads
\begin{gather}
\hspace{-3em} -\frac{1}{2}\int_\x \hat{F}_4\wedge\star \hat{F}_4 =-\frac{1}{2} \left[ \mathcal{V}\left(dC_3-B_2\wedge dA^0\right)\wedge\star_4\left(dC_3-B_2\wedge dA^0\right) \right] \nonumber\\
 +4\mathcal{V}G_{AB}\left(dA^A-b^AdA^0\right)\wedge\star_4\left(dA^B-b^BdA^0\right)\nonumber \\
-(\Im\mathcal{M})^{-1\ \hat{K}\hat{L}}\left(d\tilde{\xi}_{\hat{K}}+\mathcal{M}_{\hat{K}\hat{M}}d\xi^{\hat{M}}\right)\wedge\star_4\left(d\tilde{\xi}_{\hat{L}}+\bar{\mathcal{M}}_{\hat{L}\hat{N}}d\xi^{\hat{N}}\right)\ , \label{eq:f4}
\end{gather}
where we have used $G^{AB}:=4\mathcal{V}\int\tilde{\omega}^A\wedge\star\tilde{\omega}^B$ is the inverse metric of $G_{AB}$, as well as the period matrix defined in \eqref{eq:period_matrix} and given below for convenience
\begin{gather}
\int \alpha_{\hat{K}}\wedge\star\alpha_{\hat{L}}=-\left[\text{Im}\ \mathcal{M} + (\text{Re}\ \mathcal{M})(\text{Im}\ \mathcal{M})^{-1}(\text{Re}\ \mathcal{M})\right]_{\hat{K}\hat{L}} \ , \\
\int \beta^{\hat{K}}\wedge\star\beta^{\hat{L}}=-\left(\text{Im}\ \mathcal{M}\right)^{-1\ \hat{K}\hat{L}} \ , \\
\int \alpha_{\hat{K}}\wedge\star\beta^{\hat{L}}=-\left[(\text{Re}\ \mathcal{M})(\text{Im}\ \mathcal{M})^{-1}\right]_{\hat{K}}^{\hat{L}} \ .
\end{gather}
Finally the topological term is
\begin{gather}
\hspace{-5em}\int_\x  \hat{B}_2\wedge d\hat{C}_3\wedge d\hat{C}_3 - (\hat{B}_2)^2\wedge d\hat{C}_3\wedge d\hat{C}_1  + {1\over 3} (\hat{B}_2)^3\wedge d\hat{C}_1\wedge d\hat{C}_1 = \label{eq:top} \\
B_2\wedge d\left(\tilde{\xi}_{\hat{K}}d\xi^{\hat{K}}-\xi^{\hat{L}}d\tilde{\xi}_{\hat{L}}\right) + \mathcal{K}_{ABC}b^A\left[dA^B\wedge dA^C -b^BdA^0\wedge\left( dA^C - {1\over3} b^C dA^0\right) \right]\ . \nonumber 
\end{gather}
The action can now be written in terms of the gauge-coupling matrices defined in \eqref{eq:gc} 
\begin{gather}
\text{Re} \mathcal{N} = \begin{pmatrix}
-\frac{1}{3}\mathcal{K}_{ABC}b^A b^B b^C & \frac{1}{2}\mathcal{K}_{ABC}b^A b^B \\
 \frac{1}{2}\mathcal{K}_{ABC}b^A b^B & -\mathcal{K}_{ABC} b^C
\end{pmatrix}\ ,\\
\text{Im} \mathcal{N} = -\mathcal{V}\begin{pmatrix}
1+4G_{AB}b^A b^B & -4G_{AB}b^B\\
-4G_{AB}b^B & 4G_{AB}
\end{pmatrix}\ ,\\
(\text{Im} \mathcal{N})^{-1} = -\frac{1}{\mathcal{V}}\begin{pmatrix}
1 & b^A\\
b^A & \frac{1}{4}G^{AB}+b^A b^B
\end{pmatrix}\ .
\end{gather}
Before doing so, we will first dualise the $B_2$ and $C_3$ fields into a scalar $a(x)$ and a constant flux $e_0$, respectively. This will allow us to put the $\mathcal{N}=2$ action in its gauged form \cite{hep-th/0202168}, turning a double tensor multiplet containing $B_2$ into a hypermultiplet in terms of $a$.

\subsection{Dualising the $B_2$ and $C_3$ fields}
Let us write the Lagrangians we are trying to dualise explicitly and work from there. Starting from the $C_3$ gauge field, we collect all pieces containing $C_3$ in
\begin{gather}
\mathcal{L}_{C_3}= -\frac{\mathcal{V}}{2}\left(dC_3-B_2\wedge dA^0\right)\wedge\star_4\left(dC_3-B_2\wedge dA^0\right) \nonumber \\
:=-\frac{\mathcal{V}}{2}\left(dC_3-J_4\right)\wedge\star_4\left(dC_3-J_4\right)- e_0\left(dC_3- 0\right)\ ,
\end{gather}
where we have introduced $J_4$ for notational convenience. The Lagrange multiplier $e_0$ enforces the (trivial) Bianchi identity\footnote{Note that if one tries to simply find the equation of motion for $\star C_3$ and substitute back, because $dC_3$ is a top-form then the equations of motion would contain a derivative of a top-form which must vanish identically and this method does not work. One must introduce the Lagrange multiplier in the case of a top-form field strength.}. The equation of motion for $dC_3$ becomes
\be
\frac{\mathcal{V}}{2}\star_4\left(dC_3-J_4\right)=-e_0 \quad \Rightarrow \quad dC_3=J_4+\frac{e_0}{\mathcal{V}}\star_4 {\bf1} \label{eq:app_dc3}
\ee
and substituting back
\begin{gather}
\mathcal{L}_{e_0}=- \frac{e_0^2}{2\mathcal{V}} \star_4\mathbf{1} -e_0 J_4 = - \frac{e_0^2}{2\mathcal{V}} \star_4\mathbf{1} -e_0 \left(B_2\wedge dA^0\right)  \ .
\end{gather}
Next, we can substitute the expression back into the Lagrangian $\mathcal{L}_{C_3}$ to obtain the dual description. We also choose to group the field strength fields present in the theory as
\be
F^{\hat{A}}:=\{dA^0,dA^A\} \ ,
\ee
for later convenience. Finally, we would like to dualise the $dB_2$ field strength into the scalar $a(x)$. This follows from our results in \secref{app:dual_axions}, which we recall below for convenience. In this case, the Lagrangian for the $H_3$ field strength reads
\begin{gather}
\mathcal{L}_{H_3}=-\frac{1}{4} dB_2\wedge\star dB_2 + \frac{1}{2}dB_2\wedge\left(\tilde{\xi}_{\hat{K}}d\xi^{\hat{K}}-\xi^{\hat{L}}d\tilde{\xi}_{\hat{L}}\right)\ ,
\end{gather}
the equation of motion for $B_2$ implies
\be
d\left[{1\over 2}\star H_3 -\left(\tilde{\xi}_{\hat{K}}d\xi^{\hat{K}}-\xi^{\hat{L}}d\tilde{\xi}_{\hat{L}}\right) \right]=0 \quad \Rightarrow \quad da=\frac{1}{2}\star dB_2 -  \left(\tilde{\xi}_{\hat{K}}d\xi^{\hat{K}}-\xi^{\hat{L}}d\tilde{\xi}_{\hat{L}}\right)\ ,
\ee
which, together with \eqref{eq:f2}, \eqref{eq:f4}, \eqref{eq:top}, \eqref{eq:app_dc3}, can be substituted in \eqref{eq:iia_n2} to find the four-dimensional type IIA effective action\footnote{Where a similar Weyl rescaling has been done to put the action in Einstein frame, similar to the one done on the IIB side.} \cite{hep-th/0202168}
\begin{gather}
\hspace{-2.5em} \mathcal{S}_{IIA}^{(4)}=\int- \frac{1}{2}R\star \mathbf{1} + \frac{1}{2}\Im \mathcal{N}_{\hat{A}\hat{B}} F^{\hat{A}}\wedge \star F^{\hat{B}} + \frac{1}{2}\Re \mathcal{N}_{\hat{A}\hat{B}}F^{\hat{A}}\wedge F^{\hat{B}}  \nonumber \\
- G_{AB} dT^A\wedge\star d\bar{T}^B - h_{u v} d\tilde{q}^u\wedge\star d\tilde{q}^v - V\star {\bf 1}\ , \label{eq:iia_action_2}
\end{gather}
with
\be
V = {e^{4\phi}\over 2\vol} e_0^2 \ , \label{eq:app_scalar_e0}
\ee
and
\begin{gather}
h_{u v} d\tilde{q}^u\wedge\star d\tilde{q}^v := d\phi\wedge\star d\phi + G_{K\bar{L}}dz^K\wedge \star d\bar{z}^L + \nonumber\\
\frac{e^{4\phi}}{4}\left(Da +\tilde{\xi}_{\hat{K}}d\xi^{\hat{K}}-\xi^{\hat{L}}d\tilde{\xi}_{\hat{L}}\right)\wedge \star \left(Da + \tilde{\xi}_{\hat{K}}d\xi^{\hat{K}}-\xi^{\hat{L}}d\tilde{\xi}_{\hat{L}}\right)\nonumber \\
-\frac{e^{2\phi}}{2}(\Im\mathcal{M})^{-1\ \hat{K}\hat{L}}\left(d\tilde{\xi}_{\hat{K}}+\mathcal{M}_{\hat{K}\hat{M}}d\xi^{\hat{M}}\right)\wedge\star\left(d\tilde{\xi}_{\hat{L}}+\bar{\mathcal{M}}_{\hat{L}\hat{N}}d\xi^{\hat{N}}\right)\ ,
\end{gather}
where $Da:=da+2e_0A^0$. We have dropped the subindex on the four-dimensional Hodge star, since there are no more ambiguities. Notably, the scalar potential of the type IIA $\mathcal{N}=2$ is not flat due to the presence of the Freund-Rubin {\it flux} $e_0$, which is the gauge charge for the axion $a$ dual to the external piece of the Kalb-Ramond two-form. The spectrum of the theory is given in \tabref{tab:app_field_content}.
\begin{table}[h!]
\centering
\begin{tabular}{|c|c|c|}
\hline
Multiplet & Dimension    & Field content                                           \\ \hline
Gravity multiplet       & 1 & $\left(g_{\mu\nu},\ A^0\right)$                                      \\ \hline
Vector multiplet        & $h^{1,1}$    & $\left(A^A,v^A,b^A\right)$                              \\ \hline
Hypermultiplet        & $h^{2,1}+1$ & $\left(\xi^K,\tilde{\xi}_K,z^K\right)+\left(a,\phi,\xi^0,\tilde{\xi}_0\right)$ \\ \hline
\end{tabular}
\caption{Multiplets for the 4-dimensional type IIB supergravity spectrum.}
\label{tab:app_field_content}
\end{table}

\section{Orientifolding and $\mathcal{N}=1$ type IIA supergravity}
Similar to the type IIB case, we must reduce the supergravity spectrum by halving the number of supersymmetry generators through an orientifold projection.

We write down the orientifold projection operator as \cite{hep-th/0507153}
\be
\mathcal{O}=\Omega_p\left(-1\right)^{\text{F$_\text{L}$}}\sigma\ .
\ee
If one wishes to preserve $\mathcal{N}=1$ supersymmetry, then the involution must act on the \kah form and the holomorphic 3-form as
\be
\sigma^*J=-J\ , \qquad \sigma^*\Omega=e^{2i\theta}\bar{\Omega}\ , \label{eq:orient_conds}
\ee
for a constant phase $e^{2i\theta}$ and where $\sigma^*$ is the pullback of $\sigma$, an anti-holomorphic involution acting on \x. This in contrast with the type IIB orientifold conditions where the involution is holomorphic. In type IIA supergravity, the above conditions are consistent with the introduction of $O6$-planes in the theory.

We have seen that the effect of the involution on the cohomology of \x is to split them in even and odd cohomologies
\be
H^{p,\,q}=H^{p,\,q}_+ \oplus H^{p,\,q}_-\ ,
\ee
such that an element $\omega_{\pm}\in H^{p,\,q}_\pm$ transforms like $\sigma^*\omega_\pm=\pm\omega_\pm$. From the type IIA orientifold conditions \eqref{eq:orient_conds}, we can further see that
\begin{itemize}
\item $\sigma^*J=-J$ implies $h^{3,\,3}_-=1$, since $\vol\propto\int J\wedge J \wedge J$ and $h^{0,\,0}_+=1$, by Hodge duality. Alternatively, this implies that $h^{3,\,3}_+=h^{0,\,0}_-=0$.
\item Due to the anti-holomorphicity of $\sigma$ and Hodge duality, we find $h^{1,\,1}_{\pm}=h^{2,\,2}_{\mpl}$ and $h^3_+=h^3_-=h^{2,\,1}+1$.
\end{itemize}
In \tabref{tab:app_bases_orient} we summarise the notation for the truncated harmonic bases after orientifolding, which largely follows the notation of the type IIB analogues,
\begin{table}[h!]
\centering
\begin{tabular}{|c|c|c|}
\hline
Cohomology group & Dimension & Basis  \\ \hline
$H^{1,\,1}_+$     & $h^{1,\,1}_+ $ & $\omega_\alpha$                                   \\ \hline
$H^{1,\,1}_-   $   & $h^{1,\,1}_- $     & $\omega_a$                             \\ \hline
$H^{2,\,2}_+     $ & $h^{1,\,1}_- $  & $\tilde{\omega}^a$   \\ \hline
$H^{2,\,2}_-    $  & $h^{1,\,1}_+ $  &$\tilde{\omega}^\alpha$    \\ \hline
$H^{3}_+     $  & $h^{2,\,1}+1$  & $\left(\alpha_{\hat{k}}, \beta^\lambda\right)$    \\ \hline
$H^{3}_-     $  & $h^{2,\,1}+1 $  & $\left(\alpha_\lambda, \beta^{\hat{k}}\right)$    \\ \hline
\end{tabular}
\caption{Harmonic bases for the orientifolded \x.}
\label{tab:app_bases_orient}
\end{table}

where $\hat{k}=0,...,\tilde{h}$ and $\lambda=\tilde{h},...,h^{2,\,1}$. The non-trivial intersection numbers are given by
\be
\int_\x\omega_a\wedge\tilde{\omega}^b=\delta_a^b\ , \qquad \int_\x\omega_\alpha\wedge\tilde{\omega}^\beta=\delta_\alpha^\beta\ , \qquad \int_\x \alpha_{\hat{k}}\wedge \beta^{\hat{l}}=\delta_{\hat{k}}^{\hat{l}}\ , \qquad \int_\x \alpha_{\kappa}\wedge \beta^{\lambda}=\delta_{\kappa}^{\lambda}\ .
\ee
The action of the parity operator and the fermion number operator on the 10-dimensional fields is summarised in \tabref{tab:app_orient_1} \cite{hep-th/9804208}. It is clear from it that the invariant orientifolded states must obey
\begin{gather}
\sigma^*\hat{\phi}=+\hat{\phi}\ ,\qquad \sigma^*\hat{g}=+\hat{g}\ ,\qquad \sigma^*\hat{B_2}=-\hat{B_2}\ , \nonumber \\
\sigma^*\hat{C_1}=-\hat{C_1}\ , \qquad\sigma^*\hat{C_3}=+\hat{C_3}\ .
\end{gather}
\begin{table}[h!]
\centering
\begin{tabular}{|c|c|c|c|}
\hline
Field & Under $\Omega_p$    & Under $(-1)^{\text{F}_\text{L}}$       & Under  $\Omega_p\,(-1)^{\text{F}_\text{L}}$  \\ \hline
$\hat{\phi}$      & $+$ & $+$    & $+$                                      \\ \hline
$\hat{g}$        & $+$    & $+$      & $+$                            \\ \hline
$\hat{B_2}$       & $-$ & $+$ & $-$    \\ \hline
$\hat{C_1}$       & $+$ & $-$  & $-$   \\ \hline
$\hat{C_3}$       & $-$ & $-$ & $+$    \\ \hline
\end{tabular}
\caption{Actions of the parity and fermion number operators on the 10-dimensional spectrum.}
\label{tab:app_orient_1}
\end{table}

One can truncate the $\mathcal{N}=2$ spectrum as follows. For the \kah structure we obtain
\be
J=v^a\omega_a\ , \quad B_2=b^a\omega_a\ , \quad J_c:=J-iB_2=T^a\omega_a:=(v-ib)^a\omega_a\ .
\ee
Note, in particular, that the orientifold has projected out the external $B_2(x)$ piece, which before we had dualised into a gauge scalar charged under $e_0$. The complex structure deformations are given by applying the orientifold constraint to obtain the conditions
\be
\Im(C\mathcal{Z}^{\hat{k}})=\Re(C\mathcal{F}_{\hat{k}})=0\ , \quad \Re(C\mathcal{Z}^{\lambda})=\Im\left(C\mathcal{F}_{\lambda}\right)=0\qquad C:=e^{-\phi-i\theta}e^{K^{cs}/2} \label{eq:orient_comega}
\ee
so that the expansion of the holomorphic three-form with the regulator function $C$ is
\be
C \Omega=\Re(C\mathcal{Z}^{\hat{k}})\alpha_{\hat{k}} + i\Im(C\mathcal{Z}^{\lambda})\alpha_{\lambda}-\Re(C\mathcal{F}_{\lambda})\beta^{\lambda} - i\Im(C\mathcal{F}_{{\hat{k}}})\beta^{{\hat{k}}}\ . \label{eq:comega_n1}
\ee
One can still choose the \kah gauge so that $\mathcal{Z}^0=1$, however it is most convenient to keep the gauge freedom for now while keeping in mind that there are only $h^{2,1}$ degrees of freedom in the complex structure sector. Finally, let us look at the one and three-form gauge field in the RR sector.

The involution constraints impose that
\be
\sigma^* \hat{A}_1=-\hat{A}_1 \ , \quad \rightarrow \quad \hat{A}_1\in H^{1,0}(Y)\ .
\ee 
If \x is a Calabi-Yau threefold, there are no non-trivial one-cycles and $\hat{A}_1$ should at most be an external field. However, the orientifold acts trivially on any external fields, thus the $\mathcal{N}=2$ graviphoton $A_0$ must be projected out.

Finally,
\be
\sigma^* \hat{C}_3=\hat{C}_3 \ , \quad \rightarrow \quad \hat{C}_3\in H^{3}_+(Y)\ .
\ee 
In this case, $\hat{C}_3$ admits an external form $C_3$, that will dualise once again into the Freud-Rubin flux $e_0$, such that the gauge field can be expanded into
\be
\hat{C}_3=C_3(x)+A^\alpha\wedge\omega_\alpha+\xi^{\hat{k}}\alpha_{\hat{k}}-\tilde{\xi}_\kappa \beta^\kappa:=C_3(x)+A^\alpha\wedge\omega_\alpha+C^i_3(x)
\ee
defining a complexified three-form
\be
\Omega_c:=C^i_3+2i\Re\left(C\Omega\right)\ , \qquad C:=e^{-\phi-i\theta}e^{K^{cs}/2} \ .
\ee
Including the dilaton through the compensator function $C$ is equivalent to trading the irrelevant scale factor of the three-form $\Omega$ by the dilaton, so that all $h^{2,\,1}+1$ fields in the definition of $\Omega_c$ are physical. This will make it so introducing coordinates to the quaternionic manifold later on can be done directly through $\Omega_c$.

Much like in type IIB analogue, locally the moduli space can still be written in a product form of two \kah manifolds
\be
\tilde{\mathcal{M}}^{ks} \times \tilde{\mathcal{M}}^Q\ ,
\ee
with $\tilde{\mathcal{M}}^{ks}$ and $\tilde{\mathcal{M}}^Q$ subspaces of the $\mathcal{N}=2$ \kah and quaternionic manifolds, respectively. In contrast with the type IIB, the complex structure manifold is now part of the quaternionic manifold, $\mathcal{M}^{cs}\in \mathcal{M}^Q$. This would be expected from the usual duality arguments, where the \kah sector of type IIB maps to a dual complex structure sector in IIA and vice versa (for explicit details of the mirror symmetry see, for example, \cite{hep-th/9403096, hep-th/0507153}, and references therein). Now we would like to study the effects of the orientifold projection onto the \kah and quaternionic sector following the conditions laid out before to obtain the $\mathcal{N}=1$ effective action.
\subsection{The \kah structure of $\tilde{\mathcal{M}}^{ks}$}
The number of \kah moduli $h^{1,1}$ is reduced in half to $h^{1,1}_-$, {\it i.e.} 
\be
t^A \, \rightarrow \, t^a \ , \qquad \omega_A \, \rightarrow \, \omega_a \ ,  \qquad a=1,..., h^{1,\,1}_-.
\ee
Note that these projections imply some vanishing of the triple intersection numbers. Indeed, recall that $d\mathcal{V}\in H^6_-$ which means that
\be
\mathcal{K}_{ABC}v^Av^Bv^C:=\int_\x v^Av^Bv^C \omega_A\wedge\omega_B\wedge\omega_C \propto \int_\x d\mathcal{V}\ ,
\ee
so the integrand must be odd under the action of $\sigma$ and thus only intersection numbers with one or three odd-legs are non-trivial after the projection. Similarly, one can apply the same reasoning to the {\it contractions} of the intersection numbers, {\it i.e.} $\mathcal{K}_{AB}$ and $\mathcal{K}_{A}$, together with the fact that $J\in H^{1,1}_-$ to find
\be
\mathcal{K}_{\alpha\beta\gamma}=\mathcal{K}_{\alpha a b}=\mathcal{K}_{\alpha a}=\mathcal{K}_{\alpha}=0\ .
\ee
Then, the $\mathcal{N}=1$ \kah metric can be obtained directly by using the conditions above on the $\mathcal{N}=2$ \kah metric in \eqref{eq:ks_metric}. The orientifolded metric takes the block-diagonal form
\begin{gather}
G_{ab}=-\frac{1}{4\mathcal{V}}\left(\mathcal{K}_{ab}-\frac{\mathcal{K}_a\mathcal{K}_b}{4\mathcal{V}}\right)\ , \qquad G_{\alpha\beta}=-\frac{\mathcal{K}_{\alpha\beta}}{4\mathcal{V}}\ , \\
G_{\alpha a}=0\ .
\end{gather}
The same constraints can be applied to the gauge-kinetic coupling matrix $\mathcal{N}_{\hat{A}\hat{B}}$ to get
\begin{gather}
\Re \mathcal{N}_{{\alpha}{\beta}}= -\mathcal{K}_{\alpha\beta a} b^a\ , \qquad \Im \mathcal{N}_{{\alpha}{\beta}}=\mathcal{K}_{\alpha\beta} \ ,\\
\mathcal{N}_{0\alpha}=\mathcal{N}_{a\alpha}=0\ ,
\end{gather}
plus the components $\mathcal{N}_{\hat{a}\hat{b}}$ found by substituting $\left(\hat{A},\hat{B},\hat{C}\right)\, \rightarrow\, \left(\hat{a},\hat{b},\hat{c}\right)$, with $\hat{a}=0,a$, in the formulae of \eqref{eq:gk_coupling1}, \eqref{eq:gk_coupling2} and \eqref{eq:gk_coupling3}. Finally, the \kah potential is simply given by
\be
K^{ks}=-\ln\left[\frac{1}{6}\mathcal{K}_{abc}\left(T+\bar{T}\right)^a\left(T+\bar{T}\right)^b\left(T+\bar{T}\right)^c\right]=-\ln\left(\bar{T}^{\hat{a}}\partial_{T^{\hat{a}}} F-{T}^{\hat{a}}\partial_{\bar{T}^{\hat{a}}} \bar{F}\right)\ ,
\ee
with $\hat{a}=0,1,...,h^{1,\,1}_-$ the adapted coordinates of $\mathcal{M}^{ks}$ and the prepotential function that generates the \kah potential is $F=\frac{1}{3}\mathcal{K}_{abc}{T^aT^bT^c\over T^0}$. The $\mathcal{N}=1$ \kah sector for the orientifolded effective action reads
\be
\mathcal{S}_{IIA}^{(4)}=\int -G_{ab}\,dT^a\wedge\star d\bar{T}^b+ \frac{1}{2}\Im \mathcal{N}_{\alpha\beta} F^{\alpha}\wedge \star F^{\beta} + \frac{1}{2}\Re \mathcal{N}_{\alpha\beta}F^{\alpha}\wedge F^{\beta} \ , \label{eq:kah_sec}
\ee
with $F^{\alpha}:=dA^{\alpha}$. 

\subsection{The quaternionic structure of $\tilde{\mathcal{M}}^Q$}
The quaternionic sector is more involved. The complex structure deformations complying with the orientifolds constraints are given by considering infinitesimal variations of $\Omega$ and making use of the Kodaira identity \eqref{eq:kodaira_form} as
\be
\Omega\left(z+\delta z\right)=\Omega(z) + \delta z^K\left(\partial_{z^K}\Omega\right)(z)=\Omega(z) - \delta z^K\left(\partial_{z^K}K^{cs}\Omega-\chi^K\right)(z)\ .
\ee
We require that both $\Omega(z)$ and $\Omega(z+\delta z)$ comply with the orientifold constraints \eqref{eq:orient_conds}. Also notice that the complex conjugate to the holomorphic three-form is given by
\be
\hspace{-5mm}\bar{\Omega}(\bar{z}+\delta \bar{z})=\bar{\Omega}(\bar{z}) - \delta \bar{z}^K\left(\partial_{\bar{z}^K}\bar{K}^{cs}\bar{\Omega}-\bar{\chi}^K\right)(\bar{z})=\bar{\Omega}(\bar{z}) - \delta \bar{z}^K\left(\partial_{\bar{z}^K}K^{cs}\bar{\Omega}-\bar{\chi}^K\right)(\bar{z})\ ,
\ee
since $K^{cs}:=-i\ln\left(i\int \Omega\wedge \bar{\Omega}\right)=\bar{K}^{cs}$ by the odd properties of the external product on odd-forms. This implies that locally
\be
\delta z^K \partial_{z^K}K^{cs}=\delta \bar{z}^K \partial_{\bar{z}^K}K^{cs}\ , \qquad \delta z^K \sigma^*\chi^K=e^{2i\theta}\delta \bar{z}^K \bar{\chi}^K\ , \label{eq:loc_cond}
\ee
where we have used that the involution acts trivially on external pieces so that
\begin{gather*}
\sigma^*\left[\delta z^K \left(\partial_{z^K}K^{cs}\right)\Omega\right]=\delta z^K \left(\partial_{z^K}K^{cs}\right)\left(\sigma^*\Omega\right)\ , \\
\sigma^*\left(\delta z^K \chi^K\right) = \delta z^K \left(\sigma^*\chi^K\right)\ ,
\end{gather*}
From the first equality of \eqref{eq:loc_cond} and recalling that $K^{cs}$ is a \kah potential, {\it i.e.} $\partial_{z^K}K^{cs}\neq0$, we find that for each $\delta z^K$ either its real or its imaginary part has to be zero. The number of degrees of freedom has been halved and one may choose any basis $z^K=\left(z^k,z^\lambda\right)$ with $\Im z^k = \Re z^\lambda = 0$ for any splitting of $h^{2,1}+1$ in $k$-components and $\lambda$-components. We will keep this splitting generic for now.

These coordinates $z^K=\left(z^k,z^\lambda\right)$ can now be mapped to adapted coordinates of $\mathcal{M}_{\mathbb{R}}^{\text{cs}}$ by the embedding \cite{hep-th/0507153}
\begin{eqnarray*}
\rho: \qquad\qquad \mathcal{M}^{cs}_{\mathbb{R}} \qquad&\hookrightarrow& \qquad\mathcal{M}^{cs} \\
q^K=(z^{\hat{k}}, z^\lambda) \qquad&\mapsto&\qquad z^K=\left(z^{\hat{k}},i z^\lambda \right)
\end{eqnarray*}
and its metric is given by
\be
\rho^*\left(G_{K\bar{L}}dz^K\wedge\star d\bar{z}^L\right)=G_{KL}(q)dq^K\wedge\star dq^L\ , \qquad \rho^*\left(i G_{K\bar{L}}dz^K\wedge d\bar{z}^L\right)=0\ .
\ee
with the period matrices given by restricting 
\be
\mathcal{M}_{\hat{K}\hat{L}}:=\bar{\mathcal{F}}_{\hat{K}\hat{L}}+2i\frac{(\text{Im}\ \mathcal{F})_{\hat{K}\hat{M}}\; \mathcal{Z}^{\hat{M}}(\text{Im}\ \mathcal{F})_{\hat{L}\hat{N}}\; \mathcal{Z}^{\hat{N}}}{\mathcal{Z}^{\hat{M}}(\text{Im}\ \mathcal{F})_{\hat{M}\hat{N}}\mathcal{Z}^{\hat{N}}}\ ,
\ee
to $\tilde{\mathcal{M}}^{cs}_{\mathbb{R}}$ and applying the constraints
\be
\Im(C\mathcal{Z}^{\hat{k}})=\Re\left(C\mathcal{F}_{\hat{k}}\right)=0\ , \quad \Re(C\mathcal{Z}^{\lambda})=\Im\left(C\mathcal{F}_{\lambda}\right)=0\ ,
\ee
we find
\be
\Re \mathcal{M}_{\sigma\lambda}(q)=\Re \mathcal{M}_{\hat{k}\hat{l}}(q)=\Im \mathcal{M}_{\hat{k} \sigma}(q)=0\ ,
\ee
Finally, the \kah potential for the quaternionic manifold is \cite{hep-th/0507153, 0804.1248}
\begin{gather}
K^Q=-2\ln\left[2\int_Y\Re\left(C\Omega\right)\wedge\star \Re\left(C\Omega\right)\right]\nonumber \\=-2\ln 2\left[\Re (CF_\lambda)\Im (CZ^\lambda)-\Im (CF_{\hat{k}})\Re (CZ^{\hat{k}})\right]= -\ln e^{-4\phi}\ .
\end{gather}
For future reference, when writing the supergravity action in term of chiral supermultiplets, it will be convenient to define the coordinates of the quaternionic manifold $\mathcal{M}^{Q}$ in terms of half of the periods of $\Omega_c$ \cite{hep-th/0507153}
\begin{gather}
N^{\hat{k}}:= \int_\x \Omega_c\wedge \beta^{\hat{k}}=\xi^{\hat{k}} + 2i \Re (CZ^{\hat{k}}) \ , \\
U_\lambda := \int_\x \Omega_c\wedge\alpha_\lambda=\tilde{\xi}_\lambda+2i\Re (C\mathcal{F}_\lambda) \ . 
\end{gather}

\section{The type IIA $\mathcal{N}=1$ effective action}\label{sec:n1_action}
The orientifolded effective action is given by \cite{hep-th/0507153}
\begin{gather}
\hspace{-2.5em} \mathcal{S}_{IIA,\, \mathcal{N}=1}^{(4)}=\int -\frac{1}{2}R\star \mathbf{1}+ \frac{1}{2}\Im \mathcal{N}_{\alpha\beta} F^{\alpha}\wedge \star F^{\beta} + \frac{1}{2}\Re \mathcal{N}_{\alpha\beta}F^{\alpha}\wedge F^{\beta} \nonumber \\
-G_{ab}\,dT^a\wedge\star d\bar{T}^b - h_{uv} dq^u\wedge\star dq^v - V\star\mathbf{1}\label{eq:n1_action}\ ,
\end{gather}
with the same Freund-Rubin potential
\be
V = {e^{4\phi}\over 2\vol} e_0^2 \ ,
\ee
and where
\begin{gather}
h_{u v} d\tilde{q}^u\wedge\star d\tilde{q}^v = d\phi\wedge\star d\phi + G_{KL}(q)dz^K\wedge\star dz^L - \frac{e^{2\phi}}{2}\left(\Im \mathcal{M}\right)_{\hat{k}\hat{l}}  d\xi^{\hat{k}}\wedge\star d\xi^{\hat{l}} \nonumber \\
- \frac{e^{2\phi}}{2}\left(\Im \mathcal{M}\right)^{-1\, \sigma\lambda}\left(d\tilde{\xi}_\sigma-\Re \mathcal{M}_{\hat{k}\sigma}d\xi^{\hat{k}}\right)\wedge\star \left(d\tilde{\xi}_\lambda-\Re \mathcal{M}_{\hat{k}\lambda}d\xi^{\hat{k}}\right)\ ,
\end{gather}
is now the metric adapted to the orientifolded quaternionic manifold, of which $\mathcal{M}_{\mathbb{R}}^{cs}\in \mathcal{M}^Q$.

Analogously to the type IIB case, we are now presented with the problem of defining appropriate coordinates for the quaternionic manifold. This can be done in a couple different ways. In \cite{hep-th/0507153}, the author defines coordinates based on the holomorphic periods of the truncated theory. Another approach is to think of the appropriate coordinates as those dual\footnote{There are some subtleties due to the broken symplectic invariance in an orientifolded type IIA theory and the choice of orientifolds planes in the type IIB dual, see \cite{hep-th/0507153}.} to the \kah sector of the mirror of \x, like in \cite{0804.1248}. Below we present the results
\begin{itemize}
\item Axio-dilaton: $N^0:=S= s+i\sigma=\xi^0+2i\Re (CZ^0)=\int\Omega_c\wedge\beta^0$.
\item \kah moduli: $T^a= v^a + i b^a$, with $a=1,...,h^{1,\,1}_-$
\item Even complex structure moduli: $U_\lambda:=u_\lambda+i\nu_\lambda=\tilde{\xi}_\lambda+2i\Re (C\mathcal{F}_\lambda)=\int\Omega_c\wedge\alpha_\lambda$.
\item Odd complex structure moduli: $N^k:=n^k+i\eta^k:=\xi^k + 2i \Re (CZ^k)=\int\Omega_c\wedge\beta^k$.
\end{itemize}
The odd complex structure sector is the dual to the odd \kah sector in type IIB. To make contact with the same type of theories, we finally make use of the symplectic freedom to choose $N^k=0$. This implies that $H^3(\x)=H^{2,\,1}_+\oplus H^{3,\,0}_+$, and $\lambda=1,...,h^{2,1}_+$. We also remark that the \kah moduli have been defined in terms of the two-cycle volume moduli $v^a$ in contrast with type IIB case, which was defined in terms of the four-cycle volume moduli $\tau_\alpha$ \eqref{eq:T_coords}.

With this splitting, the \kah potential of the truncated theory is given by
\begin{gather}
K=K^{ks}+K^{Q}\ , \label{eq:app_kah_1}\\
K^{ks}= -\ln\left[\frac{1}{6}\mathcal{K}_{abc}\left(T+\bar{T}\right)^a\left(T+\bar{T}\right)^b\left(T+\bar{T}\right)^c\right] \ , \\
K^{Q} = -2\ln\left[2\int_\x \Re (C\Omega)\wedge \star \Re (C\Omega)\right]\ ,
\end{gather}
and the moduli space is, once again, given locally by the direct product\footnote{Notice that the moduli space will always be given by a direct product of the \kah sector and the quaternionic sector, no matter the choice of symplectic gauge.}
\be
\mathcal{M}^{ks}_{h^{1,\,1}_-}\times \mathcal{M}^{Q}_{h^{2,1}_+ + 1}\ .
\ee
Finally, we would like to put this in the $\mathcal{N}=1$ gauged supergravity form of \cite{hep-th/0412277}
\be
\hspace{-3em}\mathcal{S}^{(4)}_{IIA,\,\mathcal{N}=1}=-\int {1\over2} R\star{\mathbf 1} + K_{I\bar{J}} DM^I\wedge\star D\bar{M}^{\bar{J}} +{1\over 4} \text{Re}\ \mathcal{N}_{\alpha\beta} F^{\alpha}\wedge F^{\beta}+{1\over 4} \text{Im}\ \mathcal{N}_{\alpha\beta} F^{\alpha}\wedge\star F^{\beta} + V\star\mathbf{1}\ , \label{eq:iia_1_W}
\ee
where the scalar potential is given by
\be
V=e^K\left(K^{I\bar{J}}D_IWD_{\bar{J}}\bar{W}-3\left|W\right|^2\right)\ , \label{eq:app_scalar_pot_1}
\ee
and we have followed the common notation for the chiral fields and \kah derivatives described below \eqref{eq:iib_1_W}. In this case, the scalar potential is non-zero but only contains a cosmological constant-like contribution from the Freund-Rubin flux $e_0$ \eqref{eq:app_scalar_e0}. Like in the type IIB case, we would like to introduce background fluxes that will allow us to stabilise some moduli.

\section{Type IIA flux compactifications \label{app:fluxes}}
In this section we will provide the effective action in the case where NSNS and RR background fluxes are turned on. Furthermore, we will consider the case of massive type IIA supergravity by allowing for a non-zero Romans mass parameter $m$ \cite{Romans:1985tz}. The massive type IIA ten-dimensional action is 
\begin{gather}
\mathcal{S}_{IIA-M}=-\int\left({1\over2}\hat{R} \star \mathbf{1} + {1\over 4}d\hat{\phi}\wedge\star d\hat{\phi}+{1\over4}e^{-\hat{\phi}} \hat{H}_3\wedge\star\hat{H}_3\right)\\
-\frac{1}{2}\int\left(e^{3\hat{\phi}/2}\hat{F}_2\wedge\star\hat{F}_2+e^{\hat{\phi}/2}\hat{F}_4\wedge\star\hat{F}_4+e^{5\hat{\phi}/2}m^2\star\mathbf{1}+\mathcal{L}_{\text{top}}\right)\ ,\nonumber
\label{eq:iia_n2_m}
\end{gather}
where the topological term now has Romans mass dependent pieces
\begin{gather}
\mathcal{L}_{\text{top}} = \hat{B}_2\wedge d\hat{C}_3\wedge d\hat{C}_3 - (\hat{B}_2)^2\wedge d\hat{C}_3\wedge d\hat{C}_1+{1\over 3}(\hat{B}_2)^3\wedge d\hat{C}_1\wedge d\hat{C}_1  \\
m \left[{1\over3}(\hat{B}_2)^3\wedge d\hat{C}_3 - {1\over4}(\hat{B}_2)^4\wedge d\hat{C}_1  -{1\over3}(\hat{B}_2)^5\right]\nonumber \ .
\end{gather}
Similarly, the field strengths are now given by
\be
\hat{H}_3=d\hat{B}_2\ , \quad \hat{F}_2=d\hat{C}_1+m\hat{B}_2\ , \quad \hat{F}_4=d\hat{C}_3-\hat{C}_1\wedge\hat{H}_3 - {m\over2} (\hat{B}_2)^2\ ,
\ee
and the orientifold invariant fluxes will be given by
\be
H_3^{(F)}=h_0\beta^0-h^\lambda\alpha_\lambda\ , \quad F^{(F)}_2=q^a \omega_a\ , \quad F_4^{F}=e_a\tilde{\omega}^a\ .
\ee
In \cite{hep-th/0412277}, it was shown that the scalar potential found from dimensional reduction can be obtained in terms of a superpotential 
\begin{gather}
W=W_K+W_Q=\int_{\mathcal{X}} e^{-i J_c}\wedge F^{(F)}_{RR} + \Omega_c\wedge H^{(F)}_3= \\
e_0-i e_aT^a+\frac{1}{2}\mathcal{K}_{abc}q^aT^bT^c-\frac{im}{6}\mathcal{K}_{abc}T^aT^bT^c+Sh_0+U_\lambda h^\lambda\ , \nonumber
\end{gather}
and the \kah potential \eqref{eq:app_kah_1}. Here $F^{(F)}_{RR}$ is the poly-form containing all RR background fluxes. The fluxed scalar potential is given by \cite{1507.06793}
\begin{gather}
V={R\over s^4}\left[\frac{\rho_0^2}{2\mathcal{V}}+\frac{G^{ab}\tilde{\rho}_a\tilde{\rho}_b}{8\mathcal{V}}+{2\mathcal{V}G_{ab}\rho^a\rho^b}+\frac{\mathcal{V}}{2}\rho_m^2+\frac{1}{2\mathcal{V}}\left(s^{2}h_0^2-\frac{1}{3}\left(u_\lambda h^\lambda\right)^2\right)\right]\ ,  \label{eq:sca_pot_nof4}
\end{gather}
where $R=16e^{2K^{cs}}\Re(Z^0)^4$, and we define the combination $\varphi:=e_0-\xi^0h_0+\tilde{\xi}_\lambda h^\lambda$ with
\begin{gather}
\rho_0=\varphi + e_a b^a + \frac{1}{2} \mathcal{K}_{abc}q^ab^bb^c -{m\over 6}\mathcal{K}_{abc}b^ab^bb^c, \label{eq:app_dual_scalars}\\
-\rho_a= e_a + \mathcal{K}_{abc}q^bb^c -{m\over 2}\mathcal{K}_{abc}b^bb^c, \quad \tilde{\rho}^a= q^a-mb^a\ , \quad \rho_m=m\ , \nonumber
\end{gather}
the dual scalars in the democratic prescription of type IIA \cite{1606.00508} and use in \chpref{chap:clockwork}.

In contrast with the type IIA case, all moduli appear in the \kah potential and superpotential. Thus, one might be hopeful that all fields could be stabilised at tree-level. However, on duality grounds we would expect some obstruction to this goal since the type IIB theory required access to non-tree-level corrections to stabilise the \kah sector. We will study the stabilisation of the type IIA theory in \secref{app:stab_iia}.

\section{Moduli stabilisation in type IIA flux supergravity\label{app:stab_iia}}

In this section we follow the classic analysis of \cite{hep-th/0505160}. We will concentrate in studying the stabilisation of supersymmetric vacua configurations of type IIA. We will find that we are able to stabilise the dilaton, a linear combinations of axions in $\varphi$, the complex structure saxions $u_\lambda$ and the \kah sector, leaving $h^{2,\,1}$ complex structure axions flat. To begin, we look for a solution to the F-term equations
\be
D_{T}W=D_SW=D_UW=0\ .
\ee
Let us start with the quaternionic sector. The F-term for the axio-dilaton is
\be
D_S W = h_0 - {2iW\over e^{\sqrt{K^{Q}}}} \Im \left(CF_0\right)= h_0 -2i W e^{2\phi} \Im \left(C F_0\right) = 0\ ,  \label{eq:fterm_s}
\ee
and for the complex moduli we find
\be
D_{U^{\lambda}} = h^\lambda + {2iW\over e^{\sqrt{K^{Q}}}} \Im (CZ^\lambda)= h^\lambda +2i W e^{2\phi} \Im (CZ^\lambda) = 0\ ,  \label{eq:fterm_u}
\ee
where we have used the definitions of the chiral superfields in terms of the periods of $\Omega_c$ so that
\begin{gather}
\partial_S = \partial_{2\Re \left(CZ^0\right)} {\partial 2\Re \left(CZ^0\right)\over \partial S} = -i \partial_{2\Re \left(CZ^0\right)}\ , \\ \partial_{U_\lambda} = \partial_{2\Re \left(CF_\lambda\right)} {\partial 2\Re \left(CF_\lambda\right)\over \partial U^\lambda} = -i \partial_{2\Re \left(CF_\lambda\right)}\ .
\end{gather}
Noting that the dilaton is real-valued, the imaginary parts of the quaternionic F-terms lead to the same condition
\be
\Re W=0 \quad \rightarrow \quad h_0\sigma+\nu_\lambda h^\lambda+\Re W_K=0 \quad \text{or} \quad \varphi=-\Re W_K\ .
\ee
These identical equations allows us to stabilise the dilaton and $\varphi$ field at the minimum of the potential once we solve for the \kah and complex structure fields. The degeneracy in these F-terms can be seen from the fact that we have $h^{2\,1}+1$ complex equations and only $h^{2,\,1}+1$ real fluxes in $\left(h_0,h^\lambda\right)$, so there is not info degrees of freedom to stabilise the whole system.

Furthermore, we note that, after solving the previous F-terms 
\be
W = \Re W + i \Im W = i \Im W\ ,
\ee
and thus $\Im W=0$ implies a vanishing superpotential, which is inconsistent with the presence of any fluxes. This implies that the only solution to the real part of \eqref{eq:fterm_s} is given by
\be
h_0 + 2e^{2\phi} \Im W\ \Im (CF_0) = 0 \quad \rightarrow \quad e^{-\phi} = - 2{\Im W\over Q_0}\ ,
\ee
with $Q_0:=e^{-\sqrt{K^{cs}}} {h_0\over \Im (F_0)}$. Similarly, for the complex structure sector we find $\lambda$ equations
\be
h^\lambda - 2e^{2\phi} \Im W\ \Im (CZ^\lambda)= 0  \quad \rightarrow \quad {e^{-\phi}\over 2 \Im W} =e^{-\sqrt{K^{cs}}} {h^{\lambda}\over\Im (Z^\lambda) }:=-Q_0\ . \label{eq:dil_stab}
\ee
Then the chain of $h^{2,\,1}$ equations
\be
e^{-\sqrt{K^{cs}}} {h^{1}\over\Im (Z^1) }=e^{-\sqrt{K^{cs}}} {h^{2}\over\Im (Z^2) }=...=e^{-\sqrt{K^{cs}}} {h^{h^{2,\,1}}\over\Im (Z^{h^{2,\,1}}) }:= -Q_0\ ,
\ee
fixes the complex structure sector. On top of these, \eqref{eq:dil_stab} will stabilise the ten-dimensional dilaton $\hat{\phi}$ after the \kah sector is fixed, given that the four-dimensional dilaton $\phi$ contains a volume factor in its definition.

Before turning to the \kah moduli, let us note that when the previous equations are satisfied the superpotential can be written in terms of its \kah moduli only as
\be
W(T^a,S,U^k)=-i\Im W_K\left(T^a\right)\ ,
\ee
meaning the \kah sector completely decouples from the complex structure sector and axio-dilaton. We can see this by multiplying \eqref{eq:fterm_s} by $\Re (CZ^0)$ and \eqref{eq:fterm_u} by $\Re (CF_\lambda)$ and adding them to obtain
\begin{gather}
-{1\over 2}\left(h_0\sigma + h^\lambda \nu_\lambda\right) - iWe^{2\phi}\left[2\left(\Re (CF_\lambda)\Im (CZ^\lambda)-\Im (CF_{\hat{k}})\Re (CZ^{\hat{k}})\right)\right]= \nonumber \\
-{1\over 2} \Im W^Q - i W = 0\ .
\end{gather}
The F-terms for the \kah moduli are then
\begin{gather}
\hspace{-11em}D_{T^a}W=-i e_a+\mathcal{K}_{abc}q^bT^c-\frac{im}{2}\mathcal{K}_{abc}T^bT^c +i\frac{\Im W_K}{2\mathcal{V}}\mathcal{K}_{abc}\left(T+\bar{T}\right)^b\left(T+\bar{T}\right)^c \nonumber\\
=-i e_a+\mathcal{K}_{abc}q^b(v-ib)^c-\frac{im}{2}\mathcal{K}_{abc}(v^bv^c -b^bb^c - 2i b^bv^c) +  i\frac{2\Im W_K}{\mathcal{V}} \mathcal{K}_{abc} v^bv^c=0 \ . 
\end{gather}
The real part leads to
\be
\mathcal{K}_{abc}v^c\left(q^b-m b^b\right)=0\ .
\ee
Since $\mathcal{K}_{abc}v^c\propto G_{ab}$, the \kah metric, for any given row there must at least be a non-vanishing component so that the metric is invertible. Thus, the object inside the parentheses must vanish yielding the stabilisation for the \kah axions
\be
b^a=\frac{q^a}{m}\ . \label{eq:b_stab}
\ee
The imaginary part of the F-term for the \kah sector gives
\begin{gather}
\mathcal{K}_{def}v^d v^e v^f\left(4me_a+2\mathcal{K}_{abc}q^bq^c+3m^2\mathcal{K}_{abc}v^bv^c\right)+\nonumber \\
\mathcal{K}_{abc}v^bv^c\left(6me_dv^d + 3\mathcal{K}_{def}q^dq^ev^f\right)=0\ ,
\end{gather}
this gives $h^{1,1}$ real-equations that are quadratic in $v^a$ and allow us to stabilise them. We can also rewrite this equation in terms of the dual scalars from \eqref{eq:app_dual_scalars}, after  tracing over $v^a$, as
\be
v^a\left(10 m e_a + 5\mathcal{K}_{abc} q^b q^c + 3m^2 \mathcal{K}_{abc} v^b v^c\right)= v^a\left(10\rho_a + 3m^2 \mathcal{K}_{abc} v^b v^c\right)= 0 \ ,\label{eq:v_stab}
\ee
Thus, the \kah sector is fully stabilised, as opposed to the complex structure sector where only the saxions are fixed at the minimum. One can also understand this as the \kah sector having $2\cdot h^{1,1}_-$ fluxes, $(e_a,q^a)$, compared to the $h^{2,1}_+$ fluxes of the complex structure sector, $h^\lambda$. 
 
In contrast to the type IIB moduli stabilisation on \secref{sec:fluxes}, we have been able to stabilise the \kah sector, the axio-dilaton and the saxions of the complex structure sector. By duality arguments, however, we see that the complex structure saxions map to 4-volume cycles in the dual type IIB \cite{hep-th/9403096, hep-th/0507153}, which had flat profiles at the same level of the perturbative expansion. In \cite{hep-th/0607223}, it was shown that turning on any of the $h^\lambda$ fluxes required to stabilise the complex structure sector leads to a half-flat manifold and to non-geometric fluxes in the dual type IIB setting. This fluxes are not fully understood and present a problem regarding the trustability of solutions that employ them as means to stabilise moduli. 

In \secref{sec:no_go}, we present a no-go theorem for the embedding of clockwork into perturbative type IIA theories. We use a form of the potential that formally contains this fluxes, but the argument is entirely focused on the \kah sector and the impossibility of generating the clockwork mechanism through those. The argument only requires a decoupling of the \kah and complex structure sector, which is true at tree-level in any case, to go through. Therefore, the trustability issues are not a factor for our discussion.

\bibliography{bib}
\end{document}